\documentclass [a4paper,11pt]{book}
\usepackage{style_thesis}
\usepackage{newcommands}
\pdfoutput=1

\begin{document}

\dominitoc

\frontmatter
\begin{titlepage}
\ifx\pdfoutput\undefined 
\else
\pdfbookmark{Title}{title}
\fi

\newlength{\centeroffset}
\setlength{\centeroffset}{0cm}
\setlength{\centeroffset}{-0.5\oddsidemargin}
\addtolength{\centeroffset}{0.5\evensidemargin}
\thispagestyle{empty}

\noindent\vspace*{-10ex}\hspace*{\centeroffset}\makebox[\textwidth]{%
\begin{minipage}{\textwidth}
\begin{center}
\noindent\textsc{\fontfamily{ppl}\selectfont Ph.D. thesis}\\[1ex]
\noindent\textsc{\footnotesize\fontfamily{ppl}\selectfont in Theoretical Physics}\\
\end{center}
\end{minipage}}

\noindent\hspace*{\centeroffset}\makebox[\textwidth]{%
\begin{minipage}{\textwidth}
\includegraphics[height=0.15\textwidth]{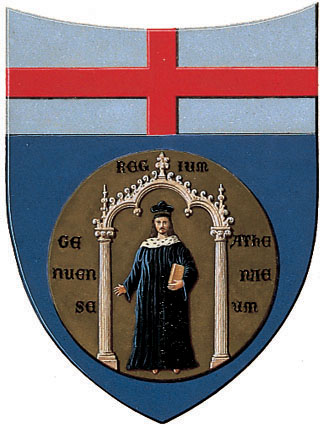}
\hspace{\stretch{1}}
\includegraphics[height=0.15\textwidth]{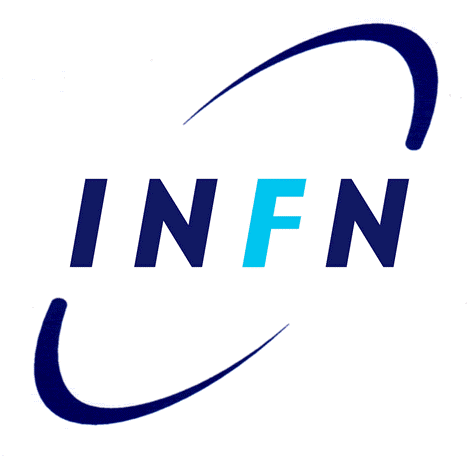}
\end{minipage}}

\vspace{5cm}

\noindent\hspace*{\centeroffset}\makebox[0pt][l]{%
\begin{minipage}{\textwidth}
\begin{center}
{\setstretch{3}
\noindent{\fontfamily{ppl}\fontsize{24.88pt}{0pt}\selectfont \mytitle}\\[2cm]
}
\noindent{\fontfamily{ppl}\selectfont by}\\[3ex]
\noindent{\Large\bfseries\fontfamily{ppl}\selectfont Marco Bonvini}
\end{center}
\vspace*{0.5cm}
\end{minipage}}

\vspace{4cm}

\noindent\hspace*{\centeroffset}\makebox[0pt][l]{%
\begin{minipage}{\textwidth}
\begin{flushright}
\begin{tabular}{rr}
{\fontfamily{ppl}\selectfont Supervisors:}
&{\bfseries\fontfamily{ppl}\selectfont Giovanni Ridolfi}\\
&{\bfseries\fontfamily{ppl}\selectfont Stefano Forte}\\[2ex]
{\fontfamily{ppl}\selectfont Exam date:}
&{\bfseries\fontfamily{ppl}\selectfont February 29, 2012}\\
\end{tabular}
\end{flushright}
\end{minipage}}

\vspace{\stretch{1}}

\noindent\hspace*{\centeroffset}\makebox[0pt][l]{%
\begin{minipage}{\textwidth}
\begin{center}
\textsc{\fontfamily{ppl}\selectfont \small Dipartimento di Fisica dell'Università di Genova,\\ via Dodecaneso 33, 16146 Genova.}
\end{center}
\begin{center}
{\color{white}\tiny Last modified: \today}
\end{center}
\end{minipage}}

  
\end{titlepage}

\cleardoublepage
\thispagestyle{plain}
\ifx\pdfoutput\undefined 
\else
\pdfbookmark{Abstract}{abstract}
\fi
\begin{abstract}
\indent
This thesis arises in the context of precision measurements at hadron colliders.
The Tevatron and the LHC provide very accurate measurements of many Standard Model
processes, such as the production of a lepton pair (Drell-Yan) of high invariant mass.
An accurate theoretical prediction of such processes is crucial to be able to
distinguish Standard Model physics from possible new physics signals.
QCD effects in the computation of the cross-sections at hadron colliders
are usually sizable; in particular, in some kinematical regimes
they behave in a non-perturbative way.
In these cases the resummation of the whole perturbative series is needed 
for accurate phenomenological predictions.
In this thesis the impact of threshold and high-energy resummations
for the production of high invariant mass systems (Drell-Yan, Higgs)
are studied in detail.
In particular, in the threshold case a prescription to deal with the divergent
nature of the perturbative series based on Borel summation is presented,
and compared with the other prescriptions in the literature.
Results for the invariant mass distributions and rapidity distributions are presented.
The high-energy resummation formalism is reviewed and improved,
and its impact in phenomenological applications at hadron colliders
is investigated.
In particular, a possible interaction between the two resummation
regimes is studied in some detail.

\end{abstract}
\cleardoublepage

\ifx\pdfoutput\undefined 
\else
\pdfbookmark{Contents}{contents}
\fi
\tableofcontents

\mainmatter
\cleardoublepage
\phantomsection
\addstarredchapter{Introduction}
\chapter*{Introduction}
\markboth{Introduction}{}

\begin{figure}[tb]
  \centering
  \includegraphics[width=381px]{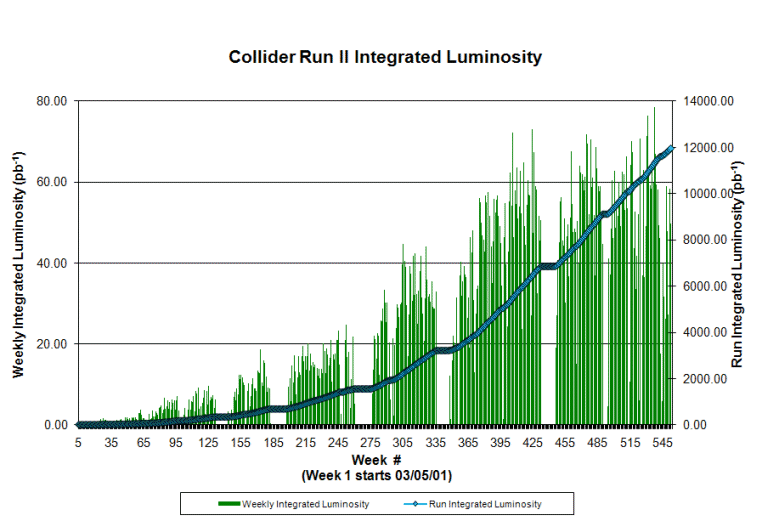}
  \caption{Tevatron integrated luminosity}
  \label{fig:Tevatron_luminosity}
\end{figure}

\begin{figure}[tb]
  \centering
  \includegraphics[width=250px]{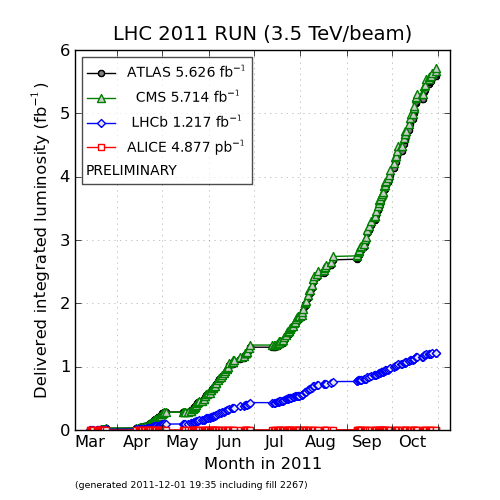}
  \caption{LHC integrated luminosity in the various experiments.}
  \label{fig:LHC_luminosity}
\end{figure}

This is a very exciting moment for particle physics.
After many years of planning and building, LHC and related experiments finally started;
in 2010, March 30, the first collision at center-of-mass energy $\sqrt{s}=7$~TeV took place,
setting the beginning of a new era for particle physics.
In April 2011, LHC set a new record in collider luminosity (roughly speaking, the number of
events per second): the communication by CERN director Rolf Heuer was
\begin{quote}
Geneva, 22 April 2011. Around midnight this night CERN's Large Hadron Collider set a new world record
for beam intensity at a hadron collider when it collided beams with a luminosity of $4.67 \cdot 10^{32}$~cm$^{-2}$s$^{-1}$.
This exceeds the previous world record of $4.024 \cdot 10^{32}$~cm$^{-2}$s$^{-1}$, which was set by the
US Fermi National Accelerator Laboratory's Tevatron collider in 2010, and marks an important milestone in LHC commissioning.
\end{quote}
Now, the luminosity peak is $3.65 \cdot 10^{33}$~cm$^{-2}$s$^{-1}$, one order of magnitude
higher than Tevatron.
Today, Tevatron has collected an integrated luminosity (the integral of the luminosity
over the run time) of about\footnote{The barn is a measure of
surface, used in particle physics for the quantity called cross-section,
and $1$~b~$=10^{-24}$~cm$^2$.}
$L_{\rm int}\simeq 12$~fb$^{-1}$ in more than $10$~years (Fig.~\ref{fig:Tevatron_luminosity});
LHC, instead, has already collected almost $L_{\rm int}\simeq 6$~fb$^{-1}$,
most of which just in the 2011 run (Fig.~\ref{fig:LHC_luminosity}).
The integrated luminosity is a measure of how many events are collected by the experiments:
a process with cross-section $\sigma$ is produced roughly $\sigma L_{\rm int}$ times.
Hence, the largest the integrated luminosity is, the more events can be registered:
this is useful when looking for some new particle, which typically has a very small cross-section.

Along the LHC ring, there are four main experiments:
ATLAS (A Toroidal Lhc ApparatuS),
CMS (Compact Muon Solenoid),
ALICE (A Large Ion Collider Experiment) and
LHCb (LHC Beauty).
The last two are devoted to the study of high-density hadron matter
(quark-gluon plasma) and the physics of $b$-hadrons (CP-violation),
respectively.
ATLAS and CMS, conversely, are general purpose experiments, and their
main goals are searches of the Higgs boson (the last missing particle to
make the Standard Model consistent) and of any kind of signals of new physics beyond
the Standard Model (SUper SYmmetry, extra-dimensions, \ldots).

To be able to see such hints of new physics beyond the Standard Model,
it is crucial to have very accurate theoretical predictions of what
is expected by the Standard Model, in order to be able to distinguish
deviation from the expectations with the maximal significance.
This thesis arises in this context.

The cross-sections in particle physics processes are typically computed using perturbation theory.
Perturbation theory provides a very powerful tool to predict observable
quantities from a quantum field theory. It is based on the assumption that
every observable can be defined by a power series in the coupling constant of the theory:
then, if such coupling is small, the computation of the first few orders
of the power series is sufficient to accurately describe the observable.
However, this assumption needs some care:
it can be proved very generally that perturbative series are divergent.
It is possible to interpret the perturbative expansion in the sense of asymptotic series:
the inclusion of higher orders improves the estimate of the physical quantity under study, up
to some finite order, thereby saving perturbation theory from a dramatic failure.
However, there are situations in which the growth of the series already starts at the level
of the first terms in the series: in these cases, a truncation of the series is of no meaning
and only a resummed result is reliable.

In QCD this situation quite often appears.
A cross-section generically depends on many energy scales,
and the dependence is typically in the form of logarithms of ratios of energies.
In some kinematical regimes, when two of such scales become very different each other,
the logs of their ratio become large: in these cases the coefficients of
the perturbative series grow fast, destroying the perturbativity of the series.
Then, the entire series of these enhanced terms has to be resummed in order
to have an accurate prediction for the observable.

This thesis faces with the problem of resummation of perturbation series in QCD.
The processes that will be discussed are the production of high invariant mass systems
at hadron colliders, such as the Drell-Yan pair production or the Higgs production.
Since at hadron colliders the initial state particles are hadrons, the cross-sections
are typically computed using the parton model, which describes the interaction of the hadron
via its partons (quarks and gluons):
the parton-level cross-section is then computed in QCD using perturbation theory,
and the hadronic cross-section is found from it by convolution with the parton
distribution functions, which are non-perturbative objects extracted from data
describing the distribution of partons momenta in the hadron.
At parton level, the relevant scales are the invariant mass $M$ of the final state
and the partonic center-of-mass energy $\sqrt{\hat s}$: their ratio $z=M^2/\hat s$
appears in the perturbative coefficients of the partonic cross-section.
If only the relevant final state were produced, $z$ would be $1$;
however, even at parton level, also the emission of other particles (in particular, gluons)
must be considered, and therefore $z\neq1$ in general.
Gluon emission produces large logarithms in the partonic cross-section:
the energy squared carried by the gluons is $\hat s-M^2=(1-z)\hat s$, and in the partonic cross-section
powers of $\log(1-z)$ and $\log z$ appear. These logs are large in the two opposite
limits $z\to1$, when the gluons have small energy (they become soft), and
$z\to0$, when the energy of the gluons is large (hard gluons).
The soft $\log(1-z)$ are also referred to as threshold logarithms,
as the limit $z\to1$ is the threshold for the production of the system with mass $M$
having an energy $\sqrt{\hat s}=M/\sqrt{z}$ available.
The hard $\log z$ are also referred to as high-energy logarithms, as they are large
when the energy $\hat s$ is large compared to $M^2$.
Then, as $z\to1$ or $0$, the partonic cross-section needs to be resummed.

The thesis is divided in three parts: the first accounts for the resummation of
threshold logarithms,
the second treats the resummation of high-energy logarithms, and in the last part the effect of
both resummations for phenomenology is discussed and some phenomenological results are shown.

Concerning threshold resummation, the main result of the work presented here
is the extension and the improvement of a prescription necessary to extract
a finite sum from the divergent perturbative series. Such prescription
is based on the Borel summation of divergent series, and provides an alternative
to another one present in the literature and widely used, called the minimal prescription.
The Borel prescription presents some convenient features, both theoretically and practically.
In particular, it is much more suitable than the minimal prescription for
phenomenological applications: one of the results presented here is indeed
a fast numerical implementation of the Borel prescription by means of Chebyshev polynomials.
Another new result is the extension of threshold resummation formalism
to the case of rapidity distributions: this is phenomenologically very useful because
rapidity distributions are measured with high accuracy at hadron colliders.

In the second part, the high-energy resummation is discussed in some detail.
Some improvements are proposed, mainly directed to an efficient numerical implementation.
In fact, the most relevant result concerning high-energy resummation is the realization
of a fast and stable code which implements the complex procedure of resummation
of high-energy logarithms. Such a code was hitherto not available, and it
will be useful to increase the accuracy of the parton distribution functions.

Finally the relevance of threshold resummation is discussed quantitatively.
As sketched above, the variable which governs threshold resummation is the
partonic ratio $z$, but $z$ is not fixed by the hadron-level kinematics.
Conversely, the hadronic process is governed by the variable $\tau=M^2/s$,
where $\sqrt{s}$ is the hadronic center-of-mass energy.
Then it is not obvious if, for a given $\tau$, the threshold region $z\sim1$
gives a sizable contribution, thereby determining the need of resummation.
An argument based on a saddle-point approximation of the integral defining the hadronic
cross-section is then presented, which provides a quantitative way to establish 
for which values of $\tau$ threshold resummation should be included.
Such values of $\tau$ are found to be unexpectedly small, thereby entering in the region
of relevance of high-energy resummation.
Then a discussion on the interplay of the two resummations is presented:
it turns out that the Drell-Yan process is not strongly affected by
high-energy resummation, while the impact on the Higgs production is potentially sizable.
Finally, some phenomenological predictions for the Drell-Yan process
at the Tevatron and the LHC are presented and compared to data.

Before discussing all these aspects in details, in the first Chapter
some basic ingredients concerning QCD are introduced.

\chapter{QCD and the Parton Model}
\label{chap:parton_model}

\minitoc

\noindent
We will review some basics of QCD, introduce the parton model and
describe in more details the GLAP evolution equations.
This Chapter is by no means intended as a complete review;
in particular we assume the Reader knows the quantum theory of fields.
We intend this Chapter as a short introduction useful to fix some notations.

\section{Basics of QCD}
\label{sec:QCD_basics}

QCD is a gauge field theory with gauge group ${\rm SU}(3)$. 
The gauge bosons of the theory are called gluons, and are massless
(no Higgs-like mechanism takes place for them).
The fermions which carry a ${\rm SU}(3)$ charge are called quarks, and in the SM they have
fractional electric charge which is either $2/3$ or $-1/3$ (and, of course, the opposite sign
for the anti-quarks).
We know so far six quarks, i.e.\ three pairs (families) of different charge,
as in the following table:
\begin{center}
\begin{tabular}[c]{cccc}
  electric & \multicolumn{3}{c}{family}\\
  charge & $1$ & $2$ & $3$ \\
  \midrule
  2/3 & $u$ & $c$ & $t$ \\
  -1/3 & $d$ & $s$ & $b$
\end{tabular}
\end{center}
The different kind of quarks are also usually called flavours.
The masses and names of the quarks are as in the following table:
\begin{center}
\begin{tabular}[c]{l|cccccc}
  flavour & $d$ & $u$ & $s$ & $c$ & $b$ & $t$\\
  \midrule
  name & down & up & strange & charm & bottom & top \\
  mass ($\overline{\rm MS}$) & $\sim2.5$~MeV & $\sim5$~MeV & $0.1$~GeV & $1.3$~GeV & $4.2$~GeV & $173$~GeV
\end{tabular}
\end{center}
The top quark $t$, the heaviest one, has been observed directly
only recently at Tevatron~\cite{top_cdf,top_d0}.
The number of quark flavours in the theory is called $n_f$, and as far as we know $n_f=6$.

Since the three lightest quarks have very small masses, the QCD lagrangian has an
approximate global ${\rm U}(3)$ symmetry (called a flavour symmetry):
it is from this symmetry that the ancient quark model has been built
(without knowing anything about QCD).
This approximate symmetry is quite rough, but if we consider only the up and down quarks
(whose mass are really very small) we get a very good approximate ${\rm U}(2)$ symmetry.
Actually, since the two chiral components of quark fields are completely independent
in the massless limit, there are two independent ${\rm U}(2)$ symmetries for each chiral component,
and the global approximate symmetry is ${\rm U}(2)_{\rm L} \times {\rm U}(2)_{\rm R}$,
or, considering vector and axial combinations, ${\rm U}(2)_{\rm V} \times {\rm U}(2)_{\rm A}$.
Now, ${\rm U}(2)_{\rm V} = {\rm SU}(2)_{\rm I}\times {\rm U}(1)_{\rm B}$ is a good symmetry
that we see in Nature, since it corresponds to isospin and baryon number.
Instead, ${\rm U}(2)_{\rm A} = {\rm SU}(2)_{\rm A}\times {\rm U}(1)_{\rm A}$ is spontaneously
broken: hence we expect to find in the spectrum of hadrons the vestiges of the four Goldstone bosons,
but only three (the pions, associated with ${\rm SU}(2)_{\rm A}$) are present.
The absence of a Goldstone boson associated with ${\rm U}(1)_{\rm A}$ was known as
the ${\rm U}(1)_{\rm A}$ problem. Its solution relies on the non-trivial topology of the QCD vacuum.
Indeed, the axial current associated with ${\rm U}(1)_{\rm A}$ has an anomaly proportional
to $\epsilon_{\mu\nu\rho\sigma}G^{\mu\nu}G^{\rho\sigma}$; this term is a total derivative,
and then would classically vanish, but due to instanton effects, this term contributes at quantum level,
breaking explicitly the ${\rm U}(1)_{\rm A}$ symmetry.
The axial anomaly induces an effective term in the QCD lagrangian proportional to the anomaly,
which clearly violates CP in the strong sector.
However we don't see any violation of the CP symmetry in QCD, and
why this is the case is still an open question, known as the strong CP problem.
An elegant solution comes from Peccei and Quinn~\cite{PQ},
who introduced a new axial scalar field (the axion) which couples
to the CP-violating term in the QCD lagrangian, dynamically solving the strong CP problem.
However, the axion has not been discovered so far.

\subsection{The running of the QCD coupling constant}

Because of renormalization, the QCD coupling constant $\as=\frac{g_s^2}{4\pi}$ runs.
The Callan-Symanzik equation (or renormalization-group equation) for $\as$ is
\beq\label{eq:QCD_RG}
\mu^2 \frac{d}{d\mu^2}\as(\mu^2) = \beta\(\as(\mu^2)\)
\eeq
where the $\beta$-function is
\beq\label{eq:betafunctiondef}
\beta(\as) = -\as^2\, \( \beta_0 + \beta_1 \as + \beta_2 \as^2 + \ldots\)
\eeq
and the $\beta_i$ coefficients are known up to $4$ loops ($i=3$).
More details on analytic solutions of the equation will be given in App.~\ref{chap:QCD_running_coupling}.
The leading coefficient is
\beq
\beta_0 = \frac{11C_A - 2n_f}{12\pi}
\eeq
and it is positive as long as $n_f<17$. Because of the minus sign in front of the $\beta_0$ term,
Eq.~\eqref{eq:betafunctiondef}, as $\mu^2$ increases $\as$ decreases:
QCD is asymptotically free, i.e.\ it decouples at high energies.%
\footnote{This statement requires a remark: in computing the solution of Eq.~\eqref{eq:QCD_RG}
the variable flavour number scheme is generally used, in which $n_f$ is the number of \emph{active}
flavours, i.e.\ flavours whose mass is lower of the current energy $\mu$, see App.~\ref{chap:QCD_running_coupling}.
If the fermion families are not only the three we know
(and the corresponding quarks have larger masses), in this scheme
they will enter step by step while increasing the energy.
If at some point the number of active flavours crosses $n_f=17$,
we would discover that the theory is actually not asymptotically free.
}

Conversely, at low energies the coupling grows, exiting the perturbative regime:
then, at such energies the perturbative $\beta$-function is no longer good,
and we no longer believe the solution of the perturbative renormalization group equation.
If we ignore for a moment this fact, and compute the perturbative solution
of the renormalization group equation even at low energies, we discover that
the running coupling $\as$ has a singularity at some $\mu^2=\Lambda^2$,
called the Landau pole.
The scale $\Lambda$, sometimes indicated $\Lambda_{\rm QCD}$,
is a scale at which QCD is non-perturbative, and it is typically
of the order of some hundreds MeV; the exact value depends on the order of the
$\beta$-function used and on the initial condition for the evolution (typically $\as(m_Z^2)$ at the $Z$ mass).
The Landau pole would probably not be there in a complete non-perturbative solution
of the evolution equation: it is something non-physical.
However, we will see that it plays an important role in resumming the perturbative
series of QCD (see Chap.~\ref{chap:soft-gluons}).
The non-perturbative region of QCD starts at higher scales, when $\as$ becomes
too large to justify a perturbative expansion, typically around $1$~GeV.

Even if we are not able to predict the behaviour of the coupling constant
at low energies, it is clear from the renormalization group equation that QCD is
strongly coupled at low energies. In Nature, indeed, we don't ever see isolated
quarks or gluons, but only hadrons, i.e.\ composite objects made of quarks and gluons
which belong to the singlet representation of the gauge group ${\rm SU}(3)$.
This fact is known as \emph{confinement}, and it cannot be explained by use of
perturbative QCD (pQCD for short); from lattice simulations, there is some evidence
that quarks confine, but a complete non-perturbative explaination is still missing.
Of course, being composite objects, the description of the interaction between hadrons
and other particles is more complicated than for elementary particles:
such a description, called the parton model, will be addressed in the following Section.

\section{The parton model}

As just said, hadrons are made of quarks and gluons, generically called \emph{partons},
being the parts of the hadron.
Therefore, for studying high-energy processes involving hadrons,
a model which describes how a hadron interacts via its partons
is generally adopted: the \emph{parton model}.
For definiteness, in the following, we will concentrate
on protons, but what we will say can be in principle applied to any other hadron.

The original (or naive) parton model was proposed by Feynman, and it is formulated
in the \emph{infinite momentum frame}, i.e.\ a reference frame
in which the proton is ultrarelativistic.
In such frame, we can neglect the mass of the proton ($m_p\sim0$) and, a fortiori,
the masses of the partons. Then, the basic assumption of the model is that
each parton $i$ carries a fraction $z_i$ of the proton momentum $p$,
\beq\label{eq:pm_assumption}
\hat p_i = z_i p,
\qquad 0\leq z_i \leq 1,
\eeq
where $\hat p_i$ is the momentum of the parton (we will use often a hat
$\hat{\phantom p}$ to indicate partonic quantities).
Note that such a relation can be defined only in the infinite momentum frame,
since otherwise the mass of the parton would vary with $z_i$, as one sees squaring
Eq.~\eqref{eq:pm_assumption}.
The exact momentum fraction $z_i$ for each parton is not fixed by the model;
instead, each parton can be picked up from the proton with a given momentum fraction
$z_i$ following the distribution
\beq
f_i(z_i),
\eeq
called \emph{parton distribution function} (PDF for short).
As another assumption, the interaction between a proton and an elementary particle
is the incoherent sum of the interaction between each parton and the elementary particle
(described by the field theory), weighted with the PDFs. Then, denoting
by $\hat\sigma_i(\hat p)$ the cross section for the partonic process, the
hadron level cross-section is
\beq\label{eq:parton_model}
\sigma(p) = \sum_i \int_0^1 dz\, f_i(z)\, \hat\sigma_i(zp).
\eeq
where $p$ is the proton momentum.

The PDFs are defined in such a way that the probability to pick up a parton $i$
with momentum fraction between $z$ and $z+dz$ is $f_i(z) dz$.
Some properties (sum rules) follow:
\begin{itemize}
\item the difference between quarks and antiquarks PDFs, integrated in $z$,
counts the number of constituents quarks of the proton:
\beq
\int_0^1 dz\, \[ f_u(z) - f_{\bar u}(z) \] = 2,
\qquad
\int_0^1 dz\, \[ f_d(z) - f_{\bar d}(z) \] = 1,
\eeq
and all the other partons give zero.
\item the sum of the momenta of all partons must equal the proton momentum:
\beq\label{eq:PDF_momentum_conservation}
\sum_i \int_0^1 dz\, z\, f_i(z) = 1.
\eeq
\end{itemize}
In Eq.~\eqref{eq:parton_model} the partonic cross-section $\hat\sigma_i(zp)$
is computable perturbatively from the field theory;
the PDFs, instead, are intrinsically non-perturbative objects,
and have to be extracted from experimental measures. The sum rules provide
important constraints on the extraction process.

This formulation of the parton model is very naive. QCD perturbative
corrections induce changes in the model, leading to what is sometimes called
the \emph{improved parton model}.
In particular, the PDFs acquire a dependence on an energy scale:
to see how this happens, we consider now in some detail a prototype process.

\subsection{Deep-inelastic scattering}
\label{sec:DIS}

\begin{figure}[th]
  \centering
  \includegraphics[width=0.6\textwidth]{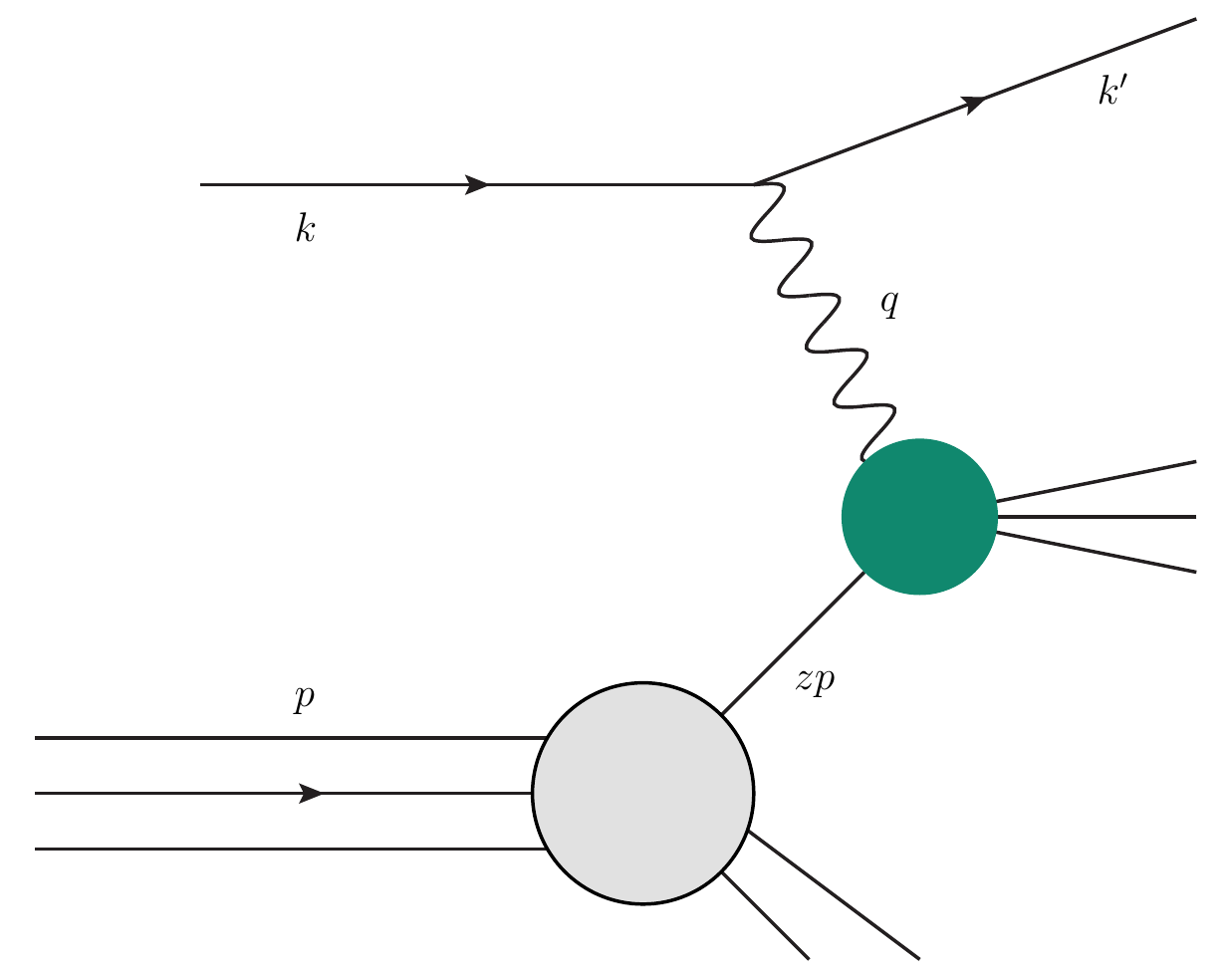}
  \caption{The DIS process. The green ball represents the interaction between a parton and the photon,
    and contains any QCD correction to the process, including real emissions.}
  \label{fig:DIS}
\end{figure}

The typical process for which the parton model has been built is the
deep-inelastic scattering (DIS), the collision of a proton with a lepton.
We will concentrate on the case the lepton is charged, see Fig.~\ref{fig:DIS}.
The process is called inelastic because the energy of the collision
is such that the proton changes nature, breaking up into pieces which will form
something else (possibly other hadrons).

Here we will review some results, in order to fix notations which will be useful
in the following. For a complete review, see for example~\cite{esw}.
Very generally, we can write the amplitude for the process as
\beq
\M = e \, \bar u(k')\, \gamma^\mu\, u(k)\, \mathcal{P}_\mu(p,q)
\eeq
where the momenta are as in Fig.~\ref{fig:DIS}, and $\mathcal{P}_\mu(p,q)$
contains the information on the hadronic part of the process, including the photon propagator.
The amplitude squared, after sum over final states and mean over initial states, can be written as
\beq
\frac{1}{N_{\rm in}}\sum_{\rm in,\, fin}\abs{\M}^2 = L_{\mu\nu} W^{\mu\nu}
\eeq
where $L_{\mu\nu}$ and $W_{\mu\nu}$ are called, respectively, leptonic
and hadronic tensors: the first is simply
\beq
L_{\mu\nu} = e^2\, \tr\!\[ \slashed{k}'\gamma_\mu \slashed{k} \gamma_\nu \],
\eeq
while the second contains all the information on the hadronic process.
Skipping the computations, the final result for the cross-section can be written as
\beq\label{eq:DIS_sigma}
\frac{d\sigma}{dx\, dQ^2} = \frac{4\pi\alpha^2}{Q^4}\[
\(1+(1-y)^2\) F_1(x,Q^2) + \frac{1-y}{x}\(F_2(x,Q^2) - 2x\,F_1(x,Q^2)\)
\]
\eeq
where we have defined the variables
\beq
x=\frac{Q^2}{2pq}, \qquad
y=\frac{qp}{kp}, \qquad
Q^2 = -q^2
\eeq
and introduced the so called \emph{structure functions} $F_1$ and $F_2$.
Such structure functions can be obtained via suitable projectors on the
hadronic tensor.
Note that this result is older than QCD: indeed, we haven't said anything so far
about the structure of the hadronic interaction, confining any hadronic information
in the structure functions, which could be measured experimentally.
It is sometimes useful to define the longitudinal structure function
\beq
F_L(x,Q^2) = F_2(x,Q^2) - 2x\,F_1(x,Q^2);
\eeq
as can be seen from Eq.~\eqref{eq:DIS_sigma}, the term with $F_L$ corresponds
to the absorption of a longitudinally polarized virtual photon,
while the term with $F_1$ corresponds to the absorption of a transversely
polarized virtual photon. Since the quarks are spin $1/2$, they cannot absorb
longitudinal virtual photons, and then they would lead to a vanishing $F_L$
(Callan-Gross relation): this is indeed observed in the limit $Q^2\to\infty$ with $x$ fixed
(Bjorken limit), confirming the spin $1/2$ nature of the quarks.

The hadronic tensor (and, then, the structure functions) can be
computed in the parton model framework. We can write
\beq\label{eq:DIS_structure_functions}
F_i(x,Q^2) = x \sum_j \int_x^1 \frac{dz}{z}\, f_j(z) \, C^{(0)}_{ij}\(\frac{x}{z}\),
\eeq
where the functions $C^{(0)}_{ij}$ are called \emph{coefficient functions}.
The superscript ${}^{(0)}$ indicates that we are ignoring QCD,
i.e.\ we are not considering QCD corrections: we will discuss them in the next Section.
Then, in the naive parton model, we have
(choosing $F_2$ and $F_L$ as the two independent structure functions)
\begin{subequations}\label{eq:DIS_C0_naive}
\begin{align}
&C^{(0)}_{2q}(z) = e_q^2 \, \delta(1-z) &
&C^{(0)}_{2g}(z) = 0 \\
&C^{(0)}_{Lq}(z) = 0 &
&C^{(0)}_{Lg}(z) = 0
\end{align}
\end{subequations}
where $e_q$ is the quark charge in fractions of the electric charge.
Then the structure functions are
\beq
F_2(x,Q^2) = x \sum_q e_q^2\, f_q(x), \qquad F_L(x,Q^2) = 0,
\eeq
where there is no actual dependence on $Q^2$: this is known as
\emph{scaling}, and was observed (in the Bjorken limit) in DIS experiments.

As a final comment, we would like to note that Eq.~\eqref{eq:DIS_structure_functions}
(divided by $x$) has the form of a Mellin convolution
\beq
(g\otimes h)(x) = \int_x^1 \frac{dz}{z}\, g(z)\, h\(\frac xz\).
\eeq
It is symmetric, i.e.\ it has the property that, changing variable $z\to x/z$,
it remains in the same form with the arguments of the two functions exchanged.
The Mellin convolution diagonalizes under Mellin transform
\beq
\Mell[g](N) = \tilde g(N) = \int_0^1 dz \, z^{N-1} \, g(z),
\eeq
i.e.\ in Mellin space we have
\beq
\Mell[g\otimes h](N) = \tilde g(N)\, \tilde h(N).
\eeq
The Mellin transformation is related to a Laplace transformation by a change
of variable; hence, the inverse Mellin transform is
\beq
\Mell^{-1}\[\tilde g\] (z) = \frac1{2\pi i}\int_{c-i\infty}^{c+i\infty} dN \, x^{-N}\, \tilde g(N)
\eeq
where $c$ has to be to the right of the rightmost singularity of $\tilde g(N)$
(it always exists because a Mellin (Laplace) transform always has a convergence abscissa).
For more details about Mellin transformation and inversion, see App.~\ref{chap:Mellin}.
Note that, in the following, by an abuse of notation
we will omit the $\tilde{\phantom g}$  to indicate a Mellin transform,
and the Reader can recognize in which space the function is by its argument.

\subsection{Radiative corrections and factorization}
\label{sec:PM_radiative_corr}

The coefficient functions $C^{(0)}_{ij}(z)$, Eqs.~\eqref{eq:DIS_C0_naive},
can be considered as the contribution to the coefficient functions at LO in pQCD.
In pQCD, a coefficient function has a perturbative expansion
\beq
C_{ij}(z,\as) = C_{ij}^{(0)}(z) + \as\, C_{ij}^{(1)}(z) + \as^2\, C_{ij}^{(2)}(z) + \ldots
\eeq
in powers of $\as$.
When considering QCD corrections to the partonic process, i.e.\ when computing $C^{(k)}(z)$ with $k\geq1$,
we have to consider both loop corrections and additional emissions of quarks and gluons.
The reason for this is that in hadronic process we usually consider inclusive quantities:
whatever happens to the proton, we don't care, we simply integrate over it.
Indeed the parton model does not give us any information on what happens to the other partons
which do not interact in the hard process we are considering.
Then, if the ``hard parton'' emits, let's say, a gluon, how can we distinguish it from
the mess of other things coming from the proton? We can't, and that's why we need to consider
also emissions to the QCD corrections of the hard process.

The QCD corrections diagrams have both UV and IR divergences: the second ones,
called also mass singularities (because they would not be there if partons were massive)
appear in two forms, called respectively soft and collinear.
Let us trace the origin of these in detail.

The loop diagrams diverge:
\begin{itemize}
\item in the UV, and such divergences are treated with renormalization, as usual;
\item in the IR, due to the fact that the particles in the loops are massless.
\end{itemize}
Fortunately, the IR singularities cancel when virtual diagrams are combined with real emission
diagrams: when an emitted particle becomes soft (its energy tends to zero) the diagram presents
a divergence which is regulated by the divergence of one of the virtual diagrams.
This fact is known as Kinoshita-Lee-Nauenberg theorem~\cite{KLN1,KLN2}, and can be recast in the sentence that
soft divergences always cancel.
However, real emission diagrams introduce other singularities,
when the transverse momentum of the emitted particle tends to zero (collinear singularities).
Such singularities are not canceled by anything, and we have to deal with them.

The way these singularities are treated is similar to what happens in renormalization:
since the PDFs are quantities which should be measured, we can imagine that in the
naive formulation of the parton model they are bare objects, and that they can be redefined
in such a way to reabsorb the collinear divergences: these new PDFs are what we actually measure
and they must be finite.
Schematically, we have (in $N$-space for simplicity)
\beq
C_{ij}(N,\as) = C_{ij}(N,\as,\mu^2) \, c_j^{\rm divergent}(N,\as,\mu^2),
\eeq
where, as usual in any regularization scheme, we have introduced a dependence on a new energy scale $\mu$;
then we can construct the new PDFs as
\beq
f_j(N,\mu^2) = f_j(N) \, c_j^{\rm divergent}(N,\as,\mu^2),
\eeq
which acquire a dependence on the new scale $\mu$.
Then, the structure functions
\beq\label{eq:DIS_structure_functions_N}
F_i(N-1,Q^2) = \sum_j f_j(N,\mu^2) \, C_{ij}(N, \as(Q^2),\mu^2)
\eeq
are well defined to all orders in perturbation theory.

In this discussion, we have tacitly assumed three important facts:
\begin{itemize}
\item the divergent part of the coefficient functions factorizes;
\item it is independent on the observable (independent on the index $i$) and on the process;
\item QCD corrections apply to the parton level process only, i.e.\ diagrams with additional
  lines connecting the partonic part of the process directly to the non-perturbative
  hadronic part (interference terms) do not count.
\end{itemize}
The first fact is known as \emph{factorization of collinear singularities}
and the second represents the \emph{universality} of collinear divergences.
The last point is valid for a large class of processes, where in particular
such terms are suppressed by powers of the hard scale $Q^2$ (higher twists).
Together, these facts are known as the \emph{(collinear) factorization theorem},
valid at leading twist for all the processes we are interested in.
In particular, it allows to reabsorb the collinear divergent terms into the PDFs,
and to do it independently on the process or the observable,
making the definition of the new PDFs really universal:
the hadronic cross-section is then completely factorized into
a perturbative part and a non-perturbative part.

Note that in Eq.~\eqref{eq:DIS_structure_functions_N} there is a dependence on $\mu$
on the right-hand-side which however is not present on the left-hand-side: this is correct, since the scale
$\mu$ (called \emph{factorization scale}) has been introduced arbitrarily by a regularization,
and hence physical quantities must not depend on $\mu$.
The $\mu$ dependence of the coefficient function is fixed by the choice of the
\emph{factorization scheme}, analogous to the subtraction scheme in renormalization,
i.e.\ the choice of which finite parts are put into the divergent coefficient $c_j(N,\as,\mu^2)$ and which
are left into the finite coefficient $C_{ij}(N,\as,\mu^2)$.
For example, we could choose a scheme in which the finite coefficient function is
\beq
C_{2q}(N,\as,\mu^2) = e_q^2,
\eeq
which is called the DIS factorization scheme: in such scheme the structure function $F_2$
corresponds (up to a factor) to the weighted sum of quark PDFs to all orders in perturbation theory.
A more used scheme is the \MSbar\ scheme, analogous to the same in renormalization,
which collects into the divergent term only the $\epsilon$ poles of a $d=4-2\epsilon$
dimensional computation and some selected terms; in the following we will always
present results in  the \MSbar\ scheme.
Once the $\mu$ dependence of the coefficient function is fixed, one can compute
the $\mu$ dependence of the PDFs by imposing $\mu$ independence of the structure functions,
order by order in perturbation theory.
We will elaborate on that in the next Section.

\section{GLAP evolution equations}
\label{sec:glap_eq}

Once the factorization scheme is fixed, the $\mu$ dependence of the coefficient function
is computable in perturbation theory. Moreover, such dependence is again process-independent
and observable-independent, since it is strictly related to the divergent piece.
Then, $\mu$ independence of any physical quantity gives an evolution equation for the PDFs:
considering for example the structure functions, the equation
\beq
\frac{d}{d\mu^2}F_i(N,Q^2) = 0
\eeq
allows us to extract an equation for the PDFs
\beq
\mu^2\frac{d}{d\mu^2} f_j(N,\mu^2) = \sum_k \gamma_{jk}\(\as(\mu^2),N-1\) f_k(N,\mu^2),
\eeq
which is called the \emph{Altarelli-Parisi} or \emph{Gribov-Lipatov-Altarelli-Parisi (GLAP) evolution equation}.
Actually the derivation is not so trivial, but a rigorous proof of the GLAP equation can be done
by means of the operator product expansion (OPE).
Note in particular that $\as$ is computed at the scale $\mu$.
The functions $\gamma_{jk}(\as,N)$, called the \emph{Altarelli-Parisi anomalous dimensions},
are computable in perturbation theory as
\beq
\gamma(\as,N) = \as \[ \gamma^{(0)}(N) + \as \gamma^{(1)}(N) + \as^2 \gamma^{(2)}(N) + \Ord(\as^3) \]
\eeq
where we have used a matrix notation (i.e.\ we have suppressed the indeces).
In this notation, the anomalous dimensions are $(2n_f+1)$-dimensional matrices,
acting on a $(2n_f+1)$-dimensional vector of PDFs of the form
\beq
f(x,\mu^2) = \left\{ f_g(x,\mu^2), f_{q_i}(x,\mu^2) \right\}.
\eeq
The $x$-space version of the evolution equation is
\beq\label{eq:GLAP_eq_integral}
\mu^2\frac{d}{d\mu^2} f(x,\mu^2) = \int_x^1 \frac{dz}{z}\, P\(\as(\mu^2),\frac xz\) \, f(z,\mu^2)
\eeq
where $P\(\as(\mu^2),x\)$ is a $(2n_f+1)$-dimensional matrix of Altarelli-Parisi
\emph{splitting functions}, whose expansion in powers of $\as$ is
\beq
P(\as,x) = \as \[ P^{(0)}(x) + \as P^{(1)}(x) + \as^2 P^{(2)}(x) + \Ord(\as^3) \].
\eeq
The LO ($1$-loop) and NLO ($2$-loops) splitting funtions (and anomalous dimensions)
are known for a long time, while the NNLO ($3$-loops) ones have been computed recently~\cite{mvv_ns,vmv}.
It has to be noted that the splitting functions are distribution, while the 
anomalous dimensions are ordinary functions; however, the knowledge
of the anomalous dimensions for complex values of $N$ is needed to compute
the inverse Mellin transform and get back the splitting functions.

Note that we use a notation which is slightly different to the common one used in the literature,
and which is more suitable for small-$x$ physics: our anomalous dimensions are defined as
the Mellin transform of $x$ times the splitting functions,
\beq\label{eq:anom-dim_Mellin_xP}
\gamma(\as, N) = \Mell\[xP(\as,x)\](N).
\eeq
With this choice, the argument of the anomalous dimensions is shifted by a unity:
the usual anomalous dimensions are recovered as $\gamma(\as, N-1)$.

The rank of the evolution matrix is not maximal, and we could find several combinations
of PDFs which decouple. Such combinations are called non-singlet and evolve independently.
There are $2n_f-1$ independent non-singlet combinations of PDFs. The remaining two degrees
of freedom do not decouple and form a rank $2$ system of equations, and are called
singlet PDFs.
Because of ${\rm SU}(n_f)$ flavour symmetry (remember that in the parton model all partons
are treated as massless), the splitting functions satisfy
\begin{subequations}\label{eq:splitting_functions_decomp}
\begin{align}
  P_{g q_i} = P_{g \bar q_i} &\equiv P_{gq}\\
  P_{q_i g} = P_{\bar q_i g} &\equiv P_{qg} /(2n_f)\\
  P_{q_i q_j} = P_{\bar q_i \bar q_j} &\equiv \delta_{ij} P^V_{qq} + P^S_{qq}\\
  P_{q_i \bar q_j} = P_{\bar q_i q_j} &\equiv \delta_{ij} P^V_{q\bar q} + P^S_{q\bar q}
\end{align}
\end{subequations}
where we have defined six of the seven independent splitting functions (the seventh is $P_{gg}$).
At LO, of the four quark splitting functions introduced here, only $P^V_{qq}$ is non-zero.

\subsection{The non-singlet sector}

The non-singlet combinations of PDFs are differences of quarks PDFs, like for example
any $f_q-f_{\bar q}$. Of course, there are $n_f$ of such combinations, but we can construct
other $n_f-1$ independent combinations whose evolution is decoupled (for more details, see Ref.~\cite{esw}).
The splitting functions governing the evolution of these two kind of non-singlet combinations
are called $P^-$ and $P^+$ respectively, with
\beq\label{eq:splitting_functions_pm}
P^\pm = P^V_{qq} \pm P^V_{q\bar q}
\eeq
in terms of the non-singlet splitting functions defined in \eqref{eq:splitting_functions_decomp}.
At LO we have $P^{V(0)}_{q\bar q}=0$ and then
\beq\label{eq:Pqq0}
P^{+(0)}(x) = P^{-(0)}(x) = P^{V(0)}_{qq}(x) = \frac{C_F}{2\pi} \plus{\frac{1+x^2}{1-x}}
\eeq
where the plus-distribution is defined in App.~\ref{sec:plus_distribution}.
The NLO and NNLO splitting functions and anomalous dimension can be found in~\cite{mvv_ns}.

\subsection{The singlet sector}
\label{sec:glap_singlet_sector}

The remaining degrees of freedom are the gluon PDF and the so called singlet quark PDF
defined as
\beq
f_S = \sum_q \(f_q + f_{\bar q}\).
\eeq
The evolution of these two PDFs is governed by an evolution matrix
\beq
\(
\begin{array}{cc}
P_{gg} & P_{gq}\\
P_{qg} & P_{qq}
\end{array}
\)
\eeq
where we have defined
\beq
P_{qq} = P^+ + n_f\(P^S_{qq} + P^S_{q\bar q}\)
\eeq
in terms of the splitting functions defined in Eqs.~\eqref{eq:splitting_functions_decomp} and
\eqref{eq:splitting_functions_pm}.
At LO, $P_{qq}$ is given by Eq.~\eqref{eq:Pqq0} and
\begin{subequations}\label{eq:P0}
\begin{align}
  P_{gg}^{(0)}(x) &= \frac{C_A}{\pi}\left[ \frac{x}{\plus{1-x}} + \frac{1-x}{x} + x(1-x) \right]
  + \frac{11 C_A -2 n_f}{12\pi}\delta(1-x) \\
  P_{gq}^{(0)}(x) &= \frac{C_F}{2\pi} \left[ \frac{1+(1-x)^2}{x} \right] \\
  P_{qg}^{(0)}(x) &= \frac{n_f}{2\pi} \left[ x^2 + (1-x)^2 \right].
\end{align}
\end{subequations}
In $N$ space, the GLAP equation for the singlet sector reads (omitting for ease of notation the $\as$ and $N$ dependencies)
\beq\label{eq:glap_eq_singlet}
\mu^2\frac{d}{d\mu^2} \dvec{f_g}{f_S} =
\(
\begin{array}{cc}
\gamma_{gg} & \gamma_{gq}\\
\gamma_{qg} & \gamma_{qq}
\end{array}
\)
\dvec{f_g}{f_S}
\eeq
where at LO
\begin{subequations}\label{eq:gamma0}
\begin{align}
  \gamma_{gg}^{(0)}(N) &= \frac{C_A}{\pi}\left[ \frac{1}{N} -\frac{1}{N+1} +\frac{1}{N+2} -\frac{1}{N+3}
  -\psi(N+2)-\gammae \right] + \frac{11 C_A -2 n_f}{12\pi} \\
  \gamma_{gq}^{(0)}(N) &= \frac{C_F}{2\pi} \left[ \frac{2}{N} -\frac{2}{N+1} +\frac{1}{N+2} \right] \\
  \gamma_{qg}^{(0)}(N) &= \frac{n_f}{2\pi} \left[ \frac{1}{N+1} -\frac{2}{N+2} +\frac{2}{N+3} \right]  \\
  \gamma_{qq}^{(0)}(N) &= \frac{C_F}{2\pi} \left[ \frac{3}{2} + \frac{1}{N+1} - \frac{1}{N+2} -2\psi(N+2)-2\gammae \right].
\end{align}
\end{subequations}
The NLO and NNLO splitting functions and anomalous dimensions $P_{ij}^{(1,2)}$, $\gamma_{ij}^{(1,2)}$ can be found in~\cite{vmv}.

\subsubsection{Eigenvalues of the singlet anomalous dimension matrix}
\label{sec:glap:projectors}

\begin{figure}[thb]
  \centering
  \includegraphics[width=0.6\textwidth]{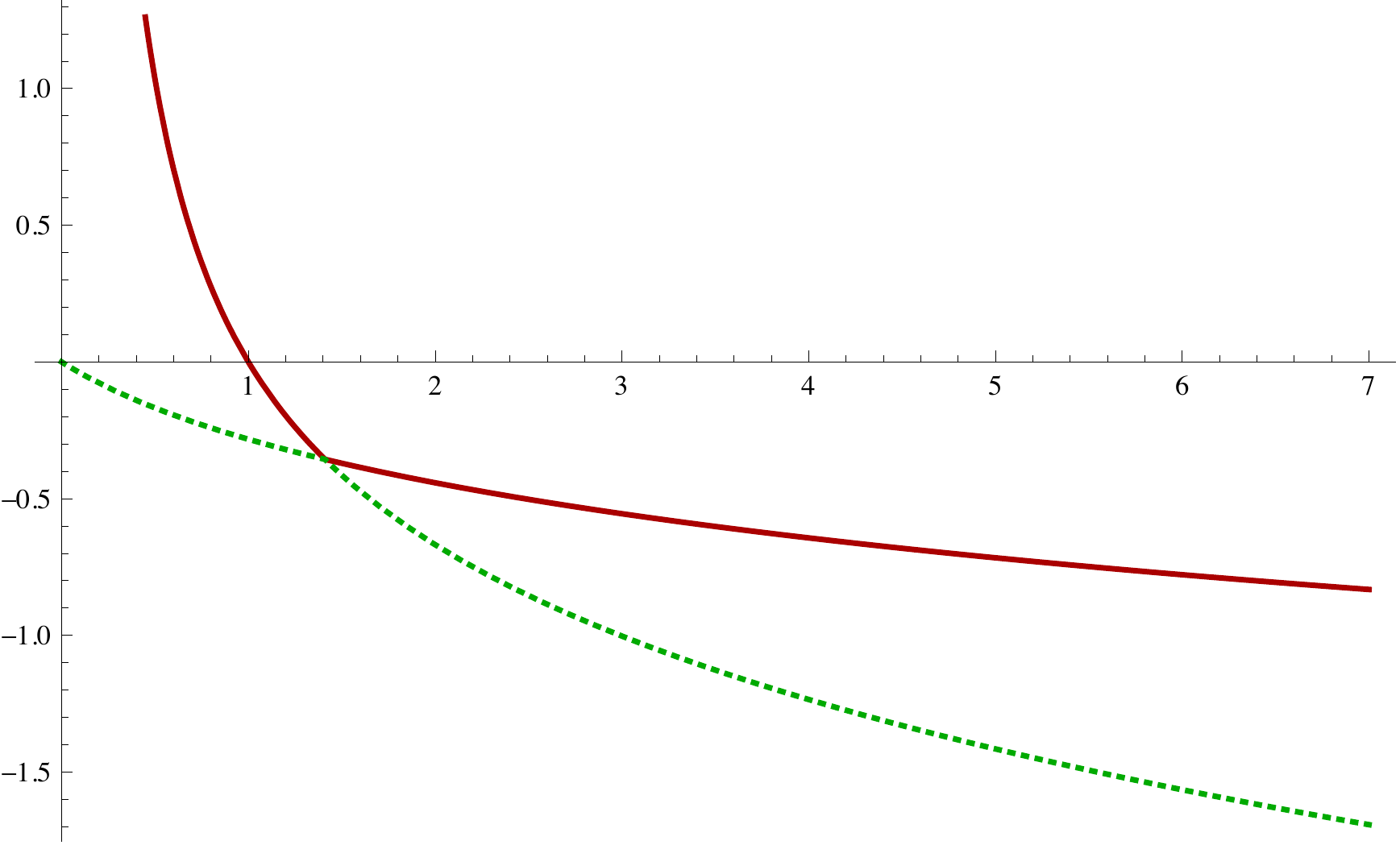}
  \caption{The eigenvalues $\gamma_+^{(0)}$ (blue upper curve) and $\gamma_-^{(0)}$
    (green lower curve) as functions of $N$ for $n_f=0$.}
  \label{fig:glap_eigen_nf0}
\end{figure}

We now study the diagonalization of the evolution matrix:
for instance, this can be useful to solve the singlet equations at LO
(see Sect.~\ref{sec:GLAP_solve_diagonalization}).
Let's call the evolution matrix $\Gamma$:
\beq\label{eq:Gamma}
\Gamma = 
\(
\begin{array}{cc}
\gamma_{gg} & \gamma_{gq}\\
\gamma_{qg} & \gamma_{qq}
\end{array}
\).
\eeq
The eigenvalues of this matrix are found as solutions of the secular equation
\beq
\gamma_\pm^2 - \gamma_\pm\, \tr\Gamma + \det \Gamma = 0
\eeq
leading to
\beq\label{eq:gamma_pm}
\gamma_\pm = \frac{1}{2} \left[ \tr\Gamma \pm \sqrt{(\tr\Gamma)^2-4\det\Gamma} \right].
\eeq
Obviously, $\gamma_+$ is the largest eigenvalue for all $N$.
In the case $n_f=0$, the matrix is triangular ($\gamma_{qg}$ vanishes to all orders,
because it is multiplied by $n_f$) and the solution has a simpler form
\beq
\gamma_\pm \overset{n_f=0}{=} \frac{1}{2} \(
\gamma_{gg} + \gamma_{qq} \pm \abs{\gamma_{gg} - \gamma_{qq}}
\).
\eeq
In this case, the largest eigenvalue $\gamma_+$ would be equal to
$\gamma_{gg}$ or $\gamma_{qq}$ depending on which of the two is the largest;
correspondingly, $\gamma_-$ is the other of the two.
This means in particular that in the case $n_f=0$ for values of $N$ such that
$\gamma_{gg}=\gamma_{qq}$ both $\gamma_+$ and $\gamma_-$ as functions of $N$
have a discontinuity in their first derivative with respect to $N$, see Fig.~\ref{fig:glap_eigen_nf0}.
This discontinuity is an artifact, and would be absent if we take separately $\gamma_{gg}$ and $\gamma_{qq}$
as eigenvalues. Moreover, when $n_f=0$, there are no quarks in the game and the only relevant quantity
is $\gamma_{gg}$.
For this reason%
\footnote{This choice is also motivated by the fact that in Chap.~\ref{chap:small-x}
we will be interested in the largest eigenvalue at small $N$.}
in the case $n_f=0$ we will conventionally choose $\gamma_+=\gamma_{gg}$.
In other words, we choose $\gamma_+$ as the largest eigenvalue at small $N$.

The matrix $R$ that diagonalizes $\Gamma$,
\beq\label{eq:glap_diagonalize_Gamma}
R\Gamma R^{-1} = \hat\Gamma = \diag(\gamma_+,\gamma_-),
\eeq
can be written as
\beq\label{eq:rotationR}
R= \frac{1}{c_--c_+}
\sqmatr{c_-}{-1}{-c_+}1
,\qquad
R^{-1}=
\sqmatr11{c_+}{c_-}
\eeq
with
\beq
c_\pm = \frac{\gamma_\pm-\gamma_{gg}}{\gamma_{gq}} = \frac{\gamma_{qg}}{\gamma_\pm-\gamma_{qq}}.
\eeq
In the diagonal basis, the diagonal evolution matrix can be written as
\beq
\hat\Gamma = \sqmatr1000 \gamma_+ + \sqmatr0001 \gamma_-
\eeq
where the two matrices are projectors (i.e.\ idempotent operators) on orthogonal subspaces.
When going back to the physical basis, we have the decomposition
\beq\label{eq:Gamma_proj}
\Gamma = R^{-1} \hat\Gamma R = \projM_+ \gamma_+ + \projM_- \gamma_-
\eeq
where the projectors $\projM_\pm$ are given by
\beq
\projM_\pm = \pm \frac{1}{c_--c_+} \sqmatr{c_\mp}{-1}{c_+c_-}{-c_\pm} = \pm \frac{\Gamma-\gamma_\mp}{\gamma_+-\gamma_-}
\eeq
or, in terms of the eigenvalues and of $\gamma_{qq}$ and $\gamma_{qg}$,
\beq
\projM_\pm = \pm \frac{1}{\gamma_+-\gamma_-} \sqmatr{(\gamma_\pm-\gamma_{qq})}{X}{\gamma_{qg}}{(\gamma_{qq}-\gamma_\mp)},
\qquad
X= \frac{(\gamma_+-\gamma_{qq})(\gamma_{qq}-\gamma_-)}{\gamma_{qg}},
\eeq
and satisfy, being projectors, the relations
\beq\label{eq:projector_properties}
\projM_\pm\projM_\pm=\projM_\pm,\qquad
\projM_\pm\projM_\mp = 0,\qquad
\projM_+ + \projM_- = 1.
\eeq
This projector formalism proves to be useful, for instance, to solve
the evolution equations, see Sect.~\ref{sec:glap_sol}.

While at LO the eigenvalues \eqref{eq:gamma_pm} are pure order $\as$,
the NLO eigenvalues computed using \eqref{eq:gamma_pm} contain spurious
terms of order $\as^3$ and higher, due to the presence of the square root.
Hence, perturbatively we can write
\beq
\gamma_\pm = \as \gamma_\pm^{(0)} + \as^2 \gamma_\pm^{(1)} + \Ord(\as^3)
\eeq
where
\beq\label{eq:le_NLO}
\gamma_\pm^{(1)} = \frac{1}{2}\left[
\gamma_{gg}^{(1)} + \gamma_{qq}^{(1)}
\pm \frac{
\gamma_{gg}^{(0)}\gamma_{gg}^{(1)} +
\gamma_{qq}^{(0)}\gamma_{qq}^{(1)} -
\gamma_{gg}^{(0)}\gamma_{qq}^{(1)} -
\gamma_{qq}^{(0)}\gamma_{gg}^{(1)} +
2\(
\gamma_{gq}^{(0)}\gamma_{qg}^{(1)} +
\gamma_{qg}^{(0)}\gamma_{gq}^{(1)}
\)
}{\sqrt{(\tr\Gamma^{(0)})^2-4\det\Gamma^{(0)}}}
\right].
\eeq
Note that this result can be obtained also in the following way:
first, we diagonalize the evolution matrix at LO, and we construct the
matrix $R_{\rm LO}$ which makes such diagonalization. Then we use $R_{\rm LO}$
to rotate the evolution matrix at NLO: the diagonal entries are exactly
the LO eigenvalues \emph{plus} the order $\as^2$ term in Eq.~\eqref{eq:le_NLO},
without higher orders.
The difference is that with this procedure also non-diagonal entries are generated at NLO.

\subsubsection{Small-$x$ behaviour}

The small-$x$ behaviour of the singlet splitting functions is, up to NLO,
(see Ref.~\cite{esw})
\begin{subequations}
\begin{align}
P_{gg} &\simeq \frac{\as C_A}{\pi x} + \frac{\as^2 n_f}{4\pi^2} \frac{6C_F-23C_A}{9x} \\
P_{gq} &\simeq \frac{\as C_F}{\pi x} + \frac{\as^2 C_F}{4\pi^2} \frac{9C_A-20n_f}{9x} \\
P_{qg} &\simeq \frac{\as^2 n_f}{4\pi^2} \frac{20C_A}{9x} \\
P_{qq} &\simeq \frac{\as^2 n_f}{4\pi^2} \frac{20C_F}{9x}.
\end{align}
\end{subequations}
In Ref.~\cite{vmv} the small-$x$ behaviour of the order $\as^3$ splitting functions
can be also found; at that order, also terms of the kind $\frac{\log^k x}x$ appear.
In general, it can be shown (and will be discussed in details in Chap.~\ref{chap:small-x})
that at the order $\as^{k+1}$ the splitting functions contain terms
\beq\label{eq:small-x_logs}
\as^{k+1}\frac{\log^j x}{x}, \qquad 0\leq j\leq k
\eeq
for the $gg$ and $gq$ components, while the others have a power less ($j<k$).
It is accidental that at NLO the dominant term $\frac{\log x}x$ does not appear.
Also at NNLO the dominant term $\frac{\log^2 x}x$ accidentally vanishes
(while the subdominant one is present, see Ref.~\cite{vmv}).

In $N$ space, the small-$x$ behaviour is determined by the rightmost singularity
of the anomalous dimensions, see App.~\ref{chap:Mellin}.
In particular, the powers of $\log x$ of Eq.~\eqref{eq:small-x_logs} correspond in $N$ space to
multiple poles in $N=0$,\footnote{Remember our definition Eq.~\eqref{eq:anom-dim_Mellin_xP}.}
\beq
\as^{k+1} \frac{1}{N^j}, \qquad 0\leq j\leq k+1
\eeq
for the $gg$ and $gq$ and with $0\leq j \leq k$ for $qg$ and $qq$.
We can say that the tower of terms with highest power,
\beq
\(\frac{\as}{N}\)^k,\qquad 0 < k < \infty
\eeq
are the \emph{leading-log} (LL) terms; sometimes the notation LL$_x$ is used,
to underline that we are talking about small-$x$ logarithms.
If we add, at each order, one power of $\as$ more, they will be NLL$_x$ terms:
in general, the tower of terms
\beq
\as^n\(\frac{\as}{N}\)^k,\qquad 0 \leq k < \infty
\eeq
constitute the N$^n$LL$_x$.

Expanding the complete 1- and 2-loops anomalous dimensions at small $N$,
we obtain the small-$x$ contributions in $N$ space, including terms up to order $\as^2$ and up to NLL:
\begin{subequations}\label{eq:sxgammaNLO}
\begin{align}
\gamma_{gg} &= \frac\as{2\pi} \[ \frac{2C_A}{N} - \frac{11C_A+2n_f}{6} + \Ord(N) \]
+ \frac{\as^2}{4\pi^2} \[ n_f \frac{6C_F-23C_A}{9 N} + \Ord(1) \] \\
\gamma_{gq} &= \frac\as{2\pi} \[ \frac{2C_F}{N} - \frac{3C_F}{2} + \Ord(N) \]
+ \frac{\as^2}{4\pi^2} \[ C_F \frac{9C_A-20n_f}{9 N} + \Ord(1) \] \\
\gamma_{qg} &= \frac\as{2\pi} \[ \frac{2n_f}{3} + \Ord(N) \]
+ \frac{\as^2}{4\pi^2} \[ \frac{20C_An_f}{9 N} + \Ord(1) \] \\
\gamma_{qq} &= \as\,\Ord(N)
+ \frac{\as^2}{4\pi^2} \[ \frac{20C_Fn_f}{9 N} + \Ord(1) \].
\end{align}
\end{subequations}
Using the small-$N$ behaviour at NLO, Eq.~\eqref{eq:sxgammaNLO}, and at NNLO, Ref.~\cite{vmv},
we can compute the small-$N$ behaviour in the more interesting case of the largest eigenvalue,
up to order $\as^3$ and up to NLL:
\begin{subequations}
\label{eq:gamma+expansion}
\begin{align}
\gamma_+^{(0)}(N) &= \frac{C_A}{\pi}\frac{1}{N} - \frac{11 C_A+2n_f(1-2C_F/C_A)}{12\pi} + \Ord(N) \\
\gamma_+^{(1)}(N) &= -\frac{n_f(23C_A-26C_F)}{36\pi^2}\frac{1}{N} + \Ord(1)\label{eq:gamma1expansion}\\
\gamma_+^{(2)}(N) &= \frac{C_A^3\(54\zeta_3+99\zeta_2-395\) + C_An_f(C_A-2C_F)\(18\zeta_2-71\)}{108 \pi ^3 N^2}
+\Ord(N^{-1}).
\end{align}
\end{subequations}
The eigenvalue $\gamma_-$, conversely, is not enhanced at small $N$;
indeed, in Ref.~\cite{abf599,abf799} it is shown that using appropriate
factorization schemes, specifically DIS and \MSbar, the eigenvalue $\gamma_-$
is free of singularities in $N=0$ to all orders in $\as$.

\subsubsection{Large-$x$ behaviour}

At large $x$ the off-diagonal splitting functions are regular: only the diagonal
components are enhanced at large $x$.
It can be proved \cite{Korchemsky:1988si} that the enhancement at large $x$
is given to all orders by
\beq
P_{ii} \simeq \frac{A_i}{\plus{1-x}} + B_i \,\delta(1-z)
\eeq
where $i$ is either $g$ or $q$.
The coefficients $A_i$ are related order by order in $\as$ by
the colour-charge relation
\beq\label{eq:Pii_color-charge}
A_q = \frac{C_F}{C_A} A_g.
\eeq
Up to $2$-loops, we have \cite{esw}
\beq
A_g = \frac{\as C_A}{\pi} \[1+\frac{\as}{2\pi}\(C_A \( \frac{67}{18}-\frac{\pi^2}{6}\) - \frac{5n_f}{9}\) + \Ord(\as^2) \]
\eeq
and
\begin{align}
B_g &= \as \beta_0 + \frac{\as^2}{4\pi^2}\[
C_A^2\(\frac{8}{3}+3\zeta_3\)
-\frac23 C_An_f -\frac12C_F n_f
\] + \Ord(\as^3) \\
B_q &= \frac{3\as C_F}{4\pi} + \frac{\as^2}{4\pi^2}\bigg[
C_A C_F \(\frac{17}{24} + \frac{11}{3}\zeta_2 - 3\zeta_3\)
+C_F^2\(\frac38 -3\zeta_2+6\zeta_3\)\\
&\qquad\qquad\qquad\qquad
-C_F n_f\(\frac{1}{12} + \frac23 \zeta_2\)
\bigg] + \Ord(\as^3).
\end{align}
In $N$-space, the diagonal anomalous dimensions in the large-$N$ limit are given by
\beq\label{eq:lxgammaNLO}
\gamma_{ii} = A_i\,\log\frac{1}{N} + \(B_i -A_i\gammae\) + \Ord\(N^{-1}\log N\),
\eeq
while the off-diagonal entries vanish at $N\to\infty$ as $N^{-1}\log^{2n-2}N$ at order $\as^n$.
The order-$\as^3$ behaviour can be found in Ref.~\cite{vmv}.

\section{Solving the GLAP equation}
\label{sec:glap_sol}

First of all, for solving the GLAP equations it is much easier
to work in $N$-Mellin space, where convolutions are simple products.
After the solution in Mellin space has been found, an inverse Mellin
transform can be performed to obtain a physical $x$-space PDF.

For convenience, we use here the variable $t=\log\frac{\mu^2}{\mu_0^2}$,
where $\mu_0$ is some arbitrary reference scale,
because the evolution depends on $\as(\mu^2)$ and it depends logarithmically on $\mu^2$;
we may also write for convenience $\as(t)$.
We can define the evolution function $U(t,t_0,N)$ by\footnote{We suppress for simplicity flavour
indexes, which are implicitly understood.}
\beq
f(N,\mu^2) = U(t,t_0,N-1)\, f(N,\mu_0^2)
\eeq
where $f(N,\mu_0^2)$ are some input PDFs given at a starting scale $\mu_0$ ($t_0=1$).
The evolution equation becomes an equation for $U$:
\beq\label{eq:GLAP_U_evol}
\frac{d}{dt} U(t,t_0,N) = \gamma\(\as(t),N\) U(t,t_0,N), \qquad U(t_0,t_0,N)=1.
\eeq
This equation have a formal solution in terms of the path-ordered integral
\begin{align}
U(t,t_0,N) &= \mathrm{P}\exp\int_{t_0}^t dt' \,\gamma\(\as(t'),N\) \\
&= 1 + \int_{t_0}^t dt'\, \gamma\(\as(t'),N\) + \int_{t_0}^t dt'\, \gamma\(\as(t'),N\)\int_{t_0}^{t'} dt''\, \gamma\(\as(t''),N\) + \ldots
\label{eq:glap_sol_path-ordered}
\end{align}
where the path-ordering symbol accounts for the non-commutativity of the $\gamma$'s.

\subsection{Explicit running coupling}
\label{sec:GLAPsol_running_coupling}

Since the $t$-dependence of the $\gamma$'s is contained in $\as(t)$,
one can recast the evolution equation in terms of an evolution in $\as$~\cite{NNPDF1}.
In practice, using the renormalization group equation
\beq
\frac{d}{dt} \as = \beta(\as)
\eeq
we have
\beq\label{eq:t_to_as_change_of_variable}
\frac{d}{dt} = \beta(\as) \frac{d}{d\as}
\eeq
and hence the evolution equation \eqref{eq:GLAP_U_evol} for $U$ becomes
\beq\label{eq:GLAP_U_evol_as}
\frac{d}{d\as} U(\as,\as^0,N) = \frac{\gamma(\as,N)}{\beta(\as)} U(\as,\as^0,N)
\eeq
with $\as^0=\as(t_0)$
(a notational change for the arguments of $U$ is implicitly understood).

From the practical point of view, using this $\as$-evolution equations
provides a simple way to insert correctly the appropriate order for the evolution of $\as$.
Indeed, if we work at N$^k$LO we need a $(k+1)$-loop $\beta$-function:
here, we simply have to put such $\beta$-function, while using the
$t$-evolution we would need an explicit solution%
\footnote{This is the case, for example, for the discretized path-ordering
(see Sect.~\ref{sec:discretized_PO}),
or in general for all solutions which involve numerical integration.
}
for the $\as$ evolution equation, which is harder to deal with.
Note that, of course, at the end of the computation we still need
the explicit evolution for $\as$, since we have to compute the results
at $\as(t_0)$ and $\as(t)$, but in this way the two problems are separated.

Then, everything we will say in the following can be translated without issues in this formalism,
by replacing $t$ with $\as$ and $\gamma$ with $\gamma/\beta(\as)$.

\subsection{The non-singlet case}

In this case all the equations are independent, the $\gamma$'s are ordinary functions
(not matrices) and the path ordering is irrelevant. The solution is then
\beq
U^{\rm NS}(t,t_0,N) = \exp\int_{t_0}^t dt'\, \gamma\(\as(t'),N\) 
\eeq
which is well defined.

Perturbatively, the complications of this solution amount to
computing $t$-integrals of functions (integer powers in this case) of $\as(t)$.
We can then recast the $t$-evolution equation in a evolution equation in $\as$
using the technique of Sect.~\ref{sec:GLAPsol_running_coupling}.
In fact, this amounts to change integration variable from $t$ to $\as$,
using the renormalization group equation to write the differential
\beq
dt = \frac{d\as}{\beta(\as)}.
\eeq
The first two powers of $\as$ (as needed for a NLO computation) give
(keeping the $\beta$-function up to NLO, i.e.\ $\beta_0$ and $\beta_1$)
\begin{align}
  \int_{t_0}^t dt'\, \as(t') &= \int_{\as(t_0)}^{\as(t)} \frac{d\as}{-\beta_0 \as - \beta_1 \as^2}
  = -\frac{1}{\beta_0} \[ \log\as -\log\(1+\frac{\beta_1}{\beta_0}\as\) \]_{\as(t_0)}^{\as(t)} \label{eq:as_integral}\\
  \int_{t_0}^t dt'\, \as^2(t') &= \int_{\as(t_0)}^{\as(t)} \frac{d\as}{-\beta_0 - \beta_1 \as}
  = -\frac{1}{\beta_1} \[ \log\(1+\frac{\beta_1}{\beta_0}\as\) \]_{\as(t_0)}^{\as(t)}. \label{eq:as2_integral}
\end{align}
At LO we have then (neglecting $\beta_1$)
\beq
U_{\rm LO}^{\rm NS}(t,t_0,N) = \(\frac{\as(t)}{\as(t_0)}\)^{-\gamma^{(0)}(N)/\beta_0}
\eeq
and at NLO (defining for simplicity $b_1 = \beta_1/\beta_0$)
\begin{align}
U_{\rm NLO}^{\rm NS}(t,t_0,N) &=
\exp\[ \(\frac{\gamma^{(0)}(N)}{\beta_0} -\frac{\gamma^{(1)}(N)}{\beta_1}\) \log\frac{1+b_1\as(t)}{1+b_1\as(t_0)} \]
\,\(\frac{\as(t)}{\as(t_0)}\)^{-\gamma^{(0)}(N)/\beta_0}\\
&\simeq \[ 1 + \(\frac{\gamma^{(0)}(N)}{\beta_0} -\frac{\gamma^{(1)}(N)}{\beta_1}\) b_1 \big[\as(t)-\as(t_0)\big] \]
\,\(\frac{\as(t)}{\as(t_0)}\)^{-\gamma^{(0)}(N)/\beta_0}\label{eq:glap_nonsinglet_sol_NLO}
\end{align}
where in the last line we have expanded the exponential in front up to NLO
(order $\as$), to be consistent with the accuracy required.

\subsection{The singlet case}

In this case there are two coupled equations, the evolution matrix $\Gamma$
is a $2$-dimensional matrix, and the path-ordered solution is not of simple usage:
in practice, only a discretized form of the path-ordering is viable.
The natural way to solve the equations would be the diagonalization of the system,
in order to have two independent equations for two linear combinations
of gluon ($f_g$) and singlet-quark ($f_S$) densities;
however, even if in some cases this is simple, in general this solution
can be as complicated as solving the original non-diagonal equations.
We discuss these methods below.

\subsubsection{Discretized path-ordering}
\label{sec:discretized_PO}

To obtain a numerical realization of the path-ordered solution, Eq.~\eqref{eq:glap_sol_path-ordered},
we start by separating the evolution from $t_0$ to $t$ into the product of $n+1$ subsequent evolution
from $t_k$ to $t_k+1$, with $t_{k+1}>t_k$,
\begin{align}
U(t,t_0,N) &= U(t,t_n,N)\, U(t_n,t_{n-1},N) \cdots U(t_2,t_1,N)\, U(t_1,t_0,N)\nonumber\\
&= \prod_{k=n}^0  U(t_{k+1},t_k,N),\qquad  (t_{n+1}\equiv t)
\end{align}
where the order of the products is such that larger $t_k$ are to the left.
A possible sequence of $t_k$ is a linear sequence
\beq\label{eq:path-ordering_linear_sequence}
t_k = t_0 + k\, \Delta t,\qquad
\Delta t = \frac{t-t_0}{n+1},
\eeq
but different (possibly optimized) sequences are allowed.

Next we consider the single-step evolution
\beq\label{eq:single-step_evolution}
U(t_{k+1},t_k,N) = \mathrm{P}\exp\int_{t_k}^{t_{k+1}} dt' \,\Gamma\(\as(t'),N\)
\simeq 1+\int_{t_k}^{t_{k+1}} dt' \,\Gamma\(\as(t'),N\),
\eeq
where the second equality holds if the integral is small, i.e.\ when
the separation between $t_k$ and $t_{k+1}$ is small enough.%
\footnote{Even if ``small enough'' is not a mathematical statement, the important thing
is that, provided $\Gamma$ is a smooth function, it is always possible to find a finite
separation $t_{k+1}-t_k$ such that the integral is as small (compared to $1$) as some
precision we would like to reach in our approximate computation.}
In this regime we can also approximate the integral with
\begin{align}
\int_{t_k}^{t_{k+1}} dt' \,\Gamma\(\as(t'),N\) &\simeq \Gamma\(\as\(\frac{t_{k+1}+t_k}{2}\),N\) (t_{k+1}-t_k)\nonumber\\
&= \Gamma\(\as(t_k+\Delta t/2),N\) \Delta t,
\end{align}
where the second equality holds if the sequence of Eq.~\eqref{eq:path-ordering_linear_sequence}
is adopted.
In this particular case, we have finally
\beq\label{eq:discretized_path-ordering1}
U(t,t_0,N) \simeq \prod_{k=n}^0 \Big[1 + \Gamma\(\as(t_k+\Delta t/2),N\) \Delta t\Big],
\eeq
or, at the same level of accuracy,
\beq\label{eq:discretized_path-ordering2}
U(t,t_0,N) \simeq \prod_{k=n}^0  \exp\Big[\Gamma\(\as(t_k+\Delta t/2),N\) \Delta t\Big],
\eeq
which both provide good numerical implementations of the path-ordered solution.
Note, by the way, that the two expressions \eqref{eq:discretized_path-ordering1}
and \eqref{eq:discretized_path-ordering2} provide a way to estimate
the numerical error in the discretization procedure, which should be of the same order
of the difference between the two results.

We have checked numerically that the two expression converge to the same value,
and in the case in which the evolution can be solved exactly, that the asymptotic value
is the exact result. It turns out that the exponential expression, Eq.~\eqref{eq:discretized_path-ordering2},
converges more rapidly, i.e.\ it gives more accurate results than Eq.~\eqref{eq:discretized_path-ordering1}
for the same value of $n$.
Indeed, looking more carefully to the single-step evolution Eq.~\eqref{eq:single-step_evolution},
we see that the first neglected correction can be approximated as
\begin{multline}
\int_{t_0}^t dt'\, \Gamma\(\as(t'),N\)\int_{t_0}^{t'} dt''\, \Gamma\(\as(t''),N\)
\\= \Gamma\(\as(t_k+\Delta t/2),N\)\Gamma\(\as(t_k+\Delta t/4),N\) \frac{(\Delta t)^2}{2},
\end{multline}
where the two $\Gamma$'s are computed at different values of $t$.
Actually, the point at which the integrand is computed is a matter of choice;
we have adopted here a central choice, which is appropriate to reproduce
the correct factor $1/2$ for this term (however, higher terms do not grow correctly:
the next is $1/8$ instead of $1/3!$).
If instead we had chosen the right bound ($t_k+\Delta t$) the argument would be the same,
but the factor at each order would be $1$.
For a left bound choice ($t_k$), this and higher terms are zero.
We conclude that, if we compute all the $\Gamma$'s in the nested integrals
at the middle point, the discretized single-step path-ordering
is better approximated by the full exponential, confirming that
the solution Eq.~\eqref{eq:discretized_path-ordering2} is more accurate.
We have also checked that the central choice of Eq.~\eqref{eq:discretized_path-ordering2}
provides the faster convergence.

We suggest then to use Eq.~\eqref{eq:discretized_path-ordering2}; practically,
to compute the exponential in each step we need to diagonalize $\Gamma$ at the
given value of $t$ for that step. It is useful here to use the projectors
defined in Sect.~\ref{sec:glap:projectors}, to write
\beq
\exp\[\Gamma\, \Delta t\] = \projM_+ e^{\gamma_+ \Delta t} + \projM_- e^{\gamma_- \Delta t}
\eeq
where everything is computed at $t_k+\Delta t/2$ for the $k$-th step.

As a final comment, if we had used the $\as$ evolution, Eq.~\eqref{eq:GLAP_U_evol_as},
we would have found
\beq
U(\as,\as^0,N) \simeq \prod_{k=n}^0  \exp\[\frac{\Gamma\(\alpha_k+\Delta \alpha/2,N\)}{\beta\(\alpha_k+\Delta \alpha/2\)} \Delta \alpha\],
\eeq
with
\beq
\alpha_k=\as^0+k\,\Delta\alpha,\qquad \Delta\alpha = \frac{\as-\as^0}{n+1}.
\eeq
This form is easier to use because we don't need to know the
explicit solution for the running coupling, but just the $\beta$-function
up to the desired order.

\subsubsection{Diagonalization of the system of singlet equations}
\label{sec:GLAP_solve_diagonalization}
Introducing a short notation for the vector of singlet PDFs,
\beq
f = \dvec{f_g}{f_S},
\eeq
we can write the singlet system, Eq.~\eqref{eq:glap_eq_singlet}, in matrix notation
\beq
\frac{d}{dt} f = \Gamma f
\eeq
where for simplicity we have suppressed the explicit dependencies on the arguments
and we have used the matrix $\Gamma$ defined in Eq.~\eqref{eq:Gamma}.

In general, the $\as$ dependence of $\Gamma$ is not trivial,
and a matrix $R$ which diagonalizes $\Gamma$ would be $\as$-dependent,
and hence it does not commute with the $t$-derivative.
Hence the right way to proceed is to introduce some ($t$-dependent) matrix $R$
and to transform the PDF vector in
\beq
\hat f = R f;
\eeq
now, multiplying the evolution equation by $R$ on the left of both sides
and expressing everything in terms of $\hat f$, we get
\beq\label{eq:singlet_eq_diag_t}
\frac{d}{dt} \hat f = \(R \Gamma R^{-1} + \frac{dR}{dt}R^{-1} \) \hat f.
\eeq
If the matrix (we introduce for simplicity a dot for the $t$-derivative)
\beq\label{eq:Gamma_diag_t}
\tilde \Gamma = R \Gamma R^{-1} + \dot RR^{-1}
\eeq
is diagonal the system \eqref{eq:singlet_eq_diag_t} splits into two independent equations,
and each can be solved as in the non-singlet case (possibly introducing the evolution function $U$).
So now the problem is to find a matrix $R$ such that $\tilde\Gamma$ is diagonal,
that is to solve the matrix equation
\beq
\dot R = \tilde \Gamma R - R \Gamma
\eeq
where also the two (diagonal) entries of $\tilde\Gamma$ are unknowns.
Solving this matrix equation may be as complicated as solving the original system
Eq.~\eqref{eq:glap_eq_singlet}, so for practical purposes
this way may be too complicated.

Let's now consider the LO case. At LO the evolution matrix is given by
\beq
\Gamma = \as \Gamma^{(0)}
\eeq
and hence, since the $\as$ dependence factorizes, we can diagonalize it
by a $\as$-independent ($t$-independent) matrix;
we can the choose $R$ to be this diagonalizing matrix, and since $\dot R=0$ we get
\beq
\frac{d}{dt} \hat f = \hat \Gamma \hat f
\eeq
where
\beq
\hat\Gamma = R\Gamma R^{-1}
\eeq
with $\hat\Gamma$ diagonal.
An explicit form of $R$ is given in Eq.~\eqref{eq:rotationR}.
The solution can be easily written by making use of the projectors
introduced in Sect.~\ref{sec:glap:projectors}; the evolutor is
\beq\label{eq:glap_singlet_sol_LO}
U_{\rm LO}(t,t_0,N) = \projM_+(N) \(\frac{\as(t)}{\as(t_0)}\)^{-\gamma_+(N)/\beta_0}
+ \projM_-(N) \(\frac{\as(t)}{\as(t_0)}\)^{-\gamma_-(N)/\beta_0}
\eeq
where we have used the integral \eqref{eq:as_integral}.

Now let's move from LO to NLO. In this case $\Gamma$ is
\beq
\Gamma = \as \Gamma^{(0)} + \as^2 \Gamma^{(1)}
\eeq
and the $\as$-dependence now is non-trivial. Unless $\Gamma^{(0)}$ and $\Gamma^{(1)}$
are diagonalized by the same matrix (and this is not the case) we cannot simply
diagonalize $\Gamma$ but we are in the general case of Eq.~\eqref{eq:singlet_eq_diag_t}.
The NLO evolution can alternatively be solved perturbatively, diagonalizing at LO
and perturbating around the LO solution (see below).

Consider now the LL case. The matrix $\Gamma$ in this case is given by
\beq
\Gamma=\gamma_s\(\frac{\as}{N}\)\Gamma_s
,\qquad
\Gamma_s=
\(
\begin{array}{cc}
1 & C_F/C_A\\
0 & 0
\end{array}
\)
\eeq
where again the $\as$-dependence has factored out, as in the LO case.
Then also in this case we can simply diagonalize the matrix $\Gamma$;
the matrix $R$ which implements this diagonalization is the same matrix
that diagonalizes the LO matrix, computed in $N=0$.

As a final comment, we note that using instead the $\as$ evolution, Eq.~\eqref{eq:GLAP_U_evol_as},
in the LO and LL cases the $\as$ dependence is still factorized,
since the evolution matrix is, respectively,
\beq
\frac{\as \Gamma^{(0)}}{\beta(\as)} = -\frac{\Gamma^{(0)}}{\beta_0 \as},\qquad\qquad
\frac{\gamma_s(\as/N)\, \Gamma_s}{\beta(\as)} = -\frac{\gamma_s(\as/N)\,\Gamma_s}{\beta_0 \as^2},
\eeq
where we have substituted the $1$-loop $\beta$-function, as appropriate.
Note that, at NLO, the evolution matrix is instead an infinite series in $\as$.

\subsubsection{Hybrid perturbative solution}
Now assume the decomposition of $\Gamma$ into the sum of two pieces
\beq
\Gamma = \Gamma_0 + \Gamma'
\eeq
in such a way that we know the solution of the evolution equation for just $\Gamma_0$:
\beq
U_0(t,t_0,N) = \mathrm{P} \exp \int_{t_0}^t dt'\, \Gamma_0(\as(t'),N).
\eeq
This is the case, for example, when the $\as$ dependence is the same for
all the entries of $\Gamma_0$, like in the previously discussed LO and LL cases.

With this setup, we can solve the evolution equation like
we do with the Schr\"odinger equation in the interaction picture.
Let's write the full solution $U$ as
\beq
U(t,t_0,N) = U_0(t,t_0,N)\, U_I(t,t_0,N);
\eeq
putting this into the evolution equation we get an equation for $U_I$
\beq\label{eq:glap_singlet_eq_interaction_picture}
\frac{d}{dt}U_I(t,t_0,N) = \tilde\Gamma'(t,N)\, U_I(t,t_0,N)
\eeq
with
\beq
\tilde\Gamma'(t,N) =  U_0(t,t_0,N)^{-1}\, \Gamma'(\as(t),N) \,U_0(t,t_0,N).
\eeq
Once we have this expression we can either use the discretized path-ordering
solution or an approximation (if $\Gamma'$ can be considered as a perturbation of $\Gamma_0$).

This procedure can be used to find, for example, the NLO solution.
Indeed, one could identify
\beq
\Gamma_0 = \as \Gamma^{(0)}
,\qquad
\Gamma' = \as^2 \Gamma^{(1)}
\eeq
and use an approximate solution for the ``perturbation'' $\Gamma^{(1)}$.
However, we have to take care of the running of $\as$: indeed, the LO solution
would be typically computed with a $1$-loop $\beta$-function, while the
$\as$ appearing in the NLO equation should be at $2$-loops.
The easiest way to tackle with this issue is to use the $\as$ evolution,
Eq.~\eqref{eq:GLAP_U_evol_as}, and write the evolution kernel as
\beq
\frac{\as \Gamma^{(0)} + \as^2 \Gamma^{(1)}}{-\beta_0\as^2 - \beta_1\as^3}
= \frac{\Gamma^{(0)}}{-\beta_0\as} + 
\frac{\Gamma^{(1)} - b_1 \Gamma^{(0)}}{-\beta_0 - \beta_1\as},
\qquad b_1 = \frac{\beta_1}{\beta_0};
\eeq
now, we can identify the two contributions in the right-hand-side as
the pure LO evolution kernel and the perturbation to it.
We could now eventually come back to the $t$-evolution: we would then have the identifications
\beq
\Gamma_0 = \as^{\rm LO}\, \Gamma^{(0)}
,\qquad
\Gamma' = \(\as^{\rm NLO}\)^2 \[\Gamma^{(1)} - b_1 \Gamma^{(0)}\]
\eeq
where we have written explicitly the order at which $\as$ must be computed.
The pure LO solution is then given by Eq.~\eqref{eq:glap_singlet_sol_LO},
\beq
U_0(t,t_0,N) = U_{\rm LO}(t,t_0,N),
\eeq
and we could take (to NLO accuracy) the first order in the expansion in the path-ordered solution
of Eq.~\eqref{eq:glap_singlet_eq_interaction_picture}
\beq
U_I(t,t_0,N) = 1+ \int_{t_0}^t dt'\,\,U_{\rm LO}^{-1}(t',t_0,N)\, \Gamma'\(\as^{\rm NLO}(t),N\) \,U_{\rm LO}(t',t_0,N) + \ldots
\eeq
leading to a solution which is equivalent to the non-singlet one, Eq.~\eqref{eq:glap_nonsinglet_sol_NLO}.
We don't show here the full result, since we are not interested in it ---
we will use the discretized path-ordering solution for applications;
for a complete solution, equivalent to this up to higher orders in $\as$, see Ref.~\cite{NNPDF1}.

\chapter{Soft-gluon resummation}
\label{chap:soft-gluons}

\minitoc

\noindent
This Chapter is devoted to a review of some basics on soft-gluon (or threshold or large-$x$)
resummation and in particular to the problem of extracting finite results from the resummation formulae.
Indeed, we will see that soft-gluon resummation is performed in $N$-space,
where a closed resummed expression for the inclusive cross-section can be found explicitly;
however, the inverse Mellin transform does not exist, and this is related to the
divergence of the perturbative series.
We will describe then two prescription to deal with such divergence.
After that, we will extend the formalism from the inclusive cross-sections to the more interesting case of
rapidity distributions.
However, we won't cover the as interesting resummation of transverse momentum distributions,
for which we refer the Reader to Refs.~\cite{DDT,ParisiPetronzio,Curci:1979sk,CSS,bfr1}.

\section{Inclusive cross-sections resummation}
\label{sec:thr_resumm}

\begin{figure}[ht]
  \centering
  \includegraphics[width=0.8\textwidth]{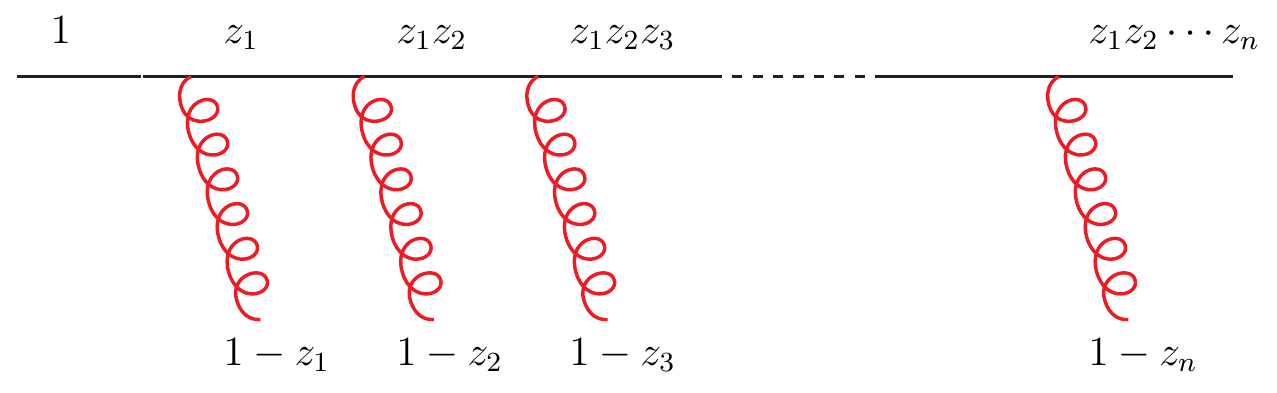}
  \caption{Emission of $n$ gluons from a quark parton line. The energy fractions of the gluons
    refer to the energy of the emitting quark, while the energy fractions in the quark line
    refer always to its initial energy.}
  \label{fig:soft-gluons}
\end{figure}

A generic partonic coefficient function $C(z,\as)$ is kinematically enhanced
when $z\to1$, because of gluon emissions.
Consider for simplicity the case of a quark parton line which emits $n$ gluons,
as in Fig.~\ref{fig:soft-gluons}.
When a gluon is emitted, it carries an energy fraction $1-z_i$ of the quark
from which it is emitted; consequently, the energy of the quark gradually decreases
becoming at the end a fraction $z=z_1 z_2\cdots z_n$ of its initial energy
(this $z$ is the argument of the coefficient function).
In the coefficient function, for each gluon emission a term enhanced when
the corresponding $z_i$ approaches one (the $i$-th gluon becomes \emph{soft}) appears.
When integrating over the emitted gluon phase spaces, these enhancements
convert into a sequence of terms
\beq
\frac{\log^k(1-z)}{1-z},\qquad
0\leq k\leq an-1
\eeq
where $a$ is a process-dependent parameter: $a=1$ for DIS and
$a=2$ for Drell-Yan and Higgs production.%
\footnote{In the Higgs production case, the parton line which emits gluons is a
gluon line, since the Higgs is predominantly produced via gluon fusion.}
For the sake of exposition, we consider just the case $a=2$.
This choice is motivated phenomenologically: the impact of soft-gluon resummation
in DIS is less relevant than in Drell-Yan or Higgs production.

Let's introduce some notations, which will be used extensively in Chap.~\ref{chap:pheno}.
The variable $\tau$ is defined as
\beq
\tau = \frac{M^2}{s},
\eeq
where $M$ is the invariant mass of the target final state (the lepton pair or the Higgs boson) and
$\sqrt{s}$ is the hadronic center-of-mass energy. The variable $z$ (the argument of the coefficient function)
is the analog of $\tau$ at parton level, and then it can be written as
\beq
z = \frac{M^2}{\hat s},
\eeq
where $\sqrt{\hat s}$ is the partonic center-of-mass energy. This explains why soft-gluon
resummation is also called threshold resummation: when $z$ approaches $1$, the partonic energy
is just sufficient to produce the final state, i.e.\ the process is close to its threshold
(same reasoning holds at hadron level).
The inclusive cross-section (differential only in $M$) can be written in the parton model framework as
\beq\label{eq:inclus_cs_conv}
\frac{1}{\tau}\frac{d\sigma}{dM^2} = \sum_{ij} \int_\tau^1 \frac{dz}{z}\,\Lum_{ij}\(\frac{\tau}{z}\) \,C_{ij}\(z,\as(M^2)\),
\eeq
where we have defined the so called \emph{parton luminosity}
\beq\label{eq:luminosity}
\Lum_{ij}(x) = c_{ij} \int_x^1 \frac{dy}{y}\,f_i^{(1)}\(\frac{x}{y}\) \,f_j^{(2)}(y),
\eeq
where the two PDFs refer to the two incoming hadrons and $i,j$ are flavour indexes.
We are omitting renormalization and factorization scale dependence, which will be restored
for the sake of phenomenology in Chap.~\ref{chap:pheno}.
In the rest of this Chapter, we will omit the flavour details (sum over flavours, indexes),
concentrating implicitly our attention to those channels which dominate at large $z$
($q\bar q$ for DY, $gg$ for Higgs).
Moreover, for convenience, we normalize the coefficient function of such dominant channel
in such a way that the LO is simply
\beq
C^{(0)}(z) = \delta(1-z) \quad \leftrightarrow \quad C^{(0)}(N) = 1;
\eeq
hence, any numerical factor appearing in the partonic cross section at LO
must be included in the definition of the coefficients $c_{ij}$
in front of the luminosity, Eq.~\eqref{eq:luminosity}.

In a fixed order computation, at LO there are no emissions, by definition;
at NLO the contribution from a single gluon emission produces (among others terms not enhanced at large $z$)
\beq
\frac{\log(1-z)}{1-z},\qquad
\frac{1}{1-z}.
\eeq
As one sees immediately, when integrating the coefficient function with the PDFs
in the convolution \eqref{eq:inclus_cs_conv} such terms produce a non-integrable singularity
in $z=1$. As already discussed in Sect.~\ref{sec:PM_radiative_corr}, such singularities
are regularized by the IR divergence of loop diagrams, giving rise to a plus distribution
(see definitions in App.~\ref{sec:plus_distribution}):
\beq
\plus{\frac{\log(1-z)}{1-z}},\qquad
\plus{\frac{1}{1-z}}.
\eeq
This mechanism takes place also at higher orders (see \cite{KLN1, KLN2}), producing at the end
a tower of terms
\beq\label{eq:log(1-z)_tower}
\as^n \plus{\frac{\log^k(1-z)}{1-z}},\qquad
0\leq k\leq 2n-1.
\eeq
It is now evident why resummation is needed: when $z$ is equal or larger than
a value $\bar z$ satisfying roughly 
\beq
\as \log^2(1-\bar z) \sim 1,
\eeq
all terms in the perturbative series are of the same order, and any
finite order truncation would be meaningless.
Being $z$ an integration variable which extends up to $z=1$,
this condition is always satisfied in a region of the
convolution integral Eq.~\eqref{eq:inclus_cs_conv},
and in that region the partonic coefficient function needs to be resummed.

\subsection{Resummation in Mellin space}

The resummation task has been achieved long ago~\cite{Sterman:1986aj,Catani:1989ne,fr}.
The details may be cumbersome because of two complications:
the running of the QCD coupling and the non-abelian nature of QCD.
It is beyond the purpose of this thesis to investigate the resummation mechanism in such detail:
we will therefore limit ourselves to a simplified exposition~\cite{Catani}
which nevertheless emphasizes the key ingredients: \emph{factorization} and \emph{exponentiation}.

In the actual computation of the contribution from $n$ gluon emissions,
one has first to compute the matrix element and then to integrate over the
phase space of the emitted gluons.
Concerning the matrix elements, the calculation can be performed in the \emph{eikonal approximation}
(see for example~\cite{Weinberg1}), which reproduces the correct soft limit.
In this limit, it can be proved that such matrix element $\M_n$ factorizes as
\beq
\M_n(z_1,\ldots,z_n) \overset{\rm soft}{\simeq} \frac{1}{n!} \prod_{i=1}^n \M_1(z_i)
\eeq
where $\M_1$ is the matrix element for the single emission;
the proof is rather easy in QED, while in QCD complications arise due to the
non-abelian couplings between gluons.
However, the phase space in $z$-space is not factorized, and is proportional to
\beq\label{eq:gluon_energy_ps}
dz_1\cdots dz_n \, \delta(z-z_1\cdots z_n).
\eeq
Nevertheless, after taking a Mellin transform, also the phase space
factorizes,
\beq
\int_0^1 \frac{dz}{z}\, z^N\, \delta(z-z_1\cdots z_n) 
= z_1^{N-1} \cdots z_n^{N-1},
\eeq
producing the factors $z_i^{N-1}$ which give, upon integration, the Mellin transforms
with respect to each individual $z_i$ variable.
In $N$-space, the threshold region $z\sim 1$ corresponds to the region
of large $N$, and in particular the tower of logarithms in Eq.~\eqref{eq:log(1-z)_tower}
converts into the tower (see App.~\ref{sec:Mellin_log})
\beq\label{eq:logN_tower}
\as^n \log^k\frac1N,\qquad
0\leq k\leq 2n.
\eeq
Hence, in $N$-space, the coefficient function at order $\as^n$ is given in the soft
limit\footnote{Including also the virtual contributions which regulate the divergence in $z=1$.} by
\beq
C^{(n)}(N) \overset{\rm soft}{\simeq} \frac{1}{n!} \[ C^{(1)}_{\rm soft}(N)\]^n
\eeq
where $C^{(1)}_{\rm soft}(N)$ is the Mellin transform of soft terms in the order $\as$ coefficient.
Considering only the leading soft term, it becomes
\beq\label{eq:C1soft}
C^{(1)}_{\rm soft}(N) = \int_0^1 dz\, z^{N-1} \,4A_1 \plus{\frac{\log(1-z)}{1-z}}
\overset{N\gg 1}{\simeq} 2A_1\, \log^2 \frac1N
\eeq
where $A_1=C_F/\pi$ for the DY case and $A_1=C_A/\pi$ for the Higgs case.

At this stage it is clear what is going to happen: the soft terms in the perturbative series
of the coefficient function can be resummed explicitly, and the results
exponentiates:
\beq\label{eq:exponentiation}
C^{\rm res}(N,\as) = \sum_{n=0}^\infty \as^n \[C^{(n)}(N)\]_{\rm soft} = \exp \[\as\, C^{(1)}_{\rm soft}(N) \].
\eeq
This result is valid only at leading-logarithmic (LL) accuracy:
only the highest power $k=2n$ in the tower of logs in Eq.~\eqref{eq:logN_tower} is resummed.
Moreover, it does not take into account effects due to the running of $\as$:
hence, in this form, it would be valid in QED,
and it would be called \emph{Sudakov} resummation~\cite{Sudakov}.
What needs to be emphasized of this simple argument is that
resummation is based on the factorization of multi-gluon emissions,
and that such factorization can take place only in Mellin space,
because the relevant phase space only factorizes in $N$-space, but not in $z$-space.

To take into account the running coupling effects, we recall that in the gluon phase space
integration there is, besides the energy fractions in Eq.~\eqref{eq:gluon_energy_ps}, the transverse
momenta of the gluons: such momenta fix the energy scale of the gluon emissions,
and then it is the proper energy scale at which $\as$ should be computed.
Then, the exponent of Eq.~\eqref{eq:exponentiation} becomes
\beq\label{eq:C1_soft_running}
\as C^{(1)}_{\rm soft}(N) \to \int_0^1 dz\,z^{N-1} \,2A_1\,\plus{\frac{1}{1-z}\int_{M^2}^{(1-z)^2 M^2}\frac{dk^2}{k^2}\,\as(k^2) };
\eeq
one can immediately verify that, if the coupling is kept fixed, the previous result
is obtained again.
Note that this expression is ill-defined: indeed, the Mellin transform
forces $z$ to take each value from $0$ to $1$, and hence $\as$ at some point has to be evaluated
in the non-perturbative region, hitting the Landau pole (see App.~\ref{chap:QCD_running_coupling}).
In actual computations, this problem is avoided by expanding in powers of $\as(M^2)$ the integrand
and computing the Mellin transform term-by-term (which is well defined), then taking the large-$N$
approximation and resumming the $\as(M^2)$ series: such procedure leads to a finite result.
In more details, the $k^2$ integral gives (using the $1$-loop $\beta$-function)
\beq
\int_{M^2}^{(1-z)^2 M^2}\frac{dk^2}{k^2}\,\as(k^2) = \frac{1}{\beta_0}\log\Big(1+\ab\log(1-z)\Big)
\eeq
having defined for future convenience the variable
\beq\label{eq:ab_def}
\ab = 2\,\beta_0\,\as
\eeq
(when written without argument, we mean $\as=\as(M^2)$).
Manifestly, the Mellin transform is divergent, because the $\log$ has a branch-cut
when $\ab\log(1-z)<-1$, which is always in the integration range.
The Mellin transform can be computed term-by-term as
\begin{align}
\Mell\[\plus{\frac{\log\[1+\ab\log(1-z)\]}{1-z}}\] &= -\sum_{k=1}^\infty \frac{(-\ab)^k}{k}\,\Mell\[\plus{\frac{\log^k(1-z)}{1-z}}\]\nonumber\\
&\simeq -\sum_{k=1}^\infty \frac{(-\ab)^k}{k(k+1)}\,\log^{k+1}\frac1N \nonumber\\
&= \(\frac1{\ab} + \log\frac1N \)\log\(1+\ab\log\frac1N\) - \log\frac1N \nonumber
\end{align}
from which we get
\beq\label{eq:exponentiation_rc}
C^{\rm res}(N,\as) = \exp \left\{\frac{2 A_1}{\beta_0\ab} \[ \(1+ \ab\log\frac1N \)\log\(1+\ab\log\frac1N\) -\ab\log\frac1N\]\right\}.
\eeq
In the second equality, we have used Eq.~\eqref{eq:Mellin_log1} in the large-$N$ limit
and kept only the highest power of $\log\frac1N$
for each term in the series, as appropriate for a LL computation; note, however, that
it is exactly this approximation which allows to avoid the problem of the Landau pole.
Nevertheless, the Landau pole problem has not disappeared, and we will come back later on it.

Eq.~\eqref{eq:exponentiation_rc}, even if it contains the proper running coupling effect, still reproduces
only the LL terms. The more general expression is~\cite{Catani:1989ne,Sterman:1986aj,Contopanagos:1996nh,fr,bfr2}
\beq\label{eq:Cres}
C^{\rm res}(N,\as) = g_0(\as) \,\exp \Sud\(\ab \log\frac{1}{N}, \ab\)
\eeq
where $g_0(\as)=1+\Ord(\as)$ collects all the constant terms, and the exponent,
called for similarity Sudakov exponent (or Sudakov form factor), has a logarithmic expansion
\beq\label{eq:S}
\Sud(\lambda, \ab) = \frac{1}{\ab}\, g_1(\lambda) + g_2(\lambda) + \ab \, g_3(\lambda) + \ab^2 \, g_4(\lambda) + \dots.
\eeq
At the next$^k$-to-leading logarithmic (N$^k$LL) level functions up to
$g_{k+1}$ must be included, and $g_0$ must be computed up to order $\as^k$.
Then, the expansion of the resummed coefficient function
Eq.~\eqref{eq:Cres} in powers of $\as$ gives the logarithmically enhanced
contributions to the fixed-order coefficient functions $C(N,\as)$
up to the same order, with, at the N$^p$LL level, 
all terms of order $\log^k\frac{1}{N}$  with  $2(n-p)\leq k\leq 2n$ correctly
predicted; note that at each order also powers of $\log\frac1N$ less
than $2(n-p)$ are generated, but their coefficients are not exact
and will change increasing the logarithmic approximation order $p$.
A summary for the first few orders is given in Tab.~\ref{tab:res_orders}.
\begin{table}[t]
  \centering
  \begin{tabular}[c]{c c c r}
    log approx. & $g_i$ up to & $g_0$ up to order & accuracy: $\as^n\log^k\frac1N$\\
    \midrule
    LL & $i=1$ & $(\as)^0$ & $k=2n$ \\
    NLL & $i=2$ & $(\as)^1$ & $2n-2\leq k\leq 2n$ \\
    NNLL & $i=3$ & $(\as)^2$ & $2n-4\leq k\leq 2n$
  \end{tabular}
  \caption{Orders of logarithmic approximations and accuracy of the predicted logarithms.}
  \label{tab:res_orders}
\end{table}

The inclusion up to the relevant order of the function $g_0$  is
necessary, despite the fact that $g_0$ contains only constants,
because of its interference with the expansion of the exponentiated 
logarithmically enhanced functions $g_i$ with $i\geq 1$.
For example, at NLL we expect, according to Tab.~\ref{tab:res_orders},
to get the correct coefficients for the three leading powers of logarithms
at each order: in particular, at order $\as$ all logarithmic terms (including a constant) should
be correct, and at order $\as^2$ the coefficients for
the powers $k=4,3,2$ of $\log\frac1N$ are expected to be correctly predicted.
By writing
\begin{align}
g_0(\as) &= 1 + \sum_{j=1}^\infty g_{0j}\,\as^j,\\
g_k(\lambda) &= (2\beta_0)^{2-k}\sum_{j=1}^\infty g_{kj}\(\frac{\lambda}{2\beta_0}\)^j ,\qquad g_{11}=0,
\end{align}
and expanding $C^{\rm res}(N,\as)$ in powers of $\as$, we get indeed ($L=\log\frac1N$)
\begin{align}
C^{\rm res}(N,\as) &= 1+ \as\[g_{12}L^2 + g_{21}L +g_{01}\]\nonumber\\
&\qquad+\as^2\bigg[\frac{g_{12}^2}{2}L^4 + \(g_{12}g_{21}+g_{13}\)L^3
+\(\frac{g_{21}^2}{2}+g_{22}+g_{12}g_{01}\)L^2 + \Ord\(L\)\bigg]\nonumber\\
&\qquad
+\Ord\(\as^3\).\label{eq:NLL_res_expansion}
\end{align}
The order $\as$, as well as the coefficients of the terms $L^4, L^3, L^2$ at order $\as^2$,
depend only on $g_1, g_2$ and $g_{01}$, i.e.\ only NLL contributions, as expected.

Details on the functions $g_i$ ($i\geq1$) will be given in App.~\ref{sec:app-resumm}.
What we want to emphasize here is that such functions depend on few coefficients:
$g_1$ depends only on the coefficient $A_1$ introduced in Eq.~\eqref{eq:C1soft},
$g_2$ on a couple of other coefficients, and so on.
One category (to which $A_1$ belongs) is fully determined by the GLAP anomalous dimensions:
for these coefficients, to compute the N$^p$LL function $g_{p+1}$, the $(p+1)$-loops
(N$^p$LO) anomalous dimension is needed.
The other (process dependent) coefficients are then fixed by matching the expansion
of $C^{\rm res}(N,\as)$ in powers of $\as$ up to order $\as^p$ and comparing with a
fixed order computation. A summary of these needs for the first few orders is given
in Tab.~\ref{tab:res_needs}.
\begin{table}[t]
  \centering
  \begin{tabular}[c]{c c c}
    & \multicolumn{2}{c}{needs}\\
    \cmidrule{2-3}
    log approx. & GLAP & fixed-order\\
    \midrule
    LL   & $1$-loop & -- \\
    NLL  & $2$-loop & NLO \\
    NNLL & $3$-loop & NNLO
  \end{tabular}
  \caption{Orders of logarithmic approximations and the needed fixed-order accuracy.}
  \label{tab:res_needs}
\end{table}

Matched resummed coefficient functions are obtained by combining
the resummed result Eq.~\eqref{eq:Cres} with the fixed-order expansion
in power of $\as$, and subtracting double-counting terms, i.e.\ the
expansion of $C^{\rm res}(N,\as(M^2))$  in powers of $\as(M^2)$ up to the
same order:
\beq\label{eq:match}
C^{\text{N$^p$LO}}_{\text{N$^k$LL}}(N,\as) =
\sum_{j=0}^{p}\as^j\, C^{(j)}(N) 
+ C^{\rm res}_{\text{N$^k$LL}}(N,\as)
- \sum_{j=0}^{p}\frac{\as^j}{j!}
\[\frac{d^jC^{\rm res}_{\text{N$^k$LL}}(N,\as)}{d\as^j}\]_{\as=0}.
\eeq
This expression will be used in Chap~\ref{chap:pheno} for phenomenology.

As a final remark, we mention that recently resummation has also been performed using
soft-collinear effective theory (SCET)
techniques both in $N$-space~\cite{scetN} and in $z$-space~\cite{Becher:2006nr,bnx};
in the latter approach the soft
scale whose logarithms are resummed is not the partonic scale
$M^2(1-z)$ but rather a soft scale $\mu_s$, independent of
the parton momentum fraction $z$, but related to 
the hadronic scale $M^2(1-\tau)$.
As a consequence, the hard coefficient function depends on $\tau$
not only through the convolution variable, but also directly through
the soft scales: therefore, the
resummed result can no longer be factorized by Mellin transformation
into the product of a parton density and a hard
coefficient.
For this reason, a direct comparison between the result
obtained through the SCET approach and that based on standard
factorization Eq.~\eqref{eq:inclus_cs_conv} is not possible at the parton level~\cite{bfgr}.
A phenomenological comparison is possible~\cite{bnx}, but it
requires either assuming a specific form of
the parton distribution functions, or switching to the $N$ space version
of the SCET result.

\subsection{Resummation in physical space: the need of a prescription}

Once the coefficient function has been resummed in $N$-space,
the inverse Mellin transform of the resummed result has to be computed
to get the physical cross-section.
However, it turns out that the inverse Mellin transform of Eq.~\eqref{eq:Cres}
does not exist.
This fact is strictly related to the divergence of perturbative series:
indeed, expanding $C^{\rm res}(N,\as)$ in powers of $\as$, the inverse Mellin transform
exists order by order, but the resulting series is divergent.

This follows from the fact that the functions $g_i$ in Eq.~\eqref{eq:S}
depend on $N$ through~\cite{fr}
\beq
\label{alpharun}
\as(M^2/N^a)=\frac{\as(M^2)}{1+\ab L}\(1+\Ord(\as(M^2))\), \qquad
L\equiv\log\frac{1}{N},
\eeq
with $\ab$ defined in Eq.~\eqref{eq:ab_def}.
As a consequence, the expansion of the $N$-space resummed coefficient function
in powers of $\as(M^2)$ has a finite radius of convergence dictated by $\abs{\ab L}<1$.
In fact, $C^{\rm res}(N,\as)$ has a branch cut in the complex $N$-plane along the positive real axis,
\beq
\Re N > N_L\equiv\exp(1/\ab), \qquad \Im N=0,
\eeq
as one can see directly from the LL expression, Eq.~\eqref{eq:exponentiation_rc}.
But a Mellin transform always has a convergence abscissa, so  
$C^{\rm res}(N,\as)$ cannot be the Mellin transform of any function.
The point $N_L$ represents the position of the Landau pole in $N$-space.
Hence, the non-invertibility of the resummed coefficient function
is strictly related to the Landau pole of the QCD running coupling.

On the other hand, any finite-order truncation of the series expansion
\beq\label{eq:expsigman}
C^{\rm res}(N,\as) = \sum_{k=0}^\infty \as^k\, C^{\rm res}_k(N)
\eeq
behaves as a power of $\log N$ at large $N$ and it is thus free of
singularities for $N$ large enough. Hence, we can construct the resummed
coefficient function in $z$-space as
\beq\label{eq:expsigmaz}
C^{\rm res}(z,\as)=\sum_{k=0}^\infty \as^k \,C^{\rm res}_k(z)
\eeq
where its coefficients are computed as
\beq\label{eq:sigpertexp}
C^{\rm res}_k(z) = \Mell^{-1} \[ C^{\rm res}_k(N) \](z).
\eeq
It follows, by contradiction, that the series Eq.~\eqref{eq:expsigmaz}
must diverge~\cite{frru}.
It turns out~\cite{frru} that, if the Mellin inversion
Eq.~\eqref{eq:sigpertexp} is performed to finite logarithmic accuracy, 
the series
Eq.~\eqref{eq:expsigmaz} acquires a finite but nonzero radius of
convergence in $z$; however, this does
not help given that the convolution integral Eq.~\eqref{eq:inclus_cs_conv}
always goes over the region where the series diverges.

Hence, any resummed definition must either explicitly or implicitly
deal with the divergence of the perturbative expansion
Eq.~\eqref{eq:expsigmaz}. We now consider two prescriptions in which this
is done by constructing a resummed expression to which the divergent
series is asymptotic.

\subsection{Minimal prescription}
\label{sec:MP}

\begin{figure}[th]
  \centering
  \includegraphics[width=0.5\textwidth]{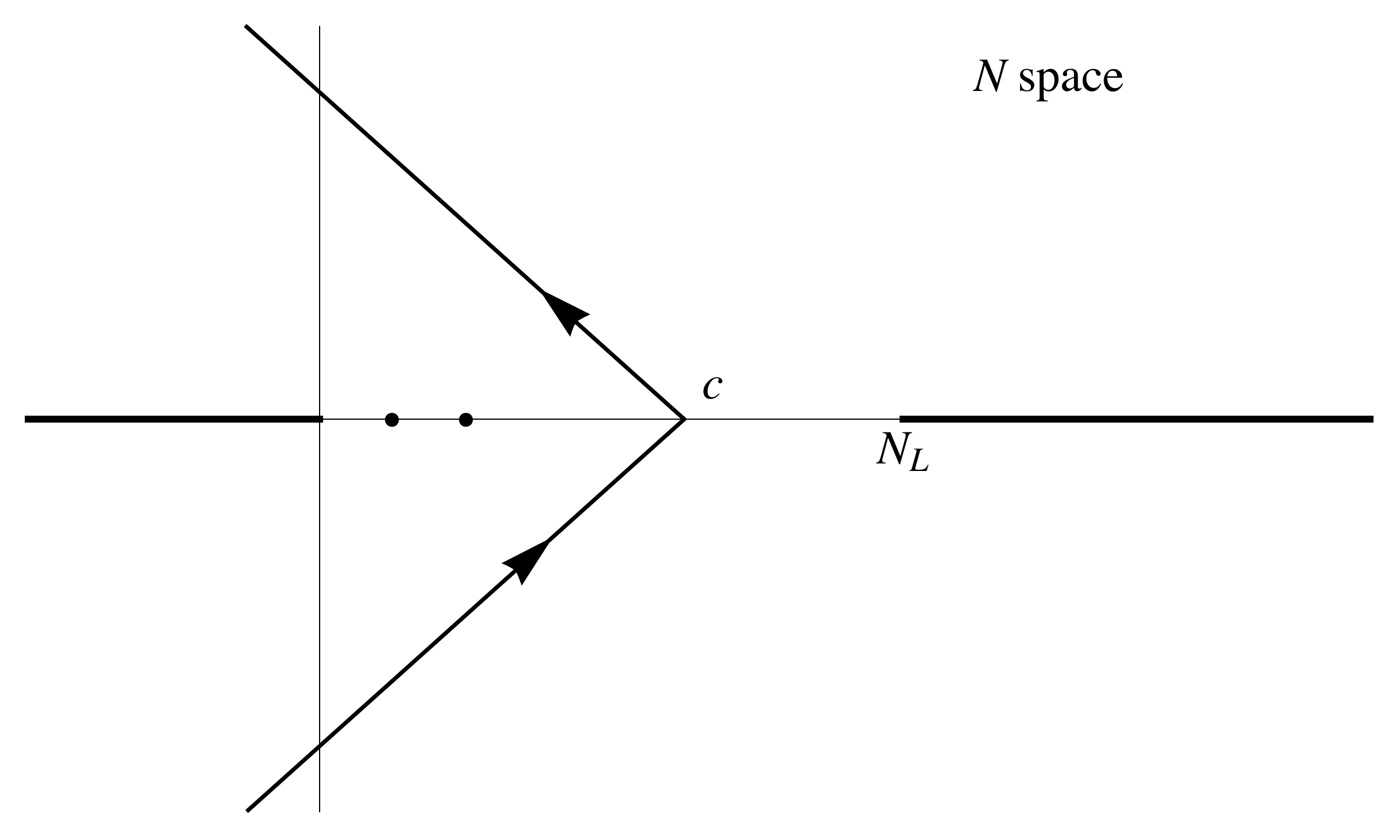}
  \caption{Minimal prescription path. In the figure the branch-cut due to the Landau pole is shown,
    as well as other singularities of the integrand.}
  \label{fig:MP_path}
\end{figure}

In Ref.~\cite{cmnt} a simple solution was proposed to avoid cure the Landau pole problem.
If the problem were not there, the hadronic cross-section could be constructed
as the inverse Mellin of the product of coefficient function and parton luminosity in $N$-space:
\beq
\frac{1}{\tau}\frac{d\sigma}{dM^2} = \frac{1}{2\pi i} \int_{c-i\infty}^{c+i\infty} dN\, \tau^{-N}\,\Lum(N) \,C(N,\as)
\eeq
with $c$, as usual, to the right of all the singularities of the integrand.
As already said above, the problem in the resummed case is that such $c$ does not exist
because of the cut.
The minimal prescription (MP)~\cite{cmnt} defines the resummed hadronic cross-section as
\beq\label{eq:MP_hadr}
\frac{1}{\tau}\frac{d\sigma^{\rm MP}}{dM^2} = \frac{1}{2\pi i} \int_{\rm MP} dN\, \tau^{-N}\,\Lum(N) \,C^{\rm res}(N,\as).
\eeq
where the integration path is chosen to pass to the left of the cut
but to the right of all the other singularities of the integrand;
moreover, the slope of the path is modified as in Fig.~\ref{fig:MP_path} in order to make
the integral convergent numerically.\footnote{The result of the integral is independent on the slope,
as proved in the original paper~\cite{cmnt}.}
It is shown in Ref.~\cite{cmnt} that the cross-section obtained in this way
is finite, and that it is an asymptotic sum of the divergent series
obtained by substituting the expansion Eq.~\eqref{eq:expsigman} in
Eq.~\eqref{eq:MP_hadr} and performing the Mellin inversion order by
order in $\as(M^2)$. Of course, if the expansion is truncated to
any finite order the MP simply gives the exact inverse Mellin transform,
namely, the truncation of Eq.~\eqref{eq:expsigmaz} to the same finite order.

Even if the MP is formulated at hadron level, we can build a sort
of partonic MP. To obtain it, we start by writing $\Lum(N)$ as the
explicit Mellin transform of $\Lum(z)$, and then exchanging the integrals:
\beq
\frac{1}{\tau}\frac{d\sigma^{\rm MP}}{dM^2}
=\int_0^1 \frac{dx}{x} \,\Lum(x) \,\frac{1}{2\pi i} \int_{\rm MP} dN\, \(\frac{\tau}{x}\)^{-N}\,C^{\rm res}(N,\as).
\eeq
We can now define the MP partonic resummed coefficient function as
\beq\label{eq:MP_part}
C^{\rm MP}(z,\as) = \frac{1}{2\pi i} \int_{\rm MP} dN\, z^{-N}\,C^{\rm res}(N,\as)
\eeq
and write, changing variable $x=\tau/z$,
\beq\label{eq:MP_part2}
\frac{1}{\tau}\frac{d\sigma^{\rm MP}}{dM^2}
=\int_\tau^\infty \frac{dz}{z} \,\Lum\(\frac{\tau}{z}\) \,C^{\rm MP}(z,\as).
\eeq
Note, however, that the integral extends to $\infty$.
If $C^{\rm res}(N,\as)$ was a real Mellin transform (i.e.\ without the cut)
as $z>1$ one could close the integration in \eqref{eq:MP_part} in the right
half $N$-plane because its contribution vanishes by virtue of the damping $z^{-N}$,
and then by residue theorem one would find that the integral vanishes for $z>1$,
thereby restoring the canonical convolution in \eqref{eq:MP_part2}.
Here, because of the cut, such procedure doesn't work and $C^{\rm MP}(z,\as)$
does not vanish in the region $z>1$.
Therefore, the MP \emph{is not a convolution}.
\begin{figure}[thb]
\begin{center}
\includegraphics[width=0.49\textwidth,page=1]{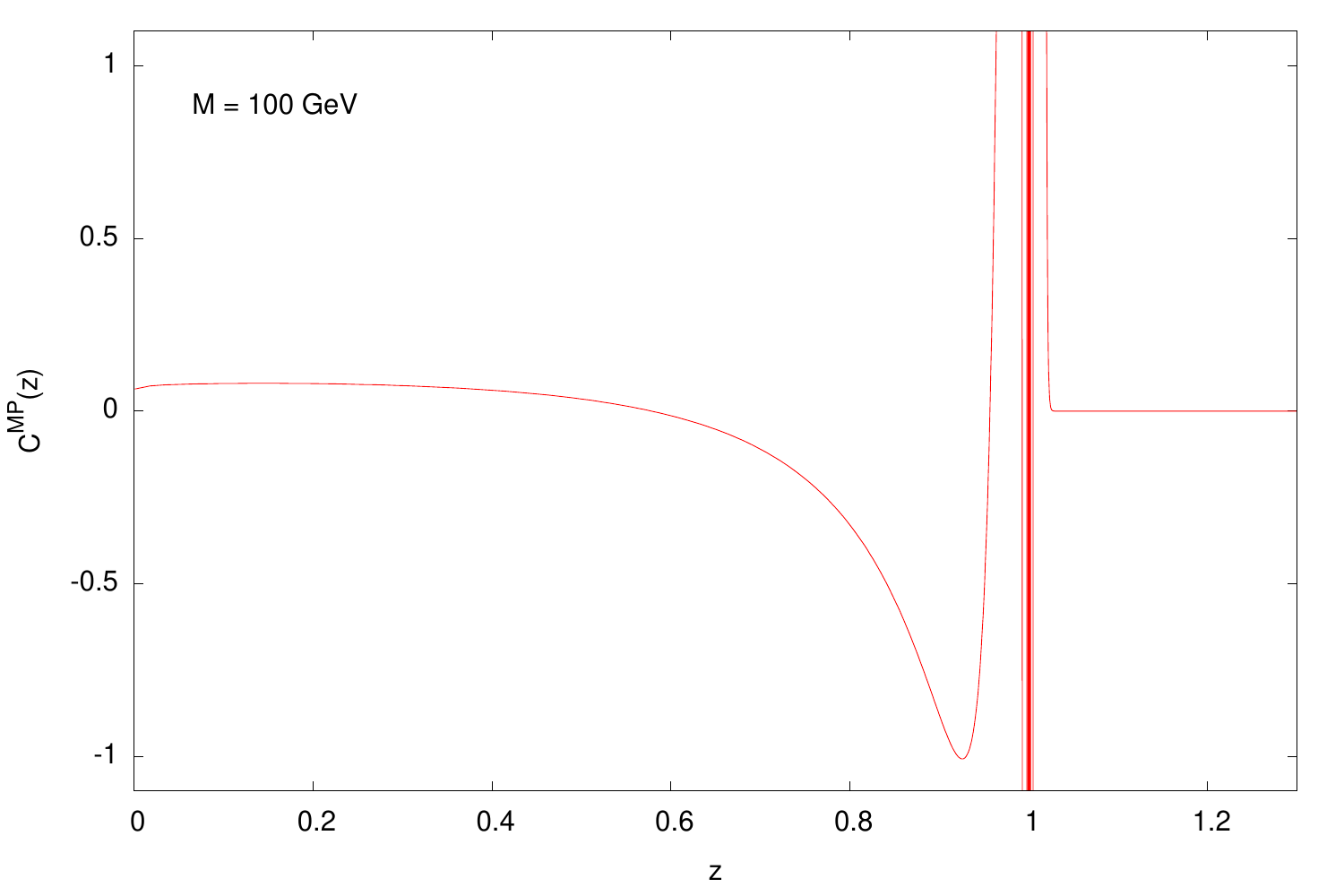}
\includegraphics[width=0.49\textwidth,page=2]{Minimal_partonic_v2}
\caption{The partonic coefficient function $C^{\rm MP}(z,\as)$
  computed using the minimal prescription, Eq.~\eqref{eq:MP_part},
  evaluated at $M=100$~GeV (left) and $M=8$~GeV (right).
  The curve shown is obtained using the NLL expression for the Drell-Yan
  coefficient function Eq.~\eqref{eq:Cres},~\eqref{eq:S} with only $g_1$ and $g_2$ included.} 
\label{fig:MP_osc}
\end{center}
\end{figure}
However, as shown in Ref.~\cite{cmnt}, the contribution from
the $z\geq 1$ region is exponentially suppressed in $\frac{\Lambda_{\rm QCD}}{M}$.
In the vicinity of $z=1$ the integral oscillates strongly, as 
shown in Fig.~\ref{fig:MP_osc}, where $C^{\rm MP}(z,\as)$ is displayed
for the Drell-Yan resummed cross-section, evaluated at two scales
which are relevant for the phenomenological discussion of
Sect.~\ref{sec:pheno-DY}. When folded with the luminosity,
the hadronic cross-section receives a contribution from
a region which is unphysical, i.e.\ when the argument of $\Lum(z)$
is less than $\tau$: this region would not contribute in any finite
order computation.
Even if the ensuing integral is finite, the oscillatory behaviour of
the partonic cross-section makes its numerical computation difficult.
A technical solution to this problem is provided in Ref.~\cite{cmnt};
we will propose in Sect.~\ref{sec:resumm_numerical} a different solution,
and use it for phenomenology in Chap.~\ref{chap:pheno}.

\subsection{Borel prescription}
\label{sec:borel}

An alternative prescription is based on the Borel
summation of the divergent series. This prescription was developed in
Refs.~\cite{frru,afr}; here we give an equivalent, but much
simpler presentation of it~\cite{bfr2}.
To this purpose, it is convenient to rewrite the resummed coefficient function
$C^{\rm res}(N,\as)$ Eq.~\eqref{eq:Cres} as
\beq\label{eq:sigdef}
C^{\rm res}(N,\as)=\Sigma\(\ab\log\frac1N,\as\),
\eeq
where the fact that $C^{\rm res}(N,\as)$ depends on $N$ via $\log\frac1N$
is made explicit:
\beq
\Sigma(\ab L,\as) = \sum_{k=0}^\infty h_k(\as) \(\ab L\)^k.
\eeq
The zeroth term $h_0(\as)$ contains $N$-constant terms, i.e.\ the function $g_0(\as)$;
however, in a matched computation, the constant term (among others) is typically subtracted.
We can then compute the inverse Mellin transform term-by-term
\begin{align}
C^{\rm res}(z,\as) &= \Mell^{-1} \[\Sigma\(\ab\log\frac1N,\as\)\](z)\\
&= \sum_{k=0}^\infty h_k(\as)\,\ab^k \, c_k(z) \label{eq:Cz_series}
\end{align}
where the $c_k(z)$ are the inverse Mellin transforms of the $k$-th power
of $\log\frac1N$ and can be written according to Eq.~\eqref{eq:inv_Mellin_logN_oint} as
\beq\label{eq:ckcoeffs}
c_k(z) = \Mell^{-1}\[\log^k\frac1N\](z)
= \frac{k!}{2\pi i} \oint\frac{d\xi}{\xi^{k+1}}\(
\frac{\plusq{\log^{\xi-1}\frac{1}{z}}}{\Gamma(\xi)} + \delta(1-z) \),
\eeq
where the integration path must enclose the origin $\xi=0$
(the $\delta(1-z)$ term survive only for $k=0$, and is therefore irrelevant
for pure logarithms).
This form, among others, is particularly useful. Indeed in this way
$\Sigma(z,\as)$ becomes, exchanging courageously%
\footnote{In principle, the exchange of integral and series can be performed
only for absolutely convergent series. In fact, since in this case we are looking for
a way to obtain a finite result from a divergent expression, such manipulations
can be allowed, provided they are interpreted as a \emph{definition} of the result they may bring to.
For a more detailed discussion, see App.~\ref{sec:divergent_series}.}
the sum and the integral,
\beq\label{eq:Sigma_xi}
C^{\rm res}(z,\as) = \frac{1}{2\pi i}\oint\frac{d\xi}{\xi} \(\frac{\plusq{\log^{\xi-1}\frac1z}}{\Gamma(\xi)} + \delta(1-z)\)
\sum_{k=0}^\infty h_k(\as)\(\frac{\ab}{\xi}\)^k k!
\eeq
where we recognize the power series of $\Sigma$ with an additional $k!$.
Since, because of the cut of $\Sigma$, its power series has finite radius
of convergence, the series in Eq.~\eqref{eq:Sigma_xi} has convergence radius \emph{zero}:
we are rediscovering, in another (explicit) form, that the inverse Mellin transform
of $C^{\rm res}(N,\as)$ does not exist because its perturbative expansion diverges.

\begin{figure}[t]
  \centering
  \includegraphics[width=0.6\textwidth]{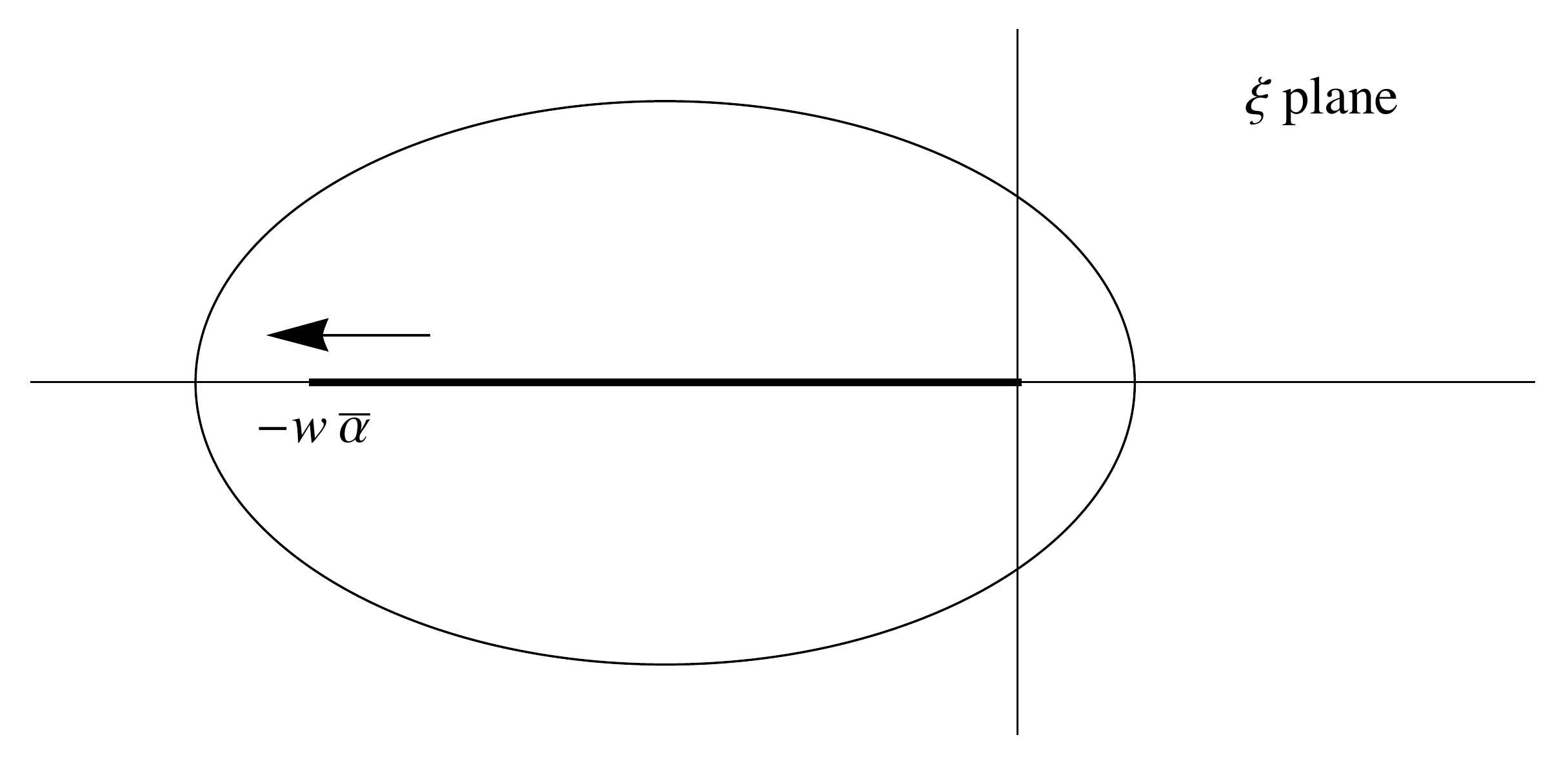}
  \caption{Branch-cut in the $\xi$ plane.}
  \label{fig:xi-cut}
\end{figure}

Being factorially divergent, we can use the $n=1$ Borel method
(see for details App.~\ref{sec:Borel}) to sum the divergent series.
Eq.~\eqref{eq:Sigma_xi} becomes then
\begin{align}
C^{\rm res}(z,\as)
&= \frac{1}{2\pi i}\oint\frac{d\xi}{\xi} \(\frac{\plusq{\log^{\xi-1}\frac1z}}{\Gamma(\xi)} + \delta(1-z)\)
\int_0^\infty dw\, e^{-w} \sum_{k=0}^\infty h_k(\as)\(\frac{\ab w}{\xi}\)^k \nonumber\\
&= \frac{1}{2\pi i}\oint\frac{d\xi}{\xi} \(\frac{\plusq{\log^{\xi-1}\frac1z}}{\Gamma(\xi)} + \delta(1-z)\)
\int_0^\infty dw\, e^{-w} \,\Sigma\(\frac{\ab w}{\xi},\as\)
\label{eq:Sigma_xi_Borel}
\end{align}
where the Borel transform of the divergent series coincides with (and has been substituted by)
the function $\Sigma$.
Some comments are in order:
\begin{itemize}
\item the branch-cut of $\Sigma$ is mapped in terms of the variable $\xi$ into the region
  $-\ab w \leq \xi \leq 0$: somehow, the series of $\xi=0$ poles in the sum has been transformed in this cut;
\item since the $\xi$ integration path must encircle the poles in $\xi=0$ order by order,
  it must encircle the cut of the sum;
\item however, the $w$ integral pushes the lower branch-point to $-\infty$ (see Fig.~\ref{fig:xi-cut}), where the
  oscillatory and factorially growing behaviour of $1/\Gamma(\xi)$ destroys the convergence of the $\xi$ integral.
\end{itemize}
These considerations hint that the series Eq.~\eqref{eq:Cz_series} is \emph{not}
Borel-summable.

The Borel prescription can be now formulated. To do this,
let's rewrite the result Eq.~\eqref{eq:Sigma_xi_Borel} by changing variable
in the $w$ integral as
\beq
C^{\rm res}(z,\as) =
\frac{1}{2\pi i}\oint\frac{d\xi}{\xi} \(\frac{\plusq{\log^{\xi-1}\frac1z}}{\Gamma(\xi)} + \delta(1-z)\)
\int_0^\infty \frac{dw}{\ab}\, e^{-\frac{w}{\ab}} \,\Sigma\(\frac{w}{\xi},\as\)
\eeq
which is valid for $\ab>0$, which is physically always the case.
To make the integrals convergent, an upper cutoff $W$ is put in the $w$ integral:%
\footnote{This notation is somewhat misleading: the full coefficient function
resummed with the Borel prescription is recovered from $C^{\rm BP}$ by adding $\delta(1-z)$.
However, since in any matched result the $\delta$ term is always subtracted
(being already included in the fixed order result, see Eq.~\eqref{eq:match}),
this is a minor notation issue.}
\beq\label{eq:Borel_prescription}
C^{\rm BP}(z,\as, W) =
\frac{1}{2\pi i}\oint\frac{d\xi}{\xi} \(\frac{\plusq{\log^{\xi-1}\frac1z}}{\Gamma(\xi)} + \delta(1-z)\)
\int_0^W \frac{dw}{\ab}\, e^{-\frac{w}{\ab}} \,\Sigma\(\frac{w}{\xi},\as\).
\eeq
This results has the following properties:
\begin{itemize}
\item the divergent series Eq.~\eqref{eq:Cz_series} is asymptotic to
the BP formula, Eq.~\eqref{eq:Borel_prescription}: the proof~\cite{frru} is rather simple,
since the damping factor $D^{(1)}_k(W/\ab)$, Eq.~\eqref{eq:damping1}, tends to $1$
faster than any power as $\ab\to0^+$;
\item even if divergent, the neglected terms due to the cutoff are formally of the form
\beq
\exp\frac{W}{\ab} = \(\frac{\Lambda^2}{M^2}\)^{W/2},
\eeq
i.e.\ they are higher-twist ($\Lambda=\Lambda_{\rm QCD}$ is the Landau pole scale);
\item such terms correspond to a twist $2+W$: since the first subleading twist
is twist $4$, $W$ must be chosen according to $W\geq 2$;
\item the parameter $W$ is a degree of freedom which can be used to estimate
the ambiguity of the resummation procedure;
\item last but not least, this resummed expression is at parton level,
and then the convolution structure of the hadronic result is not spoiled
as it is in the case of the MP.
\end{itemize}
As discussed in App.~\ref{sec:Borel_truncated}, the truncation of the
Borel integral effectively cutoff the series (smoothly), making it convergent.
The truncation point is regulated by $W$ (in fact by $W/\ab$, see App.~\ref{sec:Borel_truncated})
and the larger $W$ is the more terms are included in the series.
However, since the series is not Borel summable, as $W$ gets large
some instabilities should appear: in particular, the integral varies a
lot for small changes of $W$ large. Then, a proper choice for $W$
should be a small value: in Chap.~\ref{chap:pheno} we will then choose 
$W=2$, the minimal value.

To make contact with the original formulation~\cite{afr},
the BP result can be recast by an integration by parts in the $w$ integral,
leading to
\begin{multline}
C^{\rm BP}(z,\as, W) = 
\frac{1}{2\pi i}\oint\frac{d\xi}{\xi} \(\frac{\plusq{\log^{\xi-1}\frac1z}}{\Gamma(\xi)} + \delta(1-z)\)\\
\times\[\int_0^W dw\, e^{-\frac{w}{\ab}} \,\frac{d}{dw}\Sigma\(\frac{w}{\xi},\as\)
- e^{-\frac{W}{\ab}}\,\Sigma\(\frac{W}{\xi},\as\)\].\label{eq:Borel_deriv}
\end{multline}
Being higher-twist, the second term in square brackets can be neglected,
leading to a new (equivalent) version of the BP.
Note that such integration by parts could already have been performed
before cutting the integral off: in this case, the boundary contribution
vanishes at infinity\footnote{To see this, look for example at the LL case,
Eq.~\eqref{eq:exponentiation_rc}.}
only for $\xi<0$, which is indeed the region which gives
contributions to the integral. Putting now the cutoff, we obtain
the result of Ref.~\cite{afr}, i.e.\ Eq.~\eqref{eq:Borel_deriv} without the boundary term.
Anyway, in practice, the difference between the two results is negligible;
nevertheless, in some cases (as for example that of the LL anomalous dimension),
this alternative formulation is easier to use, because of the different analytic
structure of the derivative of $\Sigma$.

\subsection{Subleading terms}
\label{sec:subl}

The Borel and minimal prescriptions differ in the way
the high-order behaviour of the divergent series is handled.
This, as discussed in Refs.~\cite{frru,afr}, makes in
practice a small difference unless the hadronic $\tau$ is close to the
Landau pole of the strong coupling, 
\beq
\tau_L=1-\frac{\Lambda^2}{M^2},
\eeq
which is seldom the case, and never for
applications at collider energies that we are mostly interested
in. This is a consequence of the fact that for values of $\as$ in the
perturbative region it is only at very high orders that the effect of
the various prescriptions kicks in.

However, prescriptions may also differ in the subleading terms in the
$z$ dependence which are introduced when performing the resummation.
To understand this, we can simply consider the effect of the prescriptions
to the generic term of the expansion of $\Sigma\(\ab\log\frac1N,\as\)$, namely
a power of $\log\frac1N$.
The minimal prescription then just gives the exact inverse Mellin transform,
since at any finite order there is no cut; then, indicating with
$\Mell^{-1}_{\rm MP}$ the inverse Mellin \`a la minimal prescription, we have
\beq
\Mell^{-1}_{\rm MP} \[\log^k\frac1N\](z) = \Mell^{-1} \[\log^k\frac1N\](z).
\eeq
From Eq.~\eqref{eq:inv_Mellin_logN_sum} we see that the $z$ dependence of such
inverse Mellin is via $\log\frac1z$, and generates terms of the form
\beq\label{eq:log_MP}
\plus{\frac{\log^j\log\frac1z}{\log\frac1z}}, \qquad 0\leq j<k.
\eeq
In the large-$z$ limit, these terms coincide\footnote{This is not strictly true:
indeed, in a hadronic convolution, even if the lower limit $\tau$ is large
the convolution receives a contribution from an integral with lower limit $0$,
because of the plus prescription. Then, even at very large $\tau$, the result would
be different from the ``correct one''.}
with the right ones,
\beq\label{eq:log_right}
\plus{\frac{\log^j(1-z)}{1-z}},
\eeq
which are the terms that are resummed.
This mismatch is not surprising: in the resummation procedure, after passing to the $N$ space,
the large-$N$ limit was taken, spoiling the exact correspondence between the log terms
in $z$ space and their $N$ space counterparts.

Moving to the BP, the inverse Mellin of the $k$-th power of $\log\frac1N$
according to Eq.~\eqref{eq:Borel_prescription} is ($k>0$)
\begin{align}
\Mell^{-1}_{\rm BP} \[\log^k\frac1N\](z)
&= \frac{1}{2\pi i}\oint\frac{d\xi}{\xi\,\Gamma(\xi)}\plusq{\log^{\xi-1}\frac1z}
\int_0^{W/\ab} dw\, e^{-w} \(\frac{w}{\xi}\)^k\\
&= D^{(1)}_k(W/\ab) \cdot \Mell^{-1} \[\log^k\frac1N\](z)
\end{align}
with $D^{(1)}_k(W/\ab)$ defined in Eq.~\eqref{eq:damping1}
(we have used Eq.~\eqref{eq:inv_Mellin_logN_oint} in the last equality).
Then, also the BP produces the same terms of the MP, but multiplied
by the damping factor.
Being interested in the $z$ dependence, let's forget about the damping factor,
since it is actually irrelevant here --- the $W\to\infty$ limit is perfectly safe.
Then, in this view, the $z$ dependence of the MP and the BP is the same.

However, one of the features of the BP is that its $z$ dependence
is completely under control, as it appears in the term
\beq
\plusq{\log^{\xi-1}\frac1z}
\eeq
which produces the logs \eqref{eq:log_MP} by derivatives with respect to $\xi$;
this is a crucial difference with respect to the MP, where there is no
obvious way to manipulate the $z$ dependence.
We can then more conveniently use
\beq
\plusq{(1-z)^{\xi-1}}
\eeq
which generates the right logarithms Eq.~\eqref{eq:log_right}.
However this simple replacement is not enough. Indeed, in such manipulations
one can play with arbitrary subleading terms, provided
logarithmically enhanced terms are always the same. In $N$-space,
the terms generated by the two distributions above differ not only by
terms of the order $1/N$, but also by a constant term, see App.~\ref{sec:Mellin_log}.
Then, to preserve an accuracy up to constant terms in $N$-space, the appropriate
choice for the generating distribution is
\beq
\plusq{(1-z)^{\xi-1}} + \frac1\xi\,\delta(1-z),
\eeq
as one can read out of Eq.~\eqref{eq:inv_Mellin_logB1}.
With this choice the BP
\begin{multline}\label{eq:BP1}
C^{\rm BP_1}(z,\as, W) =
\frac{1}{2\pi i}\oint\frac{d\xi}{\xi\,\Gamma(\xi)}\(\plusq{(1-z)^{\xi-1}} + \frac1\xi\,\delta(1-z)\)\\
\times\int_0^W \frac{dw}{\ab}\, e^{-\frac{w}{\ab}} \,\Sigma\(\frac{w}{\xi},\as\)
\end{multline}
reproduces the right logarithms Eq.~\eqref{eq:log_right} and does not introduce
additional constant terms (in $N$-space).
This is the default Borel prescription of Refs.~\cite{frru,afr};
however, there the $\delta(1-z)$ term was not taken into account,
and therefore the resulting prescription was accurate up to constant terms
(in $N$-space).

Phenomenologically (for realistic $\tau$'s) it will differ
by a sizable amount from the MP (or the ``old'' BP).
This point is quite important: indeed, provided the large-$z$ behaviour
(or large-$N$ in Mellin space) is respected, all choices of subleading terms
are theoretically equivalent, but may lead to large differences in
the region far from threshold, eventually spoiling the accuracy of the result.
In particular, phenomenologically relevant values of the hadronic $\tau$ are
typically quite small --- of the order of $10^{-3}$ or less;
in the convolution Eq.~\eqref{eq:inclus_cs_conv}, values of $z$
in the partonic resummed coefficient function down to $z=\tau$ give contribution,
then inadequate choices of subleading terms may produce unphysical
contributions to the hadronic cross-section.

To avoid such problems, it has been suggested~\cite{Bonciani} to (smoothly) truncate
the resummed part of the coefficient function below some value of $N$:
in this case, far from the threshold region, potentially large and
uncontrolled contributions from subleading terms are suppressed.
The price to pay is the introduction of a new, completely arbitrary, degree of freedom.

Then, we prefer and suggest~\cite{bfr2} to find some optimal
choice for the subleading terms, and in any case use the ambiguity related to
different choices as an estimator of the size and the uncertainty of the resummed part of a cross-section.
In Sect.~\ref{sec:res_comparison_FO} we will investigate several choices by comparing
with fixed order results. Here, we propose a more theoretical and general argument.

The key idea is to include subleading terms that we know to appear in the
perturbative computation. Coming back to Eq.~\eqref{eq:C1_soft_running},
we see that the dependence on $z$ is all contained, apart the factor $(1-z)^{-1}$,
in the upper integration limit: such limit is determined by the kinematics of gluon emission.
A careful computation~\cite{fr} shows that the upper limit is really
\beq
k^0_{\rm max}=\sqrt{\frac{M^2(1-z)^2}{4z}},
\eeq
so that in fact the resummation produces logarithmic terms of the form
\beq\label{eq:lognew}
\log^k\frac{1-z}{\sqrt{z}}.
\eeq
This suggests to define a new version of the Borel prescription
which reproduces such logarithms.
The generating distribution can be either
\beq
\plusq{z^{-\frac\xi2}(1-z)^{\xi-1}} \qquad\text{or} \qquad z^{-\frac\xi2}\plusq{(1-z)^{\xi-1}},
\eeq
the difference being constant terms in $N$-space; since for the same reason discussed above
the constant terms must be subtracted again, we use the second one and get,
according to Eq.~\eqref{eq:inv_Mellin_logB3},
\begin{multline}\label{eq:BP2}
C^{\rm BP_2}(z,\as, W) =
\frac{1}{2\pi i}\oint\frac{d\xi}{\xi\,\Gamma(\xi)}\(z^{-\frac\xi2}\plusq{(1-z)^{\xi-1}} + \frac1\xi\,\delta(1-z)\)\\
\times\int_0^W \frac{dw}{\ab}\, e^{-\frac{w}{\ab}} \,\Sigma\(\frac{w}{\xi},\as\).
\end{multline}
This form of the Borel prescription is more accurate than the default one, Eq.~\eqref{eq:BP1}:
in Ref.~\cite{bfr2} we used it for phenomenology, but there we neglected the
$\delta(1-z)$ term. In Chap.~\ref{chap:pheno} we will use here this prescription
for phenomenological applications.
We finally mention that this choice of subleading terms also arises
in a natural way in the context of soft-collinear effective theories,
and it was adopted in Ref.~\cite{bnx}.

It turns out that these two new Borel prescriptions are
closer to the MP (or equivalently to the old BP) than the default one,
Eq.~\eqref{eq:BP1}, because
\beq
\log\frac{1}{z}=\frac{1-z}{\sqrt{z}}\(1+\Ord\[(1-z)^2\]\),
\label{eq:logMPexp}
\eeq
so that
\beq
\frac{\log^k\log\frac{1}{z}}{\log\frac{1}{z}}
=\frac{\sqrt{z}}{1-z}\log^k\frac{1-z}{\sqrt{z}} \(1+\Ord\[(1-z)^2\]\).
\label{eq:allMPexp}
\eeq
Equation~\eqref{eq:allMPexp} shows that, amusingly, up to terms
suppressed by \emph{two} powers of $1-z$, the minimal prescription
effectively also reproduces the logarithms Eq.~\eqref{eq:lognew},
though at the cost of also introducing an overall factor
$\sqrt{z}$ which is absent in the known perturbative contributions.

\subsubsection{A discussion on constant terms}
\label{sec:discussion_constants}

We have stressed that the $\delta(1-z)$ term in Eqs.~\eqref{eq:BP1} and \eqref{eq:BP2}
should be there to make the results accurate up to constants in $N$-space.
We would like to show here that in any matched result, the difference
between our expression and those used in Refs.~\cite{frru,afr,bfr2}
is in fact subleading, and therefore does not constitute a real defect.

To prove this, we take the Mellin transform of the Borel prescription.
In the three cases of Eqs.~\eqref{eq:Borel_prescription}, \eqref{eq:BP1} and \eqref{eq:BP2}
we get
\begin{align}
C^{\rm BP}(N,\as, W) &=
\frac{1}{2\pi i}\oint\frac{d\xi}{\xi}\,N^{-\xi}
\int_0^W \frac{dw}{\ab}\, e^{-\frac{w}{\ab}} \,\Sigma\(\frac{w}{\xi},\as\)\\
C^{\rm BP_1}(N,\as, W) &=
\frac{1}{2\pi i}\oint\frac{d\xi}{\xi}\,\frac{\Gamma(N)}{\Gamma(N+\xi)}
\int_0^W \frac{dw}{\ab}\, e^{-\frac{w}{\ab}} \,\Sigma\(\frac{w}{\xi},\as\)\label{eq:BP1N}\\
C^{\rm BP_2}(N,\as, W) &=
\frac{1}{2\pi i}\oint\frac{d\xi}{\xi}\,\frac{\Gamma(N-\xi/2)}{\Gamma(N+\xi/2)}
\int_0^W \frac{dw}{\ab}\, e^{-\frac{w}{\ab}} \,\Sigma\(\frac{w}{\xi},\as\).\label{eq:BP2N}
\end{align}
By making use of the Stirling approximation, we immediately see that the three equations
are equivalent up to terms of order $1/N$, as stressed above.

If we had not included the $\delta(1-z)$ term in Eqs.~\eqref{eq:BP1} and \eqref{eq:BP2},
as done in Refs.~\cite{frru,afr,bfr2},
the $N$-dependence of their Mellin transforms would have been, respectively,
\beq
\[\frac{\Gamma(N)}{\Gamma(N+\xi)} - \frac1{\Gamma(1+\xi)} \],\qquad
\[\frac{\Gamma(N-\xi/2)}{\Gamma(N+\xi/2)} - \frac1{\Gamma(1+\xi)} \].
\eeq
Hence, for each power of $L=\log\frac1N$ in $\Sigma$ a constant term is produced:
\beq
L^k \to L^k + \Delta^{(k)}(1) + \Ord\(N^{-1}\),
\eeq
where $\Delta(\xi)=1/\Gamma(\xi)$.
For definiteness, consider a NLL computation in $N$ space, before the application of any prescription:
in this case the at order $\as$ can be read out of Eq.~\eqref{eq:NLL_res_expansion}, and is
\beq
g_{12} L^2 + g_{21} L + g_{01}
\eeq
where $g_{01}$ contains all the constants appearing in the Mellin transform of the
fixed-order coefficient function $C^{(1)}(z)$; at higher orders, no constants are generated.
Indeed, at NLL, the three leading powers of $L$ are correctly predicted at each order,
as shown in Tab.~\ref{tab:res_orders}.
Using the prescriptions of Refs.~\cite{frru,afr,bfr2} we would have obtained instead
\beq
g_{12} L^2 + g_{21} L + \[g_{01} + g_{12} \Delta''(1) + g_{21} \Delta'(1) \] + \Ord\(N^{-1}\),
\eeq
and now the constant term (in square brackets) is no longer correctly predicted.
However, a NLL resummed result is usually matched with a NLO computation,
thereby substituting the order $\as$ with the correct one: hence, at the end,
the three leading powers of $L$ are in fact correctly included in the full result.
The same reasoning holds for higher-order resummation: therefore,
after matching the generated constant terms are always subleading, and more
subleading than other neglected logs.

Having understood that, after matching, the prescriptions of Refs.~\cite{frru,afr,bfr2}
are as accurate as the ones described here, we now want to stress
that the subleading constant terms induced to all orders in $\as$ by those prescriptions
are actually not random: they are exactly the constants coming from
the Mellin transform of the distributions.
Of course, since not all logarithms are included, not all the constant terms are produced;
moreover, these constants are subleading with respect to other logarithmic terms
already not included in the resummed expression.
Nevertheless, since these constants are there, their inclusion could
help somehow the convergence of the perturbative-resummed expansion.
We will investigate this at the hadron level in Sect.~\ref{sec:pheno_Borel}.

\subsection{Comparison with fixed-order}
\label{sec:res_comparison_FO}

In order to assess the impact of these different choices of subleading terms,
we compare the complete fixed-order results with the terms produced
order by order by a truncation of the resummed expressions and with the full
resummed result, both in $N$-space and in $z$-space.
For definiteness, we consider the case of the Drell-Yan process,
see Sect.~\ref{sec:pheno-DY} and App.~\ref{sec:DY_fixed-order}.

\subsubsection{$N$-space comparison}
At NLO, the coefficient function for the $q\bar q$ channel
(the logarithmically enhanced one) is given by Eq.~\eqref{eq:C1qqDY},
\beq
C^{(1)}(z) =
\frac{2C_F}{\pi}\left\{
2\plus{\frac{\log(1-z)}{1-z}}
-\frac{\log z}{1-z}-(1+z)\log\frac{1-z}{\sqrt{z}}
+\(\zeta_2 -2\)\delta(1-z)
\right\}
\eeq
or, in $N$-space (see App.~\ref{sec:Mellin_log} for details on the Mellin transforms),
\begin{multline}\label{eq:c1}
C^{(1)}(N) = \frac{2C_F}{\pi} \Bigg\{ \psi_0^2(N) + 2 \gammae \psi_0(N)
+ 2\zeta_2 -2 + \gammae^2
+ \frac{\psi_1(N+2)- \psi_1(N)}2
\\
+ \frac{\gammae + \psi_0(N+1)}{N}
+ \frac{\gammae +\psi_0(N+2)}{N+1} \Bigg\}.
\end{multline}
According to Tab.~\ref{tab:res_orders}, a NLL resummed expression,
expanded up to order $\as$, reproduces all the logarithmically enhanced terms,
including constants.
Then, consider first the NLL (or higher): the order $\as$ expansion of the resummed
expression in $N$-space is
\beq\label{eq:NLO_log_MP}
C^{(1)}_{\rm MP}(N) = \frac{2C_F}{\pi} \[ \log^2\frac1N - 2 \gammae \log\frac1N + 2\zeta_2 -2 + \gammae^2 \]
\eeq
and coincides to what the minimal prescription would reproduce.
The Borel prescription produces different terms: using directly Eqs.~\eqref{eq:BP1N}
and \eqref{eq:BP2N} in the $W\to\infty$ limit we would get
\begin{align}
C^{(1)}_{\rm BP_1}(N) &= \frac{2C_F}{\pi} \[ \psi_0^2(N) - \psi_1(N) + 2 \gammae \psi_0(N) + 2\zeta_2 -2 + \gammae^2 \]\\
C^{(1)}_{\rm BP_2}(N) &= \frac{2C_F}{\pi} \[ \psi_0^2(N) + 2 \gammae \psi_0(N) + 2\zeta_2 -2 + \gammae^2 \]
\end{align}
Of course, they all coincide at large $N$, but the finest choice of subleading terms
is the one which is more accurate even when $N$ is not so large.
\begin{figure}[tb]
\begin{center}
\includegraphics[width=0.85\textwidth,page=1]{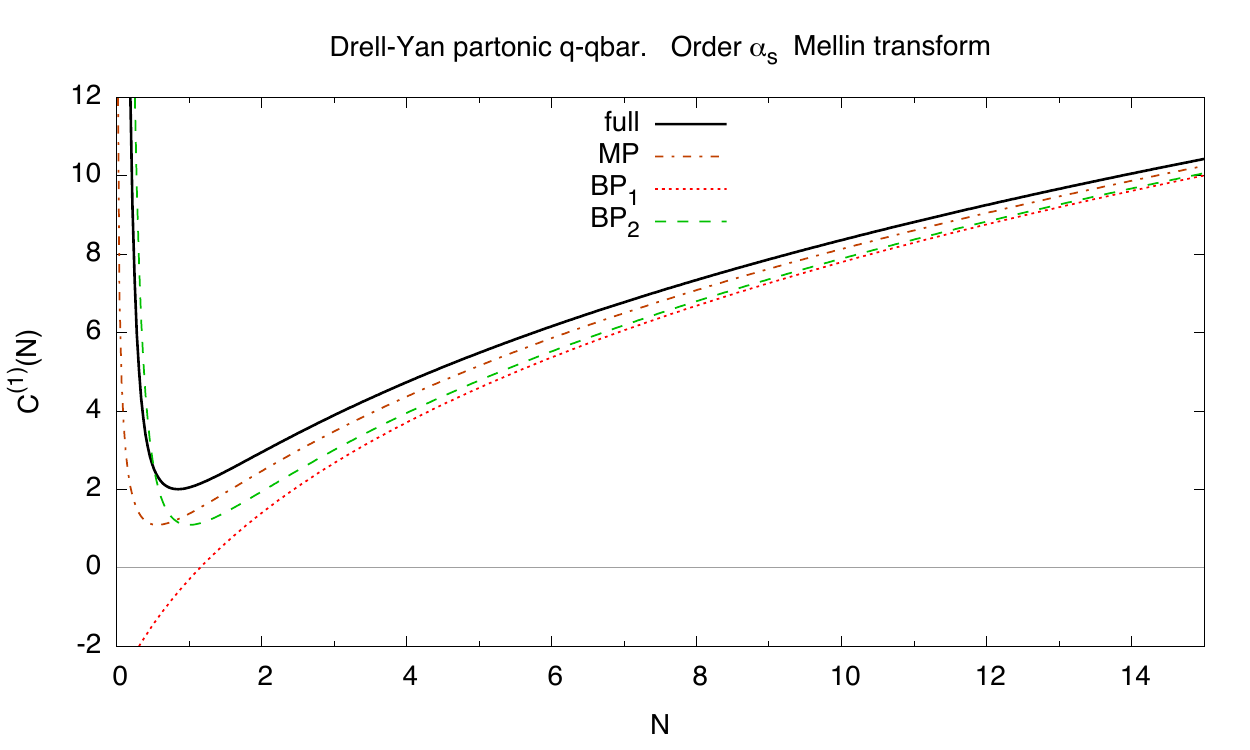}
\caption{The order $\as$ Drell-Yan partonic coefficient function
  $C^{(1)}(N)$, Eq.~\eqref{eq:c1}, plotted as a function of $N$ (solid curve)
  and its logarithmic approximations.}
\label{fig:DY_order_as}
\end{center}
\end{figure}
In Fig.~\ref{fig:DY_order_as} we compare these three large-$N$ approximations
with the full NLO coefficient for positive real values of $N$.
The MP curve is the closest one to the full order $\as$, while the BP$_1$ is the more distant.
To justify this, we expand at large $N$ the differences
\begin{align}
\frac{C^{(1)}_{\rm MP}(N) - C^{(1)}(N)}{2C_F/\pi} &= \frac{\log\frac1N-\gammae}N + \Ord\(N^{-2}\)\\
\frac{C^{(1)}_{\rm BP_1}(N) - C^{(1)}(N)}{2C_F/\pi} &= 2\,\frac{\log\frac1N-\gammae-\frac12}N + \Ord\(N^{-2}\)\\
\frac{C^{(1)}_{\rm BP_2}(N) - C^{(1)}(N)}{2C_F/\pi} &= 2\,\frac{\log\frac1N-\gammae}N + \Ord\(N^{-2}\)
\end{align}
and note that, actually, the difference between MP and full order $\as$ is half than the
difference between BP$_2$ and the full order $\as$.
Anyway, for $N\gtrsim 2$ all choices of subleading terms are quite close to the full result,
the difference being less than $50\%$ of the contribution.
Below $N\sim2$, the BP$_1$ takes its way, far from fixed order: this confirms
that such choice of subleading terms is the worst, and we therefore discard it.
The other two curves at small $N$ show a singularity in $N=0$: the complete
order $\as$ behaves as
\beq
C^{(1)}(N) = \frac{C_F}{\pi N^2} + \Ord\(N^0\)
\eeq
while the MP has a logarithmic singularity and the BP$_2$ approximation behaves as
\beq
C^{(1)}_{\rm BP_2}(N) = \frac{2C_F}{\pi N^2} + \Ord\(N^0\).
\eeq
Then, apart from a factor of $2$, the BP$_2$ reproduces quite well also the behaviour
at small $N$.

The conclusion we may extract from this comparison is that both the MP and the BP$_2$
have good properties, and reproduce quite well the behaviour of the full fixed-order
even far from the threshold region $N\gg 1$.

However, this is of course not enough for saying that the coefficient function
is well predicted by the logarithms produced by threshold resummation. Indeed,
according to Tab.~\ref{tab:res_orders} only a tower of dominant logarithms is
correctly produced by resummation, while the other subdominant logarithms will
have wrong coefficients.
For example, at LL only the highest power of log is produced:
at order $\as$ we would then have
\begin{align}
C^{(1)}_{\rm MP,\, LL}(N)  &= \frac{2C_F}{\pi} \,\log^2\frac1N\\
C^{(1)}_{\rm BP_1,\,LL}(N) &= \frac{2C_F}{\pi} \[ \psi_0^2(N) - \psi_1(N) \]\\
C^{(1)}_{\rm BP_2,\,LL}(N) &= \frac{2C_F}{\pi} \, \psi_0^2(N).
\end{align}
However, by construction, the difference between these three approximations
is never logarithmic, and then no choice of subleading terms can reproduce
any of the subleading logarithmic terms.
The subleading difference (suppressed by at least one inverse power of $N$)
is subleading with respect to the neglected logs, and therefore a comparison
is not that helpful.

The next non-trivial comparison is between the logs produced by a NNLL resummation
(or higher) at order $\as^2$ (NNLO) and the full NNLO.
Without showing the lengthy expression (which can be easily found expanding the expressions
in App.~\ref{sec:app-resumm}) we present the numerical results.
\begin{figure}[tb]
\begin{center}
\includegraphics[width=0.85\textwidth,page=1]{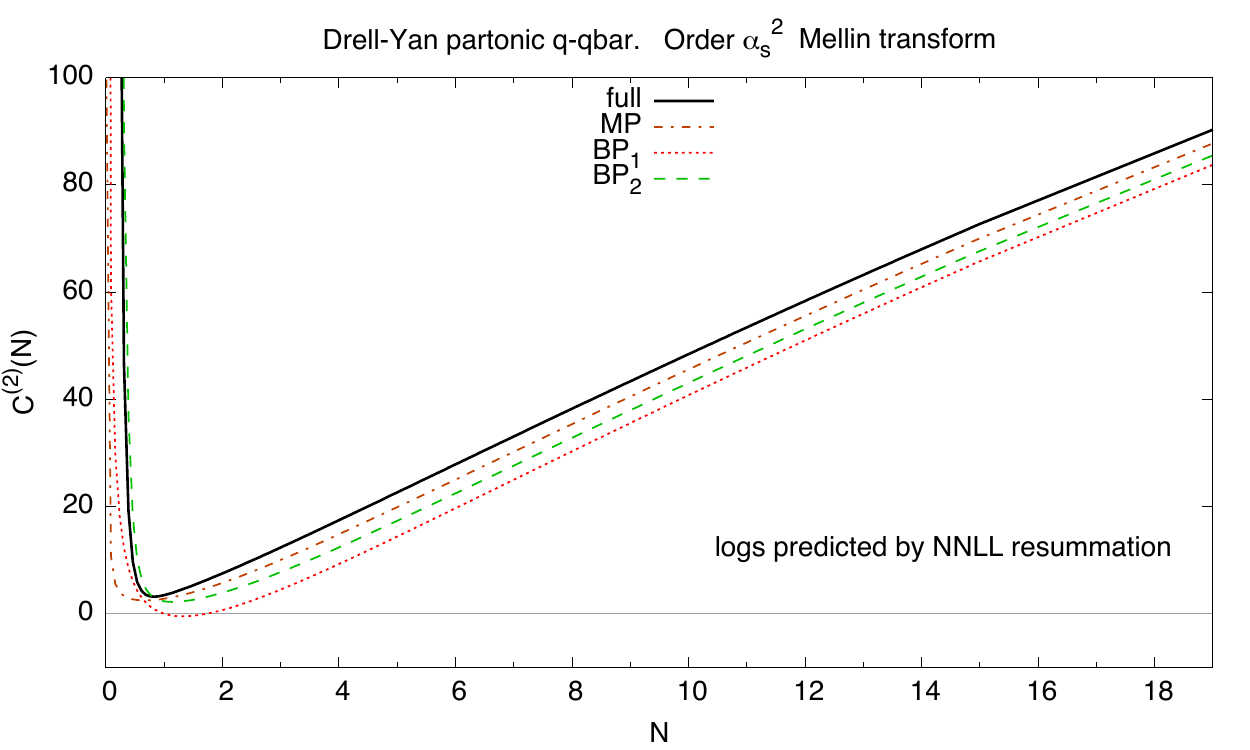}
\caption{The order $\as^2$ Drell-Yan partonic coefficient function
  $C^{(2)}(N)$ plotted as a function of $N$ (solid curve)
  and its logarithmic approximations.}
\label{fig:DY_order_as2}
\end{center}
\end{figure}
In Fig.~\ref{fig:DY_order_as2}, the same curves as in Fig.~\ref{fig:DY_order_as}
where the various log approximations are computed form NNLL resummation.
The leading terms which differ from the full result are logarithms of $N$
suppressed by a factor $1/N$.
The first observation we may do is that such terms are quite relevant,
because even at quite large $N$ the different logarithmic curves
are still quite far from the full NNLO curve.

Concerning the specific difference between each choice of subleading terms
in the logarithmic terms, the conclusions drawn at NLO are unchanged.
The MP choice is the best at large $N$, even down to such a small value as $N\sim1$.
At very small $N$ (less than about $1$), the BP$_2$ performs better,
reproducing at least qualitatively the small-$N$ behaviour.
The BP$_1$ is the worst at large and intermediate $N$, confirming
our choice of discarding it.
Then, for the MP and the BP$_2$, we can say that for $N\sim 2$
the logarithmic contribution is already about $50\%$ of the full result,
and it rapidly increases as $N$ gets larger.
This suggests that indeed the logarithmic contribution provides a sizable
or even dominant contribution to it for $N\gtrsim 2$.

As a final comment we note that the region in which we have found
logarithmic effects to be important is in fact rather wider than the
region in which $\as\log^2 N\sim 1$. In this region even though
logarithmically enhanced terms may lead to a substantial contribution,
they behave in an essentially perturbative way, in that $(\as\log^2
N)^{k+1}<(\as\log^2 N)^k$, and therefore the all-order behaviour of the
resummation is irrelevant. In this intermediate region, the
resummation may have a significant impact, but with significant
ambiguities related to subleading terms.

\subsubsection{$z$-space comparison}
We now turn to a comparison in $z$-space. Since the conclusion
would not be different from those in $N$-space, we perform here a comparison
between the fixed order result and the full resummed expressions.

\begin{figure}[tbh]
\begin{center}
\includegraphics[width=0.495\textwidth,page=1]{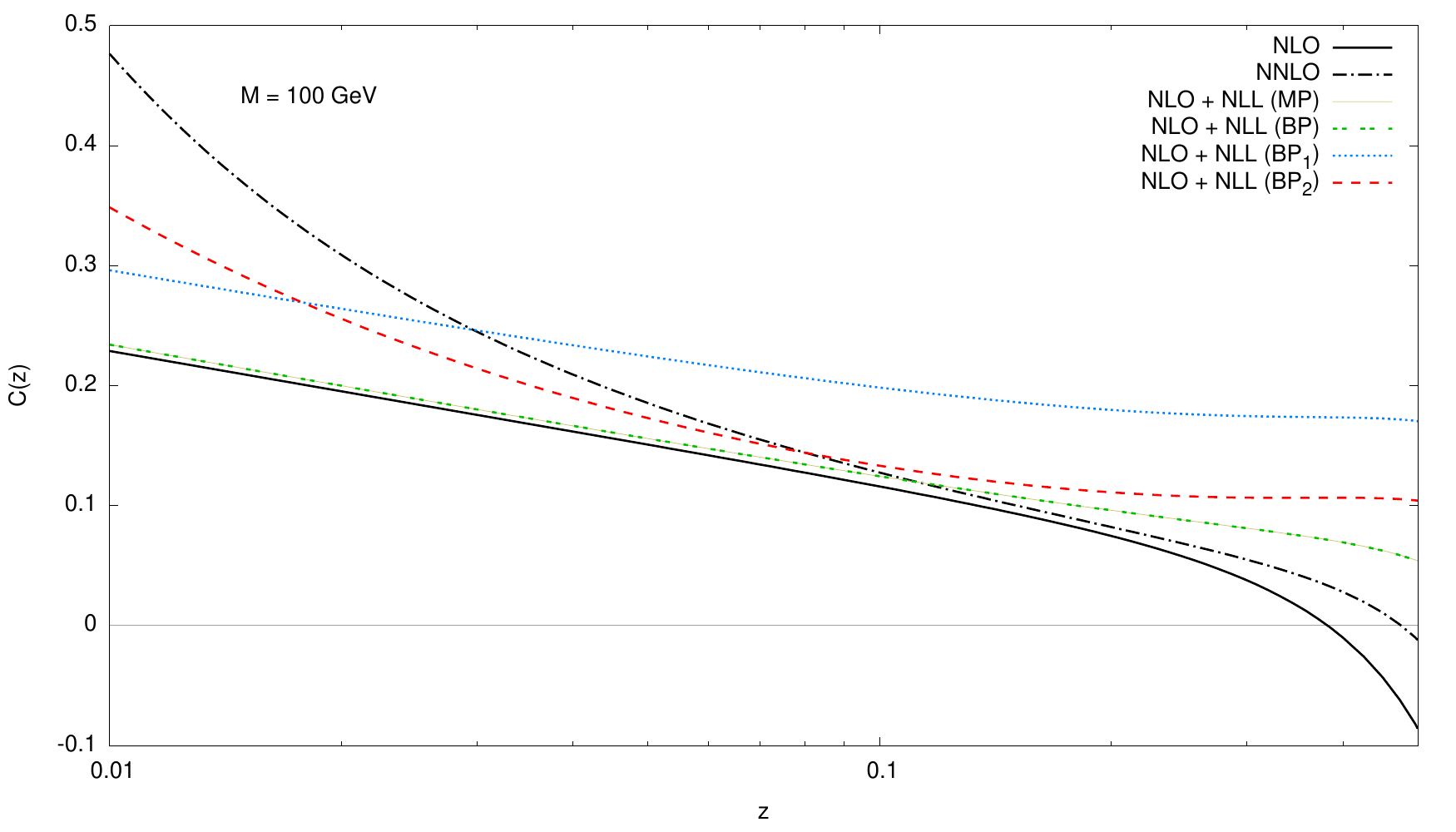}
\includegraphics[width=0.495\textwidth,page=2]{graph_partonic_thesis}\\
\includegraphics[width=0.495\textwidth,page=3]{graph_partonic_thesis}
\includegraphics[width=0.495\textwidth,page=4]{graph_partonic_thesis}
\caption{The next-to-leading log (NLL) resummed  Drell-Yan matched to
  the fixed next-to-leading order (NLO, top row) or
  next-to-next-to-leading order (NNLO bottom row), with various
  resummation prescription. The right plots show the very large $z$
  region.
}
\label{fig:subl}
\end{center}
\end{figure}

In Fig.~\ref{fig:subl} we compare the matched result Eq.~\eqref{eq:match}
for the Drell-Yan coefficient function in the quark-antiquark channel
obtained by including terms up to NLL in the resummed expression, and
either up to order $\as$ (NLO) or $\as^2$ (NNLO) in the
fixed-order expansion, with various resummation prescriptions.
First, we note that the MP and the BP Eq.~\eqref{eq:Borel_prescription}
are essentially indistinguishable (for values of $z$ less than about 0.9,
where the oscillatory behaviour of the minimal prescription sets in).
This is what one expects, since they contain the same subleading terms
and they only differ in the treatment of the high-order divergence.

However, the BP$_1$ Eq.~\eqref{eq:BP1} is seen to differ
by a non-negligible amount from the minimal prescription: at large
$z\gtrsim 0.3$, where the resummation kicks in, it is small in
comparison to the size of the resummation itself, while at small
$z\lesssim 0.03$, where the resummation just leads to unreliable
subleading contributions, it is smaller than the typical higher order
correction, as it is seen by comparing results matched to the NLO and
NNLO. However, in the intermediate $z$ region the subleading terms
introduced by this prescription are uncomfortably large.

The BP$_2$ Eq.~\eqref{eq:BP2} as expected differs less from the MP
both at large and at small $z$; also, it does not introduce
unnaturally large subleading terms for any value of $z$. The
difference between MP and this last version of the Borel prescription
is not negligible but small in comparison to the size of resummation
effects in the region where the resummation is relevant, and it is
smaller than typical higher-order terms 
in the region in which the resummation is not relevant. It can thus be
taken as a reliable estimate of the ambiguity in the resummation.
It is interesting to observe that the subleading terms which are
introduced by the replacement Eq.~\eqref{eq:lognew} grow at small
$z$, and in fact are thus reproducing part of the small-$z$ growth of
perturbative coefficient functions. The fact that inclusion of these
terms is important in keeping the ambiguities of large $z$ resummation
under control suggests that the large $z$ and small $z$ resummation
regions are not well separated, and that there  
might be an interplay between small- and
large-$z$ resummation.

\section{Rapidity distributions resummation}
\label{sec:rap_resumm}

The resummed expression for rapidity distributions was only derived
relatively recently in Refs.~\cite{Mukherjee:2006uu,bolz}, confirming
a conjecture of Ref.~\cite{Laenen:1992ey}.
In this Section, after introducing some basics formulae for rapidity
distributions, we will briefly review this result, and also compare it
to a somewhat different expression later derived in Ref.~\cite{bnx}.

The hadronic rapidity $Y$ distribution for a Drell-Yan pair or a Higgs boson
of invariant mass $M$ produced in hadronic collisions at center-of-mass
energy $\sqrt{s}$ is given by
\beq\label{eq:rap_distr}
\frac{1}{\tau}\frac{d\sigma}{dM^2 dY} =
\sum_{i,j} c_{ij} \int_{x_1^0}^1 \frac{dx_1}{x_1} \int_{x_2^0}^1 
\frac{dx_2}{x_2} \, f_i^{(1)}(x_1)\, f_j^{(2)}(x_2) \,
C_{ij} \( \frac{\tau}{x_1x_2}, y, \as\),
\eeq
where
\beq
x_1^0=\sqrt{\tau}e^Y,\qquad
x_2^0=\sqrt{\tau}e^{-Y},\qquad
\tau=\frac{M^2}{s}\,\qquad
y =Y-\frac{1}{2}\log\frac{x_1}{x_2}
\eeq
and $c_{ij}$ are suitable combination of couplings to make the LO
coefficient function simply equal to $C_{ij}^{(0)}(z,y)=\delta(1-z)\,\delta(y)$
(or zero for channels which do not contribute at LO).
The result is conveniently expressed in terms of new variables $z,u$, defined by
\beq\label{eq:x1x2tozu}
x_1 = \sqrt{\frac{\tau}{z}}  e^Y \sqrt{\frac{z+(1-z)u}{1-(1-z)u}}, \qquad
x_2 = \sqrt{\frac{\tau}{z}}  e^{-Y} \sqrt{\frac{1-(1-z)u}{z+(1-z)u}}
\eeq
with inverse
\beq\label{eq:zutox1x2}
z = \frac{\tau}{x_1 x_2}, \qquad
u = \frac{e^{-2y}-z}{(1-z)\(1+e^{-2y}\)}, \qquad
e^{-2y} = \frac{x_1}{x_2}e^{-2Y};
\eeq
the cross-section then becomes
\beq\label{eq:rap_distr_zu}
\frac{1}{\tau}\frac{d\sigma}{dM^2 dY} =
\sum_{i,j} \int_\tau^1 \frac{dz}{z} \int_0^1 du\;
\Lrap_{ij}(z,u)\,
\bar C_{ij}(z,u,\as)
\eeq
having defined the ``rapidity'' luminosity
\beq\label{eq:Lrap}
\Lrap_{ij}(z,u) = c_{ij}\, f_i^{(1)}(x_1)\, f_j^{(2)}(x_2)
\eeq
(the change of variables Eq.~\eqref{eq:x1x2tozu} is understood implicitly)
and the coefficient function
\beq
\bar C_{ij}(z,u,\as)=\abs{\frac{\partial(\log x_1,\log x_2)}{\partial(\log z,u)}}
C_{ij}\(z,y(z,u),\as\).
\eeq
Note that $\Lrap_{ij}(z,u)$ hides a dependence on $\tau$ and $Y$,
according to Eq.~\eqref{eq:x1x2tozu}.
The partonic threshold region $M^2\to \hat s=x_1x_2s$ corresponds to $z\to 1$.

As already mentioned, large threshold logs only appear in the
quark-antiquark channel for Drell-Yan and gluon-gluon channel for Higgs,
while the contributions from other channels are
suppressed by at least one more power of $1-z$ as $z \to 1$.
Therefore, we suppress flavour indexes, thereby implicitly
understanding that we are considering the logarithmically enhanced channel
(for the $q\bar q$ channel, we also include in $\Lrap$ a sum over quark flavours).

Threshold resummation of rapidity distributions is based on the
observation~\cite{Mukherjee:2006uu,bolz} (conjectured in
Ref.~\cite{Laenen:1992ey}) that at large $z$
the coefficient function $C(z,y,\as)$ factorizes as
\beq
C(z,y,\as)=C(z,\as)\,\delta(y)\[1+\Ord(1-z)\],
\eeq
where $C(z,\as)$ is the rapidity-integrated coefficient.
This is easily proved by rewriting $C(z,y,\as)$ in terms of its
Fourier transform with respect to $y$:
\beq\label{eq:fourier}
\tilde C(z,\omega, \as) = \int_{-\infty}^{+\infty}dy\,e^{i\omega y}\,C(z,y,\as).
\eeq
The integration range in Eq.~\eqref{eq:fourier} is restricted by kinematics to 
$\log\sqrt{z}\leq y\leq-\log\sqrt{z}$.
Hence, one may expand the exponential $e^{i\omega y}$ in powers
of $y$,
\beq
\tilde C(z,\omega,\as)
=\int_{\log\sqrt{z}}^{-\log\sqrt{z}}dy\,C(z,y,\as)\[1+\Ord(y)\]
=C(z,\as)\[1+\Ord(1-z)\],
\eeq
where $C(z,\as)$ is the rapidity-integrated coefficient function.
Hence, $\tilde C(z,\omega,\as)$ is independent of $\omega$ up to terms
which as $z\to1$ are suppressed by powers of $\abs{\log z}=1-z+\Ord((1-z)^2)$,
and one immediately gets the desired factorized form:
\beq
C(z,y,\as)=
\int_{-\infty}^{+\infty}\frac{d\omega}{2\pi}\,e^{-i\omega y}\,\tilde C(z,\omega,\as)
=C(z,\as)\,\delta(y)\[1+\Ord(1-z)\].
\eeq
According to Eq.~\eqref{eq:x1x2tozu}, $y=0$ corresponds to $u=1/2$,
and noting that
\beq
\abs{\frac{\partial(\log x_1,\log x_2)}{\partial(\log z,u)}}
\abs{\frac{\partial u}{\partial y}}=
\frac{z^2}{\tau}\abs{\frac{\partial u}{\partial y}
\(
\frac{\partial x_1}{\partial z}\frac{\partial x_2}{\partial u}
-
\frac{\partial x_2}{\partial z}\frac{\partial x_1}{\partial u}\)}
=1
\eeq
we have immediately
\beq\label{eq:barC_log}
\bar C(z,u,\as) = C(z,\as)\,\delta\(u-\frac12\)\[1+\Ord(1-z)\]
\eeq
with the same integrated coefficient $C(z,\as)$.
Up to power-suppressed terms we can thus write
\begin{align}
\frac{1}{\tau}\frac{d\sigma^{\rm res}}{dM^2 dY}
&= \int_\tau^1 \frac{dz}{z} \int_0^1 du\; \Lrap(z,u)\,\delta\(u-\frac12\) C^{\rm res}(z,\as)\\
&= \int_\tau^1 \frac{dz}{z} \; \Lrap\(z,\frac12\)\, C^{\rm res}(z,\as).
\label{eq:sigma_rap_res}
\end{align}
Because Eq.~\eqref{eq:sigma_rap_res} only depends on the rapidity-integrated
coefficient function, threshold resummation is simply performed by using for the
latter the resummed expressions discussed in Sect.~\ref{sec:thr_resumm}.
Moreover, by inspection of Eq.~\eqref{eq:x1x2tozu}, we see that, for $u=1/2$,
$x_1,x_2$ depend on $z$ through the ratio $\tau/z$.
Therefore, Eq.~\eqref{eq:sigma_rap_res} has the form of a convolution
product: this greatly simplify its treatment, which is then analogous
to that of the resummed integrated cross-section, with the replacement 
\beq\label{eq:lumireplacement}
\Lum\(\frac{\tau}{z}\)\to \Lrap\(z,\frac{1}{2}\).
\eeq
Note that the integration range in Eq.~\eqref{eq:sigma_rap_res} actually
starts from $z=\tau e^{2\abs{Y}}$, because  $\Lrap(z,1/2)$ vanishes for $z<\tau e^{2\abs{Y}}$:
\beq
\frac{1}{\tau}\frac{d\sigma^{\rm res}}{dM^2 dY} 
= \int_{\tau e^{2\abs{Y}}}^1 \frac{dz}{z}\,C^{\rm res}(z,\as)\,\Lrap\(z,\frac{1}{2}\).
\label{eq:sigma_rap_res1}
\eeq
This underlines that at large rapidity the threshold region always gives
the main contributions, even when the $\tau$ is small.

Now we discuss briefly a different way of relating the resummation of
rapidity distribution to that of the inclusive cross-section which
has been more recently presented in Ref.~\cite{bnx}.
Starting from Eq.~\eqref{eq:rap_distr_zu}, the authors 
observe that at NLO the logarithmically enhanced terms in $\bar C(z,u,\as)$
in the partonic threshold limit $z\to 1$ appear as coefficients of
the combination $\delta(u)+\delta(1-u)$ (see Eq.~\eqref{Fzu} to check explicitly).
At higher orders, logarithmic terms in general multiply non-trivial functions
of $u$, but Eq.~\eqref{eq:x1x2tozu} implies that the $u$ dependence of $x_1,x_2$
is of order $1-z$, so
\beq\label{Cbar}
\bar C(z,u,\as) = \[\delta(u)+\delta(1-u)\] \frac{C(z,\as)}{2} \[1+\Ord(1-z)\],
\eeq
where the factor $1/2$ has been fixed integrating in $u$ from $0$ to $1$
and comparing with the same integral of Eq.~\eqref{eq:barC_log}.
It follows that
\beq\label{eq:sigma_rap_res_scet}
\frac{1}{\tau}\frac{d\sigma^{\rm res}}{dM^2\,dY} =
\int_\tau^1 \frac{dz}{z}\, \frac{\Lrap(z,0) +\Lrap(z,1)}{2}\, C^{\rm res}(z,\as).
\eeq
Eq.~\eqref{eq:sigma_rap_res_scet} differs by power suppressed terms
from the resummed result previously derived Eq.~\eqref{eq:sigma_rap_res},
as it is easy to check explicitly.
Indeed, using Eq.~\eqref{eq:x1x2tozu} and expanding
$\Lrap(z,0)$, $\Lrap(z,1)$ and $\Lrap(z,1/2)$ 
in powers of $z$ about $z=1$ it is easy to check that
\beq
\frac{\Lrap(z,0)+\Lrap(z,1)}{2} = \Lrap\(z,\frac{1}{2}\) + \Ord\[(1-z)^2\].
\eeq
Because the difference is suppressed by two powers of $1-z$ one expects
it to be small, and indeed we have checked that (using the Borel prescription)
the difference between Eqs.~\eqref{eq:sigma_rap_res} and \eqref{eq:sigma_rap_res_scet}
is negligible, and specifically much smaller than the difference between
different choices of subleading terms (see Sect.~\ref{sec:subl}).

We note that, because $\Lrap(z,0)$ and $\Lrap(z,1)$ are not functions
of $\tau/z$, the form  Eq.~\eqref{eq:sigma_rap_res_scet} of the resummed
result does not have the structure of a convolution and thus would
require a separate numerical implementation. 
For the same reason, a comparison between
Eqs.~\eqref{eq:sigma_rap_res} and~\eqref{eq:sigma_rap_res_scet} using the minimal
prescription cannot be performed, because 
Eq.~\eqref{eq:sigma_rap_res_scet} cannot be expressed in terms
of the Mellin transform of $C^{\rm res}(z)$.
We will disregard  the
form Eq.~\eqref{eq:sigma_rap_res_scet} of rapidity distributions henceforth.

\section{Numerical implementation}
\label{sec:resumm_numerical}

For the sake of phenomenology, Chap.~\ref{chap:pheno}, an efficient numerical implementation 
of resummed results using the various prescriptions is necessary.
Such an implementation was hitherto not available and it will be
discussed here.

\subsection{Minimal prescription}
The minimal prescription
involves the numerical evaluation of the complex integral
\beq
\label{eq:main2}
\frac{1}{2\pi i} \int_{c-i\infty}^{c+i\infty}dN\,\tau^{-N} 
\Lum(N) \, C^{\rm res}(N,\as)
\eeq
where $\Lum(N)$ is the Mellin transform of either $\Lum(z)$, Eq.~\eqref{eq:luminosity},
for the inclusive cross-section or $\Lrap(z,\frac12)$, Eq.~\eqref{eq:Lrap},
for the rapidity distribution.
The integration path is usually chosen as in Fig.~\ref{fig:MP_path}
in order to make the integral absolutely convergent.
However, parton densities obtained from data analysis are commonly
available  as functions of $z$ in
interpolated form through common interfaces such as
LHAPDF~\cite{Bourilkov:2006cj}, and the numerical 
evaluation of their Mellin transform does not converge
along the path of integration (specifically, for 
${\rm Re}\;N<0$) and must be defined by analytic
continuation.
The option of  applying the MP
to the partonic cross-section, and then convoluting the result with
the parton luminosity in momentum space is not viable, because
the MP does not have the structure of a convolution: the
partonic cross-section does not vanish for $z\geq1$, and it oscillates wildly
in the region $z\sim 1$, see discussion in Sect.~\ref{sec:MP}
This problem is discussed in Ref.~\cite{cmnt}, where it is handled by
adding and subtracting the results of the minimal prescription
evaluated with a fake luminosity which allows for analytic integration.

Another possibility, adopted for example in
Ref.~\cite{bolz}, is to use parton distributions whose Mellin
transform can be computed exactly at the initial scale. This, however,
greatly restricts the choice of parton distributions, and specifically
it prevents the use of current state-of-the-art PDFs from global fits.
It is  thus not
suitable for precision phenomenology.

The method adopted here, based on an idea suggested long
ago~\cite{Furmanski:1981ja}, consists of expanding the luminosity in
$z$ space, $\Lum(z)$ or $\Lrap(z,1/2)$,
on a basis of polynomials whose Mellin transform can be computed
analytically. We have chosen Chebyshev polynomials, for which
efficient algorithms for the computation of the expansion coefficients
are available.  The details of the procedure are illustrated in
Appendix~\ref{app:cheb}. The obvious drawback of this
procedure is that it must be carried on for each value of the scale
$\muf$ and, in the case of rapidity distribution, for each value of
$\tau$ and $Y$.

\subsection{Borel prescription}

We now turn to the discussion of an implementation issue which is
specific to the Borel prescription, and has to do with the choice
of the cutoff $W$. As discussed in Sect.~\ref{sec:borel} the 
minimal choice is $W=2$; however
when $W\geq 1$
the $\xi$ integration path in Eq.~\eqref{eq:Borel_prescription}
includes values of $\xi$
with ${\rm Re}\;\xi<-1$, for which the convolution integral diverges.
As discussed  in Ref.~\cite{afr}, the integral can be nevertheless
defined by analytic continuation, by subtracting and adding back from
$\Lum(\frac{\tau}{z})/z$ its Taylor expansion in $z$ around $z=1$: the
subtracted  integrals converge, and the compensating terms can be
determined analytically and continued in the desired region.

Here we propose~\cite{bfr2} a different method which is numerically much
more efficient. The idea is to perform the convolution integral 
with the luminosity analytically, before 
the complex $\xi$-integral. To be able to compute the integral analytically,
we will use again Chebyshev polynomials for the relevant combination of luminosity.
The actual form of the result depend on whether choice of subleading
terms is adopted.
Since we have verified numerically that the naive choice of Eq.~\eqref{eq:Borel_prescription}
coincides with the MP up to very large values of $\tau$,
we concentrate first on Eqs.~\eqref{eq:BP1} and \eqref{eq:BP2}.
For $\Re \xi>0$, we can use the identity Eq.~\eqref{eq:plus_useful1} to write
\beq\label{eq:BP_dependence_z}
\plusq{(1-z)^{\xi-1}} = (1-z)^{\xi-1} - \frac{1}{\xi}\delta(1-z),
\eeq
and this formula can be analytically extended to all complex values of $\xi\neq0$.
As a result, the $\delta(1-z)$ terms cancel and the resulting $z$-dependence
is contained, respectively, in
\beq
(1-z)^{\xi-1} \qquad \text{and} \qquad z^{-\xi/2} (1-z)^{\xi-1}
\eeq
without plus distribution.
In the first case, Eq.~\eqref{eq:BP1}, the convolution integral can be written as
\beq\label{eq:conv_borel}
\int_{\tau}^1 \frac{dz}{z}\,(1-z)^{\xi-1}\, \Lum\(\frac{\tau}{z}\)
= \int_{\tau}^1 dz \, (1-z)^\xi \, g(z,\tau)
+\Lum(\tau) \int_\tau^1dz\, (1-z)^{\xi-1}
\eeq
where we have defined
\beq\label{eq:g_zx}
g(z,\tau)= \frac{1}{1-z}
\left[ 
\frac{1}{z}\,\Lum\(\frac{\tau}{z}\)-\Lum(\tau)\right].
\eeq
The subtraction introduced is justified because the function $g(z,\tau)$
is more easily approximated by an expansion on the basis
of Chebyshev polynomials than $\Lum\(\frac{\tau}{z}\)$.
Proceeding as in Appendix~\ref{sec:cheb_borel} we find
\beq\label{eq:gztau_cheb}
g(z,\tau) = \sum_{p=0}^n b_p\,(1-z)^p,
\eeq
where $n$ is the order of the approximation, and the coefficients
$b_p = b_p(\tau,\muf^2)$ can be computed by numerical methods.
We have\footnote{Note that the LO is not present in this expression.}
\begin{multline}\label{eq:BP1_final}
\frac{1}{\tau}\frac{d\sigma^{\rm BP_1}}{dM^2} = 
\frac{1}{2\pi i}
\oint \frac{d\xi}{\xi\,\Gamma(\xi)}
\int_0^W \frac{dw}{\ab} \, e^{-\frac{w}{\ab}}\,\Sigma\left(\frac{w}{\xi}\right)\\
\times\[\sum_{p=0}^n b_p\,\frac{(1-\tau)^{p+\xi+1}}{p+\xi+1}
+\Lum(\tau)\,\frac{(1-\tau)^\xi}{\xi} \];
\end{multline}
the term in square brackets has poles in $\xi=0,-1,\ldots, -n$
in correspondence of zeros of $1/\Gamma(\xi)$. Thus, the $\xi$
integrand has only a branch cut in $-w\leq\xi\leq 0$, and the contour
in the $\xi$ plane must encircle just the cut.
Then this expression is valid for arbitrarily large values of
the cutoff $W$, hence it can also be used in the case $W>2$,
which would require multiple subtractions if the method of Ref.~\cite{afr}
is used.

In fact, this result is applicable for the inclusive cross-section only,
while the rapidity distribution requires some care.
Indeed, even if the formalism can be the same, since $\Lrap(z,\frac12)$
is zero below $z=\tau e^{2\abs{Y}}$ it is convenient to compute directly
the convolution integral with such lower limit --- otherwise, the approximation
of $g(z,\tau)$ would be unstable.
In practice, it is sufficient to substitute $\Lum(\tau/z)$ with $\Lrap(z,\frac12)$
and $\tau$ with $\tau e^{2\abs{Y}}$ in the expression above, obtaining
\begin{multline}\label{eq:BP1_rap_final}
\frac{1}{\tau}\frac{d\sigma^{\rm BP_1}}{dM^2 dY} = 
\frac{1}{2\pi i}
\oint \frac{d\xi}{\xi\,\Gamma(\xi)}
\int_0^W \frac{dw}{\ab} \, e^{-\frac{w}{\ab}}\,\Sigma\left(\frac{w}{\xi}\right)\\
\times\[\sum_{p=0}^n b_p\,\frac{(1-\tau e^{2\abs{Y}})^{p+\xi+1}}{p+\xi+1}
+\Lrap\(1,\frac12\)\,\frac{(1-\tau e^{2\abs{Y}})^\xi}{\xi} \]
\end{multline}
where now $b_p=b_p(\tau,\muf^2)$ are the coefficient of the expansion
\eqref{eq:gztau_cheb} for the function
\beq
g(z,\tau)= \frac{1}{1-z}
\left[ 
\frac{1}{z}\,\Lrap\(z,\frac12\)-\Lrap\(1,\frac12\)\right]
\eeq
in the range $\tau e^{2\abs{Y}}\leq z \leq 1$ (see App.~\ref{sec:cheb_borel}).

The generalization to the other choice of subleading terms, Eq.~\eqref{eq:BP2},
is now straightforward: using the same steps for the computation of the
convolution integral we find
\begin{multline}\label{eq:BP2_final}
\frac{1}{\tau}\frac{d\sigma^{\rm BP_2}}{dM^2} =
\frac{1}{2\pi i} \oint \frac{d\xi}{\xi\,\Gamma(\xi)}
\int_0^W \frac{dw}{\ab} \, e^{-\frac{w}{\ab}}\,\Sigma\left(\frac{w}{\xi}\right)\\
\times\[\sum_{p=0}^n b_p\,\mathrm{B}\(\xi+p+1,1-\frac{\xi}{2};1-\tau\)
+\Lum(\tau)\,\mathrm{B}\(\xi, 1-\frac{\xi}{2}; 1-\tau\) \],
\end{multline}
where
\beq
\mathrm{B}(b,a;1-\tau)=\int_\tau^1 dz\, z^{a-1}(1-z)^{b-1}
\eeq
is the incomplete Beta function. The function $\mathrm{B}(b,a;1-\tau)$ is singular at $b=0$.
In Eq.~\eqref{eq:BP2_final} the first argument of the $\mathrm{B}$ functions
in the integrand vanishes for non positive integer values of
$\xi=0,-1,\dots$, in correspondence of zeros of $1/\Gamma(\xi)$.
Thus, again, the remaining singularities are only represented by the branch cut.
The rapidity distribution is obtained from this expression with the same
changes described above.

Finally, we briefly discuss the case of Eq.~\eqref{eq:Borel_prescription}.
Using for $\Re\xi>0$ the identity
\beq
\plusq{\log^{\xi-1}\frac1z} = \log^{\xi-1}\frac1z - \Gamma(\xi)\,\delta(1-z),
\eeq
we get easily
\begin{multline}\label{eq:BP0_final}
\frac{1}{\tau}\frac{d\sigma^{\rm BP}}{dM^2} = 
\frac{1}{2\pi i}
\oint \frac{d\xi}{\xi\,\Gamma(\xi)}
\int_0^W \frac{dw}{\ab} \, e^{-\frac{w}{\ab}}\,\Sigma\left(\frac{w}{\xi}\right)\\
\times\[\sum_{p=0}^n \tilde b_p\,\gamma\(\xi+p+1,\log\frac1\tau\)
+\Lum(\tau)\,\gamma\(\xi,\log\frac1\tau\) \];
\end{multline}
where $\gamma(\xi,a)$ is the truncated Gamma function (see App.~\ref{sec:Gamma}),
and with
\beq\label{eq:gztau_cheb_log1z}
\tilde g(z,\tau)= \frac{1}{\log\frac1z}
\left[ 
\frac{1}{z}\,\Lum\(\frac{\tau}{z}\)-\Lum(\tau)\right]
=\sum_{p=0}^n \tilde b_p\,\log^p\frac1z,
\eeq
see App.~\ref{sec:cheb_borel}.
It would be in principle possible to use the function $g(z,\tau)$,
Eq.~\eqref{eq:gztau_cheb}, but the integral would be complicated a bit;
the result would be
\begin{multline}\label{eq:BP0_final_code}
\frac{1}{\tau}\frac{d\sigma^{\rm BP}}{dM^2} = 
\frac{1}{2\pi i}
\oint \frac{d\xi}{\xi\,\Gamma(\xi)}
\int_0^W \frac{dw}{\ab} \, e^{-\frac{w}{\ab}}\,\Sigma\left(\frac{w}{\xi}\right)\\
\times\[\sum_{j=0}^{n+1} \hat b_j\,\gamma\(\xi,(1+j)\log\frac1\tau\)
+\Lum(\tau)\,\gamma\(\xi,\log\frac1\tau\) \];
\end{multline}
with
\beq
\hat b_j = (-)^j \sum_{p=j-1}^n b_p\, \binom{p+1}{j}.
\eeq
Extension to rapidity distributions is straightforward.

\chapter{High-energy resummation}
\label{chap:small-x}
\minitoc

\noindent
In this Chapter we will cover the resummation of high-energy logarithms.
Such resummation mainly affects the evolution of the gluon, even though also
the quark singlet gets contributions at next-to-leading log.
We will see that resummation is provided by combining the GLAP evolution equation
with the so called BFKL equation: after introducing a first leading resummation,
we will show that the result is unstable for perturbative corrections,
and it doesn't describe the observed phenomena.
Then we will go further by introducing two ingredients:
symmetrization and running-coupling resummation.
Such ingredients allow to find stable results, which are compatible with the observations.
To begin with, we introduce a factorization theorem valid in the high-energy
regime, from which the BFKL can be derived.

\section{High-energy factorization}

The mass (or collinear) factorization theorem Eq.~\eqref{eq:DIS_structure_functions_N}
is well established~\cite{Ellis:1978sf,Ellis:1978ty}.
It is a perturbative factorization, i.e.\ it is valid order by order in perturbation
theory up to terms which are of higher orders and to terms suppressed by negative powers of
the hard scale $Q^2$, called higher twists.
The perturbativity is guaranteed by the condition $Q^2\gg \Lambda_{\rm QCD}$,
in such a way that $\as(Q^2)$ is small enough.

However, this is not always enough for realistic applications.
Consider for definiteness the DIS process discussed in Sect.~\ref{sec:DIS}:
if the available partonic initial energy squared $\hat s=2\hat pq=zQ^2/x$ is large,
the convergence of the perturbative series is spoiled.
Indeed, in the coefficient function logarithms of the ratio of the energies $\sqrt{\hat s}$ and $Q$,
i.e.\ logarithms of $x/z$, appear at each order, and if $\hat s\gg Q^2$ (high-energy regime)
all such terms in the perturbative series are equally important.
Then, the perturbative result obtained in the context of
collinear factorization is not enough for the study of the high-energy regime.

\begin{figure}[tbh]
  \centering
  \includegraphics[width=0.4\textwidth]{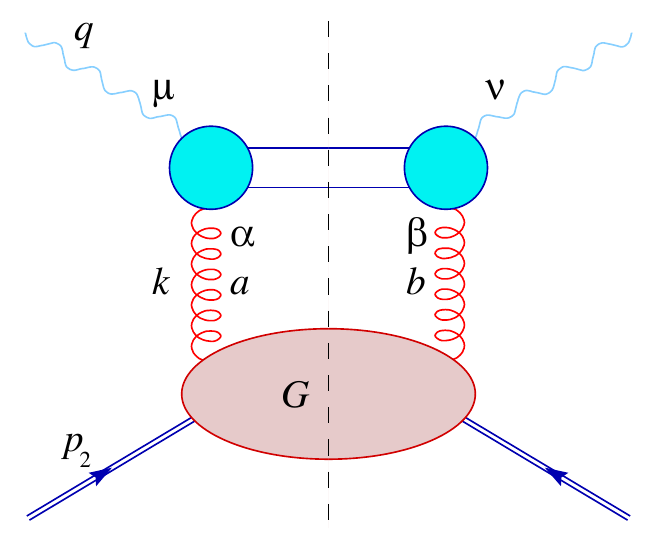}
  \caption{Two-gluon reducible diagrams contributing to the high-energy part of DIS structure functions.
    (Figure taken from Ref.~\cite{Colferai:1999em}.)}
  \label{fig:2gluons_red}
\end{figure}

A generalization of the collinear factorization valid in the high-energy regime
has been proposed long ago \cite{Catani:1990xk,Catani:1990eg,Catani:1993ww,Catani:1993rn,Catani:1994sq}.
Consider again the DIS: it can be proved that the dominant contributions
at high energy are given by diagrams in which a gluon is exchanged in the $t$-channel,
or otherwise stated, from diagrams whose forward amplitude is two-gluon reducible,
Fig.~\ref{fig:2gluons_red}.
For such diagrams, the structure function $F_2$ satisfies the fundamental factorization
\beq\label{eq:kt-factorization}
F_2(x,Q^2) = \int_x^1\frac{dz}{z} \int d^2\vk \;C\(\frac{x}{z},\frac{Q^2}{\vk^2},\as(Q^2)\)\, {\cal F}(z,\vk)
\eeq
where $C\(\frac{x}{z},\frac{Q^2}{\vk^2},\as(Q^2)\)$ is called \emph{off-shell partonic cross-section}
and represents the contribution from a partonic diagram in the $t$-channel where the gluon
is off-shell (upper part of the diagram),
and ${\cal F}(z,\vk)$ is a non-perturbative contribution analogous to the PDFs
(lower part of the diagram, including gluon propagator);
the integration limits of the $\vk$ integral are set by kinematics.
Eq.~\eqref{eq:kt-factorization} is diagonalized by a double Mellin-Laplace transform,
\beq\label{eq:HE_fact}
F_2(N,M) \equiv \int_0^1 dx\,x^{N-1} \int_0^\infty \frac{dQ^2}{Q^2}\,\(\frac{Q^2}{\Lambda^2}\)^{-M} F_2(x,Q^2),
\eeq
where $\Lambda$ is some reference scale ($F_2(x,Q^2)$ depends logarithmically on $Q^2$).
Then in double Mellin space we have
\beq\label{eq:HE_fact_NM}
F_2(N,M) = C\(N,M,\as(Q^2)\)\, {\cal F}(N,M)
\eeq
with
\begin{align}
  C\(N,M,\as(Q^2)\) &= \int_0^1 dz\,z^{N-1} \int_0^\infty \frac{d\rho}{\rho}\,\rho^{-M}\, C\(z,\rho,\as(Q^2)\)\\
  {\cal F}(N,M) &= \int_0^1 dz\,z^{N-1} \int d^2\vk \,\(\frac{\vk^2}{\Lambda^2}\)^{-M} {\cal F}(z,\vk).
\end{align}

The high-energy factorization (or $\kt$-factorization) formula Eq.~\eqref{eq:kt-factorization}
is proved to be valid in the high-energy limit
\beq
\hat s \gg Q^2 
\eeq
regardless the size of the gluon momentum $\vk^2$,
which takes all accessible values in the integral.
Conversely, the collinear factorization is valid in the limit
$Q^2\gg \vk^2$, because this region (the collinear region) dominates
the $\vk$ integral.
In the limit
\beq
\hat s \gg Q^2 \gg \vk^2
\eeq
both factorization are reliable, and then in such limit they would coincide.
Indeed, it has been proved in Ref.~\cite{bf405} that, in this regime,
standard collinear factorization can be derived in a straightforward way
from the high-energy factorization theorem.

\section{BFKL equation}

The high-energy factorization formula Eq.~\eqref{eq:HE_fact_NM} depends
on $N$ and $M$ on the same footing. This reflects a symmetry of the
factorization formula Eq.~\eqref{eq:HE_fact} if written in terms of the variables
\beq\label{eq:sx_var}
\xi = \log\frac{1}{x} , \qquad t=\log\frac{Q^2}{\Lambda^2}.
\eeq
This suggests, and in fact it is, that an evolution equation in the
variable $\xi$ could be derived. Such an equation is called
the \emph{Balitsky-Fadin-Kuraev-Lipatov (BFKL) equation}~\cite{Lipatov:1976zz,
Kuraev:1976ge,Kuraev:1977fs,Balitsky:1978ic}
\beq\label{eq:BFKL_eq_integral}
\frac{d}{d\xi}f(\xi,q^2) = \int_0^\infty \frac{dk^2}{k^2} \, K\(\as,\frac{q^2}{k^2}\)\,f(\xi,k^2),
\eeq
where $K(\as,q^2/k^2)$ is a (fixed-coupling) unintegrated kernel
(similar to the splitting functions in the case of GLAP).
In Ref.~\cite{bf405}, with the same technique used to derive the GLAP equation,
the BFKL equation is derived from the high-energy factorization.

Taking the Mellin transform with respect to $q^2$
we get another (more suitable) form of the BFKL equation
\beq\label{eq:BFKL_equation}
\frac{d}{d\xi} f(\xi,M) = \chi(\as,M)\, f(\xi,M),
\eeq
where $M$ is the variable conjugate to $t$ and
\beq\label{eq:chi_from_K}
\chi(\as,M) = \int_0^\infty \frac{dq^2}{q^2} \(\frac{q^2}{k^2}\)^{-M} K\(\as,\frac{q^2}{k^2}\)
\eeq
is the integrated BFKL kernel and
\beq
f(\xi,M) = \int_0^\infty \frac{dq^2}{q^2} \(\frac{q^2}{\Lambda^2}\)^{-M} f(\xi,q^2).
\eeq
Note that, since a Mellin transform with respect to $t$ is taken,
the BFKL equation written in the form of Eq.~\eqref{eq:BFKL_equation}
is only valid for fixed coupling $\as$.
Running coupling effects will be discussed in Sect.~\ref{sec:BFKL_running_coupling}.

As the GLAP equation describes the evolution of the PDFs in the $t$ direction,
the BFKL equation describes the evolution of $f$
in the direction of the variable $\xi$ (related to $x$ by Eq.~\eqref{eq:sx_var}).
Then, the BFKL equation resums the small-$x$ (large-$\xi$) logarithms,
as the GLAP equation resums the large-$t$ dependence.

Hence, the PDF appearing in the BFKL equation has to be identified with the largest
eigenvector $f_+$, since it contains all the small-$x$ singularities.
For practical reasons, we have chosen
\beq
f(\xi,q^2) \equiv x \, f_+(x, q^2),
\eeq
since with this factor of $x$ the GLAP equation in Mellin space becomes simply
\beq\label{eq:sx_GLAP}
\frac{d}{dt}f(N,q^2) = \gamma\(\as(q^2),N\)\, f(N,q^2),
\eeq
where $\gamma(\as,N)$ is
\beq
\gamma(\as,N) \equiv \gamma_+(\as,N).
\eeq
To further simplify the notations we will write the perturbative expansion of the anomalous dimension as
\beq
\gamma(\as,N) = \as \,\gamma_0(N) + \as^2 \,\gamma_1(N) + \Ord(\as^3)
\eeq
and that of the BFKL kernel as
\beq
\chi(\as,M) = \as \,\chi_0(N) + \as^2 \,\chi_1(N) + \Ord(\as^3).
\eeq
The LO and the NLO contributions to the BFKL kernel are known~\cite{FL};
the LO is
\beq\label{eq:chi0}
\chi_0(M) = \frac{N_c}{\pi} \[2\psi(1) - \psi(M) - \psi(1-M)\]
\eeq
where $\psi$ is the DiGamma function (see App.~\ref{sec:Gamma}),
and the NLO is given in App.~\ref{sec:NLO_BFKL_kernel}.

\section{The Double-Leading approximation}

As we have seen in Sect.~\ref{sec:glap_singlet_sector}, the singlet splitting functions
are unstable at small $x$,\footnote{At large $x$, instead, we have seen that the diagonal
terms are enhanced, but such enhancement is the same to all orders in $\as$,
leaving the perturbative expansion stable at large $x$.}
in the sense that in the generic order $\as^n$ term a tower
of powers of $\log x$ up to $n$ appears: then, if $\as \log\frac1x\sim1$, each term
of the perturbative expansion of the splitting functions is equally important and
the series must be summed.
Note that the condition $\as \log\frac1x\sim1$, the high-energy regime,
is often satisfied by the kinematic conditions of HERA.

The leading singularities at small $x$ are contained in the gluon singlet splitting functions
$P_{gg}$ and $P_{gq}$, while the quark splitting functions have singularities suppressed by
one power of $\as$ more, i.e.\ they are next-to-leading.
The coefficient functions are next-to-leading as well:
we then conclude that the most significant effect of small-$x$ resummation
appears in the gluon splitting functions, and hence in the gluon PDF.
Moreover, it is proved in Ref.~\cite{Catani:1994sq} that resummation of quark
splitting functions and of coefficient functions is related to the gluon resummation.
We then concentrate our attention to the gluon splitting functions, and we will discuss
in Sect.~\ref{sec:res_anom_dim} the next-to-leading effect of quarks.

To be more precise, we could work in the limit $n_f=0$, where there are only gluons in the game,
and everything would be clean and perfectly well defined, but the inclusion of quarks would
be difficult.
Then, it is better to work with the eigenvectors of the singlet sector: indeed, in schemes
like \MSbar\ and DIS, the ``smallest'' eigenvalue is suppressed at small $x$
by one power of $x$ with respect to the ``largest'', to all orders in $\as$.
In the following we will then use the largest eigenvector instead of the gluon:
in the limit $n_f=0$ the two would coincide.

\subsection{Duality relation}

In the region where both GLAP evolution and BFKL evolution are reliable
(large $t$ and large $\xi$ respectively), the solution of both equations should
coincide (at leading twist). Then, we can take the double Mellin transform
and get the equations
\begin{subequations}
\label{eq:GLAP-BFKL_dm}
\begin{align}
  \[ M - \gamma(\as,N) \] f(N,M) &= F_0(N) \label{eq:GLAP_double_mellin}\\
  \[ N - \chi(\as,M)   \] f(N,M) &= \tilde F_0(M)\label{eq:BFKL_double_mellin}
\end{align}
\end{subequations}
where we have defined the boundary conditions
\beq
F_0(N) = \left[ e^{-Mt}f(N,t) \right]_{t=-\infty}^{t=+\infty} ,\qquad
\tilde F_0(M) = \left[ e^{-N\xi}f(\xi,M) \right]_{\xi=0}^{\xi=+\infty}.
\eeq
Note that both Eqs.~\eqref{eq:GLAP-BFKL_dm} are valid only
in the fixed-coupling case; the extension of these equations and
of the duality we are going to derive will be treated in Sect.~\ref{sec:BFKL_running_coupling}.

The solutions of Eqs.~\eqref{eq:GLAP-BFKL_dm} are, respectively,
\beq\label{eq:GLAP-BFKL_sol}
f(N,M) = \frac{       F_0(N)}{M-\gamma(\as,N)},\qquad
f(N,M) = \frac{\tilde F_0(M)}{N-  \chi(\as,M)}.
\eeq
The leading twist behaviour of these solutions is given by the position of the perturbative pole.
Indeed, inverting the $M$-Mellin transform we get, respectively,
\begin{align}
  f(N,t) &= \int_{c-i\infty}^{c+i\infty}\frac{dM}{2\pi i}\, e^{Mt} \frac{F_0(N)}{M-\gamma(\as,N)}
  = F_0(N) \, e^{\gamma(\as,N)t}\\
  f(N,t) &= \int_{c-i\infty}^{c+i\infty}\frac{dM}{2\pi i}\, e^{Mt} \frac{\tilde F_0(M)}{N-\chi(\as,M)}
  = \frac{\tilde F_0(\bar M)}{-\chi'(\as,\bar M)} \, e^{\bar M t} 
+ \text{higher twist}
\end{align}
where $\bar M$ is the position of the rightmost\footnote{Rightmost in the
``collinear'' region, i.e.\ $M<\frac12$.}
pole of the BFKL solution Eq.~\eqref{eq:GLAP-BFKL_sol} given by
\beq\label{eq:pole_condition}
\chi(\as, \bar M) = N.
\eeq
Hence the consistency of the two solutions at leading twist requires the validity of the duality relation \cite{bf405}
\beq\label{eq:duality}
\chi\(\as,\gamma(\as,N)\) = N
\qquad \leftrightarrow \qquad
\gamma\(\as,\chi(\as,M)\) = M
\eeq
(position of the pole) and the relation
\beq
F_0(N) = -\frac{\tilde F_0\(\gamma(\as,N)\)}{\chi'\(\as,\gamma(\as,N)\)}
\qquad \leftrightarrow \qquad
\tilde F_0(M) = -\frac{F_0\(\chi(\as,M)\)}{\gamma'\(\as,\chi(\as,M)\)}
\eeq
(pole coefficient matching).

Note that the anomalous dimension and BFKL kernel, if computed both at fixed order,
are never dual.
To see this we note that, for example, momentum conservation
Eq.~\eqref{eq:PDF_momentum_conservation} requires
\beq
\gamma(\as,N=1) = 0
\eeq
to all orders, which in turn implies, by duality
\beq
\chi(\as,M=0) = 1;
\eeq
however, the BFKL kernel at fixed order is bad behaved in the $M\sim0$ region,
and in particular it has in $M=0$ a simple pole at LO, and a double pole with opposite sign at NLO
(in general, at order $\as^k$ it behaves as $(-M)^{-k}$).
Indeed, the dual of, let's say, the LO anomalous dimension contains all powers of $\as$,
and therefore it cannot coincide with any fixed-order BFKL kernel.

Therefore, the duality Eq.~\eqref{eq:duality} is intended to be valid provided all the relevant
contributions in the regime of validity (small $N$ and $M$) are included.
Then, in practice, it would be valid only if the small-$N$ contributions to the anomalous dimension
are resummed and/or the small-$M$ contributions to the BFKL kernel are resummed.
In the following, we will revert this argument and \emph{use} the duality to perform such resummation.

\subsection{Double-Leading approximation}
\label{sec:DL-approx}

The duality Eq.~\eqref{eq:duality} allows us to construct
an expansion of $\gamma$ and $\chi$ in powers of $\as$ at fixed $\as/N$ and $\as/M$ respectively \cite{Altarelli:1999vw}:
\begin{align}
\gamma(\as,N) &= \gamma_s(\as/N) + \as \,\gamma_{ss}(\as/N) + \ldots \\
\chi(\as,M) &= \chi_s(\as/M) + \as \,\chi_{ss}(\as/M) + \ldots \label{eq:chis_exp}
\end{align}
Let's consider $\chi$: all we say applies to $\gamma$ as well.
To derive the relations between $\chi_s, \chi_{ss}$ and $\gamma_0, \gamma_1$
we insert the expansion \eqref{eq:chis_exp} into the duality relation \eqref{eq:duality},
\beq
\gamma_0\[\chi_s\(\frac{\as}{M}\)\]
+ \as \gamma_0'\[\chi_s\(\frac{\as}{M}\)\] \chi_{ss}\(\frac{\as}{M}\)
+ \as \gamma_1\[\chi_s\(\frac{\as}{M}\)\]
+ \Ord\[\as^2 \(\frac{\as}{M}\)^{\forall k}\] = \frac{M}{\as},
\eeq
from which we get
\begin{align}
\chi_s\(\frac{\as}{M}\) &= \gamma_0^{-1}\(\frac{M}{\as}\) \label{eq:chis_def}\\
\chi_{ss}\(\frac{\as}{M}\) &= - \frac{\gamma_1\(\chi_s(\as/M)\)}{\gamma_0'\(\chi_s(\as/M)\)}\label{eq:chiss_def}
\end{align}
where $\gamma_0^{-1}$ is the inverse function of $\gamma_0$.
Note that, while at LO $\as\gamma_0$ and $\chi_s$ are exactly dual, at NLO
$\as\gamma_0+\as^2\gamma_1$ and $\chi_s+\as\chi_{ss}$ are dual up to order $\as^2$
corrections (at fixed $\as/M$). For this reason, for numerical application it may be better
to consider the exact dual also at NLO (or beyond), the difference being subleading.
Anyway, for the sake of theoretical discussion, we will continue to consider the
singular expansion Eq.~\eqref{eq:chis_exp}.

\begin{figure}[t]
  \centering
  \includegraphics[width=0.4\textwidth]{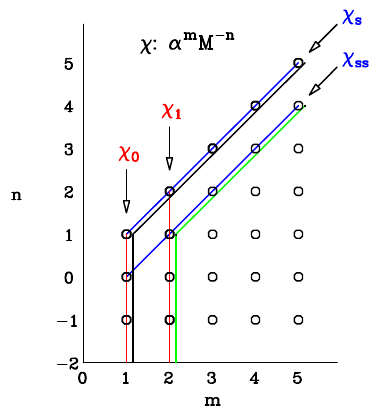}
  \caption{Structure of the Double-Leading approximation for the BFKL kernel $\chi(M,\as)$.
    (Figure taken from Ref.~\cite{Altarelli:1999vw}.)}
  \label{fig:DL_structure}
\end{figure}

The singular expansion \eqref{eq:chis_exp} allows to produce a Double-Leading (DL)
expansion for the BFKL kernel:
\begin{align}
\chi_{\rm DL}(\as,M) &= \as \chi_0(M) + \chi_s\(\frac{\as}{M}\) - \as\frac{\chi_{0,1}}{M} \nonumber\\
&\quad + \as \[ \as \chi_1(M) + \chi_{ss}\(\frac{\as}{M}\) - \as \( \frac{\chi_{1,2}}{M^2} + \frac{\chi_{1,1}}{M} \) - \chi_{0,0} \]\nonumber\\
&\quad + \ldots
\label{eq:DL_approx}
\end{align}
where the first row represents a LO DL expansion, the second row a NLO DL expansion, and so on.
The $\chi_{k,j}$ terms are subtracted to avoid double counting; these coefficients
can be read from Eqs.~\eqref{eq:chi_kj}. The structure of the terms which enter this expansion
are shown pictorially in Fig.~\ref{fig:DL_structure}.
A completely analogous expression can be obtained for the anomalous dimension.

The LO DL result resums the leading logarithms of $q^2$ in the BFKL kernel:
therefore, it represents a matched LO+LL resummed kernel
(the same for the anomalous dimension, the resummed logarithms being
logs of $1/x$).
Again, the NLO DL result resums also the next-to-leading logarithms,
and therefore represents a matched NLO+NLL resummed kernel.
Hence, concerning the double-leading expansion, the notation LO, NLO and so on
represents in fact LL, NLL, and so on. The reason for this notation
relies on the fact that to accomplish resummation to N$^k$LL within
the DL expansion one has to simply match a N$^k$LO BFKL kernel with a N$^k$LO
anomalous dimension.
We will then continue to talk about LO and NLO resummation, meaning
always LO+LL and NLO+NLL matched expressions.

\begin{figure}[t]
  \centering
  \includegraphics[width=\textwidth]{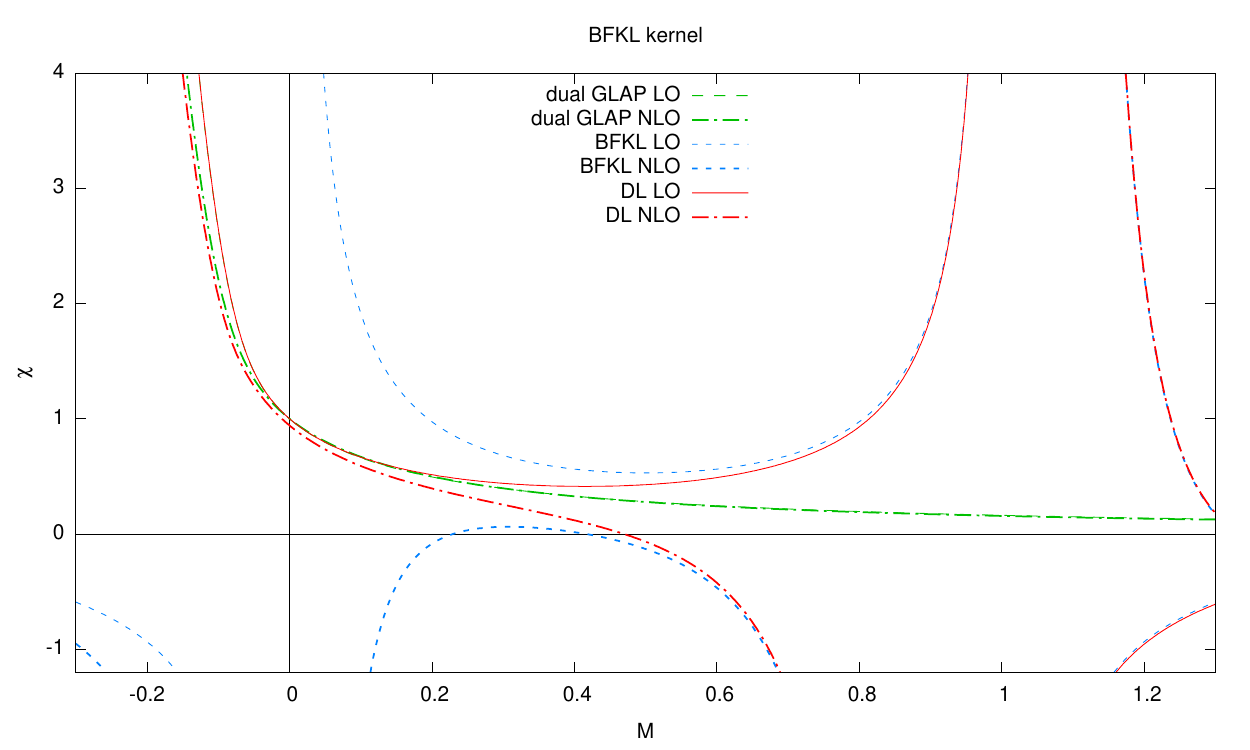}
  \caption{Double-Leading approximation for the BFKL kernel $\chi(M,\as)$. Here we have used $\as=0.2$ and $n_f=4$.}
  \label{fig:chi_DL}
\end{figure}

The results for $\as=0.2$ and $n_f=4$ are shown on Fig.~\ref{fig:chi_DL}.
From the plot, we immediately see that such result is not acceptable:
indeed the DL curves (red curves), when mirrored with respect to the main diagonal to give
the anomalous dimension, manifest a perturbative instability at small $N$.
The LO DL curve, when inverted, presents a square-root branch-point at some small positive value of $N$,
while the NLO DL curve extends smoothly to all negative values of $N$.
Moreover, form a phenomenological point of view, the LO and NLO anomalous dimensions
(which have simple poles in $N=0$), describe quite well the data, while the DL LO
approximation gives a too strong rise at small $x$, and the DL NLO does not rise at all \cite{Altarelli:1999vw}.

Such instability is related to the bad behaviour of the BFKL kernel around $M=1$:
therefore, in order to obtain a stable DL approximation, also the $M=1$ singularities must be resummed.
We will treat such resummation in the next Section.

\section{Symmetrization}
\label{sec:BFKL_symmetrization}
The first ingredient needed to improve the DL result is related to the intrinsic
symmetry of the BFKL kernel. We discuss this now.

\subsection{The symmetry of the BFKL kernel}

Since the diagrams involved in the computation of the BFKL kernels are symmetric
for the exchange of the incoming gluon momenta, the BFKL kernel should satisfy
\beq\label{eq:BFKL_kernel_symmetry}
\frac{1}{q^2} K\(\as,\frac{q^2}{k^2}\) = \frac{1}{k^2}K\(\as,\frac{k^2}{q^2}\)
\eeq
or, in terms of its Mellin transform,
\beq
\chi(\as, M) = \chi(\as,1-M).
\eeq
However, while the LO kernel Eq.~\eqref{eq:chi0} satisfies this symmetry,
the NLO kernel $\chi_1(N)$, Eq.~\eqref{eq:chi1}, is no longer symmetric.
In fact, this symmetry is broken by two effects:
\begin{itemize}
\item the running of the coupling constant $\as$
\item an asymmetric choice for the variable $\xi$.
\end{itemize}
The running of the coupling and the resummation of its effects
will be treated in Sect.~\ref{sec:BFKL_running_coupling};
so far, in $\chi_1$ only a perturbative contribution was included,
Eq.~\eqref{eq:chiNLO_ordering}.
The second point is the argument of this Section.

In the computation of the BFKL kernel from gluon-gluon scattering,
the value of the variable $\xi$ depends symmetrically on $q^2$ and $k^2$~\cite{FL,Catani:1994sq};
however, in DIS, the variable $\xi$ is defined as
\beq
\xi = \log\frac1x = \log\frac{2pq}{-q^2}\quad \overset{\text{small $x$}}{\simeq} \log\frac{s}{-q^2}
\eeq
which depends only on $q^2$, and hence has no symmetry under the exchange $q^2\leftrightarrow k^2$.
The symmetric choice for the variable $\xi$ is obtained substituting
$q^2$ with $\sqrt{q^2k^2}$ in the last equality.
Then, the BFKL evolution equation for the DIS choice of $\xi$ has to be modified
accordingly~\cite{FL}, thereby breaking the symmetry.
The transition from one choice to the other is obtained by recasting the double Mellin 
transform of the BFKL equation: using $s$ as integration variable for the $N$-Mellin we have
(from DIS to symmetric)
\beq
\int\frac{ds}{s}\int\frac{dq^2}{q^2} \(\frac{s}{q^2}\)^N \(\frac{q^2}{k^2}\)^{-M}
=
\int\frac{ds}{s}\int\frac{dq^2}{q^2} \(\frac{s}{\sqrt{q^2k^2}}\)^N \(\frac{q^2}{k^2}\)^{-M-\frac{N}2}.
\eeq
Because of the pole condition Eq.~\eqref{eq:pole_condition}, $N$
has to be identified with the current kernel. Therefore, calling for definiteness
$\chi_\sigma$ the kernel for the symmetric variable choice
and $\chi$ that for the usual DIS variables, the two kernels
are related by \cite{FL,abf742}
\begin{subequations}\label{eq:ker_symm_rel}
\beq
\begin{cases}\label{eq:ker_symm_rel3}
  \chi\(\as,M+\frac N2\) = \chi_\sigma(\as,M) \\
  N = \chi_\sigma(\as,M),
\end{cases}
\eeq
or, implicitly,
\beq\label{eq:ker_symm_rel1}
\chi\(\as, M+\frac{1}{2} \chi_\sigma(\as,M) \) = \chi_\sigma(\as, M),
\eeq
or equivalently
\beq\label{eq:ker_symm_rel2}
\chi_\sigma\(\as, M-\frac{1}{2} \chi(\as,M) \) = \chi(\as, M).
\eeq
\end{subequations}
The relation \eqref{eq:ker_symm_rel} can be solved iteratively to the desired order.
For example, given $\chi_\sigma$, one can compute $\chi$ by writing
\beq
\chi(\as,M) = \chi_\sigma\(\as,M-\frac12\chi_\sigma\(\as,M-\frac12\chi_\sigma\(\as,M-\ldots\)\) \)
\eeq
and stopping at the desired level of accuracy. For instance, at LO, it is easy to
see that the two kernels are equal. Indeed $\chi_\sigma(\as,M) = \as \chi_0(M)$ and hence
\beq
\chi(\as,M) = \as \chi_0\(M-\frac\as2\chi_0\(M+\Ord(\as)\)\) = \as \chi_0(M) + \Ord(\as^2).
\eeq
At fixed NLO, $\chi_\sigma(\as,M) = \as \chi_0(M)+\as^2\chi_1^\sigma(M)$ and
\begin{align}
\chi(\as,M) &= \as \chi_0\(M-\frac\as2\chi_0\(M+\Ord(\as)\)\) + \as^2 \chi_1^\sigma\(M+\Ord(\as)\)\nonumber\\
&= \as \chi_0(M) + \as^2 \[ \chi_1^\sigma(M) - \frac12 \chi_0(M)\chi_0'(M) \] + \Ord(\as^3).\label{eq:symm_to_DIS}
\end{align}
Note that the order $\as^2$ symmetric term $\chi_1^\sigma(M)$ is modified by adding
$-\frac12 \chi_0(M)\chi_0'(M)$: this is (one of) the non-symmetric term appearing in Eq.~\eqref{eq:chi1}.

The relation \eqref{eq:ker_symm_rel} is much better written in terms of
inverse functions. Since, by duality, the inverse function of $\chi$ in asymmetric variables
is the anomalous dimension $\gamma$, we can introduce the ``symmetric'' anomalous
dimension $\gamma_\sigma$, dual to $\chi_\sigma$: putting these into one of Eqs.~\eqref{eq:ker_symm_rel}
we get
\beq\label{eq:ker_symm_rel_gamma}
\gamma(N)=\gamma_\sigma(N)+\frac{N}{2}.
\eeq
This relation is very useful in numerical computations, and also to
simply visualize the passage from simmetric variables to asymmetric ones and vice-versa.

\subsection{Off-shell kernels}

We now build a tool that can help us to ``solve'' the relation Eq.~\eqref{eq:ker_symm_rel}.
We introduce the concept of ``off-shell'' kernel
\beq
\bar\chi(\as,M,N)
\eeq
as a function of one more variable $N$ (the name is not a coincidence)
defined by the condition that the actual kernel (also ``on-shell'' kernel from now on)
can be obtained by the ``on-shell'' (or pole) condition%
\beq\label{eq:off-shell_kernel}
\bar\chi\(\as,M,\chi(\as,M)\) = \chi(\as,M).
\eeq
If $\chi(\as,M)$ is related by duality to $\gamma(\as,N)$,
Eq.~\eqref{eq:off-shell_kernel} can be rewritten as
\beq\label{eq:off-shell_kernel_gamma}
\bar\chi\(\as,\gamma(\as,N),N\) = N.
\eeq
More generally, we can consider the ``off-shell'' relation
\beq
\bar\chi\(\as,M,N\) = N
\eeq
and obtain either Eq.~\eqref{eq:off-shell_kernel} or Eq.~\eqref{eq:off-shell_kernel_gamma}
by putting on-shell respectively $N$ or $M$, i.e.\ by setting
$N=\chi(\as,M)$ or $M=\gamma(\as,N)$.

Geometrically, we have a simple interpretation of the off-shell kernel.
Consider a two-dimensional space with coordinates $(M,N)$: the BFKL kernel
is a curve on this plane, provided we identify $N$ with $\chi(\as,M)$.
By duality, the same curve is $\gamma(\as,N)$, provided now $N$
is considered as the independent variable and we identify $M$ with $\gamma$.
In this context, the off-shell kernel $\bar\chi(\as,M,N)$ can be viewed as a function
on the $(M,N)$-plane with the property that the zeros of
\beq
\bar\chi(\as,M,N) - N,
\eeq
which form a curve in the plane, are exactly the kernels' curve.

Given $\chi(\as,M)$, we will call any $\bar\chi(\as,M,N)$ satisfying Eq.~\eqref{eq:off-shell_kernel}
an \emph{off-shell extension} of $\chi(\as,M)$; of course, there is not a unique way to obtain
such an off-shell extension, but there are infinite possibilities.
The naive off-shell extension of $\chi$ is simply
\beq
\bar\chi(\as,M,N) = \chi(\as,M).
\eeq

Relation \eqref{eq:ker_symm_rel} becomes simple from the point of view of
off-shell kernels.
Given an off-shell extensions of $\chi$ and $\chi_\sigma$
\begin{subequations}
\begin{align}\label{eq:off-shell_kernel1}
\chi(\as,M) &= \bar\chi\(\as,M,\chi(\as,M)\), \\
\chi_\sigma(\as,M) &= \bar\chi_\sigma\(\as,M,\chi_\sigma(\as,M)\).\label{eq:off-shell_kernel2}
\end{align}
\end{subequations}
and putting these equations in Eq.~\eqref{eq:ker_symm_rel3} we get
\beq\label{eq:off-shell_relation}
\bar\chi\(\as,M+\frac{N}{2},N\) = \bar\chi_\sigma(\as,M,N).
\eeq
Eq.~\eqref{eq:off-shell_relation} tells us that given
the off-shell kernel, let's say, in asymmetric variables $\bar\chi$,
we can find immediately the off-shell kernel $\bar\chi_\sigma$ in symmetric variables,
and vice-versa, by a simple variable shift.

\subsection{Improved resummation by symmetrization}
We can use relation \eqref{eq:ker_symm_rel} and the symmetry of $\chi_\sigma(\as,M)$
under the exchange $M\to 1-M$ to cure the instability of $\chi(\as,M)$ in $M=1$.
The strategy is the following:
\begin{itemize}
\item go to symmetric variables using Eq.~\eqref{eq:ker_symm_rel};
\item resum $\chi$ in the DL approximation, Eq.~\eqref{eq:DL_approx}, to obtain $\chi_{\rm DL}$;
\item the resulting $\chi^\sigma_{\rm DL}$ is not symmetric: symmetrize it,
  thereby resumming the $M=1$ singularities;
\item use again Eq.~\eqref{eq:ker_symm_rel} with the symmetrized $\chi^\sigma_{\rm DL}$
  to go back to asymmetric variables, obtaining a kernel which we will call
  from now on $\chi_{\rm SDL}$ (simmetrized double-leading).
\end{itemize}
The kernel $\chi_{\rm SDL}$ obtained with the above procedure is not symmetric
(because of the asymmetric choice of the $\xi$ variable) but has a stable
DL perturbative expansion.
In particular, a minimum near $M=\frac12$ is present to all perturbative orders:
this is relevant for the running coupling resummation discussed
in Sect.~\ref{sec:BFKL_running_coupling}.

\subsubsection{Practical realization: a toy example}
Consider for simplicity the DL expansion Eq.~\eqref{eq:DL_approx} at LO:
\beq
\chi_{\rm DL}(\as,M) = \as \chi_0(M) + \chi_s\(\frac{\as}{M}\) - {\rm d.c.}
\eeq
In this equation the $\chi_0$ term is symmetric, while $\chi_s$ is not.
Following the strategy described above, the next step for resummation
consists in going to symmetric variables
\beq
\chi^\sigma_{\rm DL}(\as,M) = \as \chi_0(M) + \chi^\sigma_s(\as,M) - {\rm d.c.}
\eeq
where here $\chi_0$ is the same as before, since it was already symmetric at this order,
and $\chi^\sigma_s$ is the dual of $\gamma_\sigma$ at LO, Eq.~\eqref{eq:ker_symm_rel_gamma}.
Being in symmetric variables, $\chi^\sigma_s$ should be symmetric, but it is not.
A way to obtain $\chi^\sigma_s$ is the approach of Ref.~\cite{abf742}, where
symmetrization is obtained by defining first the naive off-shell extension of $\chi_s$ as
\beq
\bar\chi_s(\as,M,N) = \chi_s\(\frac{\as}{M}\),
\eeq
then going to symmetric variables
\beq
\bar\chi^\sigma_s(\as,M,N) = \chi_s\(\frac{\as}{M+N/2}\),
\eeq
now symmetrizing it as~\cite{Salam:1998tj,abf742}
\beq\label{eq:chi_s_symm_offshell}
\bar\chi^\sigma_s(\as,M,N) = \chi_s\(\frac{\as}{M+N/2}\) + \chi_s\(\frac{\as}{1-M+N/2}\).
\eeq
The symmetrization procedure, though somewhat arbitrary,
satisfies the following constraints:
\begin{itemize}
\item the kernel obtained putting on-shell has to be symmetric for $M\leftrightarrow 1-M$;
\item in the $M<\frac12$ region, at large $N$ the dual kernel must match with the original
  LO anomalous dimension;
\item no other singularities must be introduced.
\end{itemize}
However, there are still some bad features in the result: for example,
momentum conservation is no longer satisfied. To solve such problem,
an additional term~\cite{abf742}
\beq
\bar\chi_{\rm mom}(N) = c_{\rm mom} f_{\rm mom}(N)
\eeq
is added to $\bar\chi^\sigma_s$ to enforce momentum conservation, provided
\beq
f_{\rm mom}(\infty) = 0, \qquad
f_{\rm mom}(1) = 1,
\eeq
where $c_{\rm mom}$ is a constant chosen in such a way that momentum conservation is preserved.
Specific forms for $f_{\rm mom}$ can be
\beq\label{eq:f_mom}
f_{\rm mom}(N) = \frac{4^{k} N^k}{(N+1)^{2k}},\quad \frac{2 N^k}{N^{2k}+1}, \qquad k>0,
\eeq
which are in particular symmetric for the exchange $N\to 1/N$ and satisfy additionally
\beq
f_{\rm mom}(0) = 0, \qquad
f_{\rm mom}'(1) = 0.
\eeq
In symmetric variables momentum conservation constraint requires
\beq
M=-\frac12 \qquad \Rightarrow \qquad N=1
\eeq
that implies, from Eq.~\eqref{eq:chi_s_symm_offshell},
\beq
c_{\rm mom} = 1-\bar\chi^\sigma_s\(\as,-\frac12,1\) = -\chi_s\(\frac{\as}{2}\).
\eeq
With these ingredients at the hand, one can use
\beq
\bar\chi^\sigma_s(\as,M,N) = \chi_s\(\frac{\as}{M+N/2}\) + \chi_s\(\frac{\as}{1-M+N/2}\) + \bar\chi_{\rm mom}(N)
\eeq
to compute $\chi^\sigma_s(\as,M)$.

Finally, one can add back $\chi_0$ minus the double counting terms, which
now should be considered both in $M=0$ and $M=1$ for symmetry.
This subtraction breaks momentum conservation again, but a simple
modification of $c_{\rm mom}$ can cure it.

At the end, one can go back to asymmetric variables; practically, the easier
way to do this is to compute the two branches of a resummed $\gamma_\sigma(N)$
and then adding $N/2$ to obtain $\gamma_{\rm SDL}(N)$, Eq.~\eqref{eq:ker_symm_rel_gamma}.

\subsubsection{Practical realization: the approach of Ref.~\cite{abf742}}

Even if the procedure described above seems to be satisfactory,
there are still some problems with the other poles of the BFKL kernel,
which spoil the matching at large $N$ with the anomalous dimension.

To avoid this, in Ref.~\cite{abf742} an off-shell extension of $\chi_0$ is also considered
(see also Ref.~\cite{Salam:1998tj}).
This off-shell $\bar\chi_0$ is then added to $\bar\chi^\sigma_s$ (after subtracting double counting)
and the resulting function is then put on-shell.\footnote{%
Note that the sum of two off-shell kernel is \emph{not} in general the
off-shell kernel for the sum of the kernels.}

Actually, the off-shell extension of $\chi_0$ performed in Ref.~\cite{abf742} is not
really an extension, since it does not lead to the original BFKL kernel when it is put on-shell;
nevertheless, it is chosen in such a way that the relevant limits are respected.
First, the collinear and anti-collinear parts of $\chi_0$ are separated as
\beq
\chi_0(M) = \chi_0^{L}(M) + \chi_0^{R}(M),
\eeq
where
\begin{align}
  \chi_0^{L}(M) &= \frac{N_c}{\pi} \[ \psi(1) - \psi(M) \],\label{eq:chi0L}\\
  \chi_0^{R}(M) &= \chi_0^{L}(1-M);
\end{align}
then, $\chi_0^{L}$ is treated as $\chi_s$ before, i.e.\ the naive off-shell extension
of $\chi_0^{L}$ is taken in asymmetric variables, then translated to symmetric variables and symmetrized,
obtaining
\beq
\bar\chi_0^\sigma(M,N) = \frac{N_c}{\pi} \[ 2\psi(1) - \psi\(M+\frac N2\) - \psi\(1-M+\frac N2\) \],
\eeq
or, in asymmetric variables,
\beq
\bar\chi_0(M,N) = \frac{N_c}{\pi} \[ 2\psi(1) - \psi\(M\) - \psi\(1-M+N\) \].
\eeq
However, since we don't want to spoil the identification at large $N$
with the anomalous dimension, $\bar\chi_0(M,N)$ should vanish at large $N$.
To accomplish this, in Ref.~\cite{abf742} a subleading term is added
to $\bar\chi_0$ obtaining at the end
\beq\label{eq:bar_chi_0}
\bar\chi_0(M,N) = \frac{N_c}{\pi} \[ \psi(1) + \psi(1+N) - \psi\(M\) - \psi\(1-M+N\) \],
\eeq
which is safe at large $N$ because $\psi\(1-M+N\) - \psi(1+N) \sim 1/N$.

\begin{figure}[tb]
  \centering
  \includegraphics[width=\textwidth,page=1]{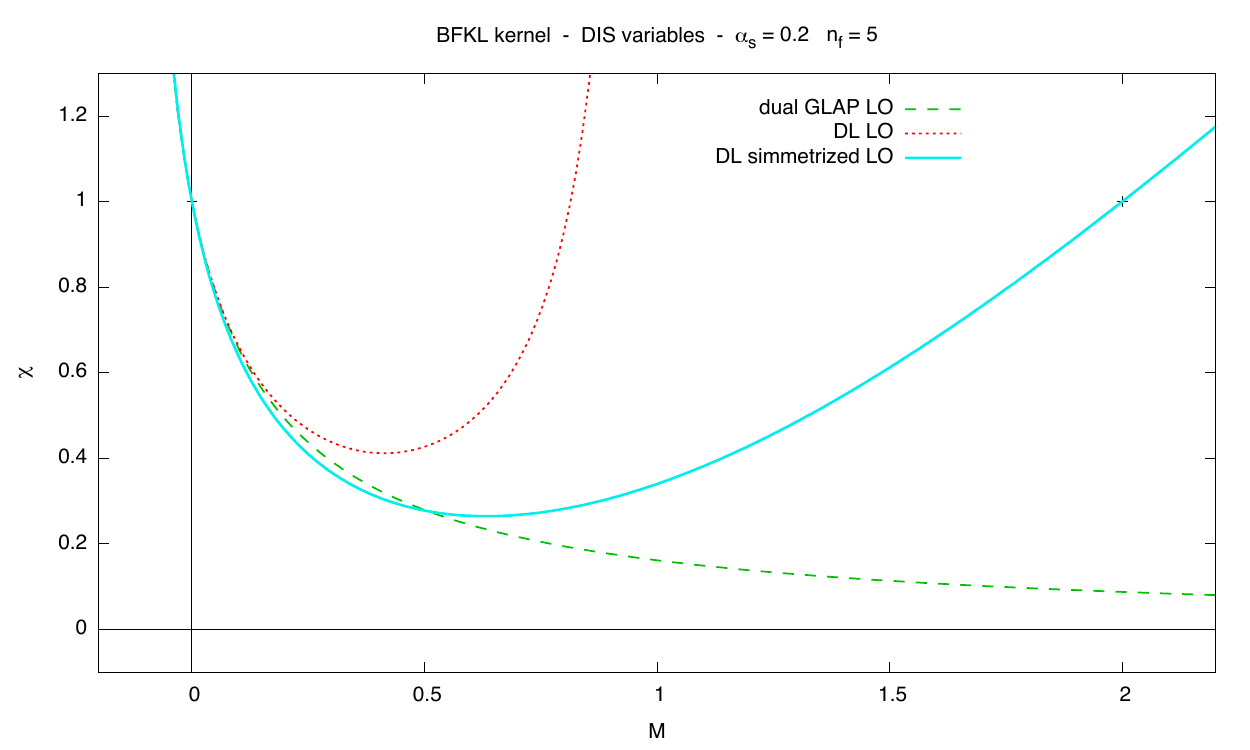}
  \caption{The simmetrized Double-Leading approximation at LO. Here we have used $\as=0.2$ and $n_f=5$.}
  \label{fig:simmDL}
\end{figure}

Putting all together, we get (in symmetric variables)
\beq
\bar\chi^\sigma_{\rm LO}(\as,M,N) = \bar\chi^\sigma_s(\as,M,N) + \as \tilde{\bar\chi}^\sigma_0(M,N) + \chi_{\rm mom}(\as,N),
\eeq
where
\beq
\tilde{\bar\chi}^\sigma_0(M,N) = \frac{N_c}{\pi} \[ \psi(1) + \psi(1+N) - \psi\(1+M+\frac N2\) - \psi\(2-M+\frac N2\) \]
\eeq
has subtracted the collinear ($M+\frac N2=0$) and anti-collinear ($1-M+\frac N2=0$) doubly counted poles,
and the momentum conservation function has been reintroduced.
The result of putting on-shell, back to DIS variables, is shown in Fig.~\ref{fig:simmDL} at LO.
The simmetrized DL curve has a minimum, thanks to the fact that the asymptotic behaviour
of the curve at large $N$ (large $\chi$) are given by the GLAP anomalous dimension.
Then, the minimum is still there also at NLO: symmetrization stabilizes the DL
expansion, as expected.
The details of the NLO symmetrization are quite cumbersome, and
we refer the Reader to App.~\ref{sec:app-SX-NLO} or to Ref.~\cite{abf742}
for further details.

\section{Running coupling corrections to the BFKL equation}
\label{sec:BFKL_running_coupling}

So far, we have supposed that the strong coupling $\as$ is fixed;
now we want to study the effect of its running.
We will find that the inclusion of running coupling effects changes
dramatically the small-$x$ behaviour of the anomalous dimension,
making the running coupling resummation a crucial ingredient for small-$x$ resummation.

The BFKL equation \eqref{eq:BFKL_eq_integral} has a kernel $K$ which depends
on the coupling $\as$.
If we let the coupling run, we can make several choices for its argument;
the most natural ones are $\as(q^2)$ and $\as(k^2)$.
In both cases, we have that under Mellin transform $\as$ becomes
a differential operator, $\ashat$, which is constructed by substituting
\beq
t \to -\frac{d}{dM}
\eeq
in the explicit form of the running $\as$. For example, at $1$-loop,
\beq\label{eq:ashat_1loop}
\ashat = \frac{\as}{1-\as\beta_0 \frac{d}{dM}},
\eeq
where $\as$ is now the value of the strong coupling at a given fixed scale $\mu_0^2$.
The difference in choosing the argument $q^2$ or $k^2$ in $\as$
is the resulting operator order.
Consider the BFKL equation~\eqref{eq:BFKL_eq_integral}
\beq
\frac{d}{d\xi}f(\xi,q^2) = \int_0^\infty \frac{dk^2}{k^2} \, \sum_{p=0}^\infty \as^{p+1}(\cdot)\,K_p\(\frac{q^2}{k^2}\)\,f(\xi,k^2),
\eeq
where we have explicitly written the BFKL kernel as a power series in $\as$, with argument of $\as$ not specified.
If we choose as argument $q^2$, the Mellin transform of the equation becomes
\beq
\frac{d}{d\xi}f(\xi,M) = \sum_{p=0}^\infty \ashat^{p+1}\,\chi_p(M)\,f(\xi,M),
\eeq
where $\ashat$ acts on everything on the right, while choosing $k^2$ as argument we get
\beq
\frac{d}{d\xi}f(\xi,M) = \sum_{p=0}^\infty \chi_p(M)\,\ashat^{p+1}\,f(\xi,M),
\eeq
where now $\ashat$ acts only on the PDF.

A unique scale choice for $\as$ inevitably breaks
the symmetry of the BFKL equation under the exchange $q^2\leftrightarrow k^2$,
unless we use $\as(\sqrt{q^2k^2})$, which however complicates somewhat the treatment.
The most convenient choice is then to compute $\as$ at the scale $q^2$ for the collinear part
and $k^2$ for the anti-collinear part:
\beq
K\(\as,\frac{q^2}{k^2}\) \to \sum_{p=0}^{\infty} \[ \as^{p+1}(q^2) K^{L}_p\(\frac{q^2}{k^2}\) + \as^{p+1}(k^2) K^R_p\(\frac{q^2}{k^2}\) \].
\eeq
If symmetric variables are used, the collinear and anti-collinear contributions
to the kernel satisfy, order by order, the symmetry relation
\beq
\frac{1}{q^2}K^L_p\(\frac{q^2}{k^2}\) = \frac{1}{k^2}K^R_p\(\frac{k^2}{q^2}\)
\eeq
which is the underlying reason for the symmetry of Eq.~\eqref{eq:BFKL_kernel_symmetry}.
This means that this choice restores the symmetry $q^2\leftrightarrow k^2$
of the BFKL equation.
The Mellin transform of the BFKL kernel with this scale choice is then
\beq\label{eq:BFKL_kernel_rc_symm}
\chi(\ashat,M) = \sum_{p=0}^{\infty} \[ \ashat^{p+1}\, \chi^L_p(M) + \chi^R_p(M) \,\ashat^{p+1} \]
\eeq
where
\beq
\chi^R_p(M) = \chi^L_p(1-M)
\eeq
and is intended that $\chi(\ashat,M)$ is an operator itself, and $\ashat$
in it will act on everything on the right, specifically on the PDF in the BFKL equation.
For example, the LO contribution is given by Eq.~\eqref{eq:chi0L},
\beq
\chi_0^L(M) = \frac{N_c}{\pi} \[\psi(1) - \psi(M)\],
\eeq
while the NLO will be given in App.~\ref{sec:NLO_BFKL_kernel}.

Note that, once we have built the running-coupling kernel Eq.~\eqref{eq:BFKL_kernel_rc_symm},
we can reshuffle the position of $\ashat$ having care of supplying
each switch with the corresponding commutator.
If we use the $1$-loop beta function Eq.~\eqref{eq:ashat_1loop}, we get easily
\beq\label{eq:commutator_as-1}
\comm{\ashat^{-1}}{M} = -\beta_0.
\eeq
From this we can compute
\beq\label{eq:commutator_as}
\comm{\ashat}{M}
= \ashat M - \ashat \ashat^{-1} M \ashat
= \ashat M - \ashat M \ashat^{-1} \ashat - \ashat \comm{\ashat^{-1}}{M} \ashat
= \beta_0\ashat^2.
\eeq
The commutator of $\ashat$ (or a function of it) with a function of $M$ can be then
built \emph{algebraically} starting from the ``fundamental''
commutators Eqs.~\eqref{eq:commutator_as-1},\eqref{eq:commutator_as}.

In general, it is quite easy to compute the commutators
of $\ashat^{-1}$ with a function of $M$, if $\as$ runs at $1$-loop.
Indeed
\beq
\comm{\ashat^{-1}}{g(M)} = -\beta_0 g'(M).
\eeq
From this, with the same computation as in \eqref{eq:commutator_as},
we get
\beq\label{eq:comm_ashat_gM}
\comm{\ashat}{g(M)} = \ashat \beta_0 g'(M) \ashat.
\eeq
If we want the $\ashat$ operator to be on the left we can
recursively commute to obtain
\beq\label{eq:comm_ashat_gM_left}
\comm{\ashat}{g(M)} = \sum_{k=1}^\infty (-)^{k+1} \beta_0^k \ashat^{k+1} g^{(k)}(M) .
\eeq

Now, suppose we want to have all the powers of $\ashat$ to the left
(as we will need it later). Then, at LO, one obtains simply
\beq
\chi(\ashat,M) = \ashat \[\chi_0^L(M) + \chi_0^R(M)\] + \Ord(\ashat^2)
\eeq
because the commutator Eq.~\eqref{eq:comm_ashat_gM_left} starts at order $\ashat^2$.
At NLO this commutator gives an effect: indeed we have
\beq\label{eq:chiNLO_ordering}
\chi(\ashat,M) = \ashat \[\chi_0^L(M) + \chi_0^R(M)\]
+ \ashat^2 \[\chi_1^L(M) + \chi_1^R(M) - \beta_0 \,\partial_M\chi_0^R(M)\]
+ \Ord(\ashat^3)
\eeq
where
\beq
\partial_M\chi_0^R(M) = \frac{N_c}{\pi} \psi_1(1-M).
\eeq

\subsection{Resummation of running coupling effects}

Perturbative inclusion of running coupling effect does not lead to stable results~\cite{abf621},
since it produces poles in the BFKL kernel at $M=1/2$, where it should have
a stationary point (in symmetric variables). Then, the leading small-$N$ singularity,
which determines the small-$x$ behaviour of the dual anomalous dimension,
changes nature whenever the running coupling effects are included at some fixed perturbative order.
For this reason, running coupling effects must be resummed to all orders in $\as$
to get a stable small-$N$ singularity.
To do this, let's take the $N$-Mellin transform of the running-coupling BFKL
equation,
\beq\label{eq:rc_BFKL_eq}
Nf(N,M) = \chi(\ashat,M)f(N,M) + \tilde F_0(M),
\eeq
which is the straightforward running-coupling extension of Eq.~\eqref{eq:BFKL_double_mellin}.
Because of the presence of the operator $\ashat$, this is a differential equation in $M$:
the solution of this equation incorporates running-coupling effects to all orders,
and from this solution we can hence compute a resummed anomalous dimension
with full running-coupling dependence.

In Ref.~\cite{abf621} it is proven that a running-coupling resummed
anomalous dimension is completely characterized by the inhomogeneus solution
of Eq.~\eqref{eq:rc_BFKL_eq}; nevertheless, up to power suppressed
terms, the inhomogeneus solution can be written in the same form of the homogeneus
solution, with an appropriate boundary condition $f(N,M_0)$,
which however is irrelevant (it cancels out) in the computation of the anomalous dimension.
Then from now on we will concentrate on the homogeneus equation
\beq\label{eq:rc_BFKL_eq_homo}
Nf(N,M) = \chi(\ashat,M)f(N,M).
\eeq
Now, suppose we manipulate $\chi(\as,M)$ in such a way that the powers of $\ashat$
are all on the left, using repeatedly the commutator Eq.~\eqref{eq:comm_ashat_gM};
if the resulting dependence on $\ashat$ is linear we can write%
\footnote{Note that this never happens exactly. Indeed, even at LO,
the anti-collinear part has $\ashat$ to the right, and the reshuffling
needed to bring it to the left produces $\Ord(\ashat^2)$ terms.
Nevertheless, being $\Ord(\ashat^2)$, they should be better included in the NLO kernel:
hence, up to NLO correction, the LO kernel can be written in the form
of Eq.~\eqref{eq:rc_linear_chi} with $\varphi(M)=\chi_0(M)$.}
\beq\label{eq:rc_linear_chi}
\chi(\as,M) = \ashat \varphi(M)
\eeq
and then, for $1$-loop beta function,
\beq\label{eq:running-coupling_eq}
\(1-\beta_0\as\frac{d}{dM}\) N f(N,M) = \as \varphi(M) f(N,M).
\eeq
The solution of this equation is
\beq\label{eq:rc_homo_sol}
f(N,M) = f(N,M_0) \,\exp\int_{M_0}^M dM'\,\frac{N-\as\varphi(M')}{N\beta_0\as}.
\eeq
The anomalous dimension which contains resummed running-coupling effects
can be found by taking the logarithmic derivative with respect to $t$
of the inverse $M$-Mellin of the solution Eq.~\eqref{eq:rc_homo_sol}:
\beq\label{eq:rc_anomalous_dimension_from_homo}
\gamma_{\rm rc}(\as(t),N) = \frac{d}{dt}\log\int_{c-i\infty}^{c+i\infty}\frac{dM}{2\pi i}\,e^{Mt}
\exp\int_{M_0}^M dM'\,\frac{N-\as\varphi(M')}{N\beta_0\as}
\eeq
where $c$ must be to the right of all the singularity of the integral.
Of course, the ability of computing the integrals depends on the explicit form of $\varphi(M)$;
in the next subsection we will show that in the case in which the kernel is quadratic
it is possible to find an analytic solution.

Another way to find a general solution for the anomalous dimension
can be obtained by taking the $N$-Mellin transform of Eq.~\eqref{eq:BFKL_eq_integral} with running coupling
\beq\label{eq:rc_homo_inverse_Mellin}
N f(N,t) = \as(t)\int^t dt' \,K_\varphi(e^{t-t'})\,f(N,t'),
\eeq
where $K_\varphi$ is the unintegrated kernel corresponding to $\varphi$,
and it is typically a distribution. In some cases we may be able
to separate the distributional part of $K_\varphi$ in terms of $\delta$
functions and its derivatives, and a finite part:
\beq
K_\varphi(e^t) = \sum_{j} k_j \delta^{(j)}(t) + K_\varphi^{\rm fin}(e^t);
\eeq
after explicit integration of the
$\delta$'s we end up with a integro-differential equation, that we can derive with respect to $t$
to obtain a genuine differential equation,
\beq
\sum_{j} (-)^j k_j f^{(j+1)}(N,t) - \frac{N}{\as(t)} f'(N,t) + \[ K_\varphi^{\rm fin}(1) -N\beta_0 \] f(N,t) =0 ,
\eeq
where we have explicitly used the $1$-loop $\beta$-function and the derivatives
are intended with respect to $t$.
This equation may be eventually easier to solve than the inverse Mellin transform
in Eq.~\eqref{eq:rc_anomalous_dimension_from_homo}.

\subsection{Quadratic approximation to the kernel}

In Ref.~\cite{abf621} it is shown that the small-$x$ behaviour of the running-coupling
resummed splitting-functions is determined by the behaviour of the kernel $\chi$
in Eq.~\eqref{eq:rc_BFKL_eq} around its minimum (which is at $M=1/2$ in symmetric variables).
This result can be simply obtained by means of a saddle point approximation \cite{Camici:1997ta}
to the $M$-Mellin inversion integral in Eq.~\eqref{eq:rc_anomalous_dimension_from_homo}.
The saddle point $M_s$ is given by
\beq
\as(t) \varphi(M_s) = N;
\eeq
from this equation it is evident that the region of small $N$
is determined by the smaller values of $\varphi$,
i.e.\ by its minimum.

This is pretty useful, because when the kernel is substituted by its quadratic approximation~\cite{Lipatov:1985uk}
\beq\label{eq:chi_quadratic}
\chi(\as,M) \quad\to\quad \chi_q(\as,M) = c(\as) + \frac12 \kappa(\as) \(M-\frac12\)^2
\eeq
we are able to analytically find a resummed anomalous dimension,
provided the $\as$ dependence of $\chi$ (or of the coefficients $c$ and $\kappa$)
belongs to the following two cases:
\begin{align}
  &\text{Airy:} & \chi(\ashat,M) &\simeq \ashat \chi_0(\as,M) \label{eq:c_approx_airy}\\
  &\text{Bateman:} & \chi(\ashat,M) &\simeq \chi(\as,M) + (\ashat-\as) \partial_{\as}\chi(\as,M) \label{eq:c_approx_bateman}
\end{align}
where $\chi_0(\as,M)$ may depend on $\as$ and is not in general the LO kernel $\chi_0(M)$.
The coefficients $c(\ashat)$ and $\kappa(\ashat)$ satisfy the same expansion.
The names Airy~\cite{abf621} and Bateman~\cite{abf742} are due to the fact that
in the respective cases the solution is expressed in terms of Airy or Bateman functions
(see App.~\ref{sec:hypergeometric}).
If the kernel is linear in $\as$ (as the LO kernel), the Airy solution coincides with the Bateman one;
otherwise, if the $\as$ dependence of $\chi$ is non-trivial,
the approximation that brings to the Airy solution is not good,
and the Bateman solution provides a better approximation.

\subsubsection{Airy anomalous dimension}
Note that a quadratic kernel $\chi$ as in Eq.~\eqref{eq:chi_quadratic}
corresponds, in the case of Airy approximation, to a unintegrated kernel $K$ of the form
\beq
K(\as,e^t) = \as(t) \[ \(c_0+\frac{\kappa_0}8\) \delta(t) - \frac{\kappa_0}2 \delta'(t) + \frac{\kappa_0}2 \delta''(t) \].
\eeq
Putting this in Eq.~\eqref{eq:rc_homo_inverse_Mellin} we get
\beq
f'' - f' = \frac{2}{\kappa_0} \[ \frac{N}{\as(t)} - c_0 - \frac{\kappa_0}{8} \] f
\eeq
where primes denote derivatives with respect to $t$.
Writing
\beq
f(N,t) \propto g(N,t) \,\exp\frac{1}{2\beta_0\as(t)}
\eeq
we get
\beq
g'' = \frac{2}{\kappa_0}\[ \frac{N}{\as(t)} - c_0 \] g.
\eeq
We recognize in this equation the equation for the Airy function,
\beq
\Ai''(z) - z\, \Ai(z) = 0,
\eeq
provided we identify $z$ with
\beq
z(\as(t),N) = \(\frac{2\beta_0N}{\kappa_0}\)^{1/3} \frac1{\beta_0}\[ \frac{1}{\as(t)} - \frac{c_0}{N} \].
\eeq
Then, we have found
\beq
f(N,t) \propto \Ai\(z(\as(t),N)\) \,\exp\frac{1}{2\beta_0\as(t)}
\eeq
from which we immediately compute the anomalous dimension
\beq\label{eq:gammaAiry}
\gamma_A(\as(t),N) = \frac12 + \(\frac{2\beta_0N}{\kappa_0}\)^{1/3} \frac{\Ai'\(z(\as(t),N)\)}{\Ai\(z(\as(t),N)\)}
\eeq
which we will call the \emph{Airy anomalous dimension}.
The same result can be found by putting the quadratic kernel
Eq.~\eqref{eq:chi_quadratic} into Eq.~\eqref{eq:rc_anomalous_dimension_from_homo},
where instead of solving a differential equation you have to compute
a inverse Mellin transformation \cite{abf621}.

For comparison with the singular expansion of the fixed-coupling anomalous dimension,
we expand Eq.~\eqref{eq:gammaAiry} in powers of $\as$ at fixed $\as/N$ and get
\beq
\gamma_A(\as,N) = \frac12 - \sqrt{\frac{N/\as-c_0}{\kappa_0/2}} - \frac{\beta_0 \as}{4(1-c_0\as/N)} + \ldots
\eeq
where the dots indicate terms of order $\as^2$ at fixed $\as/N$.
All these terms are divergent in $N=\as c_0$, but the sum is not;
it is instead divergent when the Airy function in the denominator
of Eq.~\eqref{eq:gammaAiry} goes to zero, at $z(\as(t),N)=-2.33811$
(see App.~\ref{sec:Airy}).

\subsubsection{Bateman anomalous dimension}

As already mentioned, the Bateman approximation Eq.\eqref{eq:c_approx_bateman}
is much better than the Airy one, Eq.~\eqref{eq:c_approx_airy}.
Indeed in this approximation the whole sequence
of leading-log contributions to $\as$ is correctly included in the solution~\cite{abf742}.

Defining
\begin{align}
\bar c(\as) &= c(\as) - \as c'(\as)\\
\bar\kappa(\as) &= \kappa(\as) - \as \kappa'(\as)
\end{align}
we can write, following Ref.~\cite{abf742}, the running-coupling BFKL equation
Eq.~\eqref{eq:rc_BFKL_eq} in the Bateman approximation as
\begin{multline}
\[N -\bar c(\as) - \frac12 \bar\kappa(\as)\(M-\frac12\)^2 \]f(N,M)\\
= \ashat \[c'(\as) + \frac12 \kappa'(\as)\(M-\frac12\)^2 \]f(N,M) + \tilde F_0(M).
\end{multline}
The solution to this equation can be found with techniques similar to the Airy case,
and we refer the Reader to Ref.~\cite{abf742} for further details.
The result can be written in terms of Bateman functions
$K_\nu(z)$ (see App.~\ref{sec:hypergeometric}) and reads
\beq\label{eq:gammaBateman}
\gamma_B(\as(t),N) = \frac12 - \beta_0\asb + A(\as,N)\,
\frac
{K_{B(\as,N)}'\(\frac{A(\as,N)}{\beta_0\asb}\)}
{K_{B(\as,N)} \(\frac{A(\as,N)}{\beta_0\asb}\)}
\eeq
where
\begin{align}
  \frac1{\asb} &= \frac1{\as} + \frac{\kappa'(\as)}{\bar\kappa(\as)}\\
  A(\as,N) &= \sqrt{\frac{N-\bar c(\as)}{\frac12 \bar\kappa(\as)}}\\
  B(\as,N) &= \(\frac{c'(\as)}{N-\bar c(\as)} + \frac{\kappa'(\as)}{\bar\kappa(\as)}\) \frac{A(\as,N)}{\beta_0}
\end{align}
We will call this solution the \emph{Bateman anomalous dimension}.

Expanding Eq.~\eqref{eq:gammaBateman} in powers of $\as$ at small $N$ we get
\beq
\gamma_B(\as,N) = \gamma^B_s(\as,N) + \gamma^B_{ss}(\as,N) + \ldots
\eeq
with
\begin{align}
\gamma^B_s(\as,N) &= \frac12 - \sqrt{\frac{N-c(\as)}{\frac12 \kappa(\as)}}\\
\gamma^B_{ss}(\as,N) &= \gamma^B_{ss,0}(\as) + \frac14 \as^2\beta_0\frac{c'(\as)}{c(\as)-N}\\
\gamma^B_{ss,0}(\as) &= -\beta_0\as + \frac34 \as^2 \beta_0 \frac{\kappa'(\as)}{\kappa(\as)}.
\end{align}
Note that now, after running coupling resummation, the anomalous dimension
behaves as a pole at small $N$, the position of the pole being determined by the
rightmost zero of the Bateman function in the denominator of Eq.~\eqref{eq:gammaBateman}.
This behaviour is compatible with the observations.

\subsection{On the minimum}

Since the resummation of running coupling effects depends only on the characteristics
of the minimum of the symmetric DL BFKL kernel, it is useful to make some comments
about its properties.

In symmetric variables, the minimum is always placed in $M=1/2$, because of the
symmetry $\chi_\sigma(\as,M) = \chi_\sigma(\as,1-M)$.
For this reason, the parameter $c$ is easily computed as
\beq
c(\as) = \chi_\sigma\(\as,\frac12\).
\eeq
In asymmetric variables, $M\to M+N/2$, and hence the position of the minimum is
\beq
M_{\rm min} = \frac12 + \frac{c(\as)}2,
\eeq
and the value at the minimum is the same as in symmetric variables
\beq
\chi(\as,M_{\rm min}) = \chi_\sigma\(\as,\frac12\) \equiv c(\as).
\eeq
It can be easily shown that also the curvature of the minimum is the same
in symmetric and asymmetric variables:
\beq
\kappa(\as) = \chi_\sigma''\(\as,\frac12\) = \chi''(\as,M_{\rm min}).
\eeq
In terms of the off-shell kernel $\bar\chi_\sigma(\as,M,N)$ in symmetric variables, 
the curvature can be written as
\beq\label{eq:kappa}
\kappa(\as) = \left.
\frac
{\partial_M^2\bar\chi_\sigma}
{1-\partial_N\bar\chi_\sigma}
\right|_{M=\frac12,\,N=c}
\eeq
as one can easily find by computing the second derivative of the on-shell relation
$\chi_\sigma(\as,M) = \bar\chi_\sigma(\as,M,\chi_\sigma(\as,M))$,
\beq
\de^2_M \chi_\sigma = \de^2_M \bar\chi_\sigma +\de_N\bar\chi_\sigma \,\de_M^2\chi_\sigma
+ \de_M\chi_\sigma \[ 2\de_N\de_M\bar\chi_\sigma + \de_N^2\bar\chi_\sigma\, \de_M\chi_\sigma \],
\eeq
where $N=\chi_\sigma$, and at the minimum $\de_M\chi_\sigma=0$.
The $\as$ dependence of $c$ can be found in the same way: deriving the duality relation with respect to $\as$
we get
\beq
\frac{\de}{\de{\as}}\chi_\sigma(\as,M) = \left.
\frac
{\de_{\as}\bar\chi_\sigma}
{1-\partial_N\bar\chi_\sigma}
\right|_{N=c}
\eeq
and hence
\beq
c'(\as) = \left.
\dfrac
{\de_{\as}\bar\chi_\sigma}
{1-\partial_N\bar\chi_\sigma}
\right|_{M=\frac12,\,N=c}.
\eeq
For $\kappa$, one can directly derive Eq.~\eqref{eq:kappa}, keeping in mind that $c$ depends on $\as$, and obtain
\beq
\kappa'(\as) = \left.
\dfrac
{\de_{\as}\de_M^2\bar\chi_\sigma + c'\,\de_N\de_M^2\bar\chi_\sigma + \kappa\[c'\,\de_N^2\bar\chi_\sigma+\de_{\as}\de_N\bar\chi_\sigma\]}
{1-\partial_N\bar\chi_\sigma}
\right|_{M=\frac12,\,N=c}.
\eeq
Note that in principle all these derivatives can be computed analitycally;
the more tricky one is the derivative of the $\chi_s$ part contained in the off-shell kernel,
that can be obtained by deriving duality relation
\beq
\frac{\partial}{\partial \as} \chi_s\(\frac\as M\) = -\frac{M/\as}{\as\gamma_0'\(\chi_s(\as/M)\)}.
\eeq
These relation are very useful for easily compute derivatives in a numerical code.

\section{Resummed anomalous dimensions}
\label{sec:res_anom_dim}

In the above Sections, we have introduced all the ingredients to obtain stable
resummed anomalous dimensions: we summarize the whole procedure here.
\begin{itemize}
\item The first crucial ingredient is the duality relation Eq.~\eqref{eq:duality}:
  strictly speaking, it is valid for fixed coupling, but its extension to the running coupling level
  (which amounts to the replacement of $\as$ with $\ashat$) has been proved to be valid to all orders
  in Ref.~\cite{Ball:2005mj}. The duality relates the BFKL kernel to the anomalous dimension,
  thereby providing a way to resum one of them in terms of the other.
  This procedure alone gives the Double-Leading approximation introduced in Sect.~\ref{sec:DL-approx},
  which is however unstable for perturbative corrections.
\item The way to stabilize the DL expansion is to exploit the intrinsic symmetry of the BFKL kernel,
  which is however valid for a choice of variables which is not the usual one in DIS. Then, in order
  to be able to take the advantage of such symmetry we move to symmetric variables, changing also the
  anomalous dimension via Eq.~\eqref{eq:ker_symm_rel_gamma},
  and we will get back again to asymmetric variables only at the end.
  With the help of the off-shell kernel, we symmetrize the DL result, to get a stable perturbative expansion.
\item The last ingredient is the resummation of running coupling effects.
  It produces an anomalous dimension which now has the correct resummed small-$N$ behaviour,
  which is completely determined by the value and the curvature at the minimum (and their $\as$ dependence)
  of the symmetrized DL BFKL kernel.
  Then, by matching with the symmetrized DL result, and going back to asymmetric variables,
  we finally get our resummed anomalous dimension.
\end{itemize}
The result of all this machinery can then be summarized in the following expression,
\beq\label{eq:gamma_res_final}
\gamma_{\rm res}(\as,N) = \gamma_B(\as,N) + \gamma_{\rm DL}^\sigma(\as,N) - \gamma_{B,s}(\as,N) - \gamma_{B,ss,0}(\as) + \frac{N}{2} + \gamma_{\rm mom}(N),
\eeq
where:
\begin{itemize}
\item $\gamma_B(\as,N)$ is the Bateman anomalous dimension: it resums running coupling effects and is accurate at small $N$,
  while at large $N$ is completely arbitrary;
\item $\gamma_{\rm DL}^\sigma(\as,N)$ is the symmetrized DL anomalous dimension obtained by solving
  \beq
  \bar\chi_{\rm DL}^{\sigma}(\as,\gamma_{\rm DL}^\sigma(\as,N), N) = N
  \eeq
  in symmetric variables: it has the correct large-$N$ behaviour but has a spurious square-root branch-cut at small $N$;
\item $\gamma_{B,s}(\as,N)$ subtracts double counting contributions between the two previous contributions,
  thereby removing the branch-cut in $\gamma_{\rm DL}^\sigma$ and a part of the spurious large-$N$ behaviour of $\gamma_B$;
\item $\gamma_{B,ss,0}(\as)$ removes the remaining spurious behaviour (a constant) of $\gamma_B$;
\item $N/2$ switches back to DIS (asymmetric) variables;
\item $\gamma_{\rm mom}(N) = -\bar\chi_{\rm mom}(N)$ restores momentum conservation (spoiled by running coupling resummation);
  in particular, since by construction $\gamma_{\rm DL}^\sigma(\as,N)$ preserves momentum conservation, here $c_{\rm mom}$ is given by
  \beq
  c_{\rm mom} = \gamma_B(\as,1) - \gamma_{B,s}(\as,1) - \gamma_{B,ss,0}(\as).
  \eeq
\end{itemize}
Actually this scheme is, strictly speaking, valid only at LO at $n_f=0$.

At NLO there is a mismatch between the parameters in the Bateman and DL pieces,
and a supplementary $\gamma_{\rm match}$ is needed in order to have the correct cancellation
of the square-root branch-cut and of the spurious large-$N$ behaviour, see Ref.~\cite{abf742}
for more details. Since there are many tricky aspects at NLO which are irrelevant
for the discussion, we don't show them here but we collect the results in App.~\ref{sec:app-SX-NLO}.

Already at LO, when $n_f\neq0$ some care is needed to avoid an unphysical growth in $x$-space
when Mellin-inverting the anomalous dimension matrix rotated back to the physical basis.
We discuss this now.

\subsection{The case $n_f\neq0$}
\label{sec:rat_approx_improved}

The (fixed order) eigenvalues of the anomalous dimension matrix have a branch-cut due to
the square-root of the solution Eq.~\eqref{eq:gamma_pm} of the secular equation.
This branch-cut is unphysical and indeed is not present in the matrix element of the anomalous
dimension matrix in the physical basis.

However, during the resummation procedure, we modify the largest eigenvalue,
and typically the cancellation of the branch-cut going back to the physical basis
does no longer take place. This is not acceptable because the
inverse Mellin of the branch-cut has a spurious huge growth at small $x$.
A simple way to preserve the branch-cut cancellation is to build a
\beq
\Delta\gamma_{\rm res} = \gamma_{\rm res} - \gamma_{\rm fix}
\eeq
free of branch-cut, that is to say that the branch-cut in $\gamma_{\rm res}$ has to be
the same of that appearing in the fixed-order anomalous dimension.\footnote{%
Actually this constraint is not sufficient to completely cancel the branch-cut;
in particular, the $gq$ component of the anomalous dimension would still have the cut.
We will see in Sect.~\ref{sec:back_to_physical_basis} that full cancellation
can be still achieved with a minimal modification that is subleading in both $\as$
and $\as/N$ expansions.}
However, when $n_f\neq0$, the function $\Delta\gamma_{\rm res}$ built as in
Eq.~\eqref{eq:gamma_res_final} has a cut, inherited from $\gamma^\sigma_{\rm DL}$.

In \cite{abf799} the difference between $\Delta\gamma_{\rm res}$ at a given $n_f$ (which has the cut)
and the same computed at $n_f=0$ (which is free of cut) is substituted with a rational approximation
(hence, free of cut); however, incresing the order of the approximation, some oscillations due to
the re-appereance of the cut show up, making this procedure unstable.

A better solution is to use in the computation of $\gamma^\sigma_{\rm DL}$ a fake fixed-order
$\gamma_{\rm fix}$, which should be accurate at small $N$ but can be arbitrary at large $N$.
In this way $\Delta\gamma_{\rm res}$ is still good, since at small $N$ it is accurate,
and at large $N$ it goes to zero whatever $\gamma_{\rm fix}$ we use.
Moreover, provided $\gamma_{\rm fix}$ is free of cut, the $\Delta\gamma_{\rm res}$ we get is free of cut,
as required.
The easiest way to choose the fake $\gamma_{\rm fix}$ is to use the complete $n_f=0$ part
plus the leading $n_f$ dependent pieces at small $N$. By calling this fake function
$\tilde\gamma_{\rm fix}$, we have
\beq
\tilde\gamma_{\rm fix}(\as,N) = \as \left.\gamma_0(N)\right|_{n_f=0}
+ \as \[ \as\left.\gamma_1(N)\right|_{n_f=0} + c_{10}n_f + c_{11}n_f\frac{\as}{N} \] + {\rm NNLO},
\eeq
where the coefficients
\beq
c_{10} = \frac{2C_F/C_A-1}{6\pi},\qquad
c_{11} = \frac{26C_F-23C_A}{36\pi^2}
\eeq
can be read from Eq.~\eqref{eq:gamma+expansion}. Note that at NLO $\tilde\gamma_{\rm fix}$ no longer
preserves momentum conservation.

It's a matter of choice if computing the parameters of the Bateman anomalous dimension $\gamma_B$
from the $\gamma^\sigma_{\rm DL}$ made of the fake $\gamma_{\rm fix}$ or of the real one; the difference is subleading.
However, since the small-$x$ behaviour of the splitting function is given by the Bateman contribution,
one may want to put in $\gamma_B$ the most accurate parameter as possible. In this way, however,
$\gamma_{B,s}$ no longer cancels the small-$N$ branch-cut of $\gamma^\sigma_{\rm DL}$,
because there is a mismatch between the parameters. This mismatch can be cured by a $\gamma_{\rm match}$
like that used at NLO.

\begin{figure}[tb]
  \centering
  \includegraphics[width=0.8\textwidth,page=1]{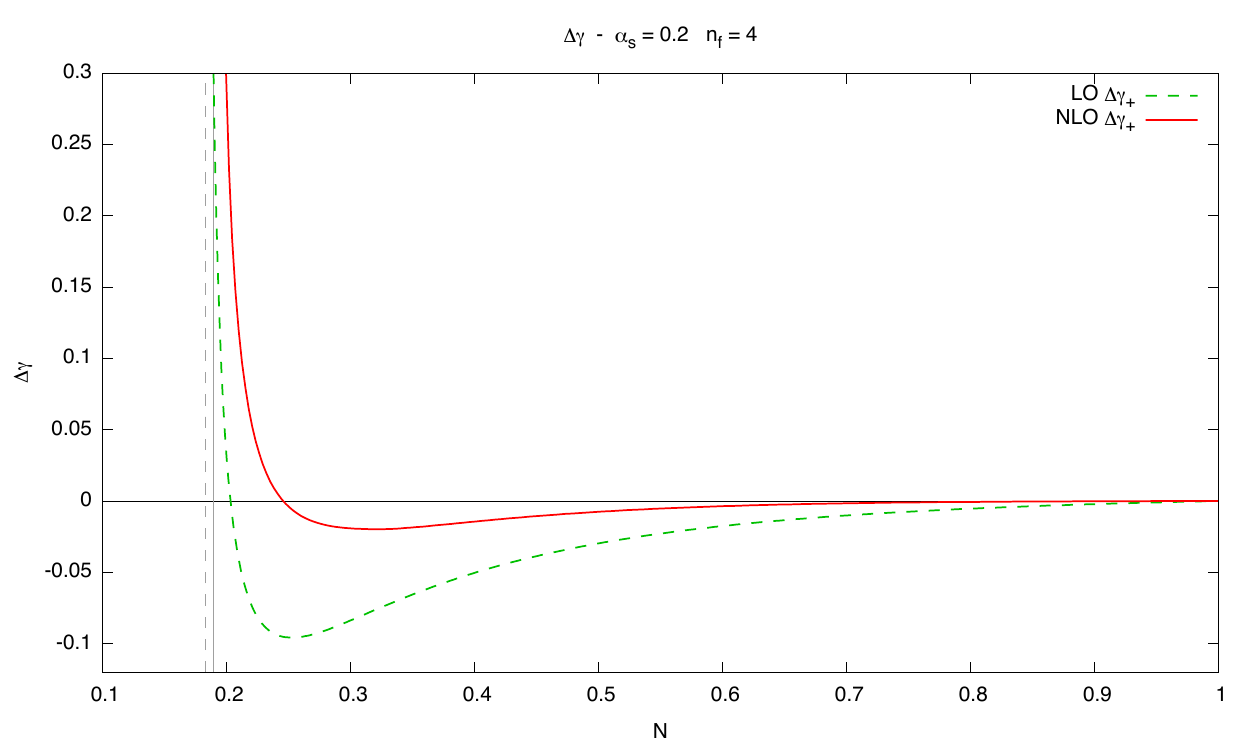}
  \caption{$\Delta \gamma_+(\as,N)$ for $\as=0.2$ and $n_f=4$ at LO and NLO.
    The Bateman pole position is shown at LO and NLO as well.}
  \label{fig:Deltagammaplus}
\end{figure}

The result for the resummed anomalous dimension at LO and NLO is shown
in Fig.~\ref{fig:Deltagammaplus}.
The Bateman pole is pushed to a somewhat larger value at NLO.
The small-$N$ rise is concentrated in the region very close to the Bateman pole.
A negative dip dominates instead the intermediate region:
this dip is softer in the NLO case.

\subsection{Resummation of quark anomalous dimensions}

At NLL, also the quark anomalous dimensions need to be resummed.
Such resummation can be expressed in terms of the resummation of the
gluon anomalous dimension~\cite{Catani:1994sq}.
In Ref.~\cite{abf799}, the running coupling effects are also taken into account:
the (NLL) singular contributions are then given by
\beq\label{eq:gamma_qg_ss_rc}
\gamma^{ss}_{qg}(\as, N) = h_{qg}\(\[\gamma_s(\as/N)\]\)
\eeq
where $h_{qg}$ is a function defined by its series expansion
\beq
h_{qg}(M) = \sum_{k=0}^{\infty} h_{qg,k} M^k
\eeq
and the square brackets mean that each power in the series must be substituted by
\beq
\[\gamma^k\] = \(\frac{\dot\gamma}{\gamma}\)^k \frac{\Gamma\(\gamma^2/\dot\gamma+k\)}{\Gamma\(\gamma^2/\dot\gamma\)}
= \(\frac{\dot\gamma}{\gamma}\)^k \(\frac{\gamma^2}{\dot\gamma}\)_k,
\eeq
where $(a)_k = \Gamma(a+k)/\Gamma(a)$ is the Pochhammer symbol.
However, the series is divergent~\cite{abf799}, and then its resummation is needed to get
a meaningful result. However, the coefficients $h_{qg,k}$ can be computed only
perturbatively, and it is hard to work many of them out.
For more details on how to perform such resummation, see App.~\ref{sec:resummation_quark_AD}.

The resummed $qg$ anomalous dimension can be written as
\beq\label{eq:gamma_qg_res}
\gamma_{qg}^{\rm res}(\as,N) = \gamma_{qg}^{\rm NLO}(\as,N) + \Delta\gamma_{qg}^{\rm res}(\as,N)
\eeq
and has to satisfy the limits
\beq\label{eq:gamma_qg_res_limits}
\begin{cases}
  \gamma_{qg}^{\rm res}(\as,N) \sim \gamma_{qg}^{\rm NLO}(\as,N) &\quad\text{as } N\to\infty \text{ or } \as\to0\\
  \gamma_{qg}^{\rm res}(\as,N) \sim \as\gamma_{qg}^{ss}(\as,N) &\quad\text{as } N\to0
\end{cases}
\eeq
which translate into analogous limits for $\Delta\gamma_{qg}^{\rm res}(\as,N)$.
These limits are trivially satisfied if we use as argument of $h_{qg}$ the
function $\gamma_s$:
\beq\label{eq:Delta_gamma_qg_res_0}
\Delta\gamma_{qg}^{\rm res}(\as,N) = \as\, h_{qg}\(\[\gamma_s(\as/N)\]\) - (\text{small-$x$})_{qg}^{\rm NLO},
\eeq
where the subtracted terms are the small-$x$ terms which are already present in the NLO
contribution of Eq.~\eqref{eq:gamma_qg_res}, i.e.
\beq
(\text{small-$x$})_{qg}^{\rm NLO} =
\as\, \frac{n_f}{3\pi} \[ 1+\frac53 \frac{C_A}{\pi} \frac{\as}{N} \].
\eeq
However, this expression inherits the branch-cut of $\gamma_s$ (which is the naive dual of $\chi_0$),
and this would give rise to a spurious small-$x$ growth.
To circumvent this, in Ref.~\cite{abf799} $\gamma_s$ is substituted by the full
NLO\footnote{In principle one could use the LO resummed anomalous dimension;
however, the use of the NLO one guarantees that the leading Bateman pole,
which gives the leading small-$x$ behaviour, is at the same position in all the entries of the
anomalous dimension matrix. An intermediate solution which preserve this property
would be to use the LO resummed anomalous dimension with the Bateman parameters taken
from the NLO: this solution coincides up to subleading terms with the NLO one,
but has a faster numerical implementation.}
resummed anomalous dimension $\gamma_+^{\rm res,\, NLO}$, which however introduces also
some spurious large-$x$ terms which would spoil the limits in Eq.~\eqref{eq:gamma_qg_res_limits}.
Then the expression in Ref.~\cite{abf799} is
\begin{multline}\label{eq:Delta_gamma_qg_res_abf799}
\Delta\gamma^{qg}_{\rm res}(\as,N) = \as \[ h_{qg}\(\[\gamma_+^{\rm res,\,NLO}(\as,N)\]\)
- h_{qg}\(\[\gamma_+^{\rm NLO}(\as,N) - (\text{small-$x$})_+^{\rm NLO} \]\)\]\\
- (\text{small-$x$})_{qg}^{\rm NLO} + \as\, h_{qg}(0),
\end{multline}
where the second term in parenthesis correct the spurious large-$x$,
and the last line eliminates the double counting, but noting that
the constant term $h(0)$ has already been eliminated by the subtraction in the first row.
The subtracted small-$x$ contribution to the largest eigenvalue is
\beq
(\text{small-$x$})_+^{\rm NLO} =
\frac{C_A}{\pi} \frac{\as}{N}
- \as \[\frac{11 C_A+2n_f(1-2C_F/C_A)}{12\pi} + n_f\frac{23C_A-26C_F}{36\pi^2} \frac{\as}{N} \],
\eeq
as can be easily extracted from Eqs.~\eqref{eq:gamma+expansion}.

We propose here a different and easier way to obtain an equivalent result,
which possibly differs by subleading contributions.
Remember that the expression in Eq.~\eqref{eq:Delta_gamma_qg_res_0} is in principle correct,
the only problem being the spurious branch-cut.
The idea of substituting $\gamma_s$ with $\gamma_+^{\rm res,\, NLO}$ solves the problem introducing
differences which are subleading: indeed, from the small-$x$ point of view,
$\gamma_+^{\rm res,\, NLO}$ has the same leading small-$x$ structure than $\gamma_s$.
In the same spirit, we propose to use as argument of $h_{qg}$ the difference
\beq
\gamma_+^{\rm res,\,NLO}(\as,N) - \gamma_+^{\rm NLO}(\as,N) + (\text{small-$x$})_+^{\rm NLO},
\eeq
which has the requested both small-$x$ and large-$x$ behaviours
(the last term restores the NLO small-$x$ terms subtracted by the second term).
With this choice we simply have
\begin{multline}\label{eq:Delta_gamma_qg_res}
\Delta\gamma_{qg}^{\rm res}(\as,N) =
\as\, h_{qg}\(\[\gamma_+^{\rm res,\,NLO}(\as,N) - \gamma_+^{\rm NLO}(\as,N) + (\text{small-$x$})_+^{\rm NLO}\]\)\\
- (\text{small-$x$})_{qg}^{\rm NLO}.
\end{multline}
This expression is also simpler for numerical evaluation, since the function $h_{qg}$
must be computed only once.

The same procedure can be used for the resummation of coefficient functions.
We do not give many details here; some minimal details are given in App.~\ref{sec:resummation_coeff_funct}.

\subsection{Back to the physical basis}
\label{sec:back_to_physical_basis}

The last step to complete the small-$x$ resummation of anomalous dimensions
consists in constructing the full anomalous dimension matrix in the physical basis.

At LO, the quark anomalous dimension do not resum, and then we can
build the anomalous dimension matrix from the quark anomalous dimensions
and the two eigenvectors, one of which is resummed.
This can be easily achieved using Eq.~\eqref{eq:Gamma_proj}, the result being
\beq\label{eq:Gamma_in_terms_of_eigen}
\Gamma = \sqmatr{(\gamma_++\gamma_--\gamma_{qq})}{X}{\gamma_{qg}}{\gamma_{qq}},
\qquad
X= \frac{(\gamma_+-\gamma_{qq})(\gamma_{qq}-\gamma_-)}{\gamma_{qg}},
\eeq
where $\gamma_+$ has to be replaced with $\gamma_+^{\rm res}$ at LL.
Then, writing
\beq
\gamma_+^{\rm res} = \gamma_+ + \Delta\gamma_+^{\rm res},
\eeq
we obtain for the gluon entries
\begin{align}
  \gamma_{gg}^{\rm res} &= \gamma_{gg} + \Delta\gamma_+^{\rm res} \\
  \gamma_{gq}^{\rm res} &= \gamma_{gq} + \frac{\gamma_{qq}-\gamma_-}{\gamma_{qg}}\,\Delta\gamma_+^{\rm res}.
\end{align}
As described above, $\Delta\gamma_+^{\rm res}$ is built to be free of branch-cuts,
in order to avoid spurious small-$x$ enhancements in the physical matrix entries.
Indeed, $\gamma_{gg}^{\rm res}$ does not have any branch-cut.
However, $\gamma_{gq}$ inherit the branch-cut of $\gamma_-$;
nevertheless, since $(\gamma_{qq}-\gamma_-)/\gamma_{qg}$ is subleading at small
$N$ with respect to $\Delta\gamma_+^{\rm res}$, we can substitute its value in $N=0$,
which at LO is $C_F/C_A$, obtaining finally
\beq
\Gamma_{\rm LO}^{\rm res} = \Gamma_{\rm LO} + \Delta\Gamma_{\rm LO}^{\rm res},\qquad
\Delta\Gamma_{\rm LO}^{\rm res} =
\sqmatr{\Delta\gamma_+^{\rm res}}{\frac{C_F}{C_A}\Delta\gamma_+^{\rm res}}00.
\eeq
Note that this result respects, in the small-$N$ limit, the colour-charge relation Eq.~\eqref{eq:colour-charge_rel}.

At NLO, also the resummation of the quark anomalous dimensions is also needed.
Then we define
\begin{align}
  \gamma_{qg}^{\rm res} &= \gamma_{qg} + \Delta\gamma_{qg}^{\rm res} \\
  \gamma_{qq}^{\rm res} &= \gamma_{qq} + \Delta\gamma_{qq}^{\rm res}
\end{align}
where the two $\Delta$ terms satisfy the colour-charge relation
\beq
\Delta\gamma_{qq}^{\rm res} = \frac{C_F}{C_A} \,\Delta\gamma_{qg}^{\rm res}
\eeq
coming from Eq.~\eqref{eq:colour-charge_rel_quark}.
Then, from Eq.~\eqref{eq:Delta_gamma_qg_res}, we have both
quark entries, and moreover we know $\Delta\gamma_+^{\rm res}$ at NLO.
In terms of these, the gluon entries become, at NLO,
\begin{align}
  \gamma_{gg}^{\rm res} &= \gamma_{gg} + \Delta\gamma_+^{\rm res} - \Delta\gamma_{qq}^{\rm res} \label{eq:Delta_gamma_gg_NLO}\\
  \gamma_{gq}^{\rm res} &= \gamma_{gq} + \Delta\gamma_{gq}^{\rm res}
\end{align}
with
\beq
\Delta\gamma_{gq}^{\rm res} = \frac{\gamma_{qq}-\gamma_-}{\gamma_{qg}}\, \Delta\gamma_+^{\rm res}
+\[\frac{C_F}{C_A}\,\frac{\gamma_{gg}-\gamma_{qq}}{\gamma_{qg}}-\frac{\gamma_{gq}}{\gamma_{qg}}\]
\Delta\gamma_{qg}^{\rm res},
\eeq
as one can find expanding $X$ in Eq.~\eqref{eq:Gamma_in_terms_of_eigen}
to first power\footnote{Actually, a term proportional to the product
$\Delta\gamma_+^{\rm res}\,\Delta\gamma_{qg}^{\rm res}$
would contribute at NLL, but such a term is not present in the expansion.}
in $\Delta\gamma_+^{\rm res}$ and $\Delta\gamma_{qg}^{\rm res}$
(or, alternatively, using the determinant condition $\gamma_+\gamma_-=\gamma_{gg}\gamma_{qq}-\gamma_{qg}\gamma_{gq}$
at the resummed level).
To avoid spurious branch-cuts, as in the LO case, we may substitute the coefficients of the $\Delta$
terms with their value in $N=0$, obtaining
\beq
\Delta\gamma_{gq}^{\rm res} = \(1+k\,\as\)\frac{C_F}{C_A}\,\Delta\gamma_+^{\rm res}
-\frac{C_F}{C_A}\, \frac{C_A+n_f}{2n_f}\, \Delta\gamma_{qg}^{\rm res}
\eeq
with
\beq
k = \frac{1}{12\pi}\[
3 C_A + \(1 - 2\frac{C_F}{C_A}\)n_f
\].
\eeq
The inclusion of the order $\as$ piece in the first term is needed because it produces
NLL terms by interference with the LL part of $\Delta\gamma_+^{\rm res}$.
Then, writing
\beq
\Gamma_{\rm NLO}^{\rm res} = \Gamma_{\rm NLO} + \Delta\Gamma_{\rm NLO}^{\rm res},
\eeq
we have finally
\beq
\Delta\Gamma_{\rm NLO}^{\rm res} =
\sqmatr
{\Delta\gamma_+^{\rm res}-\frac{C_F}{C_A}\Delta\gamma_{qg}^{\rm res}}
{\Delta\gamma_{gq}^{\rm res}}
{\Delta\gamma_{qg}^{\rm res}}
{\frac{C_F}{C_A} \Delta\gamma_{qg}^{\rm res}}.
\eeq

\subsection{Schemes}

So far, in this Chapter we didn't ever talk about the factorization scheme.
Concerning the GLAP equation, we have tacitly used the \MSbar\ scheme, since it
is the most commonly used.
However, all the results presented in this Chapter are given in a different scheme
which is usually called \QMSbar~\cite{Ciafaloni:1995bn,Ciafaloni:2005cg,abf799}.
This scheme differs to the \MSbar\ scheme only at small-$x$, and in particular
they coincide at fixed order up to NNLO, while they start differing at N$^3$LO:
therefore, all the anomalous dimension used so fare are unchanged in \QMSbar\ scheme.

The theory of small-$x$ scheme changes at the resummed level has been developed
in Ref.~\cite{Ball:1995tn,abf599}, and we don't want to discuss it here.
We simply want to emphasize that the \QMSbar\ scheme is more suitable for
small-$x$ resummation, in particular for the resummation of running-coupling
effects~\cite{abf799}. At the end, one could choose to go back to \MSbar\ scheme,
but the price to pay is the introduction of new singularities which should cancel
in a hadronic computation, but which are numerically inconvenient.
Therefore we prefer to work with the \QMSbar\ scheme.

Finally, we would like to recall that for all the schemes we are considering
the following colour-charge relations are satisfied
\begin{align}
\gamma_s^{gq}\(\frac{\as}{N}\) &= \frac{C_F}{C_A} \gamma_s^{gg}\(\frac{\as}{N}\)
\label{eq:colour-charge_rel}\\
\gamma_{ss}^{qq}\(\frac{\as}{N}\) &= \frac{C_F}{C_A} \[\gamma_{ss}^{qg}\(\frac{\as}{N}\) - \gamma_{ss}^{qg}(0) \],
\label{eq:colour-charge_rel_quark}
\end{align}
with
\beq
\gamma_{ss}^{qg}(0) = \frac{n_f}{3\pi}.
\eeq

\section{Resummed splitting function}

From $\Delta \Gamma^{\rm res}$ in $N$-space we can compute the resummed splitting functions
$\Delta P_{ij}^{\rm res}$ by taking a numerical inverse Mellin transform.
Since the resummation affects the small-$x$ region, $\Delta P_{ij}^{\rm res}$ are normal functions
rather than distributions (the distributional nature of the splitting functions regards the $x=1$ endpoint,
which is left unchanged).
Hence, they are a more suitable quantity to look at, since they are real and defined in a finite range $0<x<1$.
Moreover, the computation of $\Delta \Gamma^{\rm res}$ is not numerically robust,
since it needs several root-finding calls for the computation of $\chi_s$ and for putting
on-shell the off-shell kernels. Therefore, if the goal is the computation of the splitting functions
we can specialize the code to be stable along the inverse Mellin integration path.

The fixed Talbot algorithm described in App.~\ref{sec:fixed-talbot-algorithm}
turns out to be not the best choice, since the numerical stability decreases
going closer to the negative real axis.
The straight line of Eq.~\eqref{eq:inverse_Mellin_straight} provides a better stability.
Close to the real axis, the convergence of the root-finding algorithms
is still not optimal, and a sampling along the integration path has been adopted
to provide good initial guesses for the algorithms.

\begin{figure}[tb]
  \centering
  \includegraphics[width=0.8\textwidth,page=1]{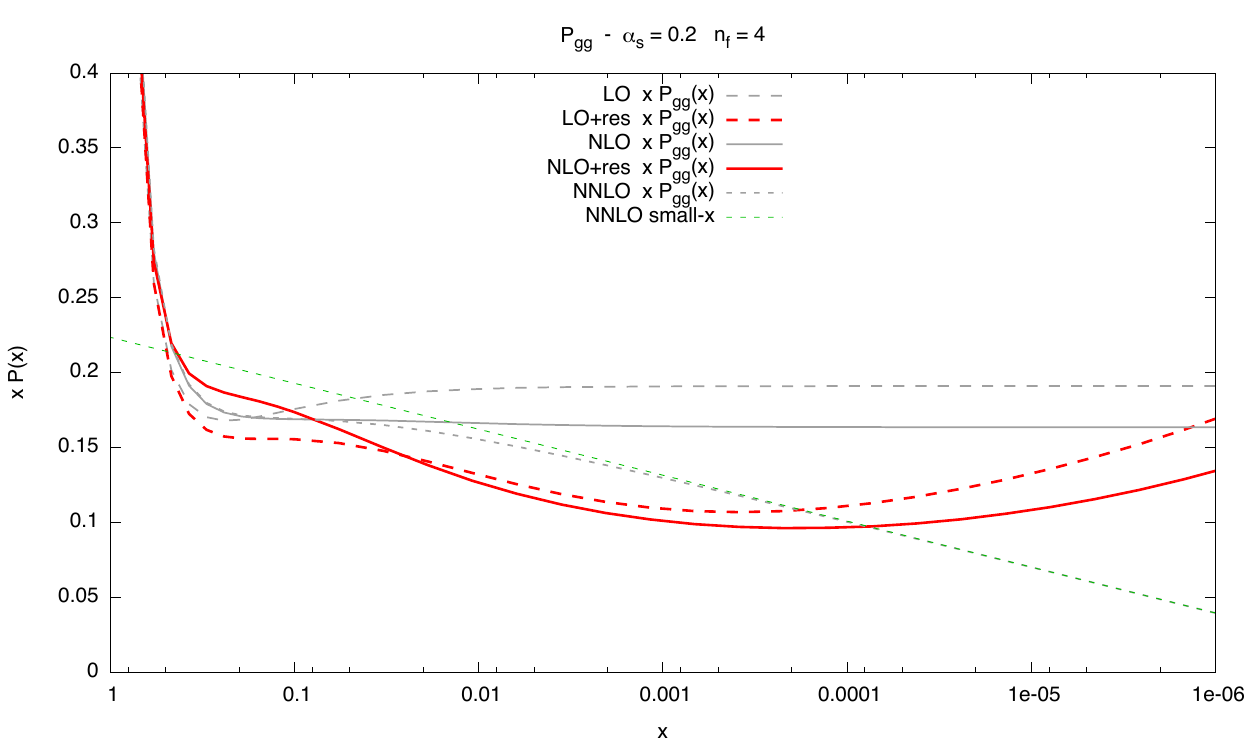}\\
  \includegraphics[width=0.8\textwidth,page=1]{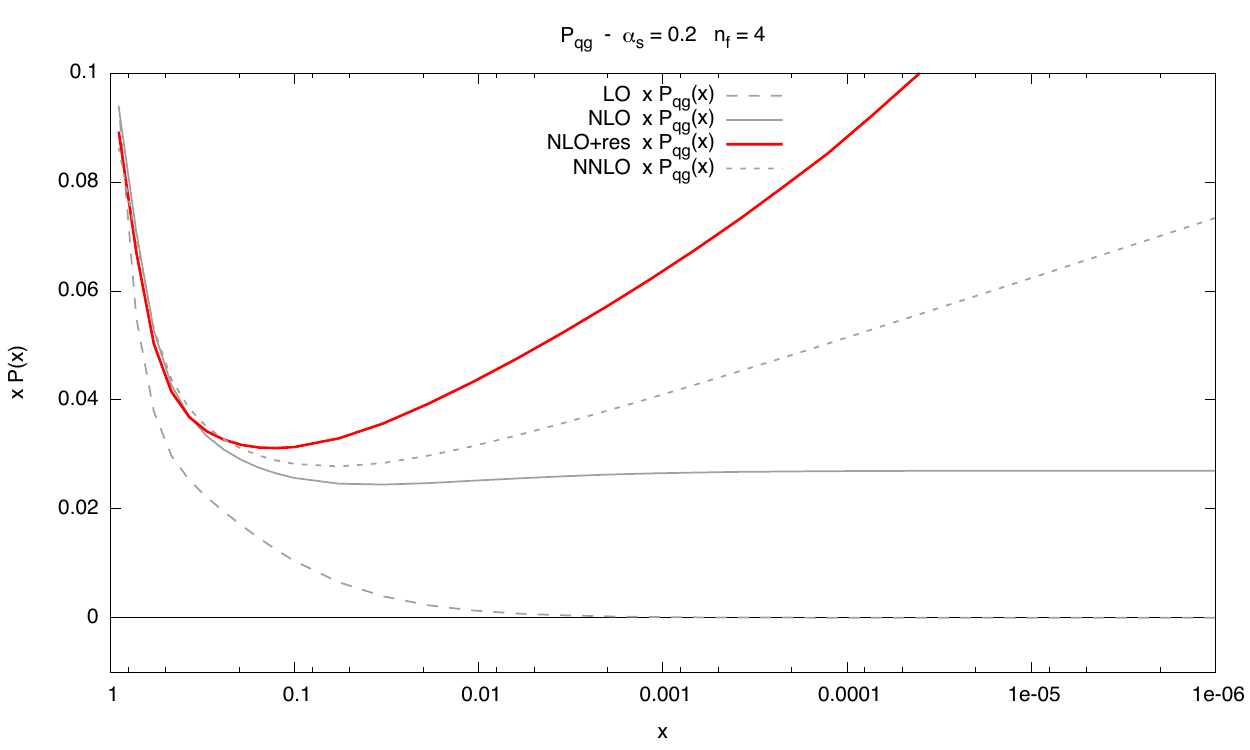}
  \caption{Unresummed and resummed $P_{gg}(\as,x)$ (upper plot) and $P_{qg}(\as,x)$ (lower plot)
    for $\as=0.2$ and $n_f=4$.
    The matching Eq.~\eqref{eq:Delta_gamma_qg_res} has been adopted, and a Pad\'e approximant $[8/8]$
    has been used to compute the Borel sum of the series $h_{qg}$.}
  \label{fig:DeltaPres}
\end{figure}

As an example, we present In Fig.~\ref{fig:DeltaPres} the results for the
resummed $gg$ and $qg$ components of the splitting function matrix,
for $\as=0.2$ and $n_f=4$, which are the values used in Ref.~\cite{abf799}.
Concerning $P_{gg}$, also the LO resummed result is shown: it agrees pretty well
to that of Ref.~\cite{abf799}, apart in the large-$x$ region, the difference being
due to the improved approximation described in Sect.~\ref{sec:rat_approx_improved}.
The NLO resummed result differs a bit more even at small $x$: however,
being $\Delta P_{gg}$ built up from $\Delta P_+$ and $\Delta P_{qg}$, Eq.~\eqref{eq:Delta_gamma_gg_NLO},
we can trace this difference in a significant difference in $\Delta P_{qg}$.
Indeed, the second plot shows a rather different behaviour for the resummed $P_{qg}$,
even if the asymptotic behaviour at small $x$ seems to be not that different.
This difference can be mainly traced in the different matching adopted to obtain
the two curves, Eq.~\eqref{eq:Delta_gamma_qg_res} here and Eq.~\eqref{eq:Delta_gamma_qg_res_abf799}
in Ref.~\cite{abf799}.
Since both matching are allowed, the difference could be considered
as an estimator of the uncertainty induced by the subleading terms included
in the resummation procedure.

\subsection{Approximations}

Once we have $\Delta \Gamma^{\rm res}$ in the complex plane we could use it to perform the GLAP evolution.
However, as described above, numerical stability is hardly achieved,
and then for fast numerical applications the use of simple analytic approximation
of $\Delta\Gamma^{\rm res}$ is required.

A good and accurate way to obtain such approximation is to fit the splitting functions $\Delta P^{\rm res}$,
computed as the inverse Mellin of $\Delta \Gamma^{\rm res}$, in terms of an appropriate parametrization.
This function appears to be a quite smooth function of $\log x$; moreover, since the convolutions
involve the splitting functions in a range that ends at $x=1$ but starts at small values of $x$ related
to the hadronic scale of the process, we practically need to know the splitting function not really down to $x=0$,
but just down to a small value, making it possible to approximate $\Delta P^{\rm res}$ in a finite $\log x$ range.

Since the matrix $\Delta\Gamma^{\rm res}$ can be built from $\Delta\gamma_+^{\rm res}$ and
$\Delta\gamma_{qg}^{\rm res}$, it is actually sufficient to fit these two functions (their inverse Mellin),
which are the primitive outputs of the resummation code,
and build analytically at the end the entire matrix according to Sect.~\ref{sec:back_to_physical_basis}.

Since the asymptotic behaviour (at small $x$) of $\Delta P^{\rm res}$ is given by the
the rightmost pole of $\Delta \Gamma^{\rm res}$,
\beq
\Delta \Gamma^{\rm res}(N) = \frac{r_B}{N-N_B} + \ldots,
\eeq
we can fix it and fit the difference (in $x$-space).
For better results, it's actually more accurate to fix the two rightmost poles $N_B$ and $N_B'$,
with residues $r_B$ and $r_B'$ respectively (these are computed numerically from $\Delta\Gamma^{\rm res}$).
Then we can fit a polynomial in $x$ and $\log\frac1x$, i.e.\ we use the parametrization
\beq
x\,\Delta P^{\rm res}(\as,x) = r_B \, x^{-N_B} + r_B' \, x^{-N_B'} + \sum_{k,\,j\geq0} c_{kj} \, x^k\, \log^j \frac1x.
\eeq
Note that $\Delta P^{\rm res}(1) = r_B+r_B'+c_{00}$ is a constant, as it should since
$\Delta \Gamma^{\rm res}(N)$ goes to zero at least as $1/N$ at large $N$.
Its Mellin transform is
\beq
\Delta \Gamma^{\rm res}(N) = \frac{r_B}{N-N_B} + \frac{r_B'}{N-N_B'} + \sum_{k,\,j\geq0} c_{kj} \, \frac{j!}{(N+k)^{j+1}}.
\eeq
For the case of the largest eigenvalue, momentum conservation constraint $\Delta\gamma_+^{\rm res}(1)=0$ imposes
\beq
\frac{r_B}{1-N_B} + \frac{r_B'}{1-N_B'} + \sum_{k,\,j} c_{kj} \, \frac{j!}{(1+k)^{j+1}} = 0,
\eeq
which fixes one of the free parameters in terms of the others.

\chapter{Combining resummations}

\minitoc

\noindent
In this Chapter we will discuss in detail the relevance of threshold
resummation. We will establish a quantitative way to assess for which values
of the hadronic variables a process is dominated by the threshold logarithms,
thereby determining is their resummation is needed or not.
We will then show how to get phenomenological implication
from the high-energy resummation formalism introduced in Chap.~\ref{chap:small-x}.
Finally, with specific reference to the Higgs boson production,
we will discuss the effect of small-$x$ resummation on the determination
of the threshold region.

\section{When is threshold resummation relevant?}
\label{sec:when_is_relevant}

First, we want to establish here a way to asses when threshold
resummation is relevant for phenomenology~\cite{bfr2}.
Of course, when the hadronic ratio $\tau=M^2/s$ is close to threshold,
$\tau\to 1$, all contributions to the cross-section come from the
threshold region and threshold resummation cannot be neglected.
But in phenomenologically interesting process, such those at the LHC and Tevatron,
$\tau$ is always very small, far from threshold.

However, it has been pointed out since long~\cite{Appell:1988ie} that
because hadronic cross-sections are found convoluting a hard
cross-section with a parton luminosity, Eq.~\eqref{eq:inclus_cs_conv},
the effect of resummation may be
relevant even relatively far from the hadronic threshold.  Indeed, in
Ref.~\cite{bolz} threshold resummation has been claimed to affect
significantly Drell-Yan production for E866 kinematics, though
somewhat different results have been found in
Ref.~\cite{bnx}.  It is important to observe that Drell-Yan
data from E866 and related experiments play a crucial role in the
precision determination of parton distributions~\cite{Ball:2010de}, so
their accurate treatment is crucial for precise LHC
phenomenology. Furthermore, threshold resummation is known~\cite{Grazzini:2010zc} to
affect in a non-negligible way standard Higgs production in
gluon-gluon fusion at the LHC, even though the process is clearly
very far from threshold.

The standard physical argument to explain why resummation may be
relevant even when the hadronic process is relatively far from
threshold goes as follows~\cite{Catani}. The quantity which is
resummed in perturbative QCD is the hard partonic cross-section, which
depends on the partonic center-of-mass energy and the dimensionless
ratio of the latter to the final state invariant mass. Therefore,
resummation is relevant when it is the partonic subprocess that is
close to threshold. The partonic center-of-mass energy in turn can
take any value from threshold up to the hadronic center-of-mass
energy, and its mean value is determined by the shape of the PDFs:
therefore, one expects threshold resummation to be more important if
the average partonic center-of-mass energy is small, i.e.\ if the
relevant PDFs are peaked at small $x$ (such as gluons or sea quarks,
as opposed to valence quarks). This for instance explains why
threshold resummation is especially relevant for Higgs production in
gluon-gluon fusion.

We will show that this can be made quantitative using a saddle-point
argument in Mellin space~\cite{bfr2}: for any given value of the hadronic
ratio $\tau$, the dominant contribution to the cross-section
comes from a narrow range of the variable $N$, conjugate to $\tau$
upon Mellin transform.
In Mellin space the cross-section factorizes in the product of a parton
luminosity $\Lum$ and a hard coefficient function $C$,
but it turns out that the position of the saddle is
mostly determined by the PDF luminosity. Moreover, the result is quite
insensitive to the non-perturbative (low-scale) shape of the parton
distribution and mostly determined by its scale dependence,
specifically by the low-$x$ (or low-$N$) behaviour of the relevant
Altarelli-Parisi splitting functions: the faster the small-$x$ growth
of the splitting function, the smaller the average partonic
center-of-mass energy, the farther from the hadronic threshold the
resummation is relevant. This is reassuring, because it means that the
region of applicability of threshold resummation is controlled by
perturbative physics. Moreover, this suggest a connection between
threshold (large-$z$) resummation and high-energy (small-$x$) resummations:
we will come back on this in Sect.~\ref{sec:joint}.

The issue of the persistence of sizable
soft gluon  emission terms even far from threshold was also addressed
in Ref.~\cite{bnx}  using methods of
soft-collinear effective theory, but in the large $\tau\gtrsim 0.2$ region 
it was related to the (non-perturbative)
shape of parton distributions, while for smaller $\tau$ values it was
also observed, but left unexplained.
The treatment of this issue in soft-collinear effective
theory issue was revisited in a quantitative way in
Ref.~\cite{Bauer:2010jv}, where it was related to a parameter
determined by the shape of parton distributions.

We now show how we can assess the impact of parton distributions
by means of a Mellin-space argument~\cite{bfr2}.
For a given process and a given value of $\tau=M^2/s$, we
determine the region of the variable $N$ which provides the
dominant contribution to the cross-section.
We show that the such region is mostly determined by the small-$x$
behaviour of the PDFs, which in turn is driven by perturbative evolution.
Then, we assess the $N$ region where threshold
resummation is relevant, and putting everything together,
we obtain the resummation region for the Drell-Yan process
and the Higgs process.

\subsection{Saddle-point argument}

A cross-section for a hadronic process with scale $M^2$ and
center-of-mass energy $s=M^2/\tau$ can be written as a sum of
contributions of the form
\beq
\label{eq:fact}
\sigma(\tau,M^2) = \int_\tau^1\frac{dz}{z} \,
\Lum(z,M^2) \,C\(\frac{\tau}{z},\as(M^2)\)
\eeq
in terms of a partonic coefficient function $C$ and a parton
luminosity, in turn determined in terms of parton distributions $f_i(x_i)$ as
\beq
\label{eq:lumi}
\Lum(z,\mu^2)=\int_z^1\frac{dx}{x}\, f^{(1)}(x,\mu^2) \, f^{(2)}\(\frac{z}{x},\mu^2\).
\eeq
Here we denote generically by $\sigma$ a suitable quantity
(in general, process-dependent) which has  the property of
factorizing as in Eq.~\eqref{eq:fact}.
Such quantities are usually related in a simple way to cross-sections
or distributions; for example, in the case of the invariant mass distribution
of Drell-Yan pairs, $\sigma$ is given by Eq.~\eqref{eq:inclus_cs_conv}.

In general, the cross-section gets a contribution like Eq.~\eqref{eq:fact}
from all parton channels which contribute to the given
process at the given order, but this is inessential for our argument.
Indeed, since we are interested on the threshold region, only one channel
is enhanced and then contributes: hence, we concentrate on one such contribution.

In Eq.~\eqref{eq:fact}, the partonic coefficient function,
which is computed in perturbation theory, is evaluated as a function
of the partonic center-of-mass energy
\beq
\label{eq:shatdef}
\hat s= \frac{M^2}{\tau/z}=x_1 x_2 s,
\eeq
where $x_1=x$ and $x_2\equiv z/x$ are the momentum fractions of the two partons.
Therefore, the threshold region, where resummation is relevant, is the 
region in which $\hat s$ is not much larger than $M^2$. However, all
values of $x_1,x_2$ between $\tau$ and $1$ are accessible, so whether or not
resummation is relevant depends on which region gives the dominant
contribution to the convolution integrals
Eqs.~\eqref{eq:fact} and \eqref{eq:lumi}. This dominant region can be determined
using a Mellin-space argument~\cite{bfr2}.

For ease of notation, we temporary omit the dependence on the energy scale
$M^2$. The Mellin transform of $\sigma(\tau)$ is 
\beq\label{eq:mt}
\sigma(N)=\int_0^1d\tau\,\tau^{N-1}\, \sigma(\tau),
\eeq
with inverse
\beq\label{eq:imt}
\sigma(\tau) = \frac{1}{2\pi i} \int_{c-i\infty}^{c+i\infty} dN \,\tau^{-N}\,\sigma(N)
=\frac{1}{2\pi i} \int_{c-i\infty}^{c+i\infty} dN\,e^{E(\tau,N)},
\eeq
where in the last step we have defined
\beq\label{eq:melexp}
E(\tau,N) \equiv N\log\frac{1}{\tau}+\log \sigma(N).
\eeq
We would like to evaluate the integral with a saddle-point approximation:
since the integration path in Eq.~\eqref{eq:imt} has to be to the
right of all the singularities of the integrand (see App.~\ref{chap:Mellin}),
we must look for a saddle in such region.
Since $\sigma(N)$ is a real function\footnote{I.e.,~$\overline{\sigma(N)}=\sigma(\bar N)$.},
we expect that if a saddle-point exists it is on the real axis.
The function $\sigma(N)$ has its rightmost (positive) singularity on the real positive axis
because of the small-$x$ behaviour of the parton luminosity;
to the right of this singularity, it is a decreasing function of $N$,
because $\sigma(\tau)$ is not a distribution and Theorem~\ref{th:bonvini1} applies.
Conversely, the $N\log\frac1\tau$ term increases at large $N$
for all $\tau<1$.
As a consequence, $E(\tau,N)$ always has at least one minimum on the real
positive $N$-axis.

We now want to understand if the saddle is unique.
If $\tau$ is small enough, the growth of the $N\log\frac1\tau$
sets in at very small $N$, leaving no space between the rightmost pole
of $\sigma(N)$ and the straight line $N\log\frac1\tau$ for more than
one minimum. When $\tau$ is larger, we need to supply this argument
with other information on the shape of $\sigma(N)$.
To clarify, let's write the saddle condition $\frac{d}{dN}E(\tau,N)=0$ explicitly:
\beq\label{eq:saddle-point_def}
\log\frac1\tau = -\frac{d}{dN}\log\sigma(N).
\eeq
The condition for the existence of a unique saddle
is that the right-hand-side is a monotonically decreasing function of $N$
extending from $+\infty$ down to $0$.
At small $N$, this is guaranteed by the positive pole of $\sigma(N)$,
as already discussed. Indeed, if $\sigma(N)$ behaves like
\beq
\sigma(N)\sim \frac{a}{(N-N_p)^\alpha}, \qquad a>0, \quad \alpha>0
\eeq
its logarithmic derivative behaves as
\beq
-\frac{d}{dN}\log\sigma(N) \sim \frac{\alpha}{N-N_p},
\eeq
which has a positive pole in $N=N_p$ and decreases as $N$ increases.
In the opposite limit, large $N$, we can compute the
behaviour of $\sigma(N)$ by knowing that $\sigma(\tau)$ vanishes
as $\tau\to1$: we may assume that
\beq
\sigma(\tau) \sim b\,(1-\tau)^\beta \qquad\beta>0
\eeq
as $\tau\to1$, whose Mellin transform is
\beq
\sigma(N) \sim b\,\frac{\Gamma(N)\Gamma(\beta)}{\Gamma(N+\beta)} 
\overset{N\to\infty}{\sim} b\,\Gamma(\beta) N^{-\beta}.
\eeq
Then,
\beq\label{eq:largeNdlogsigma}
-\frac{d}{dN}\log\sigma(N) \overset{N\to\infty}{\sim} \frac{\beta}{N},
\eeq
which is a decreasing function and goes to $0$ as $N\to\infty$.
We may guess (and we have verified numerically) that also the intermediate region
respect this behaviour: then, the saddle-point is unique for all values of $\tau$.

Then, from now on we will call the saddle-point $N_0=N_0(\tau,M^2)$,
given by the solution of Eq.~\eqref{eq:saddle-point_def}.
From the definition, Eq.~\eqref{eq:saddle-point_def}, and from the discussion above, 
we immediately see that the saddle-point $N_0$ increases with $\tau$:
in particular, this proves once again that large-$\tau$ and large-$N$ regions
are mutually related.
We will see in the following this relation explicitly for processes
like the Drell-Yan process and the Higgs production.

The position of the saddle-point gives us a quantitative way to assess
the region of relevance of threshold resummation.
Indeed, the Mellin inversion integral, Eq.~\eqref{eq:imt},
is dominated by the region of $N$ around $N_0$, and we have exploited the
relation between $N_0$ and $\tau$ (and $M^2$). Then, given $\tau$ and $M^2$
(the hadron-level parameters), we have a precise estimate of the $N$-region
which dominates the hadronic cross-section. If such region (around $N_0$)
is at values of $N$ large enough to be in the threshold region,
then we could say that for those $\tau$ and $M^2$ the threshold region
gives the dominant contribution, and hence that threshold resummation
cannot be neglected.

We have already established in Sect.~\ref{sec:res_comparison_FO}
that for $N\gtrsim 2$ the threshold logarithms give the main contribution
to the coefficient function (for the Drell-Yan case), provided the proper
choice of subleading terms is made.
Then, we can conclude that the condition $N_0>2$ can be considered as a
good estimate for the threshold region. This condition translates into
$\tau>\tau_0$, with $\tau_0$ being determined by Eq.~\eqref{eq:saddle-point_def}
with $N=2$. We have then obtained the desired result: given $M^2$,
we are able to say for which values of $\tau$ resummation is relevant.
In the following we will see explicit results.

\subsection{The impact of PDFs}
\label{sec:PDF_impact}
The position of the saddle-point $N_0$ is strongly influenced by the 
rate of decrease of the cross-section $\sigma(N)$ as $N$ grows.
Indeed, in Mellin space, the cross-section Eq.~\eqref{eq:fact} factorizes:
\beq
\label{eq:factN}
\sigma(N)=\Lum(N)\,C(N,\as).
\eeq
It is then easy to see that the decrease of $\sigma(N)$ with $N$
is driven by the parton luminosity $\Lum(N)$: in fact,
for large $N$, $C(N,\as)$ is an \emph{increasing} function of $N$,
because $C(z,\as)$ is a distribution (see Theorem~\ref{th:bonvini2}).
However, the parton luminosity always offsets this increase if the
convolution integral exists, because the cross-section $\sigma(\tau)$ is an
ordinary function and its Mellin transform decreases with $N$.
Indeed, because the partonic coefficient function rises at most as a power of $\log N$
as $N\to\infty$, it is easy to show that a sufficient condition 
for $\sigma(N)$ to decrease is that the parton luminosity $\Lum(z)$ 
vanishes at large $z$ at least as a positive power of $(1-z)$, as it usually does.

As a consequence, when $\tau$ is large, the position of the
saddle-point $N_0$ is completely controlled by the drop of the parton
luminosity: indeed, in the absence of parton luminosity, the saddle-point
would be very close to the minimum of the coefficient function
(which is around $N\simeq 1$ for Drell-Yan, see Fig.~\ref{fig:DY_order_as}).
When $\tau$ is smaller, even without PDFs the location of the saddle is
controlled by the partonic coefficient function, which in this region
is a decreasing function of $N$. However, this
decrease is much stronger in the presence of a luminosity, so the
location of the saddle is substantially larger. 
Hence, in the large $\tau$ region the effect of the resummation
is made much stronger by the luminosity, while for medium-small $\tau$  if
the luminosity decreases fast enough, $N_0$ may be quite large even if
$\tau\ll 1$, i.e.\ far from the hadronic threshold, thereby extending
the region in which resummation is relevant.

The position of the saddle-point in the various regions 
can be simply estimated on the basis of general considerations.
At the leading-log level, parton densities can be written as linear
combinations of terms of the form (see Sect.~\ref{sec:glap_sol})
\beq\label{eq:pdfform}
f_i(N,\mu^2)=\exp\[-\frac{\gamma^{(0)}_i(N-1)}{\beta_0} \log\frac{\as(\mu^2)}{\as(\mu_0^2)}\] f_i(N,\mu_0^2)
\eeq
in terms of initial PDFs $f_i(N,\mu_0^2)$ at some reference scale $\mu_0^2$.
The cross-section is correspondingly decomposed into a sum of contributions,
each of which has the form of Eq.~\eqref{eq:imt}, with 
\begin{multline}\label{eq:appexp}
E(\tau,N;M^2) = N\log\frac{1}{\tau}-\frac{\gamma^{(0)}_i(N-1)+\gamma^{(0)}_j(N-1)}{\beta_0}
\log\frac{\as(M^2)}{\as(\mu_0^2)}\\
+\log f_i(N,\mu_0^2)+\log f_j(N,\mu_0^2) +\log C(N,\as(M^2)).
\end{multline}
At large $N$, this expression is dominated by the first term, 
which grows linearly with $N$, while at small $N$ the behaviour
of $E(\tau,N;M^2)$ is determined by the singularities of the anomalous
dimensions, which are stronger than those of the initial conditions if
$M^2>\mu_0^2$, given that low-scale physics is both expected
theoretically from Regge theory~\cite{Collins:1977jy,Abarbanel:1969eh} 
and known phenomenologically from PDF
fits~\cite{NNPDF1} to produce at most poles but not essential singularities such as
those obtained exponentiating the anomalous dimensions. Indeed, assuming
a power behaviour for $f_i(z,\mu_0^2)$ both at small and large $z$,
\beq\label{eq:inpdfform}
f_i(z,\mu_0^2)=z^{\alpha_i}(1-z)^{\beta_i},
\eeq
so that
\beq\label{eq:mellinpdfform}
f_i(N,\mu_0^2) = \frac{\Gamma(N+\alpha_i)\,\Gamma(\beta_i+1)} {\Gamma(N+\alpha_i+\beta_i+1)},
\eeq
$\log f_i(N,\mu_0^2)$ behaves as $\log N$ both at large and small $N$, and is
thus subdominant in comparison to either the $\tau$ dependent term or the
anomalous dimension contribution  in Eq.~\eqref{eq:appexp}.
A similar argument holds for the partonic coefficient function term
$\log C(N,\as)$.

The position of the minimum is therefore mainly determined by the transition
from the leading small-$N$ drop due to the anomalous dimension
term and the leading large-$N$ rise due to the $\tau$-dependent
term, up to a correction due to the other contributions to
Eq.~\eqref{eq:appexp}. When $\tau$ is large, the rise in the first term
is slow, and it only sets in for rather large $N$ so the correction
due to the other contributions may be substantial. This is the
region in which resummation is surely relevant because the hadronic
$\tau$ is large.  But when $\tau$ is not so large, the rise sets in
more rapidly, in the region 
where the second term is dominant and the correction from
the initial PDFs and the partonic coefficient function is less important.

Moreover, the small-$N$ region is the one which is sensitive
to the high-energy logarithms, Chap.~\ref{chap:small-x}.
Then, the exact relation between $N_0$ and $\tau$ can be
sensibly modified by high-energy resummation: this suggests
that there can be an interplay of the threshold region
and the high-energy region. We will discuss this in more detail
in Sect.~\ref{sec:joint}.

\subsection{The resummation region for the Drell-Yan process}
\label{sec:resDY}

We now assess the relevance of resummation in the specific case
of Drell-Yan production.
We consider the case of the invariant mass distribution:
the quantity $\sigma(\tau,M^2)$ which appears in Eq.~\eqref{eq:fact} is given by
\beq\label{eq:dyfact}
\sigma(\tau,M^2) = \frac{1}{\tau} \frac{d\sigma_{\rm DY}}{d M^2}(\tau,M^2),
\eeq
see App.~\ref{sec:DY_fixed-order} for the relevant formulae.
The main contribution to the cross-section is given by the $q\bar q$
channel, and this is also the only contribution which is enhanced
in the threshold region. Then, we concentrate on that from now on ---
the other channels behaves perturbatively in the threshold region
and then there's no need to investigate them.

\begin{figure}[tb]
\begin{center}
\includegraphics[width=0.85\textwidth]{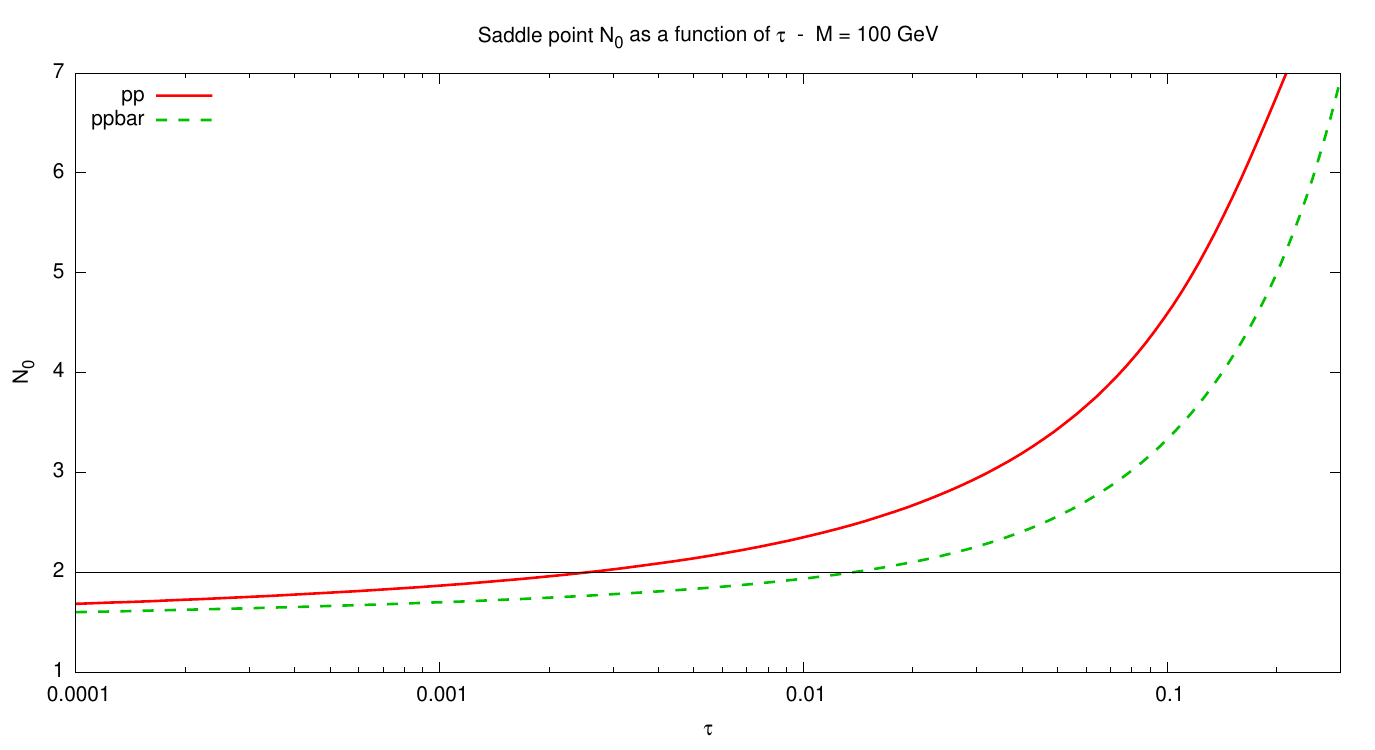}
\caption{
  The position of the saddle-point $N_0$ for the Mellin inversion
  integral Eq.~\eqref{eq:imt} as a function of $\tau$ with the cross-section
  Eq.~\eqref{eq:factN} determined  using the NLO Drell-Yan
  cross-section for neutral di-leptons and NNPDF2.0~\cite{Ball:2010de} parton
  distributions, with  $M=100$~GeV. The two upper curves refer to $pp$ and
  $p\bar p$ collisions.}
\label{fig:sd1}
\end{center}
\end{figure}
We have then determined the position of the saddle-point $N_0$ in a
realistic situation, i.e.\ using the partonic coefficient function at NLO 
and the luminosity for the production of a neutral lepton pair
of invariant mass $M=100$~GeV at a $pp$ or $p\bar p$ colliders,
using NNPDF2.0~\cite{Ball:2010de} parton distributions.
In Fig.~\ref{fig:sd1} we show the dependence of $N_0$ on $\tau$
in the two cases. The behaviour previously predicted
is indeed manifest: at large $\tau$, the saddle-point $N_0$ increases,
going in the region where threshold resummation is more and more relevant,
while it decreases at small $\tau$, but slower and slower, because of the
lower limit set by the pole of parton luminosity at small $N$.

Choosing $N=2$ as the value above which threshold logarithms are relevant,
the intersection of the line $N_0=2$ (showed in Fig.~\ref{fig:sd1}) with
the curve $N_0(\tau,M^2)$ gives the lower value $\tau_0$ above which
resummation is important. It is already surprising that, for $pp$
colliders, $\tau_0$ is as small as $0.003$, very far from hadronic threshold.
To understand better the result, it is convenient to fix the collider energy $\sqrt{s}$
instead of the mass $M$. To do this, we first write\footnote{Note, by the way,
that the dependence on $M^2$ of $\tau_0$ is completely controlled by perturbative physics,
in which it depends naively on the physical anomalous dimension.}
\beq\label{eq:tau0def}
\tau_0(M^2) = \exp\[\left.\frac{d}{dN}\log\sigma(N,M^2)\right|_{N=2}\]
\eeq
and compute it for a sequence of values of $M^2$; then, using
the relation of $\tau=M^2/s$, one can either produce a curve
$\tau_0(s)$ or even better $M_0(s)$, being $M_0$ the lower value
above which resummation is important.
\begin{figure}[tb]
  \centering
  \includegraphics[width=0.85\textwidth,page=4]{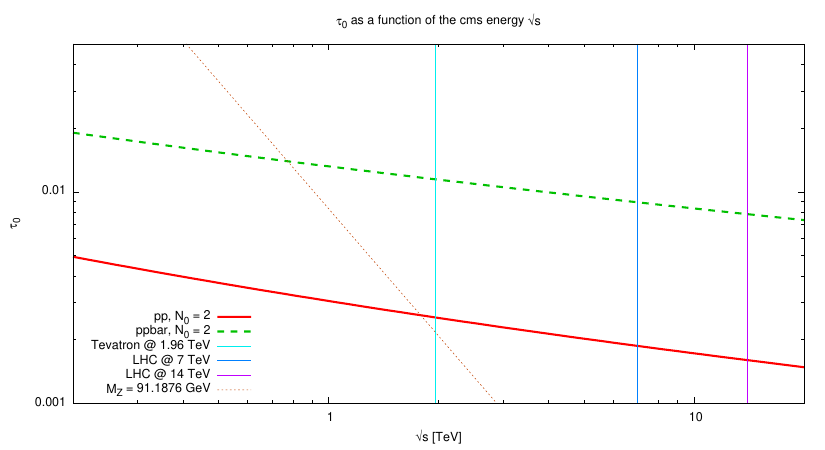}\\
  \includegraphics[width=0.85\textwidth,page=5]{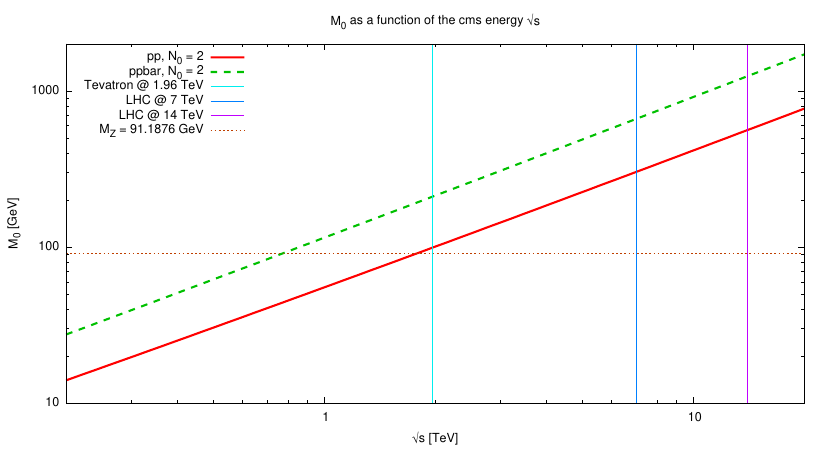}
  \caption{Dependence of $\tau_0$ (upper plot) and $M_0$ (lower plot) on the
    collider c.m.s.\ energy $\sqrt{s}$ for the Drell-Yan process.
    The $Z$ mass is shown as a reference.}
  \label{fig:sd2}
\end{figure}
The results are shown in Fig.~\ref{fig:sd2}, where the upper plot
shows $\tau_0$ as a function of $\sqrt{s}$ and the lower plot shows
$M_0$ as a function of $\sqrt{s}$.
For example, by intersection of the solid red curve with the vertical blue curve
we can conclude that at LHC at $7$~TeV a system of mass larger than about
$M_0\sim300$~GeV, corresponding to $\tau_0\sim0.002$, resummation may be relevant.
For Tevatron, we find instead $M_0\sim200$~GeV, corresponding to $\tau_0\sim0.01$.

These values are, from one side, surprisingly smaller than one could naively expect,
since the values of $\tau_0$ are very far from threshold, especially in the LHC case.
From the other side, they still tell us that, for practical purposes, threshold resummation
does not play a crucial role for phenomenologically relevant Drell-Yan masses.
We will see in a moment that in the Higgs case the situation is different.

\subsection{The resummation region for the Higgs production process}
\label{sec:resHiggs}

We now move to the case of Higgs production.
For simplicity, we consider the infinite-$m_t$ limit;
we will discuss the effect of the top mass in Sect.~\ref{sec:joint}.
Then, we consider
\beq
\sigma(\tau,M^2) = \frac{1}{\tau} \frac{d\sigma_{\rm H}}{d M^2}(\tau,M^2),
\eeq
see App.~\ref{sec:Higgs_fixed-order} for the relevant formulae.
The main (and threshold enhanced) contribution to the cross-section is given by the $gg$
channel: then, as before, we concentrate only on that.
Note that the gluon PDF is the same for the proton and an anti-proton:
therefore, the $gg$ luminosity is not different between $pp$ and $p\bar p$
colliders.

\begin{figure}[tb]
\begin{center}
\includegraphics[width=0.85\textwidth]{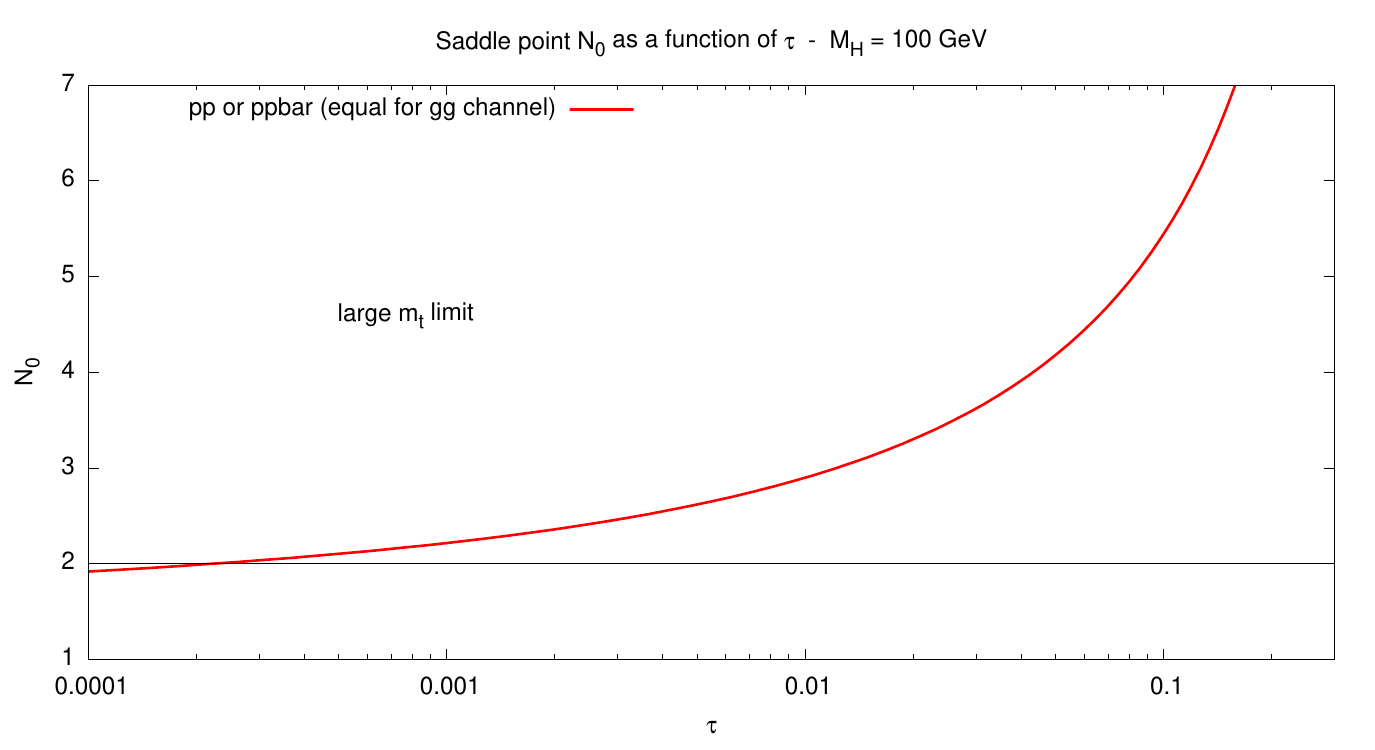}
\caption{
  The position of the saddle-point $N_0$ for the Mellin inversion
  integral Eq.~\eqref{eq:imt} as a function of $\tau$ with the cross-section
  Eq.~\eqref{eq:factN} determined  using the NLO Higgs production
  cross-section and NNPDF2.0~\cite{Ball:2010de} parton
  distributions, with  $M=100$~GeV. The two upper curves refer to $pp$ and
  $p\bar p$ collisions.}
\label{fig:sd3}
\end{center}
\end{figure}

As for the Drell-Yan case, we show in Fig.~\ref{fig:sd3} the saddle-point
$N_0$ as a function of $\tau$ using a final state mass\footnote{%
Note that this is not the Higgs mass as a parameter of the SM, but just the invariant mass
of the final state produced via a virtual Higgs (two photons, heavy quark pair, \ldots).
The dependence on the parameter $m_H$ does not play any role in the determination
of the saddle, being contained in a prefactor with no $N$ dependence.}
of $M=100$~GeV. The same PDF set NNPDF2.0 is used.
The conclusion we may draw here are pretty much the same as in Drell-Yan
case, but here the threshold region is much more extended. Indeed, at the same
final state mass $M=100$~GeV we find here $\tau_0\sim0.0002$, one order of magnitude smaller
than in the Drell-Yan case.
Note that we are assuming than $N\gtrsim2$ is a good definition for the
threshold region also for the Higgs case. To prove this, we make the
same comparison we did for the Drell-Yan case.
\begin{figure}[tb]
\begin{center}
\includegraphics[width=0.85\textwidth,page=1]{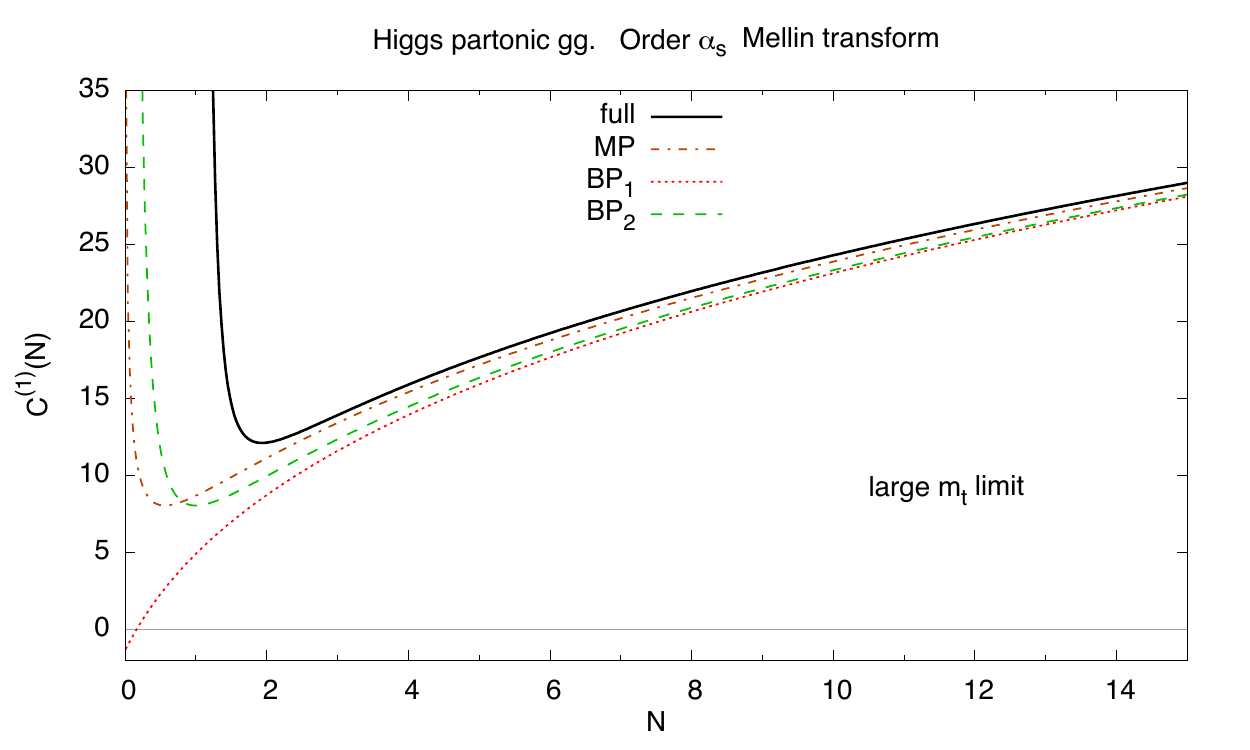}
\caption{The order $\as$ Higgs partonic coefficient function
  $C^{(1)}(N)$, Mellin transform of Eq.~\eqref{eq:Higgs_C1},
  plotted as a function of $N$ (solid curve) and its logarithmic approximations.
  The curves are obtained in large-$m_t$ limit.}
\label{fig:Higgs_order_as}
\end{center}
\end{figure}
In Fig.~\ref{fig:Higgs_order_as} we present a comparison of the
first order coefficient function and its logarithmic approximations
(same notations as in Fig.~\ref{fig:DY_order_as} are used).
We see again that the region $N\gtrsim2$ is dominated by the logarithms.
However, here the small-$N$ behaviour is no longer mimicked by the BP$_2$,
because the small-$N$ behaviour of the Higgs production in $gg$ channel
is dominated by a pole in $N=1$. However, it turns out that the small-$N$
behaviour in the large-$m_t$ limit is not correct~\cite{Bonciani:2007ex}:
therefore, we cannot trust the shape of the complete NLO curve in the region close
to $N=1$. In Sect.~\ref{sec:joint} we will investigate it more in detail.
Anyway, the point $N=2$ seems to be not affected that much from
the small-$N$ behaviour: we therefore conclude that $N_0=2$ is a
good choice also for the Higgs production case.

\begin{figure}[tb]
  \centering
  \includegraphics[width=0.85\textwidth,page=4]{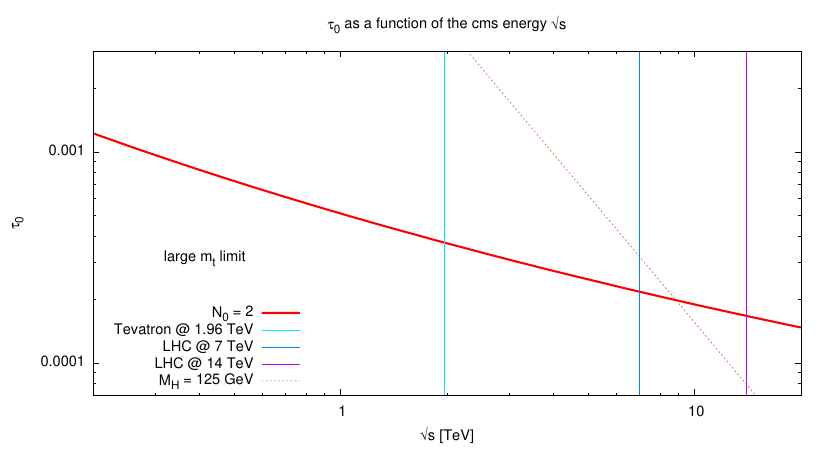}\\
  \includegraphics[width=0.85\textwidth,page=5]{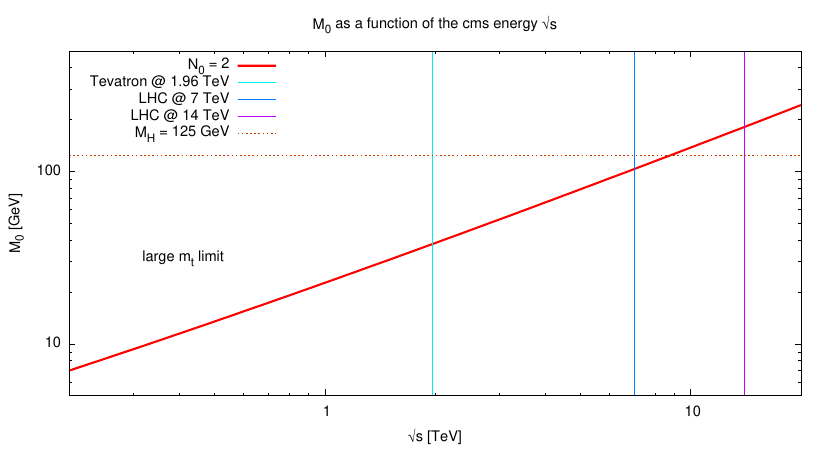}
  \caption{Dependence of $\tau_0$ (upper plot) and $M_0$ (lower plot) on the
    collider c.m.s.\ energy $\sqrt{s}$ for the Higgs production process.
    An hypothetical Higgs mass of $125$~GeV is shown as a reference.}
  \label{fig:sd4}
\end{figure}

To present the result in a more convenient way, we perform the same manipulations
as before, to obtain the curves for $\tau_0$ and $M_0$ as functions of $\sqrt{s}$
in Fig.~\ref{fig:sd4}.
The results are even more surprising than in Drell-Yan case:
for LHC at $7$~TeV, an approximately on-shell Higgs of more than
about $M_0\sim100$~GeV (corresponding to $\tau_0\sim0.0002$)
is dominated by the threshold region. It surprising how small
and far from threshold the value of $\tau_0$ is; moreover,
this result is phenomenologically relevant, because, if the Higgs boson exists,
its mass will be higher than $114$~GeV (the limit set by LEP),
and then its production is always in the threshold region at LHC $7$~TeV.
This confirms the well known observation~\cite{Grazzini:2010zc} that the 
Higgs boson production cross-section is dominated by the threshold contributions,
but here we are also giving a quantitative evaluation of where and how this is true.

These considerations are even more dramatic in the Tevatron case:
there, even an Higgs as light as $M_0\sim40$~GeV would be in the threshold region.
Then, when looking for a Higgs boson of a reasonable mass at Tevatron,
it is not a matter of choice whether include or not the threshold resummation,
since its effect would be non-negligible.

To conclude, we have to recall that, being the Higgs dominated by
the $gg$ channel, it is likely to be influenced more than other processes
by small-$x$ resummation.
In particular, since at small $\tau$ the position of the saddle is mainly
determined by the rightmost pole of the luminosity, and given that
small-$x$ resummation mainly affects the position and shape of such pole,
the inclusion of small-$x$ resummation may change quantitatively
(even if not qualitatively) our conclusions.
We will therefore investigate this in Sect.~\ref{sec:joint}.

\section{Phenomenology of high-energy resummation}
\label{sec:small-x_pheno}

Phenomenology of threshold resummation is rather easy to perform:
one simply has to compute resummed coefficient functions and, with those at
the hand, compute hadronic cross-sections.
In principle, to be perfectly consistent, one should use PDFs fitted
with resummed coefficient functions; nevertheless, the effect is expected to be small
in the kinematical regions of the data used in the fit, and therefore
``unresummed'' PDFs can be used without losses of precision.

The case of high-energy resummation is instead very different.
Beyond the coefficient functions, small-$x$ resummation affects
the anomalous dimensions, and the leading effect is in fact on the
gluon anomalous dimensions. Therefore, in this case, one cannot use
unresummed PDFs for phenomenological predictions, because in practice
most of the effect is contained in the evolution of PDFs.
Hence, small-$x$ resummed anomalous dimensions (and coefficient functions)
should be included in a PDF fit in order to have resummed PDFs
suitable for phenomenology.

However, it is beyond the purpose of this thesis to perform such
resummation task.
An approximate way to have resummed PDFs is to simply evolve PDFs from a given scale, at
which we trust the small-$x$ behaviour, to the scale we are interested in.
The smaller-$x$ DIS data ($x\sim 10^{-4}$) were taken by ZEUS and H1 for a very low energy,
$Q=2\div 3$~GeV, so we can considered this scale as the starting scale for a resummed evolution.
Such procedure should be done fixing the data at the reference scale, and not simply
the PDFs.
To clarify, the procedure should be the following:
\begin{itemize}
\item first, we take a current PDF set and evolve the PDFs to the given small reference scale;
\item then, we compute structure functions using \emph{unresummed} coefficient functions;
\item then, we extract back the PDFs using \emph{resummed} coefficient functions:%
  \footnote{For the resummation of coefficient functions see App.~\ref{sec:resummation_coeff_funct}.}
  we have now initial conditions for the resummed PDFs;
\item we evolve to the desired scale using the resummed anomalous dimensions.
\end{itemize}
Even if this procedure is not as good as a complete resummed fit,
it allows to assess in a consistent framework the impact of small-$x$ resummation.

\subsection{Resummed PDFs at the initial scale}
Usually, in PDF fits, the PDFs are extracted from DIS data using $F_2$ and its
derivative with respect to $\log\mu^2$, which gives in particular information on
the gluon PDF. We can then use these two quantities as reference at the initial scale.
Omitting for simplicity the $N$ and scale dependence (see
Eq.~\eqref{eq:DIS_structure_functions_N} for the correct arguments)
we may write
\begin{subequations}\label{eq:F2andF2p}
\begin{align}
  F_2  &= \sum_i C_{2i}\,f_i\\
  F_2' &= \sum_i C_{2i}'\,f_i + \sum_{i,j}C_{2i}\,\gamma_{ij}\, f_j\nonumber\\
  &= \sum_j\[ C_{2j}' + \sum_i C_{2i}\,\gamma_{ij} \] f_j
\end{align}
\end{subequations}
having denoted with a prime the $\mu^2\frac{d}{d\mu^2}$ derivative.%
\footnote{Note that we could also consider the prime as an $\as$-derivative,
provided we perform the substitution $\gamma_{ij}\to \gamma_{ij}/\beta(\as)$.}
We can write the two equations in a matrix form
\beq
F = C\, f
\eeq
with
\beq
F=\dvec{F_2}{F_2'},\qquad
f=\tvec{f_g}{f_S}{(f_{\rm ns})},
\eeq
and $C$ can be extracted from Eqs.~\eqref{eq:F2andF2p}.
Note that we have separated the quark singlet and non-singlet contributions:
this can be helpful because resummation affects only singlet quantities.
Indeed, we want to extract ``resummed'' initial PDFs using resummed coefficient function
from the known functions $F_2$ and $F_2'$: there are many more unknowns than equations,
but fortunately we know that the non-singlet PDFs are not affected by resummation,
and then the only two unknowns are the ``resummed'' initial $f_g$ and $f_S$.
To make this more manifest, since the singlet and non-singlet do not interfere
each other we may keep fixed just the singlet part of the structure function,
\beq
F_{\rm s} = F - F_{\rm ns},
\eeq
since the non-singlet part does not change with small-$x$ resummation.
The, our equations are simplified to
\beq
F_{\rm s} = C_{\rm s}\, f_{\rm s}
\eeq
with
\beq
C_{\rm s}=\sqmatr{C_{2g}}{C_{2q}}
{C_{2g}'+\sum_i C_{2i}\,\gamma_{ig}}
{C_{2q}'+\sum_i C_{2i}\,\gamma_{iq}},
\qquad
f_{\rm s}=\dvec{f_g}{f_S}.
\eeq
Then, using the procedure described above, the resummed PDFs at the initial energy
are obtained as
\beq\label{eq:resummed_initial_pdfs}
f_{\rm s}^{\rm res}(\mu_0^2) = C_{\rm s, res}^{-1}\(\as(\mu_0^2)\)\,
C_{\rm s}\(\as(\mu_0^2)\) \, f_{\rm s}(\mu_0^2),
\eeq
where $f_{\rm s}(\mu_0^2)$ should be computed from a fixed-order PDF set.

The fixed-order singlet coefficient functions in the \MSbar\ scheme are
given by~\cite{esw,Catani:1994sq}
\beq
C_{2i}(N,\as) = \frac{1}{n_f}\(\sum_j e_j^2\) \[ \delta_{iq} + \as\,C^{(1)}_{2i}(N) + \Ord(\as^2)\]
\eeq
where $e_j$ are the quark charges in unit of the electric charge and
\begin{align}
  C^{(1)}_{2g}(N) &= \frac{n_f}{4\pi} \bigg[
  - \frac{\gammae+\psi(N+1)}{N}
  + 2\,\frac{\gammae+\psi(N+2)}{N+1}
  + 2\,\frac{\gammae+\psi(N+3)}{N+2}
  \nonumber\\&\qquad\qquad
  + \frac1{N^2}
  - \frac2{(N+1)^2}
  - \frac2{(N+2)^2}
  - \frac1{N}
  + \frac8{N+1}
  - \frac8{N+2}
  \bigg] \\
  C^{(1)}_{2q}(N) &= \frac{C_F}{2\pi} \bigg[
  \psi^2(N) + \(2\gammae+\frac32\)\psi(N)+\gammae^2-\zeta_2+\frac32\gammae-\frac92 + \psi_1(N+2)
  \nonumber\\&\qquad\qquad
  + \frac{\gammae+\psi(N+1)}{N}
  + \frac{\gammae+\psi(N+2)}{N+1}
  +\frac3N + \frac2{N+1}
  \bigg].
\end{align}

\subsection{Resummed evolution}

Once the resummed PDFs at the initial scales $f_{\rm s}^{\rm res}(\mu_0^2)$
Eq.~\eqref{eq:resummed_initial_pdfs} has been computed, the evolution
to the desired scale must be performed.
To do this, we use for the resummed evolution of the singlet sector
the discretized path-ordering evolution described in Sect.~\ref{sec:discretized_PO}.
The resummed anomalous dimensions are given in Sect.~\ref{sec:back_to_physical_basis}.
We will see in the next Section the effect of this approximate procedure
in the determination of the Higgs threshold region.

\section{Joint effect of both resummations}
\label{sec:joint}

As already pointed out in Sect.~\ref{sec:when_is_relevant},
small-$x$ resummation of PDFs may have an effect on threshold resummation
of coefficient functions.
This fact come from the observation~\cite{Catani,bfr2} that in the convolution,
Eq.~\eqref{eq:fact},
\beq\label{eq:fact2}
\sigma(\tau) = \int_\tau^1\frac{dz}{z} \,
\Lum\(\frac\tau z\) \,C(z,\as)
\eeq
when the argument of the coefficient function is close to threshold,
that is, when $z\to1$, then the argument of the parton luminosity
is small, $\tau/z\sim\tau$.
Nevertheless, this simple observation does not tell us if this
region gives the dominant contribution to the integral.
The naive observation~\cite{Catani} is that, if the luminosity is peaked
at small $x=\tau/z$, then it will give the dominant contribution to the integral.
The saddle-point argument discussed in Sect.~\ref{sec:when_is_relevant}
had exactly the aim of quantifying this aspect; the price to pay
is that the formulation of the argument is done in $N$ space, where
the simple relation between the high-energy and the threshold regions
is no longer clear.
Indeed, in $N$ space the convolution becomes
\beq
\sigma(N) = \Lum(N)\, C(N,\as)
\eeq
where both the luminosity and the coefficient function are computed at the
same $N$. Therefore, the naive identification of the threshold and high-energy regions
with the large- and small-$N$ regions respectively seems to point out
that both functions should contribute in the same region.
This is not true, of course: we have proved that the saddle-point,
which gives the dominant contribution to the Mellin inversion integral,
is pushed to large $N$ because of the small-$N$ shape of the luminosity.
Therefore, we recover the conclusion drawn above.

Here, we want to revisit the result in momentum space, to better clarify
the relation between the two regions.
To begin with, it is convenient to rewrite Eq.~\eqref{eq:fact2} in the form
\begin{align}
\sigma(\tau) &= \int_0^1dx \int_0^1dz \;\delta(xz-\tau)\, \Lum(x)\, C(z,\as)\\
&= \int_0^1dx \int_0^1dz \;\delta(xz-\tau)\,
\int\frac{dN_1}{2\pi i}\, e^{N_1\log\frac1x + \log \Lum(N_1)}
\int\frac{dN_2}{2\pi i}\, e^{N_2\log\frac1z + \log C(N_2,\as)},
\end{align}
where in the second row we have written the luminosity and the coefficient
function as the inverse Mellin transform of their Mellin transforms.
We did it because now we want to consider the saddle-points for each
inverse Mellin separately:
\beq\label{eq:saddles_lum_C}
  \log\frac1x + \frac{\Lum'(\bar N_1(x))}{\Lum(\bar N_1(x))} = 0,\qquad
  \log\frac1z + \frac{C'(\bar N_2(z))}{C(\bar N_2(z))} = 0;
\eeq
note that, because of the $\delta$ function, $x$ and $z$ always satisfy $xz=\tau$.
At some point $\bar x$, the two saddle will coincide
\beq
\bar N_1(\bar x) = \bar N_2(\bar z) \equiv N_0(\tau),
\qquad \bar z=\frac{\tau}{\bar x},
\eeq
where we have noted that this point is completely determined by $\tau$.
It turns out that $N_0(\tau)$ is the saddle-point for $\sigma(\tau)$,
since putting together the definitions Eq.~\eqref{eq:saddles_lum_C} we have
\beq
\log\frac1\tau + \frac{\Lum'(N_0(\tau))}{\Lum(N_0(\tau))} + \frac{C'(N_0(\tau))}{C(N_0(\tau))} =0
\eeq
which is the saddle-point condition for $\sigma(\tau)$, Eq.~\eqref{eq:saddle-point_def}.
We have already discussed in Sect.~\ref{sec:when_is_relevant}
that the hadronic saddle-point gives the dominant contribution to $\sigma(\tau)$.
We conclude from this discussion that, in momentum space,
the main contributions to $\sigma(\tau)$ come from the region around
$x=\bar x$ ($z=\bar z$).
As discussed in Sect.~\ref{sec:PDF_impact}, the hadronic saddle $N_0$
is mostly determined by the luminosity, being the contribution
from the coefficient function negligible (at least when $\tau$ is small).
Then, we expect $\bar x$ to be rather close to $\tau$, thereby
implying that $\bar z$ is expected to be rather close to $1$.
This completes our argument: we have proved that, in fact,
in the convolution Eq.~\eqref{eq:fact2} the main contribution
comes from the region $z\sim1$, i.e.\ when the coefficient function
is in the threshold region and the luminosity is in the high-energy region.

Having this in mind, we now turn to the study of the effect of small-$x$
resummation to the determination of the saddle-point.
We won't cover the case of Drell-Yan, since $q\bar q$ channel (which dominates at threshold)
is not expected to get significant contribution from small-$x$ resummation,
since the quark PDFs are affected only at NLL level.
Conversely, the Higgs production being dominated at threshold by the $gg$ channel,
is expected to get sizable contribution from small-$x$ resummation,
since the gluon PDF is affected at the LL level.

\subsection{Joint effect on Higgs production}

Using the approximate procedure drawn in Sect.~\ref{sec:small-x_pheno} to implement
small-$x$ resummation, we present here the effect of resummed PDFs in the
determination of the threshold region for Higgs production.

The results presented here will differ from those of Sect.~\ref{sec:resHiggs}
in the following aspects:
\begin{itemize}
\item the full $m_t$ dependence is considered (see Ref.~\cite{Bonciani:2007ex}
  and App.~\ref{sec:Higgs_finite-mt}): this allows to have the correct small-$x$
  behaviour of the coefficient function, which is instead wrong in the large-$m_t$ approximation;
\item the gluon PDF used for the computation of the luminosity is resummed according
  to the procedure of Sect.~\ref{sec:small-x_pheno}.
\end{itemize}
In principle, we should also include the small-$x$ resummed Higgs coefficient function
\cite{Marzani:2008az}: however, since it affects the coefficient function,
the effect should be small compared to the effect of resummation on PDFs.
Also the effect of the finite top mass turns out to be almost negligible.

\begin{figure}[tb]
  \centering
  \includegraphics[width=0.85\textwidth,page=1]{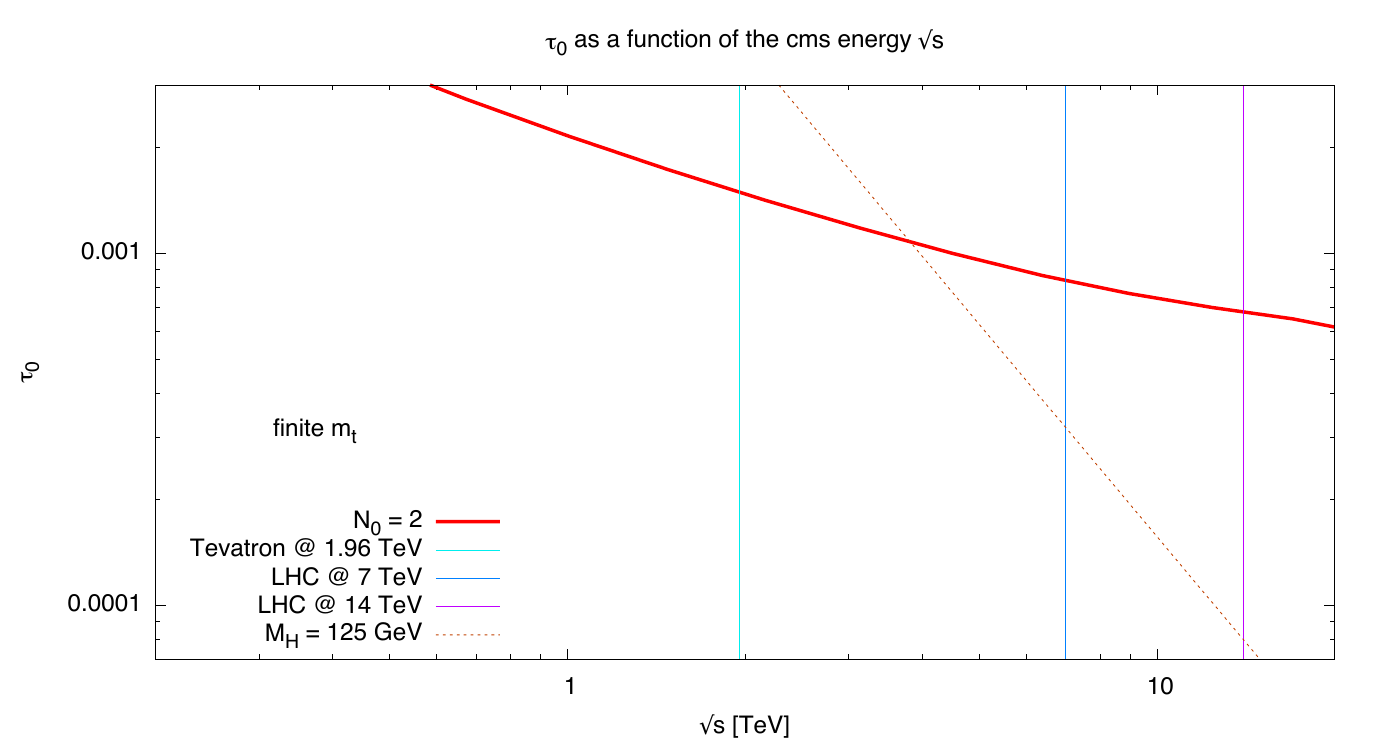}\\
  \includegraphics[width=0.85\textwidth,page=1]{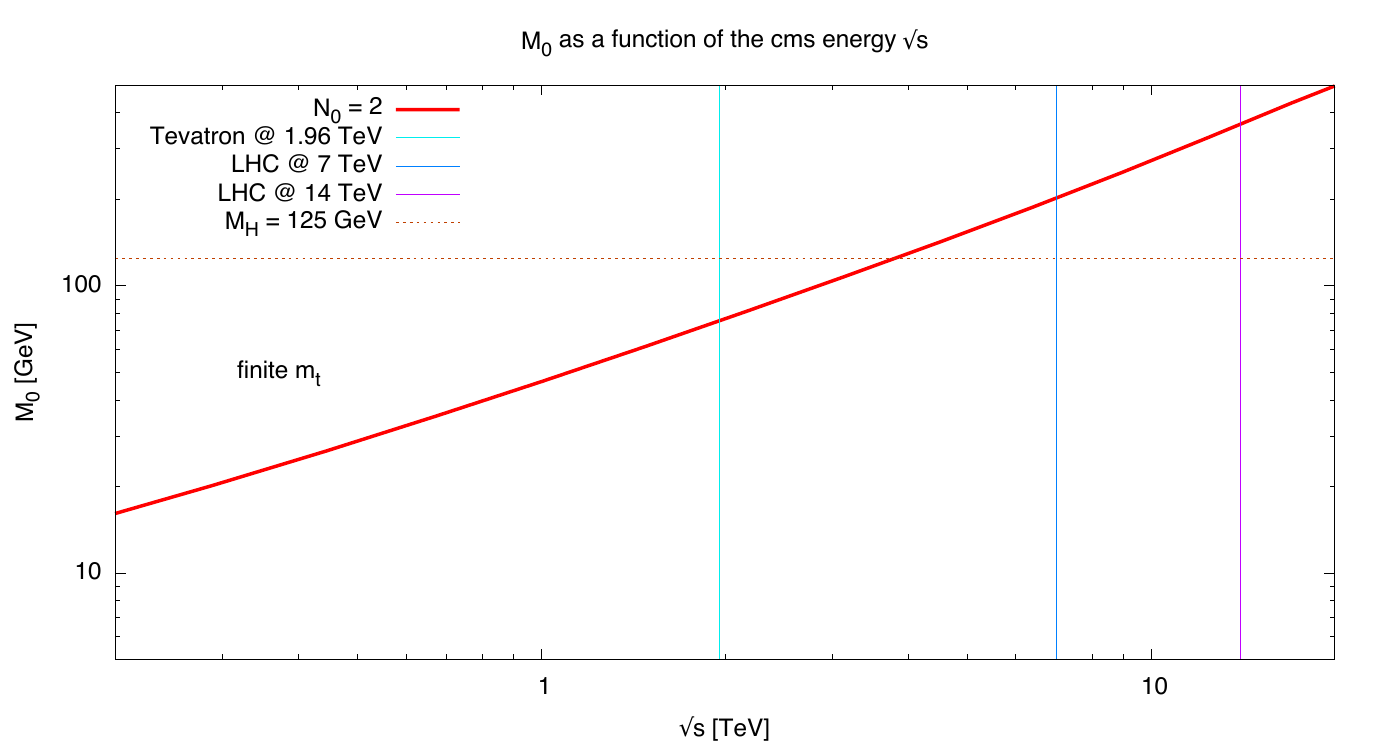}
  \caption{Dependence of $\tau_0$ (upper plot) and $M_0$ (lower plot) on the
    collider c.m.s.\ energy $\sqrt{s}$ for the Higgs production process.
    The approximate resummed PDFs described in Sect.~\ref{sec:small-x_pheno}
    are used.
    The curves are obtained with full $m_t$ dependence.}
  \label{fig:sd4_finite-mt+small-x}
\end{figure}

In Fig.~\ref{fig:sd4_finite-mt+small-x} the same plots of Fig.~\ref{fig:sd4}
are shown, with the modifications described above.
We see that the threshold region now begins at higher masses (higher $\tau$'s):
for example, from these plot we discover that actually threshold
resummation is not that important even at the LHC at $7$~TeV.

The result is rather unexpected: since small-$x$ resummation
makes the rise of the gluon PDF stronger at small-$x$,
one naively expects that the saddle-point should move to larger values of $N$.
However, the small-$x$ rise is located at very small $N$, very close to the
Bateman pole.
From Fig.~\ref{fig:Deltagammaplus} one sees indeed that the Bateman
pole is located roughly at $N\sim0.2$: remember however that the anomalous dimension
is defined as in Eq.~\eqref{eq:anom-dim_Mellin_xP}, so that its argument
is shifted by a unity with respect to the usual definition for the other quantities,
as those used here. Therefore, the Bateman pole affects the
cross-section $\sigma(N)$ in the region close to $N\sim 1.2$,
which is not relevant for the determination of the saddle-point in $N_0=2$.
Instead, the saddle-point $N_0=2$ is influenced by the region around
$N=1$ in the plot of Fig.~\ref{fig:Deltagammaplus}, where
momentum conservation constraint imposes $\Delta\gamma_+$ to be zero.
In particular, the $N$-derivative there is negative, thereby
justifying the effect of moving the saddle-point to lower values of $N$,
that is to say to move $\tau_0$ Eq.~\eqref{eq:tau0def} to larger values.

We emphasize that these results are obtained with the approximate
implementation of small-$x$ resummation. It may well be that
the proper implementation of small-$x$ resummation in a PDF fit
gives a somewhat different result.
Be this result correct or wrong, it clearly shows the importance
of a good understanding of the impact of small-$x$ resummation at hadron colliders:
this should be a strong motivation for PDF fitter community
to include resummed anomalous dimensions and coefficients functions
in the fit.

\section{Toward a joint resummation}

As last argument of this Chapter, we discuss the possibility
and the implications of a joint resummation of both
the high-energy and the threshold logarithms.

What we would like to discover is that a joint resummed
cross-section is as close as possible to the ``full result''.
This sentence is a bit misleading, because the full result,
which would correspond to the sum of the full perturbative series,
is of course unknown.
The way to interpret the sentence is just the requirement that
the remaining unresummed pieces behave perturbatively.

As a preliminary test, we consider again the Higgs production,
and in particular the NLO coefficient function.
Since we are interested in the small-$x$ behaviour, we must work at
finite $m_t$.
The large-$N$ logarithms which would be reproduced in a NLL or higher
threshold resummation are
\beq
C^{(1)}_{\rm BP_2}(N) = \frac{2C_A}{\pi}
\[ \psi_0^2(N) + 2 \gammae \psi_0(N) + 2\zeta_2 + \gammae^2 + \frac{11}{12} \],
\eeq
where the subscript BP$_2$ reflects the fact that these logs are those
generated by resummation if the Borel prescription of Eq.~\eqref{eq:BP2} is used
(see discussion in Sect.~\ref{sec:subl}).
The small-$N$ behaviour has been determined at NLO with full $m_t$ dependence
in Ref.~\cite{Marzani:2008az}: it is
\beq
C^{(1)}_{\rm sx}(N) = c\,\frac{C_A}{\pi}\, \frac{\as}{N-1},
\eeq
where the coefficient $c$ depends on the Higgs mass, and for
$m_H=125$~GeV it is $c\simeq4.522$.
The sum of these two opposite approximations provides our
NLO prototype of a joint resummed coefficient:
\beq
C^{(1)}_{\rm joint}(N) = C^{(1)}_{\rm BP_2}(N) + C^{(1)}_{\rm sx}(N).
\eeq
The difference
\beq
C^{(1)}(N) - C^{(1)}_{\rm joint}(N)
\eeq
is the NLO remainder term, i.e.\ the lowest-order term which would
be neglected in a joint resummed coefficient function.

\begin{figure}[tb]
\begin{center}
\includegraphics[width=0.85\textwidth,page=1]{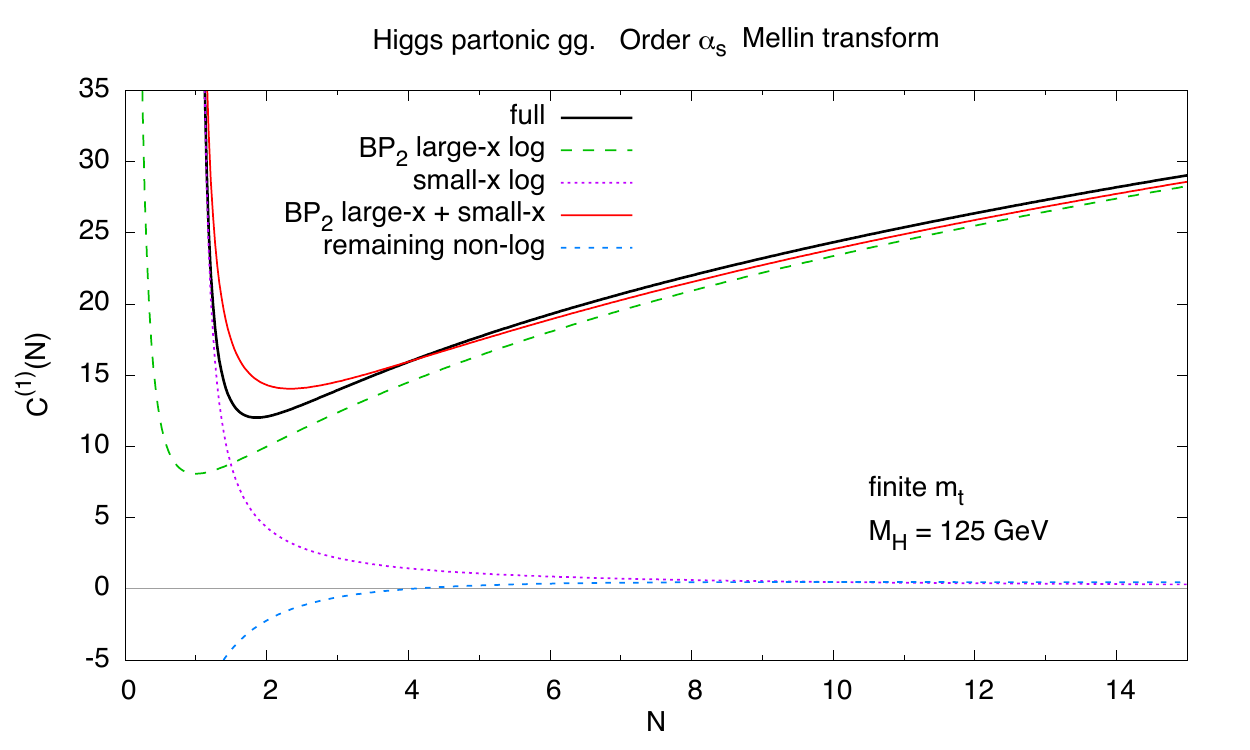}
\caption{The order $\as$ Higgs partonic coefficient function
  $C^{(1)}(N)$ plotted as a function of $N$ (black solid curve)
  and its small- and large-$N$ logarithmic approximations.
  The curves are obtained with full $m_t$ dependence.}
\label{fig:Higgs_order_as__finite-mt}
\end{center}
\end{figure}

To verify our guess, that the remainder terms behave perturbatively,
we start by inspecting the size of this term at NLO.
In Fig.~\ref{fig:Higgs_order_as__finite-mt} all the quantities
mentioned above are shown.
Indeed, the remainder term is absolutely very small for $N\gtrsim2$,
and relatively small everywhere.
In particular, at $N\sim2$ where the relative size of the remainder term
is largest, its magnitude is about $3$, which multiplied by $\as\sim 0.1$
(as appropriate at the scale $m_H=125$~GeV) gives a small contribution $0.3$.
Assuming that this behaviour is not changed at higher orders,
we would conclude that the remainder terms behave indeed perturbatively.

Moreover, the other important observation is that the small- and large-$N$
regions overlap, the first being dominant for $N\lesssim 1.5$ and the second
for $N\gtrsim 2$. Therefore, there is a very tight region in which
non-log effects may be relevant, enforcing our guess that even
at higher orders non-log contributions cannot become too large.

To conclude, we stress that jointly resummed cross-section
may be very accurate (at least for the Higgs production case),
and that the inclusion of non-logarithmically enhanced
terms from fixed-order computations may be considered
as a perturbatively stable improvement of such resummed cross-sections.

\chapter{Phenomenology}
\label{chap:pheno}

\minitoc

\noindent
In this Chapter we will present application of soft-gluon and high-energy
resummation for phenomenology. We will present results for
the Drell-Yan pair production process.
Such process is expected not to get relevant contributions from high-energy logarithms,
since the production is dominated by the quark-antiquark channel:
we will then show only results from threshold resummation, and compare them to data.
It would be more interesting to study the Higgs production process:
being dominated by the gluon-gluon channel, it is expected to
get a sizable contribution from the high-energy region; moreover
it is well known~\cite{Grazzini:2010zc} that the main contributions to the Higgs production
come from the threshold region.
Unfortunately, this is still work in progress, and we will not provide
results here.

\section{The Drell-Yan pair production}
\label{sec:pheno-DY}

\begin{figure}[thb]
  \centering
  \includegraphics[width=0.7\textwidth]{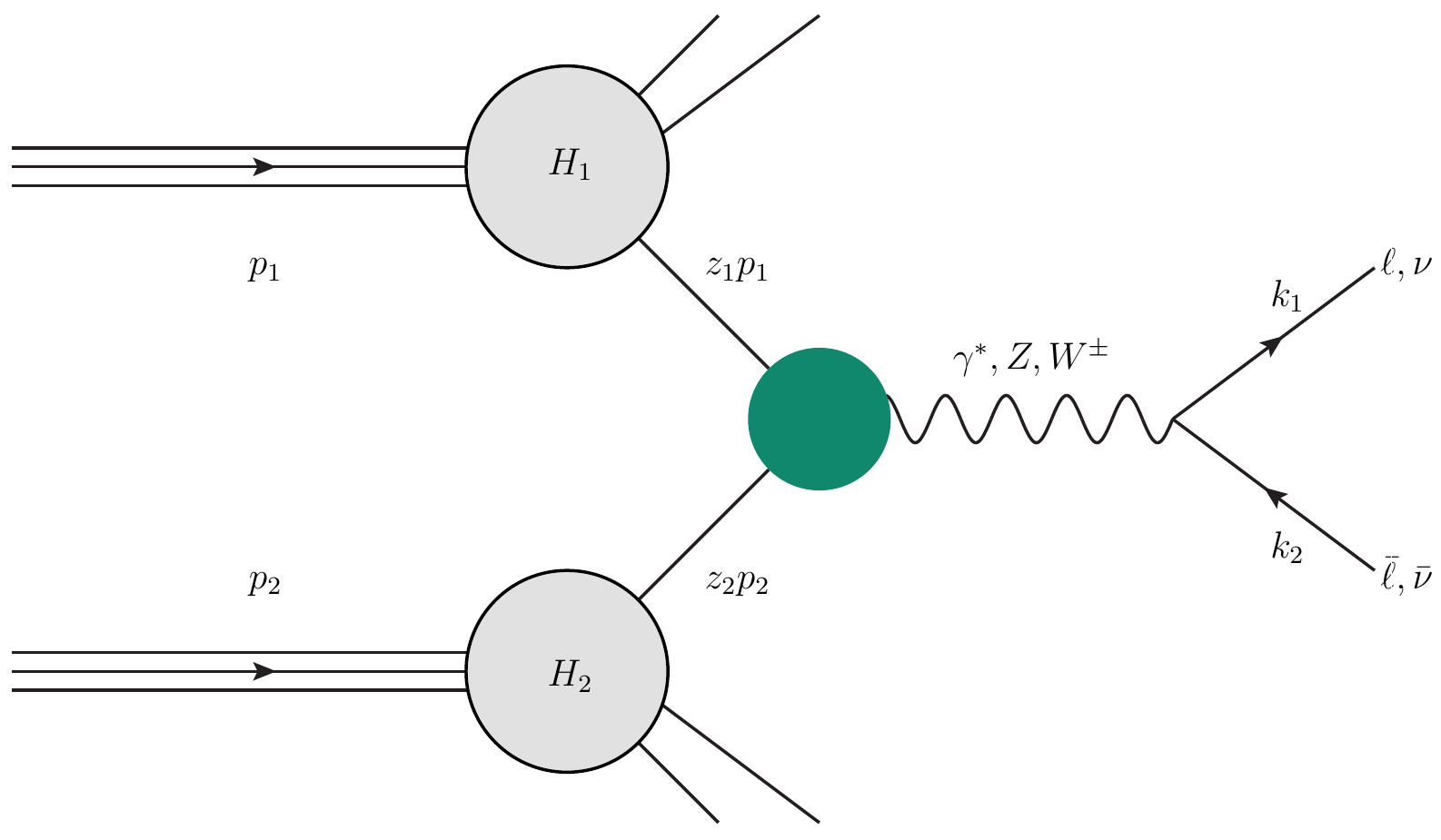}
  \caption{The Drell-Yan process at leading order in electroweak interactions.
    The green ball contains the QCD corrections to the process, including real emissions.}
  \label{fig:DY_proc_Feyn}
\end{figure}

The Drell-Yan process is likely to be the standard candle
which is both theoretically calculable and experimentally measurable
with highest accuracy at hadron colliders, in particular the LHC.
It consists in the production of a neutral or charged lepton pair,
$\ell\bar\ell$ or $\ell\bar\nu$,
in the collision of two initial hadrons $H_1$ and $H_2$:
\begin{align}
H_1(p_1) + H_2(p_2)&\to\ell(k_1)+\bar\ell(k_2)+X(q) && \text{(neutral)}\\
&\to\ell(k_1)+\bar\nu(k_2)+X(q) && \text{(charged)}
\end{align}
where $X$ is the entire set of other hadronic stuff produced in the event.
The process is sketched in Fig.~\ref{fig:DY_proc_Feyn}.
A large abundance of events collected at the LHC,
combined with a very precise theoretical determination of the process,
can be a very powerful test of perturbative QCD. Moreover, the high LHC energy
will allow for detailed measurements at a previously unexplored kinematic
domain of low parton momentum fraction at a high energy scale,
significantly improving the precision on the determination of the PDFs.

At leading order in both electroweak and QCD the partonic process
consists of a quark and an antiquark annihilating into a
virtual vector boson $\gamma$, $Z$ or $W^{\pm}$, which subsequently
decays into the lepton pair. At this order all the available
partonic center-of-mass energy $\sqrt{\hat s}$,
\begin{equation}
\hat s = z_1 z_2 s,\qquad s= (p_1+p_2)^2,
\end{equation}
goes into the lepton pair invariant mass $M$,
\begin{equation}
M^2 = (k_1+k_2)^2
\end{equation}
($z_1,z_2$ are the momentum fractions of the two partons),
and the inclusive cross-section is characterized by the variable
\beq
\tau = \frac{M^2}{s}
\eeq
which describes the amount of initial hadronic energy going into
the relevant final state (the lepton pair).
When QCD correction to the initial partons are considered,
real emission of gluons and quarks must be taken into account,
in order to remove infrared divergences coming from virtual corrections
(see discussion in Sect.~\ref{sec:PM_radiative_corr}):
in this case the available partonic center-of-mass energy is no longer
equal to the final state mass. We can then introduce the variable
\beq
z = \frac{M^2}{\hat s} = \frac{\tau}{z_1 z_2}
\eeq
which is the partonic analog to $\tau$: the Born amplitude (and its virtual corrections)
selects $z=1$, while in the case of real emissions $0\leq z \leq 1$.
As discussed in Sect.~\ref{sec:thr_resumm}, in the coefficient function
logarithms of $1-z$ appear: in the soft limit $z\to1$ the resummation
of the entire perturbative series is needed.

The current QCD theoretical accuracy for this process is
NNLO, both for integrated cross-section~\cite{vanNeerven:1991gh}
and rapidity-distributions~\cite{admp}; explicit formulae up to NLO
can be found in App.~\ref{sec:DY_fixed-order}.
Also, small effects such as those related to the coupling of the gauge boson
to final-state leptons have been studied recently~\cite{Catani:2010en}.

Whereas the resummation of contributions related
to the emission of soft gluons are routinely included in the
computation of Drell-Yan transverse-momentum
distributions~\cite{Grazzini:2009nd}, where they are mandatory in order to
stabilize the behaviour of the cross-section at low transverse momenta,
their impact on rapidity distributions, as well as inclusive cross-section,
has received so far only a moderate amount of attention.
This is partly due to the fact that even in
fixed-target Drell-Yan production experiments, such as the Tevatron
E866~\cite{e866_1,e866_2,e866_3}, let alone LHC experiments,
the available center-of-mass energy is much larger than the mass of typical
final states, thereby suggesting that threshold resummation is not relevant. 

However, we have abundantly discussed in Sect~\ref{sec:when_is_relevant}
that resummation may be relevant even far from hadronic threshold.
Indeed, in Ref.~\cite{bolz} threshold resummation has been claimed to affect
significantly Drell-Yan production for E866 kinematics, though
somewhat different results have been found in Ref.~\cite{bnx}.
It is important to observe that Drell-Yan data from E866 and
other fixed-target or collider Tevatron experiments play a crucial role
in the precision determination of parton distributions
such as MSTW08~\cite{Martin:2009iq},
NNPDF2.0~\cite{Ball:2010de} or CT10~\cite{Lai:2010vv},
so their accurate treatment is crucial for precise LHC phenomenology.
Neither threshold resummation has been so far included in predictions for the LHC,
such as those of Ref.~\cite{Balossini:2009sa}.

Then, in this Section we will present a detailed phenomenological
study~\cite{bfr2} of the impact of threshold resummation on inclusive cross-section
and rapidity distributions for the Drell-Yan process, and compare it
with an analogous study performed in the context of SCET~\cite{bnx}.

\begin{figure}[htb]
\begin{center}
\includegraphics[width=0.49\textwidth]{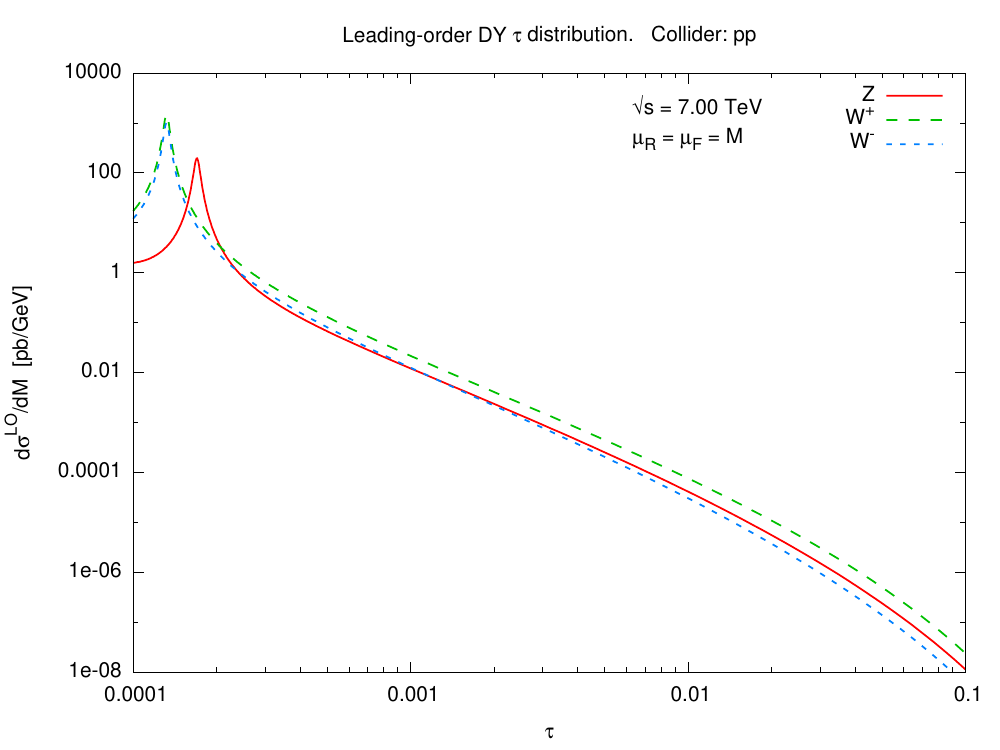}
\includegraphics[width=0.49\textwidth]{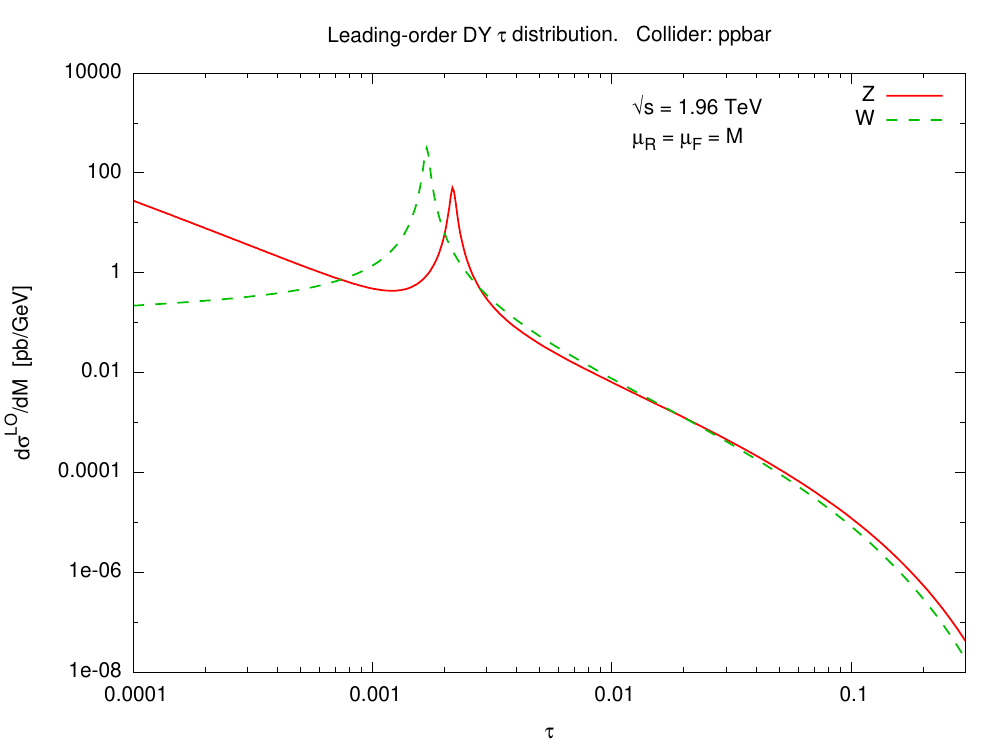}
\caption{The  
invariant mass distribution of charged and neutral Drell-Yan pairs
in $pp$ collisions at $\sqrt{s}=7$~TeV (left) and in $p\bar p$
collisions at $\sqrt{s}=1.96$~TeV (right) computed at leading order with
NNPDF2.0 parton distributions at $\as(m_Z^2)=0.118$.
}
\label{fig:born}
\end{center}
\end{figure}

We will therefore consider three cases: $pp$ collisions at a
center-of-mass energy of $38.76$~GeV, which corresponds to the case of
the experiment E866/NuSea, taken as representative of Tevatron
fixed-target experiments; $p\bar p$ collisions at the Tevatron
collider energy of $\sqrt{s}=1.96$~TeV; and the LHC case, $pp$
collisions at the intermediate energy of $\sqrt{s}=7$~TeV.
For the Tevatron and LHC configurations, we will
consider both charged ($\ell\bar\nu$) and neutral ($\ell\bar\ell$)
Drell-Yan pairs, taking into account in the latter case the
interference between virtual $Z$ and $\gamma$. Lepton masses will
always be neglected. We will show results for both the invariant mass
distribution as a function of $\tau=M^2/s$, and for the
doubly-differential distribution in invariant mass and rapidity as a
function of the rapidity for fixed values of $\tau$.
Specifically, for invariant mass distributions we will show results for the
$K$-factor defined as
\beq
K\(\tau, \frac{\muf^2}{M^2}, \frac{\mur^2}{M^2}\) =
\frac{\dfrac{d\sigma}{dM}\(\tau,\dfrac{\muf^2}{M^2}, 
\dfrac{\mur^2}{M^2}\)}{\dfrac{d\sigma^{\rm LO}}{dM}(\tau,1,1)},
\eeq
where $\mu_F$ and $\mu_R$ are the factorization and renormalization
scale, respectively. Since we will be considering
different experimental configurations, the results for the $K$-factors
will be shown for fixed value of $s$, with $M^2$ determined by the value
of $\tau$, $M^2=\tau s$.
The Born cross-section ${\dfrac{d\sigma^{\rm LO}}{dM}(\tau,1,1)}$,
which provides the scale for these plots, is
shown in Fig.~\ref{fig:born} for LHC at $7$~TeV
and Tevatron at $1.96$~TeV.

Our aim will be to assess the potential impact of inclusion of
resummation effects on cross-sections and their associate perturbative
uncertainty, both on experiments which are already used for PDF
determinations (and thus, potentially, on the PDF extraction itself),
as well as on future LHC measurements, both for real $W$ and $Z$
production as well as for high-mass $1$~TeV Drell-Yan pair
(relevant for instance
as a background to hypothetical $Z^\prime$ production). 
For each observable, we will show fixed-order predictions at leading,
next-to-leading and next-to-next-to-leading order, and,
correspondingly, leading, next-to-leading and next-to-next-to-leading
log resummed curves. 

All curves will be computed using a fixed (NLO)
set of parton distributions. In a realistic situation, parton
distributions would be refitted each time at the corresponding
perturbative order; the effect of the perturbative corrections on the
hard cross-section is then partly reabsorbed in the PDFs (with fixed
experimental data), and the effect on the Drell-Yan process 
gets tangled with the effect on other processes which are used for PDF
fitting. Hence, a comparative assessment of size of various
perturbative corrections on cross-sections and uncertainties, which is
our main aim here, can only be done with fixed PDF. It should be born
in mind, however, that our predictions will only be fully realistic
when considering the NLO case.

In order to assess perturbative uncertainties, we will perform
standard variations of factorization and renormalization scales, and
furthermore in order to assess the ambiguities related to the
resummation procedure we will compare results obtained with the
minimal and Borel prescriptions, as discussed in
Chap.~\ref{chap:soft-gluons}: specifically, for the Borel prescription
we will use the modified BP$_2$ Eq.~\eqref{eq:BP2_final}, which
provides a more moderate but realistic estimate of ambiguities as
discussed in Sect.~\ref{sec:subl}, and take $W=2$ (see Sect.~\ref{sec:borel}).
Actually, we are going to present the results published in Ref.~\cite{bfr2},
where the constant terms coming from the distributions are resummed as well,
see discussion in Sect.~\ref{sec:discussion_constants}; we will discuss it
in Sect.~\ref{sec:pheno_Borel}.
Note that we have checked explicitly that also
at the hadronic level curves obtained using the Borel
prescription Eq.~\eqref{eq:Borel_prescription}, which contains logarithms of
$\frac{1}{z}$, are indistinguishable from those obtained with the
minimal prescription, in agreement with the discussion in
Sect.~\ref{sec:subl}, provided $\tau$ itself is not too close to threshold:
we will come back on that in Sect.~\ref{sec:pheno_Borel}.

Other sources of uncertainty will be discussed briefly in
Sect.~\ref{sec:thunc}, where we will provide an overall assessment of
uncertainties related to the value of the strong coupling and to the
parton distributions, and then present and evaluate critically the use
of scale variation to assess perturbative uncertainties. In the
remainder of this Section detailed predictions for the three cases of
interest will be presented.

\subsection{Uncertainties}
\label{sec:thunc}

Theoretical predictions for the Drell-Yan process are affected by a
number of uncertainties, related to the treatment of both the strong
and electroweak interactions. 
Of course, in a realistic experimental situation further
uncertainties arise because of the need to introduce kinematic cuts,
which in turns requires comparing to fully exclusive
calculations~\cite{Catani:2009sm}.
Here, we will make no attempt to
estimate the latter, nor electroweak uncertainties 
and their interplay with strong corrections
(see e.g. Ref.~\cite{Balossini:2009sa} for a
recent discussion), and we will concentrate on uncertainties related to the treatment of the strong
interactions. 
 Before turning to an assessment of the way higher order
corrections can be estimated from scale variation, we discuss
uncertainties related to the value of the strong coupling and to the
choice of parton distributions (PDFs).

\subsubsection{Uncertainties due to the parton distributions and $\as$}
\label{sec:alphapdf}

The uncertainty due to PDFs is usually dominant in hadron
collider processes. Tevatron Drell-Yan data are used for PDF
determination, so PDF uncertainties here reflect essentially the
current theoretical uncertainty in knowledge of this process, as
well as possible tension between Drell-Yan data 
and other data which go into global PDF
fit (which however seems~\cite{Ball:2010de} to be very moderate).
Predictions for the LHC are affected by sizable PDF uncertainties
because of the need to extrapolate to a new kinematical region, and
also, in the case of Drell-Yan, because at the LHC, unlike at the
Tevatron, one of the two PDFs which enter the leading-order process
is sea-like. 

\begin{figure}[htb]
\begin{center}
\includegraphics[width=0.6\textwidth]{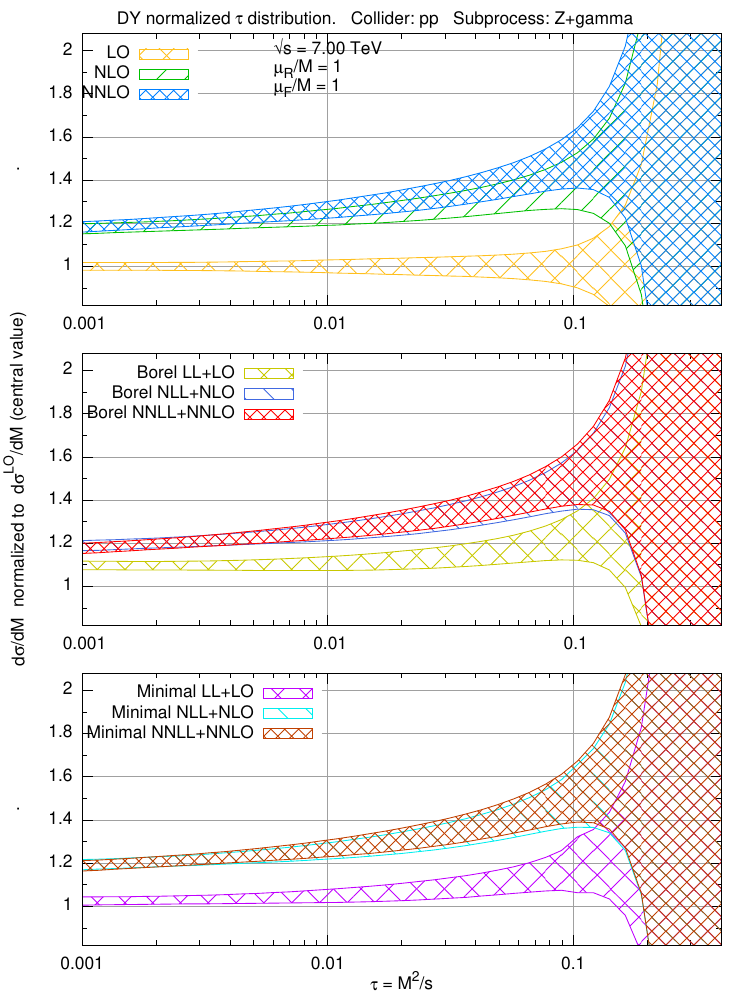}
\caption{
Invariant mass distribution of neutral Drell-Yan pairs
in $pp$ collisions at $\sqrt{s}=7$~TeV. The
band corresponds to the PDF uncertainty, using the
NNPDF2.0 set with $\as(m_Z^2)=0.118$.
}
\label{fig:totLHCpdf}
\end{center}
\end{figure}

PDF uncertainties for the invariant mass distribution of 
neutral Drell-Yan pairs at $\sqrt{s}=7$~TeV are shown in
Fig.~\ref{fig:totLHCpdf} as a function of $\tau=M^2/s$. We use NNPDF2.0
PDFs with $\as(m_Z^2)=0.118$; other PDF sets are expected to give
similar results~\cite{PDF4LHC}. Because we are using a fixed PDF set,
the  uncertainty does not depend
significantly on the perturbative order, or the inclusion of resummation.
It ranges between 5\% and 15\% for $\tau\lesssim
0.1$. For larger values of $\tau$ the cross-section becomes
essentially undetermined, because there are no data in PDF global fits
to constrain PDFs in that region: the few available large--$x$ data
are at much lower scale, and the uncertainty due to lack of
information at very large $x\gtrsim0.5$ contaminates PDFs down to
$x\gtrsim0.1$ when evolving up to the LHC scale.
Note however that the Drell-Yan cross-section at large
$\tau\gtrsim0.1$ rapidly drops to unmeasurably small values (see
Fig.~\ref{fig:born}). The fact that PDF uncertainties blow up for
$\tau\gtrsim0.1$ implies that data in this region would allow a
determination of PDFs in a region where they are currently almost
unknown; conversely, any signal of new physics in this region would
have to be validated by measurements in an independent channel (such
as for example jet production) which provides an independent
constraint on the relevant PDFs.

\begin{figure}[htb]
\begin{center}
\includegraphics[width=0.7\textwidth]{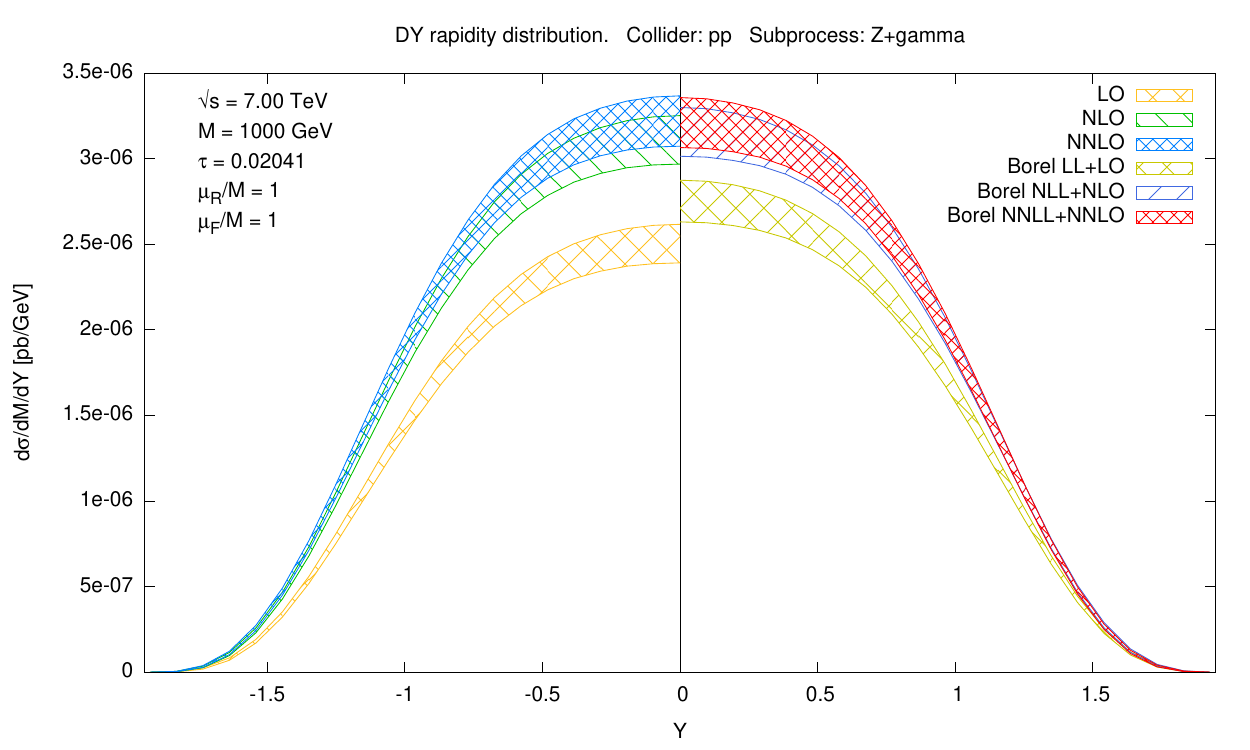}
\caption{
  Rapidity distribution of neutral Drell-Yan pairs with invariant
  mass $M=1$ TeV in $pp$ collisions at $\sqrt{s}=7$~TeV. Unresummed
  results are shown for negative $Y$ and resummed results for positive
  $Y$.  The band corresponds to the PDF uncertainty, obtained by the
  NNPDF2.0 set with $\as(m_Z^2)=0.118$.}
\label{fig:nnpdf}
\end{center}
\end{figure}

In Fig.~\ref{fig:nnpdf} we show the PDF uncertainties for the rapidity
distribution of neutral Drell-Yan pairs with $M=1$~TeV at
$\sqrt{s}=7$~TeV, again using the NNPDF2.0 set with $\as(m_Z^2)=0.118$. As in
the previous case, the PDF uncertainty does not depend significantly
on the perturbative order, and it is typically larger than 5\%.

We now turn to the uncertainty due to the value of $\as$.
The current PDG~\cite{PDG} value for $\as(m_Z^2)$ is taken from
Ref.~\cite{Bethke:2009jm} and it is $0.1184\pm 0.0007$. However, this
uncertainty seems quite small, especially when taking into account the
fluctuation in central values between the determinations that go into
it, and the dependence on the perturbative order of some of them: indeed,
current recommendations for precision LHC studies from the PDF4LHC
group~\cite{PDF4LHC} advocate a rather more conservative uncertainty
estimate. We thus take
\beq
\label{alphaband}
\as(m_Z^2)=0.118\pm 0.002
\eeq
as a reasonable current range.

\begin{figure}[htb]
\begin{center}
\includegraphics[width=0.6\textwidth]{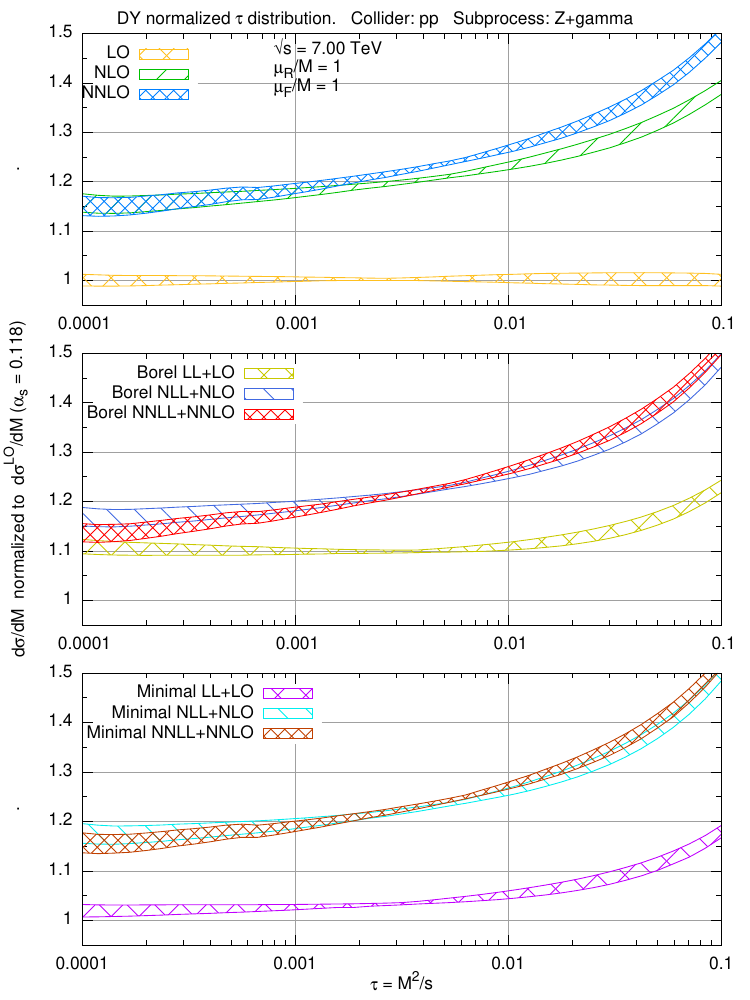}
\caption{
  Invariant mass distribution of neutral Drell-Yan pairs in $pp$
  collisions at $\sqrt{s}=7$~TeV.  The uncertainty bands corresponds
  to a variation of  $\as(m_Z^2)$ in the range $0.116$ to $0.120$ in the hard
  matrix element and in the parton distributions;  NNPDF2.0 parton
  distributions are used.}
\label{fig:tot7alpha}
\end{center}
\end{figure}

The impact of this uncertainty on $\as$ on the Drell-Yan cross-section
at $\sqrt{s}=7$~TeV can be estimated from Fig.~\ref{fig:tot7alpha},
where we show the effect on the inclusive Drell-Yan cross-section due
to variation in the range Eq.~\eqref{alphaband} of the value of
$\as(m_Z^2)$ used both in the computation of the hard matrix element and
the PDF evolution. Results are only shown for $\tau<0.1$, because for
larger value the PDF uncertainty blows up and results loose
significance,  as discussed above.
Note that this full dependence of the physical
cross-section on $\as$ is in general somewhat different from
the dependence of the hard matrix element alone, because of the
dependence on $\as$ of the relevant parton luminosity. This total
dependence might be larger or smaller according to whether the
luminosity is correlated or anticorrelated to the value of $\as$,
either of which might be the case for a quark luminosity, according to
the kinematic region~\cite{Ball:2010de}.  A priori, the size of the
uncertainty due to variation of $\as$ in the matrix element and
that due to the dependence on $\as$ of the PDFs are likely to
be comparable: after all, the Drell-Yan rapidity distribution plays a
significant role in the determination of the PDFs themselves.

It appears from Fig.~\ref{fig:tot7alpha} that the $\as$ uncertainty
increases with the perturbative order, but it is of similar size at
the resummed and unresummed level; at NNLO it is of order of $\sim
1.5$\% at LHC energies; we have checked that it is about a factor two
larger at Tevatron fixed-target experiments. The uncertainty due to
$\as$ on rapidity distributions is clearly of comparable size.

Noting that within the approximation of linear error propagation the
PDF and $\as$ uncertainties should be combined 
in quadrature~\cite{Lai:2010nw}, we conclude that PDF uncertainties
are somewhat larger than $\as$ uncertainties and the combined
effect of PDF and $\as$ uncertainties is likely to be smaller
than about 10\% but not much smaller, at least in the region in which
PDFs are constrained by presently available data. 
Once PDF uncertainties will be
reduced due to LHC data, it should be possible to keep the combined
effect of these uncertainties at the level of few percent. Therefore,
perturbative accuracies at the percent level are relevant for precision
phenomenology.

\subsubsection{Perturbative uncertainties: scale variations}
\label{scales}

\begin{figure}[tbp]
  \centering
  \includegraphics[scale=1]{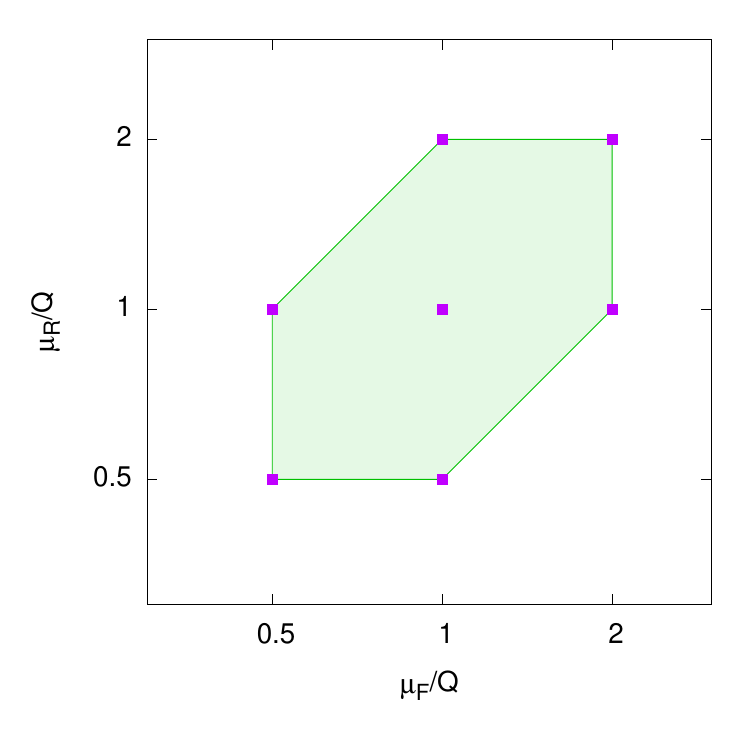}
  \caption{Scale variation grid; the cross-section is computed
    in correspondence of the purple dots.}
  \label{fig:scales}
\end{figure}

A standard way of estimating unknown higher order perturbative
corrections is to vary factorization and renormalization scales. We
perform this variation 
by writing the generic factorized cross-section as
\beq
\label{factNscal}
\sigma(N,M^2)=\Lum(N,\muf^2)\,
C\(N,\as(\mur^2),\frac{M^2}{\muf^2},\frac{M^2}{\mur^2}\),
\eeq
which is independent of $\muf^2$ and $\mur^2$ at the
order at which the partonic coefficient 
$\hat\sigma\(N,\as(\mur^2),\frac{M^2}{\muf^2},\frac{M^2}{\mur^2}\)$
is computed.
The residual scale dependence is therefore of the first neglected
order, and can be used as an estimate of the higher order terms in
the perturbative expansion.
We vary the two scales in the range
\beq\label{eq:scale_variation}
\abs{\log\frac{\muf}{M}} \leq \log 2 ,\qquad
\abs{\log\frac{\mur}{M}} \leq \log 2 ,\qquad
\abs{\log\frac{\mur}{\muf}} \leq \log 2 ,
\eeq
depicted in Fig.~\ref{fig:scales}, which guarantees that both
higher-order corrections to the partonic cross-section and to perturbative
QCD evolution are generated, with the last condition ensuring that no
artificially large scale ratios are introduced.

In the sequel, we will perform scale variation of both unresummed and
resummed cross-sections. The interpretation of results deserves a
comment. When performing scale variation of a result determined at
fixed $\Ord(\as^k)$, one generates terms of $\Ord(\as^{k+1})$:
consequently, the scale uncertainty is reduced as
one increases the perturbative order. However, terms generated by
scale variation are proportional to those which are present at the
given order: therefore, scale variation underestimates the size of higher
order corrections when these are enhanced by higher logarithmic
powers. For instance, scale variation of  the $\Ord(\as)$
Drell-Yan coefficient function $C_1(N)$ Eq.~\eqref{eq:c1} produces terms which
at large $N$ grow at most as $\log^2 N$, whereas the actual 
$\Ord(\as^2)$ $C_2(N)$ coefficient function at large $N$ grows as
$\log^4 N$. Hence, if $N$ is so large that these terms dominate the
coefficient functions and must be resummed to all orders their impact
might be rather larger than the scale variation of the fixed-order
result may suggest. Nevertheless, in this case the scale dependence of the resummed
result will still be smaller than that of the fixed order result
because the resummed result includes the dominant contributions to the
cross-section to all orders.

However, in Sect.~\ref{sec:resDY} we have seen that there is an
intermediate kinematic region in which logarithmically enhanced
contributions may provide a sizable fraction of the coefficient
function even though $\as\log^2 N\ll 1$: in this case, the resummation
improves the fixed-order result in that it includes a sizable fraction
of the higher order correction, but it still behaves in a perturbative
way, i.e.\ terms of higher order in $\as$ included through the
resummation give an increasingly small contribution. If so, 
the scale dependence of the resummed result may well be
comparable to or even larger than that of the fixed order
result, because the resummation amounts to the inclusion of large
terms in the next few higher orders, which are not necessarily seen when
performing the scale variation of the lower orders. Furthermore, 
resummation
only affects the quark channel, while fixed-order scale variation
mixes the quark and gluon channels: in an intermediate region, the
logarithmic terms in the quark channel may be sizable, but with the
gluon channel not being entirely negligible.  In such case, the scale
variation is dominated by subleading terms and thus we expect the
residual scale dependence of the resummed result to differ according to the
resummation prescription. We will see that this is indeed the case for
resummation of Tevatron rapidity distributions, with scale variation
of resummed results different according to whether the Borel or
minimal prescription is used.

\subsection{Results and comparison with data}

In this Section we present the prediction for the $K$-factors and
the rapidity distributions both at the unresummed level and at the resummed
level, using in the latter case both the minimal and Borel prescriptions.
When available, we compare our predictions with data.
We want to remember again that we use NNPDF2.0 NLO PDFs set:
then, our prediction are reliable and comparable with data
consistently only at NLO.

\subsubsection{Tevatron at fixed target: NuSea}

\begin{figure}[tb]
\begin{center}
\includegraphics[width=0.6\textwidth]{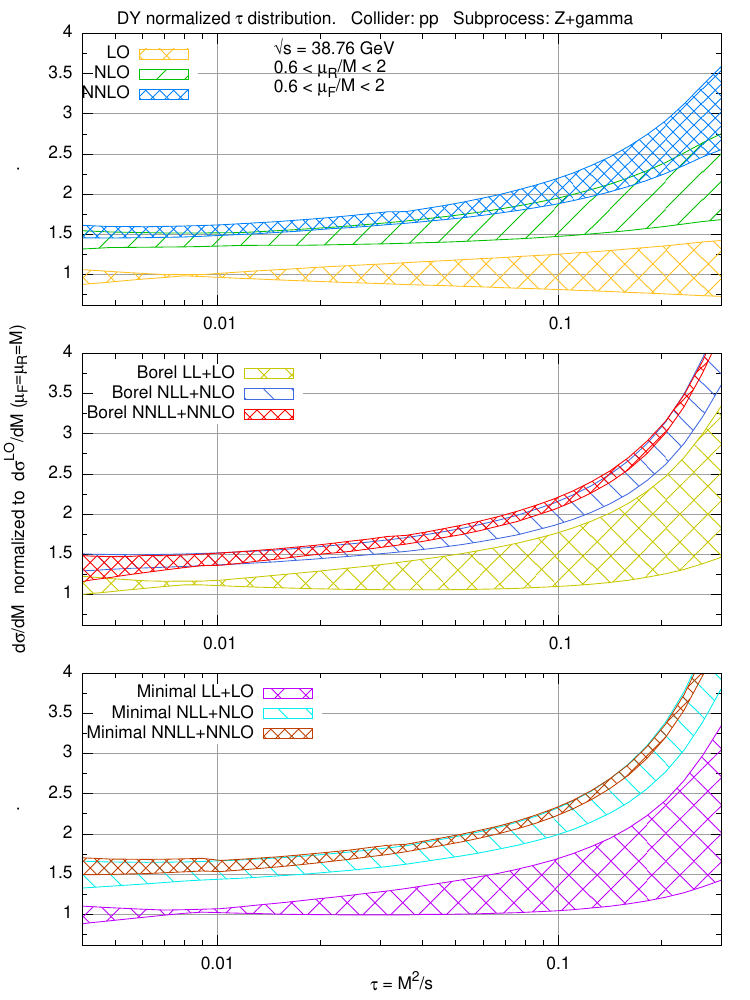}
\caption{Invariant mass distribution of neutral Drell-Yan pairs
in $pp$ collisions at $\sqrt{s}=38.76$~GeV.}
\label{fig:totft}
\end{center}
\end{figure}

We begin by studying the invariant mass distribution of lepton pairs
produced by collisions of a proton beam of energy
$E=800$~GeV on a proton or deuteron target, at rest in the laboratory
($\sqrt{s}=38.76$~GeV). This is the experimental
configuration of the experiment E866/NuSea.
We first consider the inclusive invariant mass distribution. Results
are shown in Fig.~\ref{fig:totft}. All uncertainties shown here and
henceforth are due to scale variation as described in Sect.~\ref{scales}.
As expected, the width of the error bands decreases
with increasing perturbative order. Note that for sufficiently small
$\tau$ the uncertainty blows up, due to the fact that for fixed $s$
the small $\tau$ limit corresponds to low scale: for example, at this energy
$\tau=10^{-3}$ corresponds to $M\approx 1.2$~GeV, and varying the scales
as in Eq.~\eqref{eq:scale_variation} the values $\mur,\muf\approx 0.6$~GeV
are reached.

Turning now to the
resummed results, we note that the numerical impact of
resummation is large for $\tau \gtrsim 0.1$, while for
$0.03\lesssim\tau\lesssim0.1$ it is moderate but sill not negligible.
Furthermore, starting with the NLL level,
the scale uncertainty band for resummed results is dramatically
smaller than in the case of fixed order results. This is because scale
variation of the LL result produces NLL terms which beyond the first
few orders are not contained in the fixed order result; starting with
NLL these terms are already included in the resummed result.
It is interesting to
note that in the case of the resummed cross-section (with both
prescriptions) the NNLL band is almost entirely contained in the NLL
band, while the fixed-order NLO and NNLO error bands are only
marginally compatible with each other.  The ambiguity in the
resummation, estimated from the difference
between Borel
and minimal prescription, is not negligible,
but smaller than the scale uncertainty;
moreover, it is more evident at small $\tau$,
since the different subleading terms
give a larger contribution in that region.

\begin{figure}[htb]
\begin{center}
\includegraphics[width=0.7\textwidth,page=1]{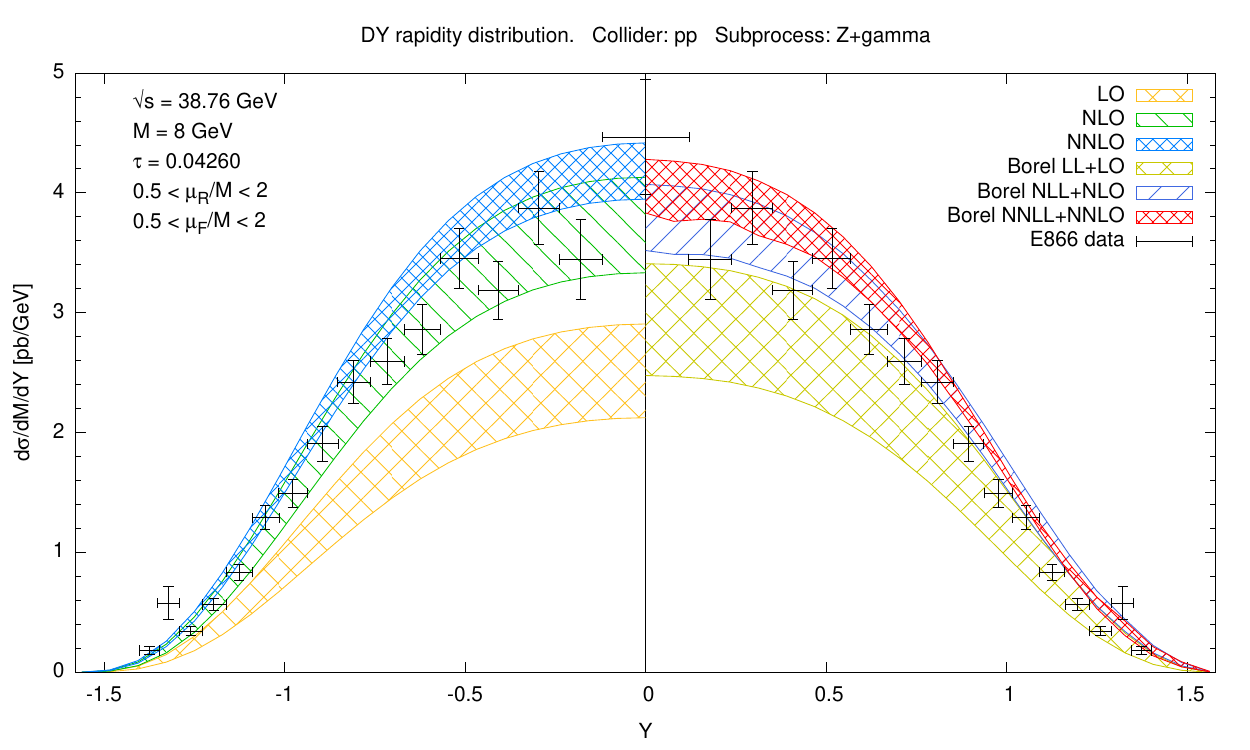}
\\
\includegraphics[width=0.7\textwidth,page=2]{graph_rap_pp_nusea}
\caption{Rapidity distribution of neutral Drell-Yan pairs of invariant
mass $M=8$~GeV in $pp$ collisions
at $\sqrt{s}=38.76$~GeV; E866 data are also shown.}
\label{fig:rapft}
\end{center}
\end{figure}
The experiment E866/NuSea~\cite{e866_1,e866_2,e866_3} has measured the $x_F$-distribution
Eq.~\eqref{xfdef} of lepton pairs with an invariant mass $M=8$~GeV.
The E866 data are displayed in Fig.~\ref{fig:rapft}, superimposed to
the QCD prediction and the corresponding scale uncertainty.
The distribution is symmetric about $Y=0$; the curves shown for $Y<0$ refer
to fixed-order calculations, and those with $Y>0$ to resummed results. 
The data agree with the NLO calculation because these data were
included in the determination of the  NLO PDFs that we are using.

The impact of the resummation is small but not negligible: for
instance the difference between NNLO and NNLL is about half of the
difference between NNLO and NLO. Furthermore, the scale uncertainty of
resummed results is somewhat smaller than that of the unresummed ones.
This is consistent with the
observation that for this experiment $\tau=0.04$, which, as 
discussed in Sect.~\ref{sec:when_is_relevant}, is in the region
in which resummation is relevant. 
However, the difference between resummed results obtained using the
Borel and the minimal prescription is almost as large as the size of
the resummation itself: in fact the NLL Borel results is somewhat
lower than the NNLO one, while the NLL minimal prescription
result is a bit higher.
Hence we conclude that the overall impact of the resummation on these
data is essentially negligible. A large and negative NLL resummed
correction to the NLO result was claimed in
Ref.~\cite{bolz}, using the minimal prescription, but we do not confirm it:
we find a positive and rather smaller correction.
The result of Ref.~\cite{bolz} was first criticized in 
Ref.~\cite{bnx}.
Our result with the Borel prescription is
in good quantitative agreement with Ref.~\cite{bnx};
however,
the minimal prescription gives a somewhat larger correction, though still
positive.

\subsubsection{Tevatron collider}

\begin{figure}[htbp]
\begin{center}
\includegraphics[width=0.496\textwidth]{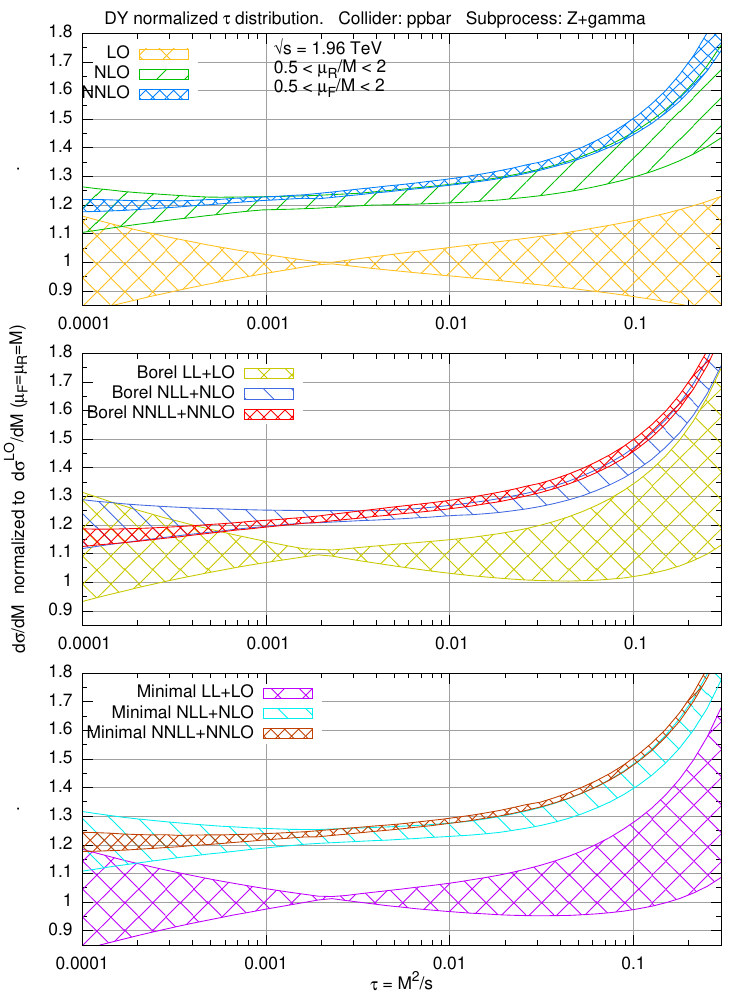}
\includegraphics[width=0.496\textwidth]{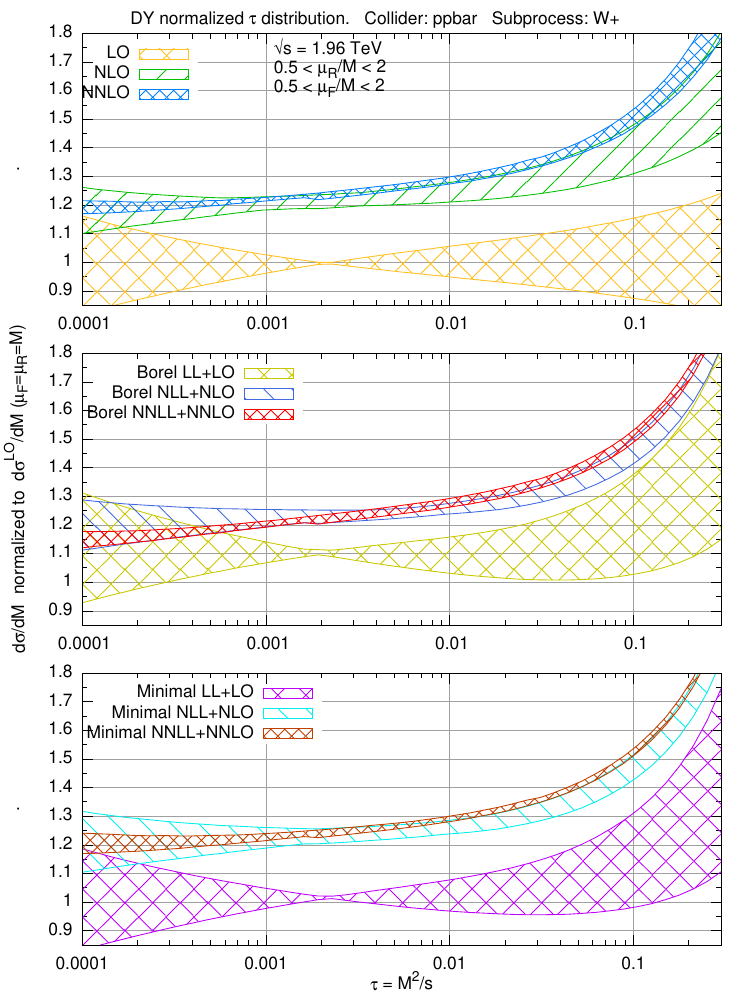}
\caption{Invariant mass distribution of neutral (left) and charged 
(right) Drell-Yan pairs
in $p\bar p$ collisions at $\sqrt{s}=1.960$~TeV.}
\label{fig:totTev}
\end{center}
\end{figure}

We now turn to Drell-Yan production in $p\bar p$ collisions
at a center-of-mass energy $\sqrt{s}=1.96$~TeV. 
Results for the invariant mass distribution  of neutral and
charged Drell-Yan pairs in this configuration are shown
in Fig.~\ref{fig:totTev}.

The behaviour of these curves is similar to that seen in the case of NuSea,
Fig.~\ref{fig:totft}, but with the impact of the resummation yet a
bit smaller, as one would expect both because of the higher energy and
because of the collider configuration, as discussed in
Sect.~\ref{sec:resDY} (in particular Fig.~\ref{fig:sd1}).
Interestingly, even when the resummation has a very small impact, it
still leads to a non-negligible reduction of the uncertainty: this 
is consistent with the expectation
that for these medium-small values of $\tau$ 
resummation is in fact a perturbative
correction, as discussed in the end of Sect.~\ref{sec:resDY}.
Note that in these plots the smallness of leading-order 
uncertainty bands  when $\tau\approx 0.002$
(i.e.\ $M\approx100$~GeV) is due to the
fact that the scale dependence of the parton luminosity, to which the
leading-order cross-section is proportional (see Eq.~\eqref{eq:sigmaNLOqqbar}), is
stationary at this scale.

\begin{figure}[tb]
\begin{center}
\includegraphics[width=0.7\textwidth,page=1]{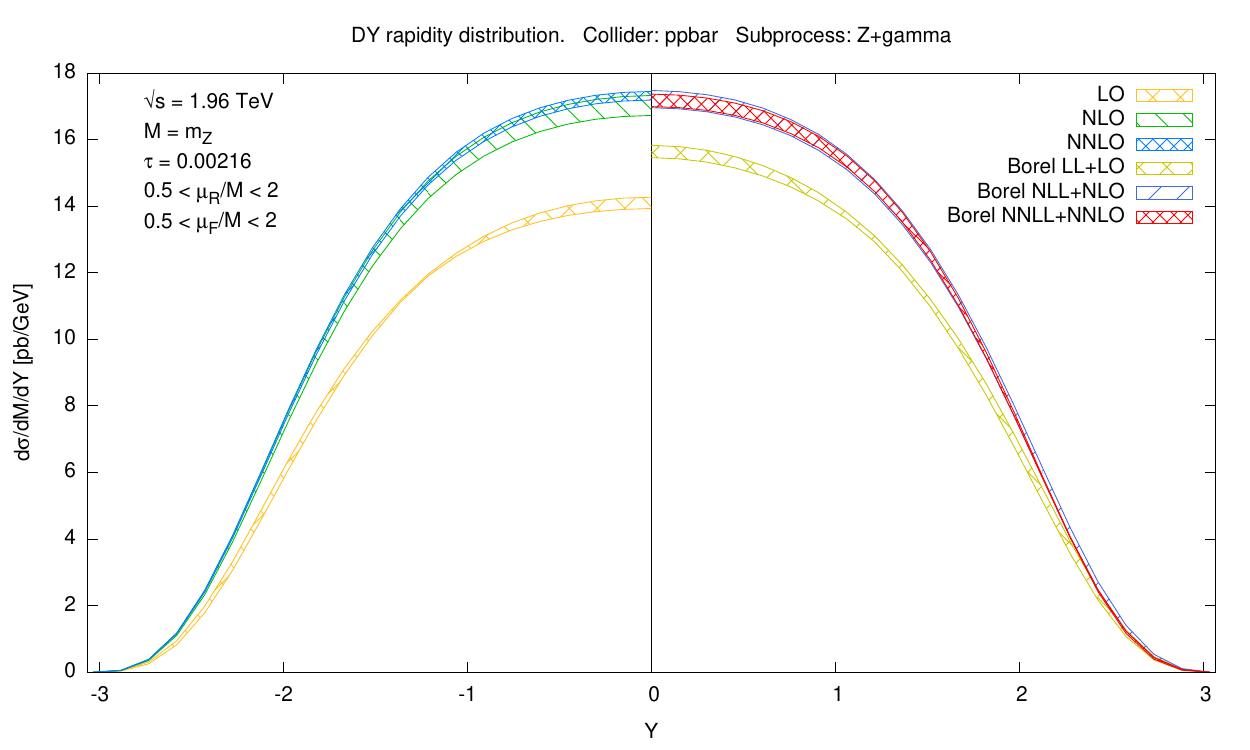}
\\
\includegraphics[width=0.7\textwidth,page=2]{graph_rap_ppbar_196_Z}
\caption{Rapidity distribution of neutral Drell-Yan pairs of invariant
  mass $M=m_Z$ in $p\bar p$ collisions at $\sqrt{s}=1.96$~TeV
  (the contribution of virtual $\gamma$ at the $Z$ peak is included).}
\label{fig:rapTEVZ}
\end{center}
\end{figure}
\begin{figure}[tb]
\begin{center}
\includegraphics[width=0.7\textwidth,page=1]{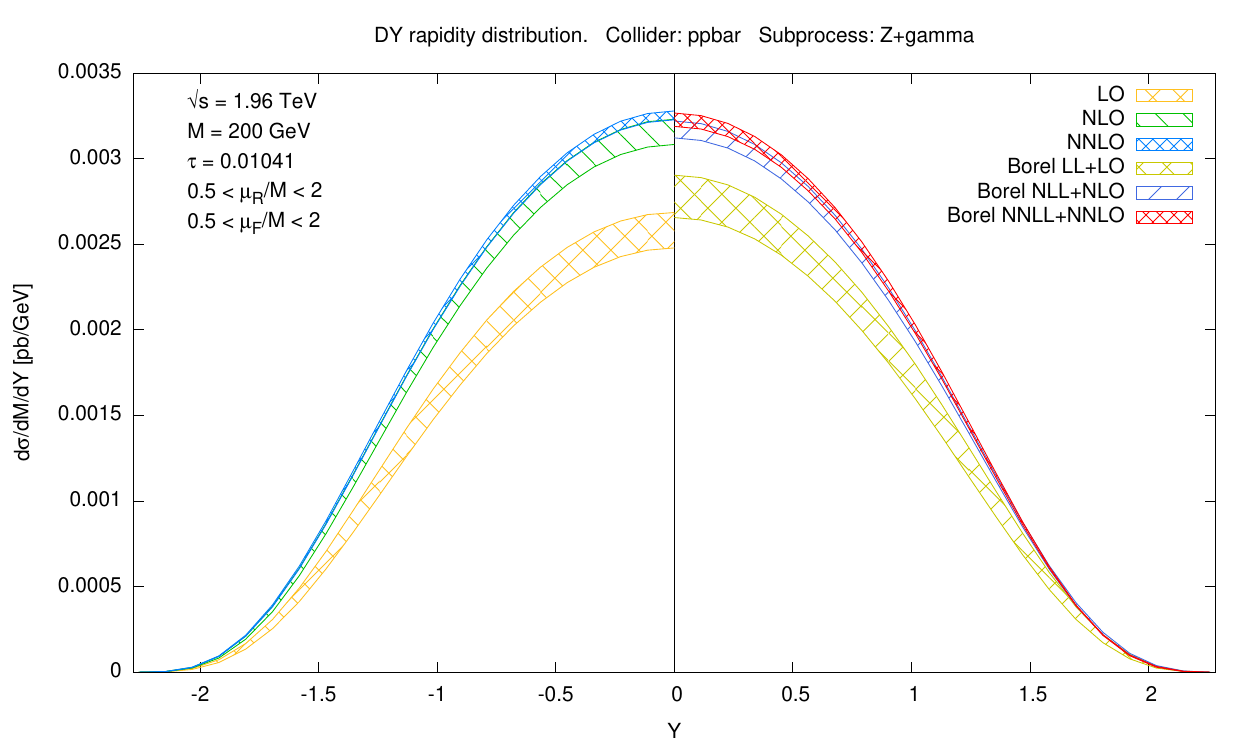}
\\
\includegraphics[width=0.7\textwidth,page=2]{graph_rap_ppbar_196_200}
\caption{Rapidity distribution of neutral Drell-Yan pairs of invariant
  mass $M=200$~GeV in $p\bar p$ collisions at $\sqrt{s}=1.96$~TeV.}
\label{fig:raptevg}
\end{center}
\end{figure}
\begin{figure}[tbp]
\begin{center}
\includegraphics[width=0.67\textwidth,page=1]{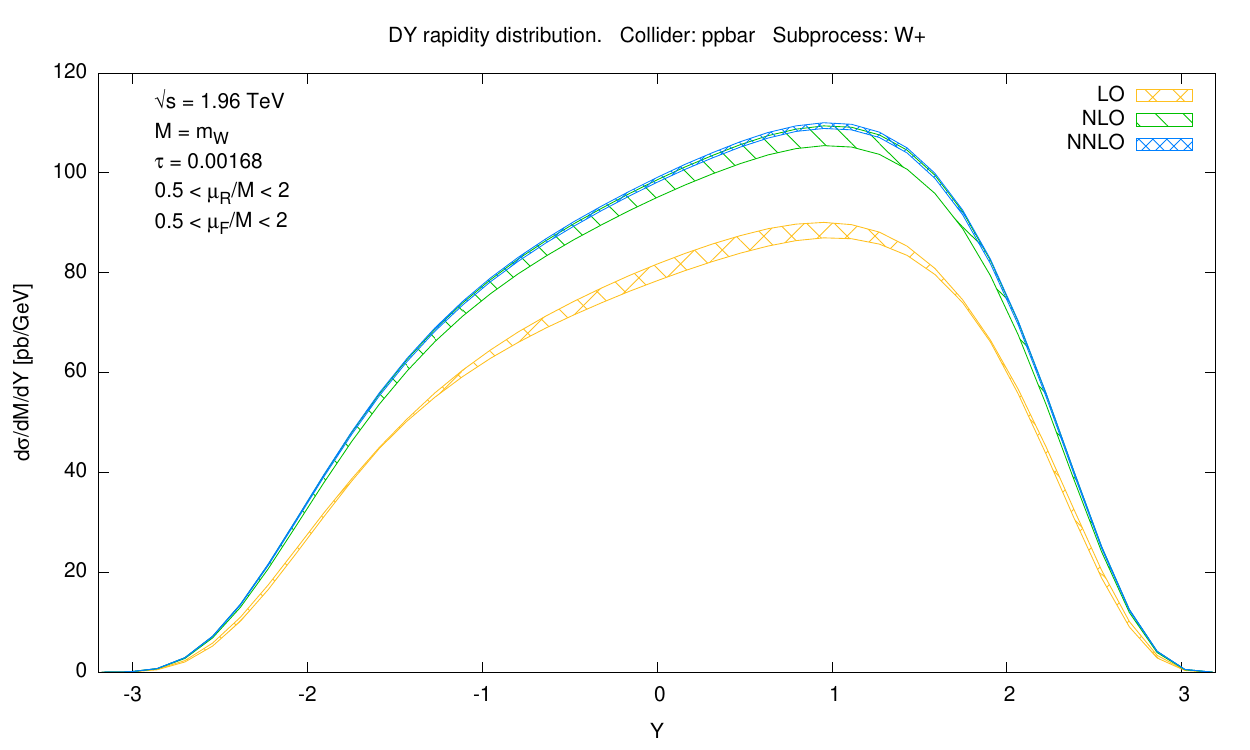}
\\
\includegraphics[width=0.67\textwidth,page=2]{graph_rap_ppbar_196_Wp}
\\
\includegraphics[width=0.67\textwidth,page=3]{graph_rap_ppbar_196_Wp}
\caption{Rapidity distribution of charged Drell-Yan pairs of invariant
  mass $M=m_W$ in $p\bar p$ collisions at $\sqrt{s}=1.96$~TeV.}
\label{fig:rapTEVWp}
\end{center}
\end{figure}

We now turn to rapidity distributions,  shown in
Figs.~\ref{fig:rapTEVZ} and~\ref{fig:raptevg} for neutral Drell-Yan
pairs of invariant mass $M=m_Z$ and $M=200$~GeV respectively. The impact
of the resummation is now very small, as one would expect given the
smallness of the relevant values of $\tau$. However, interestingly,
resummed uncertainty bands are systematically smaller than the
unresummed one, with the resummation ambiguity (i.e.\ the difference
between minimal and Borel results) now essentially negligible. Hence,
even in this small $\tau$ region the resummation leads to perturbative
improvement, while behaving of course as a perturbative correction.
Again, note that the smallness of leading-order 
uncertainty bands is due to the fact that the scale here is close to
the stationary point of 
the scale dependence of the parton luminosity already seen in
Fig.~\ref{fig:totTev}.

We can compare these uncertainties to that of typical current data,
thanks to recent measurements at the Tevatron. Specifically,
the rapidity distribution of $e^+ e^-$ pairs with
invariant mass in the range $66$~GeV $\leq M\leq 116$~GeV has
been recently measured by the CDF collaboration~\cite{Aaltonen:2010zza}.
In principle, the data should be
compared with the theoretical prediction for the full process
\beq
p+\bar p \to e^+ + e^- +X.
\eeq
For values of $M$ close to the
$Z$ mass, however, a good approximation is provided by the Breit-Wigner
approximation, which amounts to assuming that the amplitude is dominated by 
$Z$ exchange, and takes into account the finite width $\Gamma_Z$ of the $Z$
boson:
\beq
\label{BW}
\frac{d\sigma(\tau,Y,M^2)}{dM^2dY}
=\frac{2m_Z\Gamma_{\ell\bar\ell}}{(M^2-m_Z^2)^2+m_Z^2\Gamma^2_Z}
\frac{1}{2\pi}\frac{d\sigma_Z}{dY}
\eeq
where $\Gamma_{\ell\bar\ell}$ is the $Z$ decay width into
a lepton pair, and $d\sigma_Z$ is the
differential cross-section for the production of a real on-shell $Z$ boson.
Eq.~\eqref{BW} gives
\beq
\label{BW2}
\frac{d\sigma(\tau,Y,M^2)}{dM^2dY}
=\frac{m_Z^2\Gamma_Z^2}{(M^2-m_Z^2)^2+m_Z^2\Gamma^2_Z}
\frac{d\sigma(\tau,Y,m_Z^2)}{dM^2dY},
\eeq
and therefore
\beq
\int_{M^2_{\rm min}}^{M^2_{\rm max}}dM^2\,\frac{d\sigma(\tau,Y,M^2)}{dM^2 dY}
= N_Z(M^2_{\rm min},M^2_{\rm max})\,
\frac{d\sigma(\tau,Y,m_Z^2)}{dMdY},
\eeq
where
\begin{align}
N_Z(M^2_{\rm min},M^2_{\rm max})&=
\frac{m_Z \Gamma_Z^2}{2}\int_{M^2_{\rm min}}^{M^2_{\rm max}}dM^2\,
\frac{1}{(M^2-m_Z^2)^2+m_Z^2\Gamma^2_Z}\nonumber\\
&= \frac{\Gamma_Z}{2}\[\arctan\(\frac{M_{\rm max}^2-m_Z^2}{m_Z\Gamma_Z}\) + \arctan\(\frac{m_Z^2-M_{\rm min}^2}{m_Z\Gamma_Z}\)\]
\end{align}
is just a $Y$--independent multiplicative factor.
A better approximation is obtained including also the contribution from
the virtual photon exchange, and neglecting its interference with the $Z$ boson.
Schematically, we can separate the contributions as
\begin{align}
\sigma(M^2)
&= \sigma_Z(M^2) + \sigma_\gamma(M^2) + \sigma_{Z\gamma}(M^2) \nonumber\\
&\simeq \frac{m_Z^2\Gamma_Z^2}{\(M^2-m_Z^2\)^2 + m_Z^2\Gamma_Z^2}\,\sigma_Z(m_Z^2)
+ \frac{m_Z^2}{M^2}\,\sigma_\gamma(m_Z^2)
+ \frac{M^2-m_Z^2}{\(M^2-m_Z^2\)^2 + m_Z^2\Gamma_Z^2}\,\sigma^0_{Z\gamma}
\end{align}
where the first piece is the Breit-Wigner term used before,
the second is the contribution from the virtual photon and the last term
is the interference term: since it is almost odd in $M^2-m_Z^2$,
its contribution is strongly suppressed in the vicinity of the $Z$ peak and it can be neglected.
Note that at the $Z$ peak
\beq
\sigma(m_Z^2) = \sigma_Z(m_Z^2) + \sigma_\gamma(m_Z^2)
\eeq
exactly. Then, knowing the relative contributions of $Z$ and $\gamma$ at the peak
the $M^2$ integral can be estimated to a higher accuracy. Defining
\beq
R_Z = \sigma_Z(m_Z^2) / \sigma(m_Z^2)
,\qquad
R_\gamma = \sigma_\gamma(m_Z^2) / \sigma(m_Z^2)
,\qquad
R_Z + R_\gamma = 1,
\eeq
we have
\beq
\sigma(M^2) \simeq \sigma(m_Z^2)
\[ R_Z\,\frac{m_Z^2\Gamma_Z^2}{\(M^2-m_Z^2\)^2 + m_Z^2\Gamma_Z^2} + R_\gamma\,\frac{m_Z^2}{M^2} \]
\eeq
and hence
\beq
\int_{M^2_{\rm min}}^{M^2_{\rm max}}dM^2\,\frac{d\sigma(\tau,Y,M^2)}{dM^2 dY}
= \[R_ZN_Z(M^2_{\rm min},M^2_{\rm max}) + R_\gamma N_\gamma(M^2_{\rm min},M^2_{\rm max})\]
\frac{d\sigma(\tau,Y,m_Z^2)}{dMdY},
\eeq
with
\beq
N_\gamma(M^2_{\rm min},M^2_{\rm max}) = \frac{1}{2m_Z} \int_{M_{\rm min}^2}^{M_{\rm max}^2} dM^2\,\frac{m_Z^2}{M^2}
= \frac{m_Z}{2} \log\frac{M_{\rm max}^2}{M_{\rm min}^2}
\eeq
With a simple electroweak computation one can compute the $R$-factors
\beq
R_\gamma = \frac{\aem/3 m_Z}{\frac{\aem}{3m_Z} + \frac{\Gamma_{\ell\bar\ell}}{\Gamma_Z^2}}
= 0.0021,\qquad
R_Z = \frac{\Gamma_{\ell\bar\ell}/\Gamma_Z^2}{\frac{\aem}{3m_Z} + \frac{\Gamma_{\ell\bar\ell}}{\Gamma_Z^2}}
= 0.9979;
\eeq
substituting the CDF values, the full factor becomes
\beq\label{eq:factor_final}
R_ZN_Z(M^2_{\rm min},M^2_{\rm max}) + R_\gamma N_\gamma(M^2_{\rm min},M^2_{\rm max}) = 3.893 \unitm{GeV},
\eeq
which is a bit larger than that obtained with the simple Breit-Wigner approximation.

\begin{figure}[tb]
\begin{center}
\includegraphics[width=0.9\textwidth,page=4]{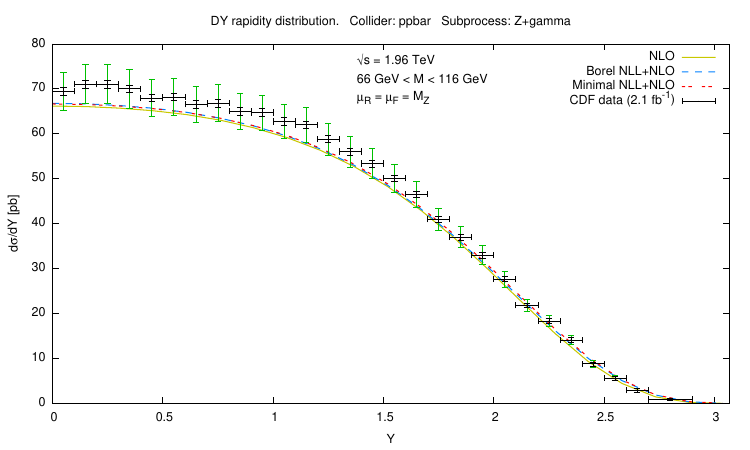}
\caption{Rapidity distribution of $Z$ bosons in $p\bar p$ collisions
at $\sqrt{s}=1.96$~TeV (the contribution of virtual $\gamma$ at the $Z$ peak
is included). Data are taken from~\cite{Aaltonen:2010zza}. The smaller black
uncertainty bands are statistical only, while the wider green bands also
include normalization uncertainties.}
\label{fig:rapTEVZdata}
\end{center}
\end{figure}
In Fig.~\ref{fig:rapTEVZdata} we show the CDF
data~\cite{Aaltonen:2010zza}, corresponding to an integrated luminosity
of $2.1 \text{ fb}^{-1}$, compared to the NLO QCD prediction with
the inclusion of threshold resummation at NLL, using both Borel
and minimal prescriptions.
The comparison shows an excellent agreement in shape between the data
and the theoretical curves; there is clearly a mismatch in
normalization of the total cross-section, which is however consistent
with the sizable $6\%$ normalization uncertainty on the cross-section.
Also in this case, as for the NuSea experiment, this simply
reflects the fact that these data are used in the determination of the
PDFs that we are using.

\begin{figure}[tbp]
\begin{center}
\includegraphics[width=0.9\textwidth,page=3]{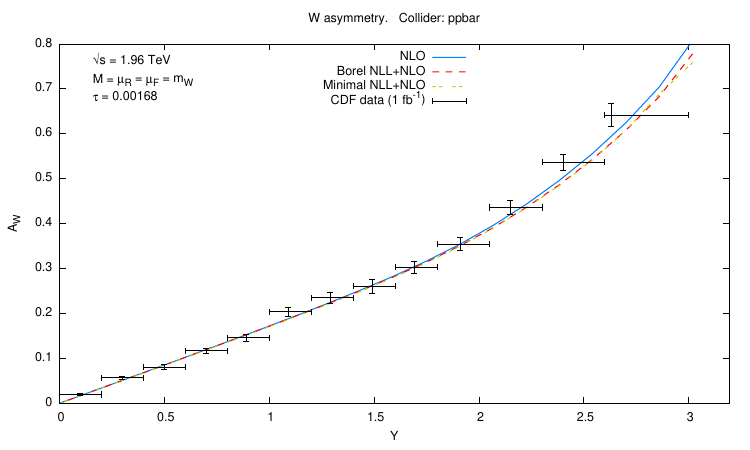}
\caption{$W^\pm$ asymmetry in $p\bar p$ collisions at $\sqrt{s}=1.96$~TeV.
Data are taken from~\cite{Aaltonen:2009ta}.}
\label{fig:rapTEVWdata}
\end{center}
\end{figure}

A similar comparison can be performed for the $W^\pm$ asymmetry, defined as
\beq
A_W(Y)=\frac
{\dfrac{d\sigma_{W^+}}{dY}-\dfrac{d\sigma_{W^-}}{dY}}
{\dfrac{d\sigma_{W^+}}{dY}+\dfrac{d\sigma_{W^-}}{dY}},
\eeq 
also measured by CDF~\cite{Aaltonen:2009ta}. In this case, normalization 
uncertainties cancel in the ratio.
In Fig.~\ref{fig:rapTEVWdata}
we show the measured CDF data~\cite{Aaltonen:2009ta} compared to the
QCD prediction at NLO and resummed NLO+NLL (Borel and minimal
prescriptions).
Clearly, the accuracy of present-day data is insufficient to
appreciate the effect of NNLO or resummation correction, and it is
rather comparable to the difference between LO and NLO predictions,
which can thus be barely appreciated. However, an improvement of
statistical accuracy by an order of magnitude would be sufficient for
NNLO and resummation corrections 
to become significant. The normalization uncertainty 
has a negligible effect on the shape of the distribution
and therefore it does not affect this conclusion

\subsubsection{LHC}
\label{sec:LHC7}

We now consider predictions for Drell-Yan production at the LHC,
with the current energy $\sqrt{s}=7$~TeV.
Prediction for collider energy $\sqrt{s}=14$~TeV can be found in Ref.~\cite{bfr2}.
Very recently, CERN announced the energy upgrade for the 2012 run
to $\sqrt{s}=8$~TeV; predictions for such energy are not yet
available.

\begin{figure}[tb]
\begin{center}
\includegraphics[width=0.6\textwidth]{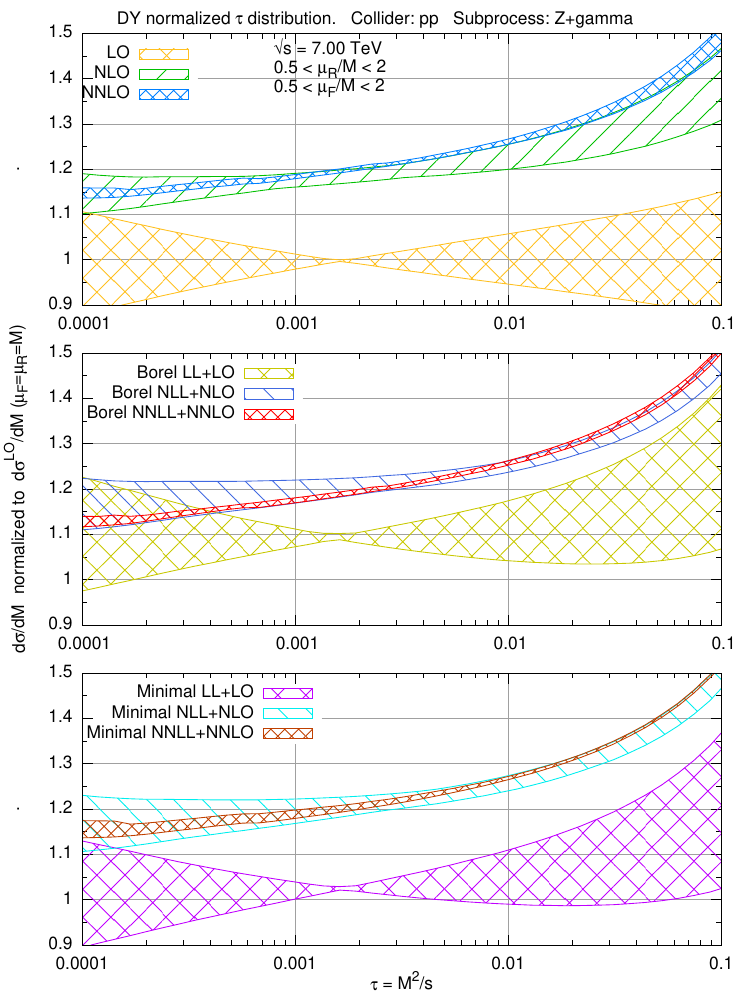}
\caption{Invariant mass distribution of neutral Drell-Yan pairs
in $p p$ collisions at $\sqrt{s}=7$~TeV.}
\label{fig:totLHC7}
\end{center}
\end{figure}
\begin{figure}[tbp]
\begin{center}
\includegraphics[width=0.496\textwidth]{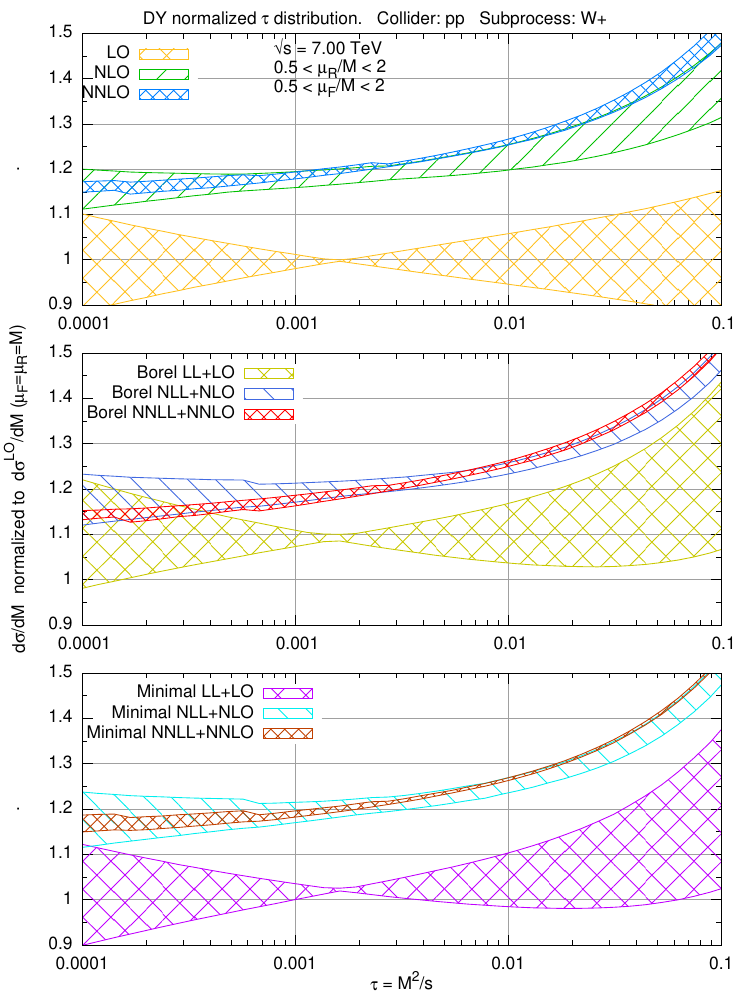}
\includegraphics[width=0.496\textwidth]{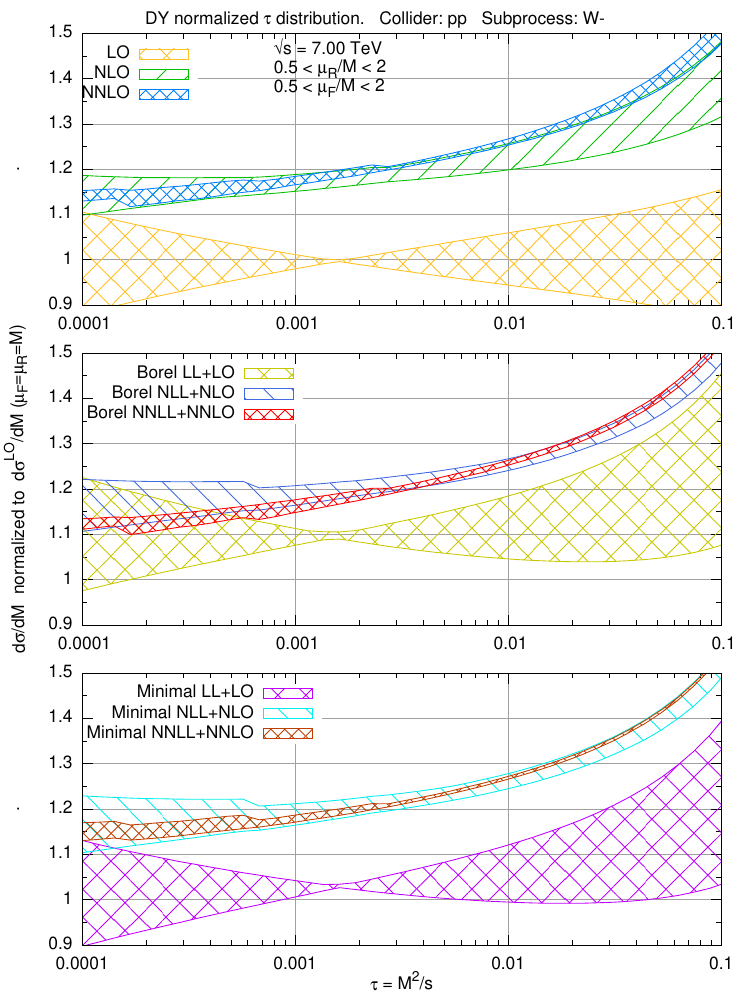}
\caption{Invariant mass distribution of positively (left) and negatively (right)
  charged Drell-Yan pairs in $p p$ collisions at $\sqrt{s}=7$~TeV.}
\label{fig:totLHC7W}
\end{center}
\end{figure}
Invariant mass distributions for both neutral and charged Drell-Yan
pairs are shown in
Figs.~\ref{fig:totLHC7},~\ref{fig:totLHC7W}. While the impact of
fixed-order perturbative corrections is unsurprisingly similar to
that at the Tevatron collider shown in
Fig.~\ref{fig:totTev}, interestingly the reduction
in uncertainty obtained thanks to the resummation is more marked at
the LHC, consistent with the expectation (recall
Sect.~\ref{sec:when_is_relevant}) that the effect of the resummation is
somewhat more significant at a $pp$ than at a $p\bar p$ collider. 
Moreover, the consistency of the NLO error band
with the NNLO prediction is improved by the inclusion of
resummation.
Of course, as in the case of Tevatron, for realistic values of $\tau\lesssim
0.1$ the impact of the resummation is mostly on the uncertainty but
very small or negligible on central values, so the resummation is
behaving as a perturbative correction.

\begin{figure}[htbp]
\begin{center}
\includegraphics[width=0.52\textwidth,page=1]{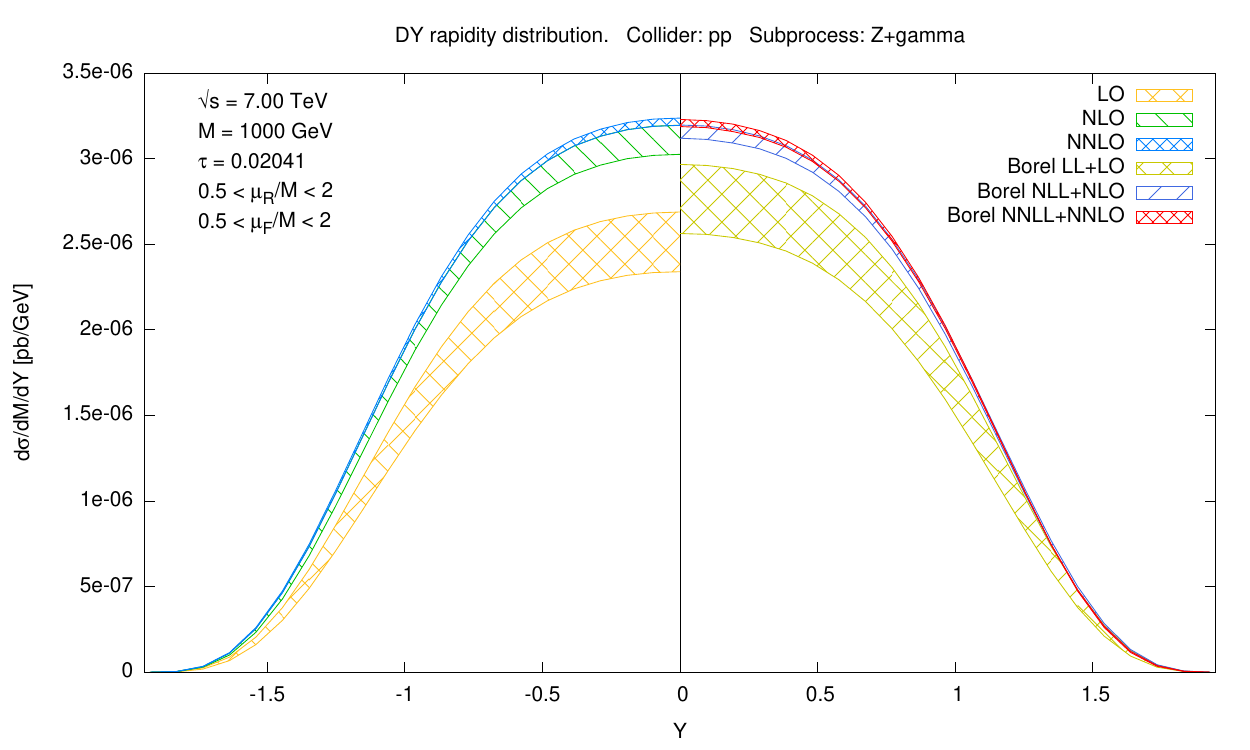}
\\
\includegraphics[width=0.52\textwidth,page=2]{graph_rap_pp_7_1000}
\caption{Rapidity distribution of neutral Drell-Yan pairs of invariant
  mass $M=1$~TeV in $pp$ collisions at $\sqrt{s}=7$~TeV.}
\label{fig:rapLHC7g}
\end{center}
\end{figure}

\begin{figure}[htbp]
\begin{center}
\includegraphics[width=0.52\textwidth,page=1]{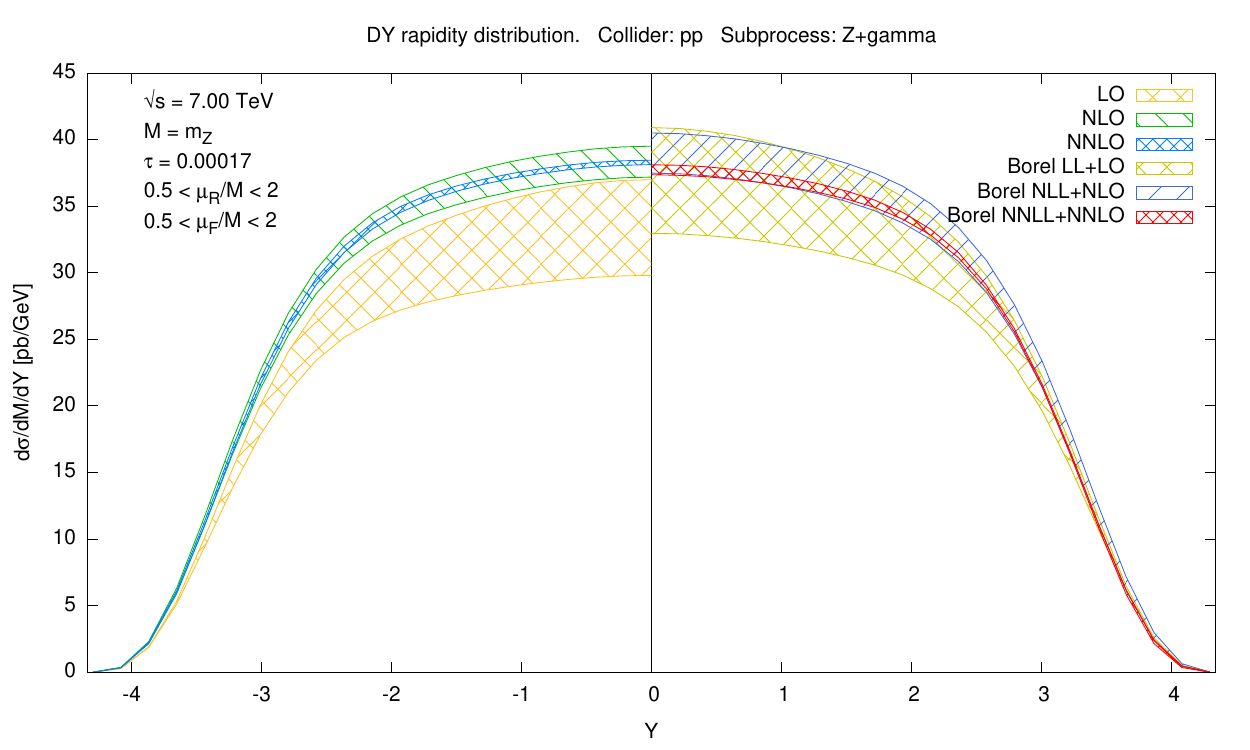}
\\
\includegraphics[width=0.52\textwidth,page=2]{graph_rap_pp_7_Z}
\caption{Rapidity distribution of neutral Drell-Yan pairs of invariant
  mass $M=m_Z$ in $pp$ collisions at $\sqrt{s}=7$~TeV
  (the contribution of virtual $\gamma$ at the $Z$ peak is included).}
\label{fig:rapLHC7Z}
\end{center}
\end{figure}

\begin{figure}[htbp]
\begin{center}
\includegraphics[width=0.495\textwidth,page=1]{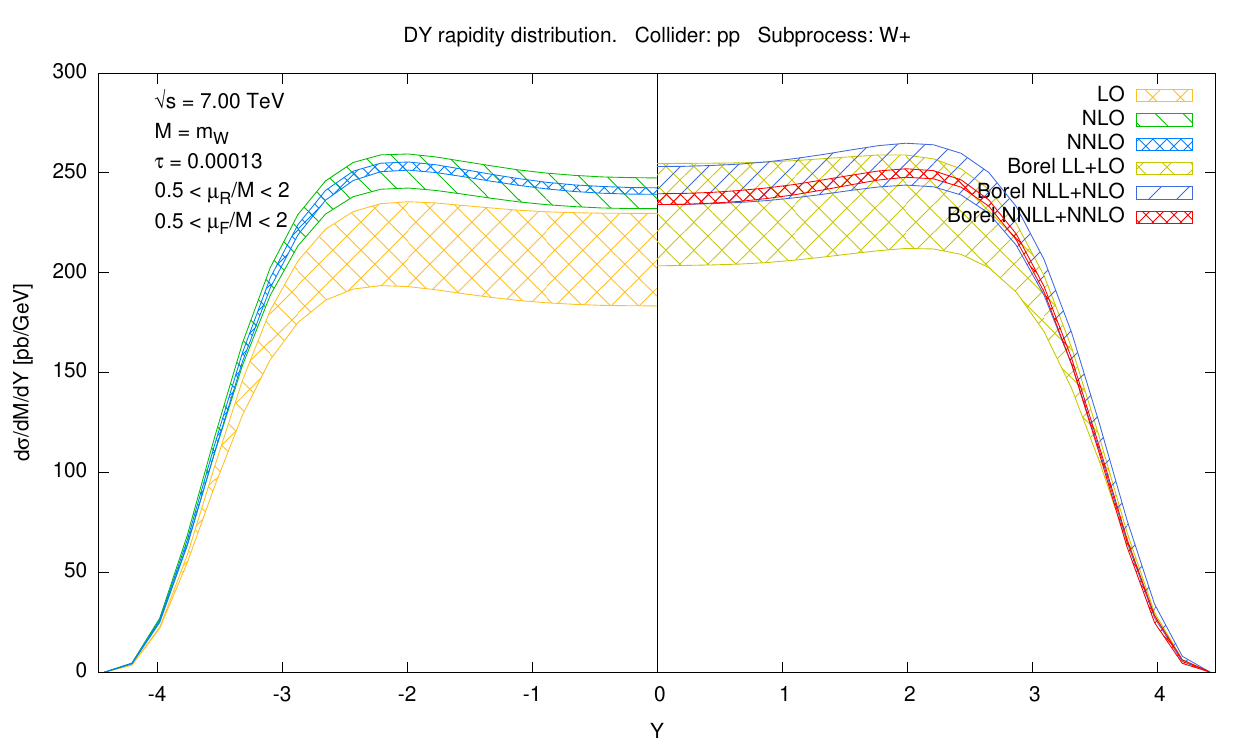}
\includegraphics[width=0.495\textwidth,page=1]{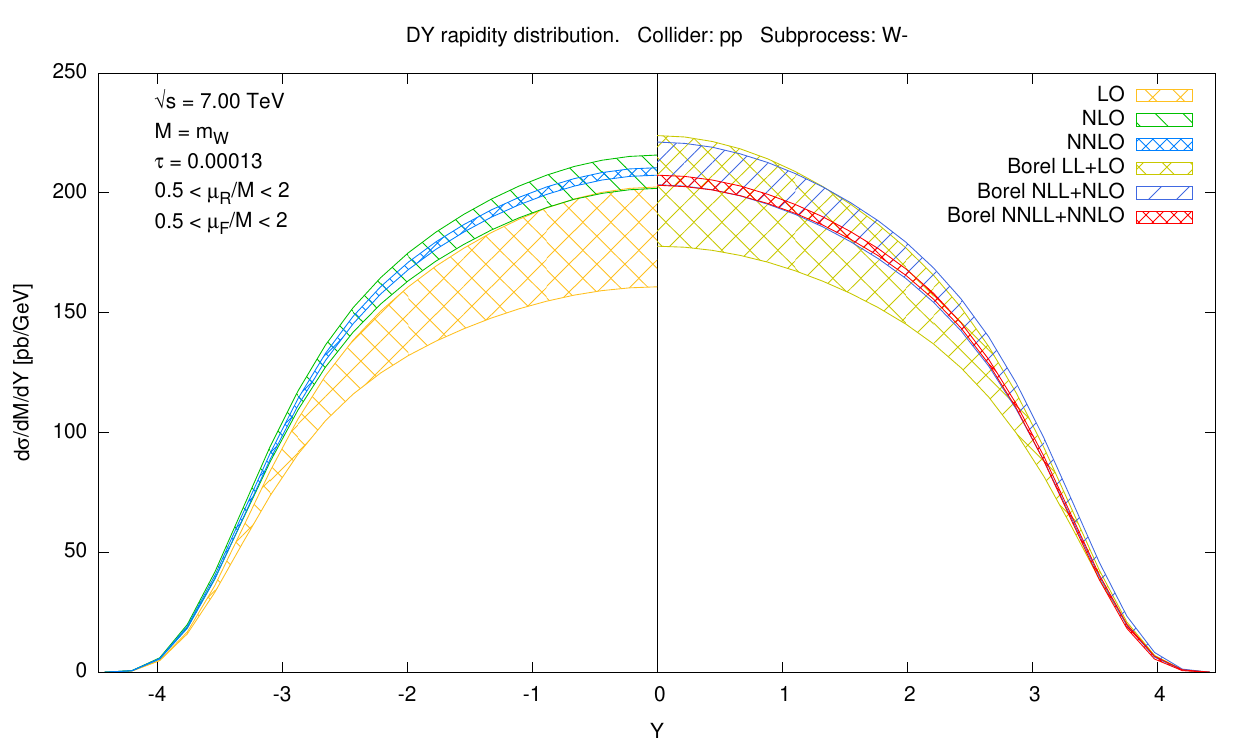}
\\
\includegraphics[width=0.495\textwidth,page=2]{graph_rap_pp_7_Wp}
\includegraphics[width=0.495\textwidth,page=2]{graph_rap_pp_7_Wm}
\caption{Rapidity distribution of positively (left) and negatively (right) charged Drell-Yan pairs of invariant
  mass $M=m_W$ in $pp$ collisions at $\sqrt{s}=7$~TeV.}
\label{fig:rapLHC7W}
\end{center}
\end{figure}

Turning to rapidity distributions, we present results for the following
observables:
\begin{itemize}
\item
neutral Drell-Yan pairs with invariant mass $M=1$~TeV, Fig.~\ref{fig:rapLHC7g};
\item
neutral Drell-Yan pairs with invariant mass $M=m_Z$, Fig.~\ref{fig:rapLHC7Z};
\item
positively charged Drell-Yan pairs with invariant mass $M=m_W$, Fig.~\ref{fig:rapLHC7W};
\item
negatively charged Drell-Yan pair with invariant mass $M=m_W$, Fig.~\ref{fig:rapLHC7W}.
\end{itemize}

The first case corresponds to $\tau\sim 0.02$, comparable to
the case of a final state with $M=200$~GeV at the Tevatron shown in 
Fig.~\ref{fig:raptevg}. As in that case, we clearly see an improvement
in uncertainty (with small resummation ambiguities)
when going to the resummed result, though also in that
case the effect on central value is moderate.
On the other hand, the other cases correspond to very small values of
$\tau$ and indeed in this case the uncertainty on resummed results is
larger than that on unresummed ones, indicating that whatever effect
is induced by the resummation is related to the inclusion of terms
which are not dominant in this region. This is also reflected in a
sizable difference between Borel and minimal results.

\subsection{Comparison of the different prescriptions}
\label{sec:pheno_Borel}

We finally turn to a comparison of the phenomenological impact of
different choices of subleading terms.

\begin{figure}[ht]
\begin{center}
\includegraphics[width=0.8\textwidth,page=1]{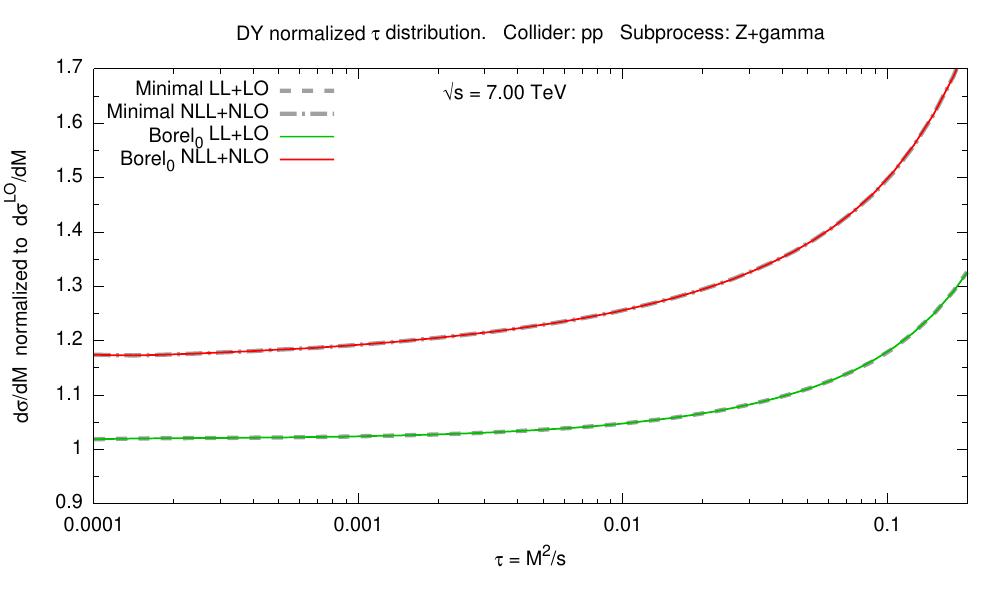}
\caption{Invariant mass distribution of neutral Drell-Yan pairs
  in $pp$ collisions at $\sqrt{s}=7$~TeV for central values of the renormalization and factorization scales.
  The original Borel prescription Eq.~\eqref{eq:BP0_final} is used.}
\label{fig:totBPMPcomp}
\end{center}
\end{figure}

First, we want to prove what we claimed several times in the text, i.e.\
that the ``natural'' Borel prescription which depends on $z$ via $\log\frac1z$,
Eq.~\eqref{eq:Borel_prescription}, gives the same result as the Minimal prescription,
provided $\tau$ is not too large.
Therefore, in Fig.~\ref{fig:totBPMPcomp} we plot the predictions for the resummed $K$-factors as obtained
with the Minimal prescription and with the Borel prescription, Eq.~\eqref{eq:BP0_final}.
We consider as a reference case the production of neutral DY pairs at the
LHC at $7$~TeV; since the NNLL+NNLO curves are very close to the NLL+NLO ones,
we don't show them in order to better read the plots.
The plot shows the impressive similarity between such Borel prescription and the Minimal prescription:
in both the LL and NLL case they are indistinguishable in the whole plotted range.%
\footnote{This plot is, by the way, a powerful check of the goodness and correctness
of the numerical implementation of resummation.}
This plot shows clearly that, for experimentally accessible values of $\tau$,
the difference in the way the Borel and Minimal prescriptions treat the divergence
of the perturbative series is not relevant at all, thereby reassuring who could be worried
either by the non-convolutive nature of the Minimal prescription or
by the higher-twists included in the Borel prescription.
This fact is a consequence of the observation that, for such not to large values of $\tau$,
the perturbative series behaves perturbatively, and the divergence of the series
(and the details of the treatment of it) does not play a role.

\begin{figure}[!t]
\begin{center}
\includegraphics[width=0.75\textwidth,page=1]{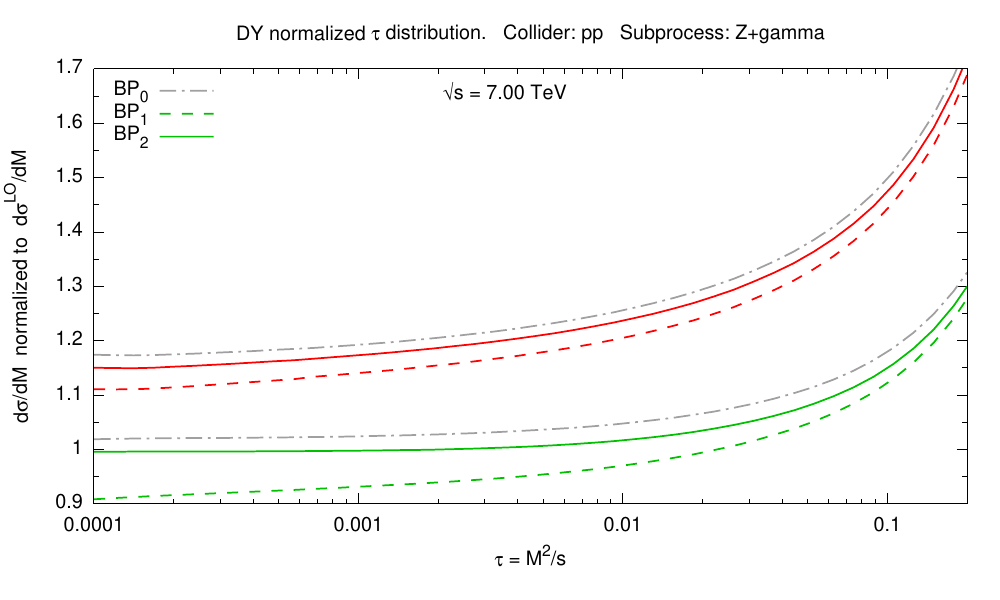}\\
\includegraphics[width=0.75\textwidth,page=1]{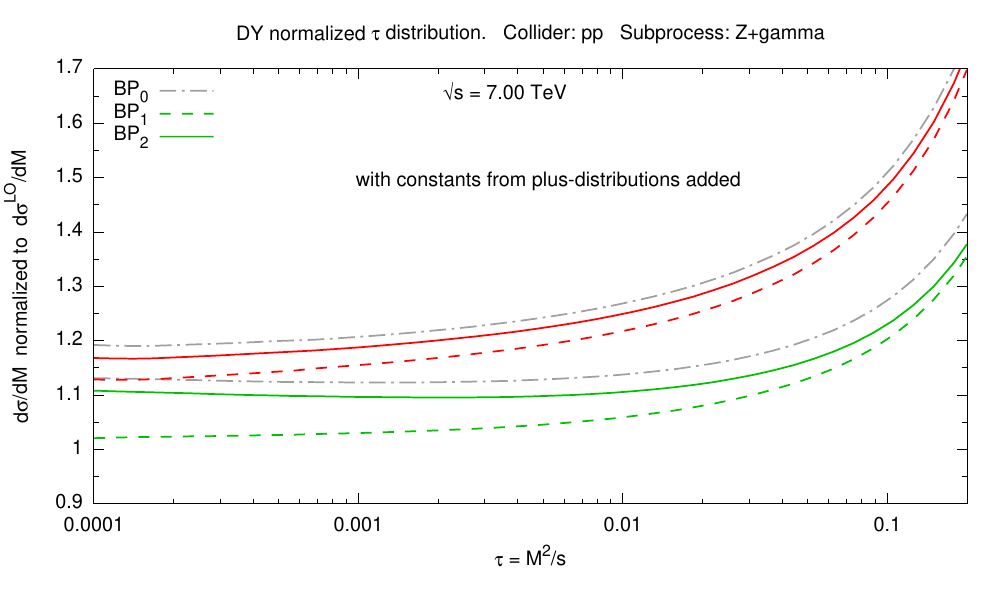}
\caption{Invariant mass distribution of neutral Drell-Yan pairs
  in $pp$ collisions at $\sqrt{s}=7$~TeV for central values of the renormalization and factorization scales.
  Different versions of the Borel prescription are used: BP$_0$ corresponds to Eq.~\eqref{eq:BP0_final},
  BP$_1$ to Eq.~\eqref{eq:BP1_final} and BP$_2$ to Eq.~\eqref{eq:BP2_final}. The upper plot shows
  the results for these prescriptions, while in the lower plot also the constants from the plus-distributions
  are added. The upper curves in the plots correspond to NLL+NLO: same style means same prescription.}
\label{fig:totBPcomp}
\end{center}
\end{figure}

Next, in Fig.~\ref{fig:totBPcomp} we compare the different versions of the
Borel prescription.
The upper plot is a direct comparison of the three choices Eqs.~\eqref{eq:Borel_prescription},
\eqref{eq:BP1} and \eqref{eq:BP2}.
First, we note that at large $\tau$, where the different subleading terms are less important,
all the curves at a given order tend to be very similar each other, as expected.
Moreover, the BP$_0$ and BP$_2$ are rather close each other in the whole range,
while the BP$_1$ is quite different especially at small $\tau$:
this is consistent with the discussion in Sect.~\ref{sec:res_comparison_FO},
and indeed we had decided not to take into account the BP$_1$ for phenomenology.

The lower plot shows the same curves with the constant terms coming
from plus-distributions added: in practice, these prescriptions are obtained by adding
\beq
-\frac1\xi\, \delta(1-z)
\eeq
to the plus-distribution terms in Eqs.~\eqref{eq:Borel_prescription},
\eqref{eq:BP1} and \eqref{eq:BP2} (see also discussion in Sect.~\ref{sec:discussion_constants}).
The BP$_2$ obtained with this modification is the prescription we used in this Chapter
for phenomenology~\cite{bfr2}.
Even if this modification is subleading and somewhat unjustified theoretically,
its effect on the Drell-Yan case is to improve the convergence of the
resummed perturbative expansion.
Indeed, while the NLL curves do not change too much between upper and lower plot,
the LL curves in the lower plot are much closer to the NLL ones.

We conclude noting that, discarding the BP$_1$,
the other prescriptions, differing for the choices of subleading terms produced,
if used as estimators of the uncertainty due to the resummation procedure would
give a rather small uncertainty (from NLL+NLO on), comparable to that from scale variation.

\cleardoublepage
\phantomsection
\addstarredchapter{Conclusions}
\chapter*{Conclusions}
\markboth{Conclusions}{}

We have presented a detailed overview of the current status of
soft-gluon and high-energy resummations in perturbative QCD.
On the theoretical side, we have obtained several results.

First, concerning soft-gluon resummation, we have presented an
alternative prescription, the Borel prescription, to deal with
the problem of the Landau pole of the strong coupling, related
to the divergence of the perturbative series.
The Borel prescription turns out to be very powerful, in which
it gives complete control on the subleading terms introduced by the resummation procedure.
This property can be either used to estimate the uncertainty 
due to such subleading terms, or to find the best choice of these terms,
in such a way that the threshold logarithms could best reproduce the
partonic coefficient even far from the threshold region.

We have then discussed the resummation of high-energy logarithms,
providing several improvements in the whole procedure.
The most important result on this subject is the development
of a code implementing all the machinery of high-energy resummation,
providing for the first time publicly available small-$x$ resummed
splitting functions and coefficient functions.

One of the main achievements is that we have provided a way to estimate the value of
hadronic variable $\tau=M^2/s$ at which resummation of threshold logarithms is expected
to provide an improvement over fixed-order calculations. This result
has been accomplished through a determination of the relevant partonic
center-of-mass energy, whose distance from threshold 
determines the impact of resummation. This
estimate relies on the singularity structure of the anomalous
dimensions which drive $M^2$ evolution of parton distributions in
perturbative QCD. Using this technique, we have shown that
resummation is expected to be relevant down to fairly small values of $\tau$,
very far from hadronic threshold.
For example, at the LHC at 7~TeV the Drell-Yan process is expected to be
influenced by threshold resummation for $\tau\gtrsim0.002$,
corresponding to $M\gtrsim300$~GeV. More interestingly,
the Higgs boson production at LHC at 7~TeV is expected to get contributions
from the threshold region down to values of $\tau$ as small as
$\tau\gtrsim 0.0002$, corresponding to $M\gtrsim100$~GeV.

Being such values of $\tau$ so small, we have investigated the
modifications in our predictions induced by high-energy resummation, 
in particular for the Higgs boson production, being it dominated by
gluon fusion.
By a preliminary analysis, based on an approximate implementation of small-$x$ resummation,
it seems that the threshold region is reduced, in the sense that threshold
logarithms are dominant only for higher masses.
This observation needs to be clarified: this request enforces the need
of a complete inclusion of small-$x$ resummation
in PDF fits for precision phenomenology at hadron colliders.

At the phenomenological level, we have shown how to use different
versions of the Borel resummation prescription and their comparison to
the minimal prescription as a means to assess the ambiguities related
to the resummation, and in particular to the treatment of the
subleading terms. The application of these tools to the Drell-Yan process
at Tevatron and LHC has shown that resummation is relevant for the production of
states of mass as light as $W$ and $Z$ vector bosons at the Tevatron,
and for the production of heavy dileptons of mass in the TeV range at
the LHC: in all these cases threshold resummation leads to a
significant improvement in perturbative accuracy. The impact of
resummation on Tevatron fixed-target rapidity distributions is less
clear, in that, despite being larger, the effect of resummation may be
marred by its ambiguities.

Our general conclusion is that the impact of threshold resummation at
hadron colliders even for significantly small values of $\tau$
is comparable to that of NNLO corrections, and it should thus be included
both in the determination of parton distributions and in precision phenomenology,
though care should be taken in also estimating carefully the ambiguity which is
intrinsic in the resummation procedure.

\appendix
\chapter{QCD running coupling}
\label{chap:QCD_running_coupling}

\minitoc

\noindent
In this Appendix we review some features of the QCD running coupling
and we compute the solutions to its renormalization-group equation.

\section{The running of $\as$}
The QCD coupling $\as$ satisfies the renormalization-group equation
\beq
\mu^2 \frac{d}{d\mu^2}\as(\mu^2) = \beta\(\as(\mu^2)\)
\eeq
where the $\beta$-function in the right-hand-side has the perturbative expansion
\begin{align}
\beta(\as) &= -\as^2\, \( \beta_0 + \beta_1 \as + \beta_2 \as^2 + \ldots\) \\
&= -\beta_0 \as^2\, \( 1 + b_1 \as + b_2 \as^2 + \ldots\)
\end{align}
where $\beta_k = b_k \beta_0$ for $k\geq1$.
The $k$-th coefficient $\beta_k$ can be computed by a $(k+1)$-loop calculation,
and so far are known up to $4$ loops \cite{vanRitbergen:1997va,Czakon:2004bu}.
Here, we present the first three, which are those that will be used for resummation:
\begin{align}
  \beta_0 &= \frac{11 C_A - 4\,T_F\, n_f}{12\pi} = \frac{33-2n_f}{12\pi}\label{eq:beta0}\\
  \beta_1 &= \frac{17\,C_A^2-(10\,C_A+6\,C_F)\,T_F\,n_f}{24\pi^2}
  = \frac{153-19\,n_f}{24\pi^2}\\
  \beta_2 &= \frac{1}{(4\pi)^3} \bigg[
  \frac{2857}{54}\,C_A^3 + \left( 2\, C_F^2 - \frac{205}{9}\,C_F\, C_A
    - \frac{1415}{27}\,C_A^2 \right) T_F\, n_f\nonumber\\
  &\qquad\qquad + \left( \frac{44}{9}\,C_F + \frac{158}{27}\,C_A \right) T_F^2\, n_f^2 \bigg]
  \nonumber\\
  & = \frac{1}{128 \pi^3}\left[ 2857-\frac{5033}{9}\,n_f 
    +\frac{325}{27}\,n_f^2 \right]
\end{align}
where in the second equality we have substituted the numerical values
\beq
C_A=N_c=3, \qquad
C_F = \frac{N_c^2-1}{2N_c}=\frac43, \qquad
T_F=\frac12,
\eeq
which are appropriate for the gauge group ${\rm SU}(N_c)$ with $N_c=3$ of QCD.

As already discussed in Chap.~\ref{chap:parton_model}, as long as $n_f<17$
(and so far we know $6$ flavours) $\beta_0$ is positive, and because of
the minus sign in front of the perturbative expansion of the $\beta$-function
the solution of the renormalization-group equation is decreasing as $\mu^2$ increases.
We now show explicit solutions where this fact, known as asymptotic freedom, is manifest.

\subsection{Solutions of the renormalization-group equation}

The renormalization-group equation can be easily solved, because the
$\mu^2$ dependence appear only as argument of $\as$, and in general we have then
\beq
\int\frac{d\mu^2}{\mu^2} \,F\(\as(\mu^2)\) = \int\frac{d\as}{\beta(\as)}\,F(\as)
\eeq
with the appropriate integration limits.
At one loop, the only coefficient of the $\beta$-function is $\beta_0$ and the solution is
\begin{subequations}\label{eq:alpha_running_LO}
\begin{align}
\as(\mu^2) &= \frac{\aq}{1+\beta_0 \aq \log\frac{\mu^2}{Q^2}}\label{eq:alpha_running_LO_1}\\
&= \frac{1}{\beta_0\log\frac{\mu^2}{\Lambda^2}},\label{eq:alpha_running_LO_lambda}
\end{align}
\end{subequations}
where in the first line we have used an initial condition for $\as$
at the scale $Q^2$, while in the second one the initial condition is
chosen at the scale $\Lambda^2$ at which $\as$ goes to infinity (Landau pole).
At this order, then, the Landau pole $\mu_L$ can be written as
\beq\label{eq:asLandauPole}
\mu_L^2 = \Lambda^2 = \mu^2\,\exp\(-\frac1{\beta_0\as(\mu^2)}\)
\eeq
valid for every choice of $\mu^2$, then in particular for $\mu^2=Q^2$,
making the connection between the two solutions Eqs.~\eqref{eq:alpha_running_LO}.

Note that the first form of the solution, Eq.~\eqref{eq:alpha_running_LO_1},
shows explicitly that the renormalization group equation resums all powers of
$\log\frac{\mu^2}{Q^2}$.
In particular, we may write an expansion of $\as(\mu^2)$ in powers of $\as(Q^2)$
at $\as(Q^2)\log\frac{\mu^2}{Q^2}$ fixed,
\beq
\label{eq:as_logexpansion}
\as(\mu^2)=\alpha_0\, f_1(\alpha_0\ell)+\alpha_0^2\,f_2(\alpha_0\ell)+\ldots,
\eeq
where we have defined for simplicity $\alpha_0=\as(Q^2)$ and $\ell = \log\frac{\mu^2}{Q^2}$.
The functions $f_1, f_2,\ldots$ contain all powers of their argument; therefore,
such expansion is valid provided only $\as(Q^2)\ll 1$, regardless the value of $\ell$.
The functions $f_j$ can be found for all $j=1,\ldots,k$ solving the renormalization-group
equation with a $k$-loop $\beta$-function, i.e.\ including all the coefficients
$\beta_i$ in the range $0\leq i\leq k-1$.

The $1$-loop solution indeed predicts $f_1$:
\beq
f_1(\alpha_0\ell)=\frac1{1+\beta_0\alpha_0\ell}.
\eeq
At two loops, we must include $\beta_1=b_1\beta_0$ and we find
\beq
\[\frac{1}{\beta_0 \as} -\frac{b_1}{\beta_0}\log\( \frac{1}{\as} +b_1 \)\]_{\as(Q^2)}^{\as(\mu^2)} = \log\frac{\mu^2}{Q^2}.
\eeq
Solving for $\as(\mu^2)$ and expanding in order to take terms up to the second order
in the expansion \eqref{eq:as_logexpansion} we then obtain
\beq\label{eq:alpha_running_NLO}
\as(\mu^2) = \frac{\aq}{X} \left[ 1-\frac{b_1 \aq}{X} \log X \right]
\eeq
where $X=1+\beta_0 \aq \log\frac{\mu^2}{Q^2}$.
With the same technique, it is easy to find also the three loops solution, which reads
\begin{multline}
\as(\mu^2) = \frac{\aq}{X} -b_1\frac{\as^2(Q^2)}{X^2} \log X\\
+ \frac{\as^3(Q^2)}{X^3} \Big[ b_1^2(\log^2X-\log X-1+X) + b_2(1-X) \Big].
\label{eq:alpha_running_NNLO}
\end{multline}
Note that the Landau pole is still present in the two- and three-loops solutions,
and it is always given by Eq.~\eqref{eq:asLandauPole}.
This is indeed a general feature: the Landau pole is always present in any
perturbative solution to the running coupling equation.

Typically, the initial condition for $\as$ is given at the $Z$ mass $Q^2=m_Z^2$,
because many measurements were made at LEP and more recently at colliders with high precision.
Then, at the $Z$ mass
\beq
m_Z = 91.18\unitm{GeV},
\eeq
the current better estimate of $\as$ is \cite{PDG,Bethke:2009jm}
\beq
\as(m_Z^2) = 0.1184\pm 0.0007.
\eeq
From this value we could compute the Landau pole; however, we have to choose
the value of $n_f$, i.e.\ the number of flavours, which appears in the coefficient $\beta_0$.
With the two extreme choices $n_f=0$ and $n_f=6$ we get, respectively,
\beq
\mu_L(n_f=0) \sim 730\unitm{MeV}, \qquad
\mu_L(n_f=6) \sim 47\unitm{MeV}.
\eeq
These two results are quite different each other, but nevertheless they set the order of magnitude
of the non-perturbative scale of QCD, which is of the order of some hundreds MeV.
The proper choice for the value of $n_f$ is a matter of renormalization scheme choice.
We are going to discuss the most common scheme in the next Section.

\subsection{Quark masses thresholds}

\begin{figure}[tbh]
  \centering
  \includegraphics[width=0.9\textwidth]{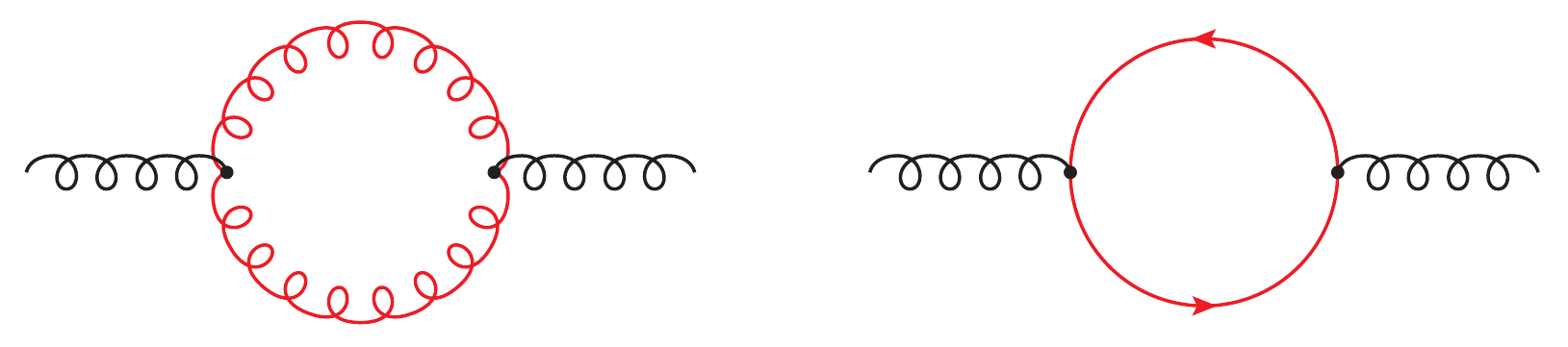}
  \caption{Diagrams contributing to the $1$-loop $\beta$-function.}
  \label{fig:1loop_beta_diag}
\end{figure}

The number of flavours $n_f$ is generically not kept fixed.
What is usually adopted is the so called \emph{variable flavour number scheme},%
\footnote{More precisely, this is the \emph{zero-mass} variable flavour number scheme;
there are several extensions of this scheme, but their discussion is beyond the purpose of this Appendix.}
a renormalization scheme in which the number of \emph{active} flavours 
depends on the energy scale $\mu$.
This choice is motivated by the computation of the $\beta$ function:
the $1$-loop coefficient $\beta_0$ is computed by the diagrams of Fig.~\ref{fig:1loop_beta_diag},
where in the second diagram the fermions running in the loop are the quarks.
If they were real, there would be a threshold for the diagram given by
\beq
\mu > 2m_q,
\eeq
being $m_q$ the mass of the quark.
Then, a natural choice would be to use as number of active flavours the numbers of quarks
which have a mass less than half of the current energy $\mu$.
Since however such procedure is somehow arbitrary (quarks in the loop are virtual
and the threshold is not really there) a simpler choice is to use just
the mass, and not twice the mass, of the quarks to set the number of active flavours.

\begin{figure}[tbh]
  \centering
  \includegraphics[width=0.9\textwidth]{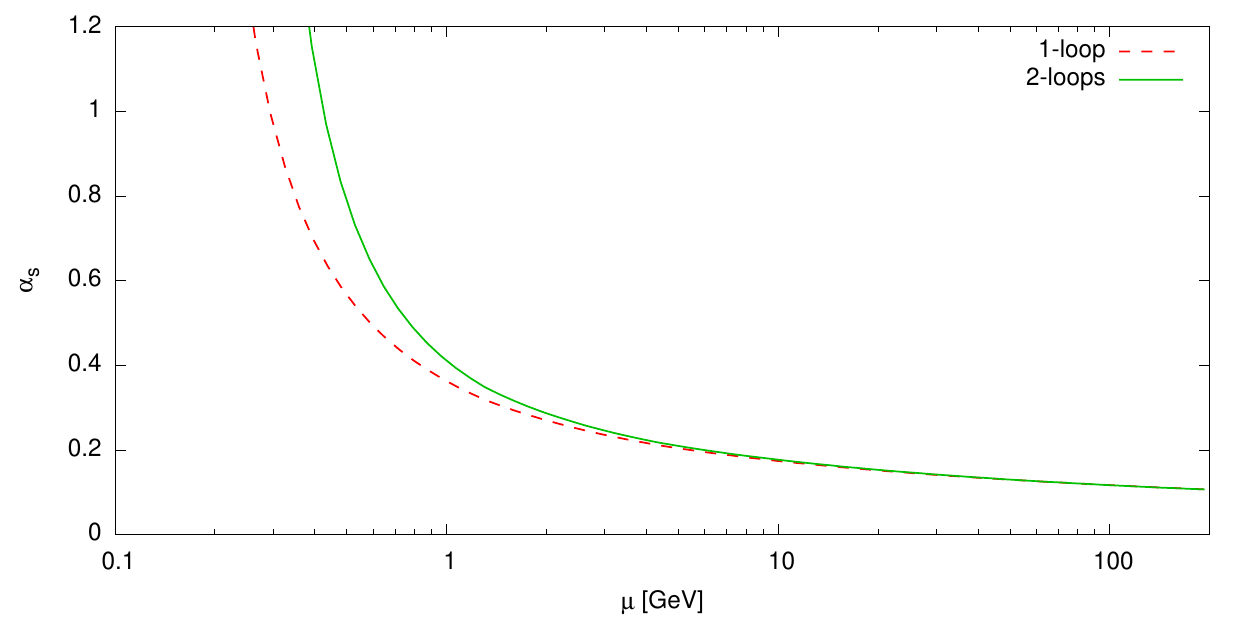}
  \caption{The running coupling $\as$ as a function of $\mu$ at $1$- and $2$-loops.}
  \label{fig:running_coupling}
\end{figure}

Then, starting as usual from $\as(m_Z^2)$ with $n_f=5$ (only the top quark is heavier than the $Z$),
one can build $\as(\mu^2)$ with $n_f$ changing with $\mu^2$, by requiring continuity
of the solution and then using as initial condition for any new branch
the value of $\as$ at the quark thresholds.
The resulting solution has a discontinuous derivative: one could imagine to
perform a smearing to eliminate this problem, but this goes beyond the purpose of
this Appendix.
In Fig.~\ref{fig:running_coupling} we show explicitly the $1$- and $2$-loops solutions;
the Landau pole at low energies and the asymptotic freedom at large energies are both
evident.

As a final comment, we want to show where the Landau pole is located using this
variable flavour number scheme. We first note that, because of the patching of $\as(\mu^2)$
at quark thresholds, the position of the Landau pole now will be different
for different orders of the solution. We then take as representative examples
the $1$- and $2$-loops solution Eq.~\eqref{eq:alpha_running_NLO} and find
\beq
\mu_L^{1\text{-loop}} \sim 150 \unitm{MeV}
,\qquad
\mu_L^{2\text{-loop}} \sim 180 \unitm{MeV}
,
\eeq
where we have used the quarks masses given in Sect.~\ref{sec:QCD_basics}.
Note that the two predictions are of the same order, even if they
are in the non-perturbative region of QCD where the running is no longer under control.
From Fig.~\ref{fig:running_coupling} it is clear that indeed
perturbation theory in QCD breaks down at a scale around $1$~GeV,
where the coupling starts growing fast and the two curves become very different.

\chapter{Mellin transformation}
\label{chap:Mellin}

\minitoc

\noindent
In this Appendix we collect some useful definitions and results
concerning Mellin transformations. In particular, we concentrate our attention
to the case of logarithms, which are the main subject of this thesis.

\section{Laplace transform}

The Mellin transform is a particular case of a Laplace transform.
Hence we first start considering some definitions and results about Laplace transforms,
which remains also in the case of Mellin transforms.

The (unilateral) Laplace transform of a function $f(t)$ is defined as
\beq\label{eq:Laplace}
\tilde f(s) \equiv \int_0^\infty dt\, e^{-st} \, f(t).
\eeq
If the function $\tilde f(s)$ exists, then it is regular for $\Re s>k$ for some value of $k$ depending on $f(t)$.
This is because, in order for the transform to exist, the function $f(t)$ can grow at most as $e^{kt}$
as $t\to\infty$, and then the integral is convergent for $\Re s>c$.

The inverse is given by ($c>k$)
\beq\label{eq:Laplace_inverse}
f(t) = \frac{1}{2\pi i} \int_{c-i\infty}^{c+i\infty} ds\, e^{st} \, \tilde f(s).
\eeq
The proof is easy:
\begin{align}
\frac{1}{2\pi i} \int_{c'-i\infty}^{c'+i\infty} ds\, e^{st} \, \tilde f(s)
&= \int_0^\infty dt'\, f(t') \, \frac{1}{2\pi i} \int_{c-i\infty}^{c+i\infty} ds\; e^{s(t-t')} \nonumber\\
&= \int_0^\infty dt'\, f(t') \, \delta(t-t') \nonumber\\
&=
\begin{cases}
f(t) & t\geq 0\\
0 & t<0.
\end{cases}
\end{align}
Note that, if we had taken the bilateral transform, i.e.\ with lower limit $-\infty$,
the inverse transform Eq.~\eqref{eq:Laplace_inverse} would have reproduced $f(t)$
in the whole range $-\infty<t<\infty$.
Hence, the unilateral transform is suitable for functions defined only for $t>0$,
while the bilateral for functions defined everywhere.

If $f(t)$ is real, then $\tilde f(s)$ is a real function, i.e.\ it satisfies
\beq\label{eq:Laplace_real}
\tilde f(s^*) = \tilde f(s)^*
\eeq
where ${}^*$ indicates complex conjugation, as one can immediately verify from the
definition Eq.~\eqref{eq:Laplace}.

\section{The Mellin transform}

When a function $f(x)$ is defined in the range $0<x<1$
(as many of the functions in this thesis) a Laplace transform can be taken
by using $x=e^{-t}$. The resulting transform,
\beq
\tilde f(N) \equiv \Mell\[f\](N) \equiv \int_0^1 dx\, x^{N-1} \, f(x),
\eeq
is called \emph{Mellin transform}.
Even if we changed name, $N$ is the same variable as $s$, and then the inverse,
according to Eq.~\eqref{eq:Laplace_inverse}, is
\beq\label{eq:Mellin_inverse}
f(x) \equiv \Mell^{-1}\[\tilde f\](x) = \frac{1}{2\pi i} \int_{c-i\infty}^{c+i\infty} dN\, x^{-N} \, \tilde f(N),
\eeq
where $c$ must be greater than the real part of the rightmost singularity
(which must exist, because again $\tilde f(N)$ has a convergence abscissa).
The integration contour can be deformed at will, provided it does not cross
any singularity --- all the singularities must be at the left of the contour.
Being $x<1$, a typical deformation consists in giving a phase to the upper and lower
parts of the integration path in such a way that the real part of $N$ be negative
and increasing as its absolute value goes to infinity.
In particular, when $\tilde f(N)$ is a real function (see definition in Eq.~\eqref{eq:Laplace_real}),
we can perform the following manipulations
\begin{align}
  f(x) &= \frac{1}{2\pi i} \int_{c-i\infty}^{c+i\infty} dN\; x^{-N} \tilde f(N) \nonumber\\
  &= \frac{1}{\pi}\; \Im \int_c^{c+i\infty} dN\; x^{-N} \tilde f(N) \nonumber\\
  &= \frac{1}{\pi} \int_0^{+\infty} dt \; \Im\[ (i-\epsilon)\, x^{-c-(i-\epsilon)t}\, \tilde f(c+(i-\epsilon)t)\]\nonumber\\
  &= \frac{x^{-c}}{\pi} \int_0^{+\infty} du \; \Im\[ \frac{\epsilon-i}{\log x}\, e^{(i-\epsilon)u}\, \tilde f\(c-(i-\epsilon)\frac{u}{\log x}\)\]
  \label{eq:inverse_Mellin_straight}
\end{align}
where in the last two steps we have deformed the integration contour
according to $N=c+(i-\epsilon) t$, $\epsilon>0$, to guarantee numerical convergence.

\subsection{Fixed Talbot Algorithm}
\label{sec:fixed-talbot-algorithm}

An alternative and numerically powerful way to compute the inverse Mellin transform
is to use the Talbot contour~\cite{talbot}, where
\beq
N = N(\theta) = r\theta\(\cot\theta+i\)
\eeq
with $r$ a free parameter.
Then the inverse Mellin integral becomes
\beq
f(x) = \frac{1}{2\pi i} \int_{-\pi}^\pi d\theta \, x^{-N(\theta)}\, \tilde f(N(\theta))\,N'(\theta)
\eeq
or, using the reality of $\tilde f$,
\beq
f(x) = \frac{r}{\pi} \int_0^\pi d\theta \, \Re\!\[ x^{-N(\theta)}\, \tilde f(N(\theta))\,[1+i\sigma(\theta)]\]
\eeq
where $N'(\theta) = i r [1+i\sigma(\theta)]$ and 
\beq
\sigma(\theta) = \theta + \(\theta\cot\theta-1\) \cot\theta.
\eeq
For numerical applications, the integral can be computed using
a trapezoidal rule using the points $\theta_k = k \pi/K$ with $K$ a constant:
the resulting expression
\beq
f(x)\simeq \frac rK \[\frac12 x^{-r} \tilde f(r) + \sum_{k=1}^K \Re\!\(x^{-N(\theta_k)}\tilde f(N(\theta_k))\,[1+i\sigma(\theta_k)]\) \]
\eeq
is known as the \emph{fixed Talbot algorithm}~\cite{talbot}.
In Ref.~\cite{talbot} an optimal choice for $r$ is suggested:
\beq
r = \frac{2K}{5\log\frac1x};
\eeq
then, the relative precision is approximately $10^{-0.6 K}$.

\subsection{Convolution}

Let us define the convolution product $\otimes$
\begin{align}
(f \otimes g)(x)
&= \int_x^1 \frac{dy}{y}\, f(y)\, g\(\frac{x}{y}\).\label{eq:conv2}\\
&= \int_0^1 dy \int_0^1 dz \, f(y)\, g(z) \, \delta(x-yz) \label{eq:conv1}
\end{align}
In the second form, it is clear that it is commutative
\beq
f\otimes g = g\otimes f
\eeq
and it can be easily extended to many functions
\beq
(f_1\otimes\ldots \otimes f_n)(x) =
\int_0^1 dy_1 \cdots \int_0^1 dy_n \, f_1(y_1)\cdots f_n(y_n) \, \delta(x-y_1\cdots y_n).
\eeq
Under Mellin transformation, the convolution product diagonalizes:
\begin{align}
  \Mell\[f\otimes g\](N) &= \int_0^1 dx\, x^{N-1} \int_0^1 dy \int_0^1 dz \, f(y)\, g(z) \, \delta(x-yz)\nonumber\\
  &= \int_0^1 dy\,y^{N-1}\,f(y) \int_0^1 dz \,z^{N-1}\,  g(z) = \tilde f(N) \, \tilde g(N)
\end{align}

\section{Plus distribution}
\label{sec:plus_distribution}

The plus distribution arises form the cancellation of soft divergences,
and is defined as
\beq\label{eq:plus_distr_def}
\int_0^1 dz \, \plusq{f(z)} g(z) = \int_0^1 dz \, f(z) \left[ g(z) - g(1) \right] \;.
\eeq
From the definition it follows that, formally,
\beq\label{eq:plus_distr_form}
\plusq{f(z)} = f(z) -\delta(1-z) \int_0^1 dz \, f(z) \;,
\eeq
but, if $f(z)$ diverges as $z\to1$, this expression makes sense only in a regularized form:
\beq
\plusq{f(z)} = \lim_{\eta\to0^+} \left[ \thetaH(1-\eta-z) \, f(z) - \delta(1-z) \int_0^{1-\eta} dz \, f(z) \right] \;,
\eeq
where the limit is intended to be performed \emph{after} the integration over
the test function $g(z)$.

The plus distribution defined in Eq.~\eqref{eq:plus_distr_def} regularizes
functions which diverge as $z\to1$ at most as
\beq
(1-y)^{-\alpha}, \qquad \alpha<2
\eeq
in the sense that the integral \eqref{eq:plus_distr_def} over any test function
$g(z)$ is finite.
In particular, the usual logarithms
\beq
\frac{\log^k(1-z)}{1-z}
\eeq
are properly regularized.

Some useful identities can be derived directly from the definition;
if $g(z)$ is a regular function as $z\to1$, then
\begin{align}
\plusq{g(z)\,f(z)} &= g(z)\,\plusq{f(z)} - \delta(1-z) \int_0^1 dy \; g(y)\,\plusq{f(y)}\label{eq:plus_useful1} \\
g(z)\,\plusq{f(z)} &= g(1)\,\plusq{f(z)} + \[g(z)-g(1)\] f(z) \;.\label{eq:plus_useful2}
\end{align}
In the last term of second line, the plus distribution is no longer necessary since
$g(z)-g(1)$ regularizes $f(z)$.
One interesting consequence is the following relation
\beq\label{eq:logz_zeta2}
\plus{\frac{\log z}{1-z}} = \frac{\log z}{1-z} + \zeta_2\, \delta(1-z)
\eeq
where we have noted that $\log 1=0$ and we have used
\begin{align}
  \int_0^1 dy \,\frac{\log z}{1-z} &= \lim_{b\to0}\frac{d}{da}\left.\int_0^1dy\; z^{a} (1-z)^{b-1}\right|_{a=0}\nonumber\\
  &= \lim_{b\to0} \left.\left[ \psi_0(1+a) - \psi_0(1+a+b)\right] \frac{\Gamma(1+a) \Gamma(b)}{\Gamma(1+a+b)}\right|_{a=0}\nonumber\\
  &= \lim_{b\to0} \frac{ \psi_0(1) - \psi_0(1+b)}{b}\nonumber\\
  &= -\psi_1(1) = -\zeta_2
\end{align}
(for details about $\Gamma$ and $\psi$ functions, see App.~\ref{sec:Gamma}).

\subsection{Convolution}

Now we want to investigate how a convolution appear when extended to distributions.
This is not trivial, because the definition of the plus distribution Eq.~\eqref{eq:plus_distr_def}
involves an integral from $0$ to $1$, while a convolution written as Eq.~\eqref{eq:conv2}
is an integral from $x$ to $1$.
The obvious way to extend the definition is to use instead Eq.~\eqref{eq:conv1}:
\begin{align}
(\plusq{f} \otimes g)(x) &= \int_0^1 dy \int_0^1 dz \, \plusq{f(y)}\, g(z) \, \delta(x-yz) \nonumber\\
&= \int_0^1 dy \int_0^1 dz \, f(y)\, g(z)  \left[ \delta(x-yz) -\delta(x-z) \right] \nonumber\\
&= \int_0^1 dy \, f(y) \left[ \int_0^1 dz \, g(z) \, \delta(x-yz) -g(x) \right] \nonumber\\
&= \int_x^1 dy\, f(y) \left[ \frac{1}{y}\,g\(\frac{x}{y}\) - g(x) \right] - g(x) \int_0^x dy\, f(y).
\end{align}
The last form could be grouped in the more natural way
\beq
\int_x^1 \frac{dy}{y}\, f(y)\, g\(\frac{x}{y}\) - g(x) \int_0^1 dy\, f(y),
\eeq
as one would have immediately obtained using Eq.~\eqref{eq:plus_distr_form},
but in this form both integrals are not convergent, and can be interpreted
only as a limit
\beq
\lim_{\eta\to0^+} \left[ \int_x^{1-\eta} \frac{dy}{y}\, f(y)\, g\(\frac{x}{y}\) - g(x) \int_0^{1-\eta} dy\, f(y) \right].
\eeq
It must be noticed that the $0$ to $x$ region contributes to the result
of the convolution: this is not surprising, because it was already intrinsically
included in the definition of the plus distribution, but it can be relevant for
applications.

\subsection{Mellin transform of plus distributions}

From the definition, the Mellin transform of a distribution is
\beq
\Mell\[\plus{f}\](N) = \int_0^1 dz\,\(z^{N-1}-1\) f(z).
\eeq
The behaviour in $N$ space of a distribution is qualitatively different
from that of an ordinary function. To see this, we prove first the following
\begin{theorem}\label{th:bonvini1}
If $f(z)$ is real, then $\abs{\Mell\[f\](N)}$ for real $N>0$ is bounded by a decreasing function
which tends to $0$ as $N\to\infty$.
\end{theorem}
\proof First, we see that $\abs{\Mell\[f\](N)}$ is bounded by
\beq
\abs{\Mell\[f\](N)} \leq \int_0^1 dz \,\abs{z^{N-1}\, f(z)} = \Mell\[\abs{f}\](N)
\eeq
where in the last equality we used the fact that $z^{N-1}$ is positive for $N$ real.
The rest of the proof consists in showing that $\Mell\[\abs{f}\](N)$ is a decreasing function:
\beq
\frac{d}{dN} \Mell\[\abs{f}\](N)
= - \int_0^1 dz\,\log\frac{1}{z}\, z^{N-1}\,\abs{f(z)}
\leq 0
\eeq
because $\log\frac1z$ is positive in the integration range.
The last part of the theorem is proved by noting that, in the $N\to\infty$ limit,
the integrand tends to $0$ almost everywhere: by using the monotone convergence theorem
we have the hypothesis.

The proof shows in particular that if $f(z)$ is positive in the whole integration range
then the Mellin transform monotonically decreases.
This is not true for distributions. For example, the Mellin transform of $\delta(1-z)$
is $1$, a constant for all values of $N$, and in particular it does not go to $0$
as $N$ gets large.
For the interesting case of plus distributions, we can prove the following
\begin{theorem}\label{th:bonvini2}
  For real $N>1$, if $f(z)$ is real and singular in $z=1$ then $\abs{\Mell\[\plus{f}\](N)}$
  is bounded by an increasing function of $N$ and it diverges as $N\to\infty$.
\end{theorem}
\proof As before, we note first that
\beq
\abs{\Mell\[\plus{f}\](N)} \leq \int_0^1 dz \,\abs{\(z^{N-1}-1\) f(z)}
\overset{N>1}{=} -\Mell\[\plus{\abs{f}}\](N)
\eeq
where now the minus sign comes form the fact that $z^{N-1}-1<0$ for $N>1$.
Then, $\abs{\Mell\[\plus{f}\](N)}$ is bounded by $-\Mell\[\plus{\abs{f}}\](N)$;
note that, in this case, the Mellin transform is positive is $f(z)<0$.
The derivative of the bounding function is
\beq
-\frac{d}{dN} \Mell\[\plus{\abs{f}}\](N)
= \int_0^1 dz\,\log\frac{1}{z}\, z^{N-1}\,\abs{f(z)}
\geq 0
\eeq
and hence it is an increasing function.
Using again the monotone convergence theorem, we can pass the limit $N\to\infty$
under the integral and see that the result is
\beq
\abs{\Mell\[\plus{f}\](N)} \overset{N\to\infty}{\to} \abs{\int_0^1 dz\,f(z)}
\eeq
which is divergent.

Note that, since the monotonically increase applies only to the bound,
this theorem does not limit the Mellin transform from being small
for a wide range in $N$; however, since
$\abs{\Mell\[\plus{f}\](N)}$ diverges in the limit $N\to\infty$,
it must start increasing at some point.
Anyway, for simple functions which have the same sign in $0<z<1$, the bounds are
saturated and the monotonically increase applies.

Then, we can conclude that the difference between the Mellin transform
of a function and that of a plus distribution of a $z=1$ singular function
is that the first is limited and goes to zero at lager $N$, the second instead
at some point increases and goes to infinity at large $N$.

\section{Mellin transformation of logarithms}
\label{sec:Mellin_log}

In this Section we compute explicitly the Mellin transforms of
logarithms, which appear in (or are related to) perturbative computations of coefficients functions.
To begin with, we consider the following distributions:
\begin{subequations}\label{eq:logarithms}
\beq
\plus{\frac{\log^k(1-z)}{1-z}},
\qquad
\plus{\frac{\log^k\log\frac{1}{z}}{\log\frac{1}{z}}},
\qquad
\plus{\frac{\log^k\frac{1-z}{\sqrt{z}}}{1-z}}
\eeq
and also
\beq\label{eq:plus_prime}
\plus{\frac{\log^k\frac{1-z}{\sqrt{z}}}{1-z}}^{\prime}
= \sum_{j=0}^k \binom{k}{j} \plus{\frac{\log^{k-j}(1-z)}{1-z}} \log^j\frac{1}{\sqrt{z}}
\eeq
\end{subequations}
where the ${}^\prime$ denotes that the plus distribution has to be put
only around $\frac{\log^{k-j}(1-z)}{1-z}$; note that in the sum only
the first term needs the plus symbol, since $\log\sqrt{z} = 0$ in $z=1$.
These distributions can be obtained, respectively, as the $k$-th $\xi$-derivative of the following
generating distributions, computed in $\xi=0$:
\beq
\plusq{(1-z)^{\xi-1}},\qquad
\plusq{\log^{\xi-1}\frac1z},\qquad
\plusq{z^{-\xi/2}(1-z)^{\xi-1}},\qquad
z^{-\xi/2}\plusq{(1-z)^{\xi-1}}.
\eeq
Their Mellin transforms can be computed easily, and are
\begin{subequations}\label{eq:Mellin_log_gen}
\begin{align}
  \Mell\[\plus{(1-z)^{\xi-1}}\] &= \Gamma(\xi)\[\frac{\Gamma(N)}{\Gamma(N+\xi)} - \frac1{\Gamma(1+\xi)}\]\\
  \Mell\[\plus{\log^{\xi-1}\frac1z}\] &= \Gamma(\xi) \[N^{-\xi}-1\]\label{eq:Mellin_log_gen2}\\
  \Mell\[\plus{z^{-\xi/2}(1-z)^{\xi-1}}\] &=
  \Gamma(\xi)\[\frac{\Gamma(N-\xi/2)}{\Gamma(N+\xi/2)} - \frac{\Gamma(1-\xi/2)}{\Gamma(1+\xi/2)}\]\label{eq:Mellin_log_gen3}\\
  \Mell\[z^{-\xi/2}\plus{(1-z)^{\xi-1}}\] &= 
  \Gamma(\xi)\[\frac{\Gamma(N-\xi/2)}{\Gamma(N+\xi/2)} - \frac1{\Gamma(1+\xi)}\].\label{eq:Mellin_log_gen4}
\end{align}
\end{subequations}
Since $\Gamma(\xi)$ has a simple pole in $\xi=0$, the limit $\xi\to0$ of these functions
needs the computation of a derivative. To be precise, we note that every Mellin transform
above is written in the form
\beq
\Gamma(\xi)\, F_N(\xi) = \frac1\xi \,\Gamma(1+\xi)\, F_N(\xi) \equiv \frac1\xi \, G_N(\xi)
\eeq
with $F_N(0)=G_N(0)=0$, which has the power expansion
\beq
\frac1\xi \, G_N(\xi) = \sum_{k=0}^\infty \frac{G_N^{(k+1)}(0)}{(k+1)!} \,\xi^k.
\eeq
Hence, the $k$-th derivative of the generic Mellin transform Eqs.~\eqref{eq:Mellin_log_gen}
is
\beq
\frac{G_N^{(k+1)}(0)}{k+1} = \frac1{k+1} \sum_{j=0}^{k}\binom{k+1}{j}\, \Gamma^{(j)}(1)\, F_N^{(k+1-j)}(0)
\eeq
where the $(k+1)$-th term in the sum has been omitted because $F_N(0)=0$.
In the case of the first and the last of Eqs.~\eqref{eq:Mellin_log_gen},
a somewhat simpler expansion can be found by noting that for them $G_N(\xi)$ has the form
\beq
G_N(\xi) = \Gamma(1+\xi)\,H_N(\xi) -1,
\eeq
and then the derivatives are
\beq
\frac{G_N^{(k+1)}(0)}{k+1} = \frac1{k+1} \sum_{j=0}^{k+1}\binom{k+1}{j}\, \Gamma^{(j)}(1)\, H_N^{(k+1-j)}(0)
\eeq
where now the $(k+1)$-th term is not null.
We can now show explicitly the Mellin transforms of Eqs.~\eqref{eq:logarithms}:
\begin{subequations}\label{eq:Mellin_log}
\begin{align}
\Mell\[\plus{\frac{\log^k(1-z)}{1-z}}\]
&=\frac{1}{k+1}\sum_{j=0}^{k}\binom{k+1}{j}\,
\Gamma^{(j)}(1)\[ \Gamma(N)\,\Delta^{(k+1-j)}(N) - \Delta^{(k+1-j)}(1)\]
\nonumber\\
&=\frac{1}{k+1}\sum_{j=0}^{k+1}\binom{k+1}{j}\,
\Gamma^{(j)}(1)\, \Gamma(N)\,\Delta^{(k+1-j)}(N)
\label{eq:Mellin_log1}
\\
\Mell\[\plus{\frac{\log^k\log\frac{1}{z}}{\log\frac{1}{z}}}\]
&=\frac{1}{k+1}\sum_{j=0}^k\binom{k+1}{j}\, \Gamma^{(j)}(1)\,
\log^{k+1-j}\frac{1}{N}
\label{eq:Mellin_log2}
\\
\Mell\[\plus{\frac{\log^k\frac{1-z}{\sqrt{z}}}{1-z}}\]
&=\frac{1}{k+1}\sum_{j=0}^{k}\binom{k+1}{j}\,
\Gamma^{(j)}(1) \,\big[\Upsilon_{k+1-j}(N,0)-\Upsilon_{k+1-j}(1,0)\big]
\label{eq:Mellin_log3}
\\
\Mell\[\plus{\frac{\log^k\frac{1-z}{\sqrt{z}}}{1-z}}^{\prime}\]
&=\frac{1}{k+1}\sum_{j=0}^{k}\binom{k+1}{j}\,
\Gamma^{(j)}(1) \[\Upsilon_{k+1-j}(N,0)-\Delta^{(k+1-j)}(1) \]
\nonumber\\
&=\frac{1}{k+1}\sum_{j=0}^{k+1}\binom{k+1}{j}\,
\Gamma^{(j)}(1) \,\Upsilon_{k+1-j}(N,0)
\label{eq:Mellin_log4}
\end{align}
\end{subequations}
where we have defined
\begin{subequations}
\begin{align}
\Delta(\xi) &= 1/\Gamma(\xi)\\
\Upsilon_0(N,\xi) &= \Gamma(N-\xi/2) \,\Delta(N+\xi/2)
\end{align}
\end{subequations}
and denoted with $\Upsilon_k(N,\xi)$ the $k$-th derivative of $\Upsilon_0(N,\xi)$ with respect to $\xi$.
The equivalence of the two expressions in the first and last of Eqs.~\eqref{eq:Mellin_log}
is encoded in the relation
\beq
-\sum_{j=0}^k\binom{k+1}{j}\,\Gamma^{(j)}(1)\,\Delta^{(k+1-j)}(1) = \Gamma^{(k+1)}(1);
\eeq
indeed, grouping all terms on the same side we recognize
the $(k+1)$-th derivative of $\Gamma(\xi)\,\Delta(\xi)=1$,
which trivially vanishes.

At large $N$, all the expressions in Eqs.~\eqref{eq:Mellin_log} are equivalent
up to constant terms or terms suppressed by inverse powers of $N$.
Indeed, the dominant terms in each first line of Eqs.~\eqref{eq:Mellin_log}
coincide at large $N$
\beq\label{eq:lagreNeq}
\Gamma(N)\,\Delta^{(k+1-j)}(N) \simeq
\log^{k+1-j}\frac{1}{N} \simeq
\Upsilon_{k+1-j}(N,0)
\eeq
up to corrections of order $1/N$.
This can be verified using the Stirling approximation Eq.~\eqref{eq:GammaStirling}
on the generating functions, i.e.\
\begin{subequations}\label{eq:generating_largeN}
\begin{align}
\frac{\Gamma(N)}{\Gamma(N+\xi)} &= N^{-\xi} + \Ord(N^{-1})\\
\frac{\Gamma(N-\xi/2)}{\Gamma(N+\xi/2)} &= N^{-\xi} + \Ord(N^{-1}).
\end{align}
\end{subequations}
The first subleading difference between Eqs.~\eqref{eq:Mellin_log} is a constant:
indeed in Eq.~\eqref{eq:Mellin_log2} there are no constants, while
in Eqs.~\eqref{eq:Mellin_log1} and \eqref{eq:Mellin_log4} there is a $\Gamma^{(k+1)}(1)/(k+1)$,
and finally in Eq.~\eqref{eq:Mellin_log3} there is a constant term given by
\beq
-\frac{1}{k+1}\sum_{j=0}^{k}\binom{k+1}{j}\, \Gamma^{(j)}(1) \,\Upsilon_{k+1-j}(1,0).
\eeq
The difference between these two non-zero constants can be better computed
by taking the difference of the generating functions Eqs.~\eqref{eq:Mellin_log_gen3}
and \eqref{eq:Mellin_log_gen4}:
\begin{align}
\Mell\[\plus{\frac{\log^k\frac{1-z}{\sqrt{z}}}{1-z}}
-
\plus{\frac{\log^k\frac{1-z}{\sqrt{z}}}{1-z}}^{\prime}\]
&=
\frac{d^k}{d\xi^k}\[\Gamma(\xi)\(\frac{1}{\Gamma(1+\xi)} - \frac{\Gamma(1-\xi/2)}{\Gamma(1+\xi/2)}\)\]_{\xi=0}
\nonumber\\
&=
\frac{d^k}{d\xi^k}\[\frac{1 - C(\xi)}\xi\]_{\xi=0}
\nonumber\\
&= -\frac{C^{(k+1)}(0)}{k+1}
\label{eq:Mellin_constant_difference}
\end{align}
having defined
\beq\label{eq:C(xi)}
C(\xi) = \frac{\Gamma(1+\xi)\, \Gamma(1-\xi/2)}{\Gamma(1+\xi/2)}.
\eeq
In threshold resummation (see Chap.~\ref{chap:soft-gluons}),
only the large-$N$ part of the coefficient function, i.e.\ powers of $\log N$, is resummed.
This is the reason why we computed also the Mellin transforms Eq.~\eqref{eq:Mellin_log2}:
even if those $z$-space logarithms never appear in perturbative computations,
their Mellin transform produces instead the resummed logarithms.
From Eq.~\eqref{eq:Mellin_log_gen2}, we can then find the inverse Mellin transforms
of powers of $\log\frac1N$, since $N^{-\xi}$ is their generating function.
We have
\beq
N^{-\xi}=1+\frac{1}{\Gamma(\xi)}\,\Mell\[\plus{\log^{\xi-1}\frac{1}{z}}\]
\eeq
and therefore
\beq
\Mell^{-1}\[N^{-\xi}\]=\delta(1-z)+
\plus{\frac{\log^{\xi-1}\frac{1}{z}}{\Gamma(\xi)}}.
\eeq
Taking $k$ derivatives with respect to $\xi$ at $\xi=0$ we get
either one of the following expressions
\begin{subequations}\label{eq:inv_Mellin_logN}
\begin{align}
\Mell^{-1}\[\log^k\frac{1}{N}\]
&=\delta_{k0}\,\delta(1-z)
+\plus{
\left.\frac{d^k}{d\xi^k}\frac{\log^{\xi-1}\frac{1}{z}}{\Gamma(\xi)}
\right|_{\xi=0}}
\label{eq:inv_Mellin_logN_der}
\\
&=\delta_{k0}\,\delta(1-z)
+\frac{k!}{2\pi i}\plus{\oint\frac{d\xi}{\xi^{k+1}}\,
\frac{\log^{\xi-1}\frac{1}{z}}{\Gamma(\xi)}
}
\label{eq:inv_Mellin_logN_oint}
\\
&=\delta_{k0}\,\delta(1-z)
+\plus{\frac{1}{\log\frac{1}{z}}\sum_{j=1}^k\binom{k}{j}\Delta^{(j)}(0)\,
\log^{k-j}\log\frac1z},
\label{eq:inv_Mellin_logN_sum}
\end{align}
where we have used again $\Delta(\xi)=1/\Gamma(\xi)$ and $\Delta(0)=0$.
The derivatives of $\Delta(\xi)$ in $\xi=0$ can be related to those
in $\xi=1$ using Eq.~\eqref{eq:Delta_rec_rel_0}; equivalently, we can
write directly $\Delta(\xi) = \xi\Delta(1+\xi)$ before computing the derivatives,
obtaining
\beq
\Mell^{-1}\[\log^k\frac{1}{N}\]
=\delta_{k0}\,\delta(1-z)
+\plus{\frac{k}{\log\frac{1}{z}}\sum_{j=0}^{k-1}\binom{k-1}{j}\Delta^{(j)}(1)\,
\log^{k-1-j}\log\frac1z}.
\eeq
\end{subequations}
Note that the result is a distribution,
consistently with the fact that $\log^k N$ is an increasing function
of $N$, see Theorem~\ref{th:bonvini2}.
In the second form, directly related to the first by Cauchy theorem,
the integration contour is any closed curve in the complex $\xi$ plane
which encircles the origin $\xi=0$.

Starting from one of the other of Eqs.~\eqref{eq:Mellin_log_gen},
we can use the same procedure to find the inverse Mellin transform of
the analogous of a power of $\log\frac1N$, according to the large-$N$
equivalence Eq.~\eqref{eq:lagreNeq}.
At the level of the generating functions, we have then, respectively,
\begin{subequations}\label{eq:inv_Mellin_gen}
\begin{align}
\Mell^{-1}\[\frac{\Gamma(N)}{\Gamma(N+\xi)}\] &= \frac1{\Gamma(1+\xi)} \,\delta(1-z) + \plusq{\frac{(1-z)^{\xi-1}}{\Gamma(\xi)}}
\label{eq:inv_Mellin_gen1}\\
\Mell^{-1}\[\frac{\Gamma(N-\xi/2)}{\Gamma(N+\xi/2)}\] &=
\frac{\Gamma(1-\xi/2)}{\Gamma(1+\xi/2)} \,\delta(1-z) + \plusq{\frac{z^{-\xi/2}(1-z)^{\xi-1}}{\Gamma(\xi)}}\\
&=
\frac1{\Gamma(1+\xi)} \,\delta(1-z) + \frac{z^{-\xi/2}}{\Gamma(\xi)}\plusq{(1-z)^{\xi-1}}. \label{eq:inv_Mellin_gen3}
\end{align}
\end{subequations}
The left-hand-side of each equation can be confused, up to terms of order $1/N$,
with the inverse Mellin of $N^{-\xi}$, Eq.~\eqref{eq:generating_largeN}.
Moreover, since the last two lines are different forms of the same inverse Mellin,
we can choose to use one and discard the other: the second, Eq.~\eqref{eq:inv_Mellin_gen3},
is somewhat simpler and then we use it.
By computing the $\xi$-derivatives in $\xi=0$ in the left-hand-side we get
powers of $\log\frac1N$ up to terms suppressed as $1/N$.
Then, up to this accuracy, using the derivative and the Cauchy notations
we have in the first case, Eq.~\eqref{eq:inv_Mellin_gen1},
\begin{subequations}\label{eq:inv_Mellin_logB1}
\begin{align}
\Mell^{-1}\[\log^k\frac1N\] &\simeq \Delta^{(k)}(1) \,\delta(1-z)
+\plus{\left.\frac{d^k}{d\xi^k}\frac{(1-z)^{\xi-1}}{\Gamma(\xi)}\right|_{\xi=0}}\\
&= \frac{k!}{2\pi i} \oint\frac{d\xi}{\xi^{k+1}\,\Gamma(\xi)}\( \plusq{(1-z)^{\xi-1}} + \frac1\xi\,\delta(1-z)\)
\end{align}
\end{subequations}
and in the second case, Eq.~\eqref{eq:inv_Mellin_gen3},
\begin{subequations}\label{eq:inv_Mellin_logB3}
\begin{align}
\Mell^{-1}\[\log^k\frac1N\] &\simeq \Delta^{(k)}(1) \,\delta(1-z)
+\left.\frac{d^k}{d\xi^k}\frac{z^{-\xi/2}\plusq{(1-z)^{\xi-1}}}{\Gamma(\xi)}\right|_{\xi=0}\\
&= \frac{k!}{2\pi i} \oint\frac{d\xi}{\xi^{k+1}\,\Gamma(\xi)}\( z^{-\xi/2} \plusq{(1-z)^{\xi-1}} + \frac1\xi\,\delta(1-z)\).
\end{align}
\end{subequations}
The presence of the $\delta(1-z)$ term in Eqs.~\eqref{eq:inv_Mellin_logB1},
\eqref{eq:inv_Mellin_logB3}, is crucial for the equalities to hold
up to power suppressed terms: indeed, in threshold resummation,
$\delta(1-z)$ terms (constants in $N$ space) are considered enhanced at large $N$,
since they don't vanish, and hence they must not be neglected.
Note that, in both cases, the $z$ dependence can be simplified
using the relation
\beq\label{eq:plus(1-z)_delta}
\plusq{(1-z)^{\xi-1}} + \frac1\xi \,\delta(1-z) = (1-z)^{\xi-1}
\eeq
which holds for $\Re \xi>0$. Since the integral involves values
of $\xi$ with negative real part, this substitution is in principle not allowed;
nevertheless, Eq.~\eqref{eq:plus(1-z)_delta} can be continued analytically
to the relevant values of $\xi$, obtaining expressions equivalent to
Eqs.~\eqref{eq:inv_Mellin_logB1}, \eqref{eq:inv_Mellin_logB3} without the $\delta(1-z)$
term and without the plus distribution symbol.
This can be used, for instance, if one performs a convolution analytically \emph{before}
the $\xi$ integration; otherwise, if the $\xi$ integral is computed before, the convolution
would not be convergent anymore.

We now move to the Mellin transform of suppressed logarithms.
In particular we consider
\beq
(1-z)^p \log^k(1-z),\qquad p\geq0;
\eeq
its Mellin transform is
\begin{align}
\Mell\[ (1-z)^p \log^k(1-z) \]
&= \frac{d}{d\xi}\,\Mell\[ (1-z)^{p+\xi}\]_{\xi=0}\\
&= \frac{d}{d\xi}\[ \frac{\Gamma(N) \,\Gamma(p+1+\xi)}{\Gamma(N+p+1+\xi)} \]_{\xi=0}\\
&= \Gamma(N) \sum_{j=0}^k\binom{k}{j}\Gamma^{(j)}(1+p)\,\Delta^{(k-j)}(N+1+p).
\end{align}
The prefactor can be written as
\beq
\Gamma(N) = \frac{\Gamma(N+1+p)}{N(N+1)\cdots(N+p)}
\eeq
from which we have a clear large-$N$ interpretation in terms
of powers of $\log\frac{1}{N}$:
\beq
\Mell\[ (1-z)^p \log^k(1-z)\] \overset{N\to\infty}{=}
\frac{1}{N^{p+1}} \sum_{j=0}^k\binom{k}{j}\Gamma^{(j)}(1+p)\,\log^{k-j}\frac{1}{N}.
\eeq
With the same technique we can compute the Mellin transform of a power of
$\log z$ divided by $1-z$, which turns out to be
\beq
\Mell\[\frac{\log^k z}{1-z}\]
= \frac{d^k}{d\eta^k}\frac{d}{d\epsilon}\[\frac{\Gamma(N+\eta)}{\Gamma(N+\eta+\epsilon)}\]_{\eta=\epsilon=0}
= -\psi_k(N).
\eeq
Finally, the Mellin transform of a power of $\log z$ is
\beq
\Mell\[\log^k z\] = \psi_k(N+1)-\psi_k(N) = (-)^k\frac{k!}{N^{k+1}}.
\eeq

\subsection{Explicit computation of Mellin transforms for the first few orders}
\label{sec:exact-Mellin-logs}
We now turn to the explicit computation of the Mellin transforms Eqs.~\eqref{eq:Mellin_log}
in terms of elementary special functions.
Using the recursion
\beq
\Delta^{(k+1)}(N) = -\sum_{j=0}^{k} \binom{k}{j} \Delta^{(j)}(N)\, \psi_{k-j}(N)
\eeq
we get for the first few terms of Eq.~\eqref{eq:Mellin_log1}
\begin{subequations}
\begin{align}
\Gamma(N)\, \Delta^{(1)}(N) &= - \psi_0(N)\\
\Gamma(N)\, \Delta^{(2)}(N) &= - \psi_1(N) + \psi_0^2(N)\\
\Gamma(N)\, \Delta^{(3)}(N) &= - \psi_2(N) + 3\psi_1(N)\psi_0(N) - \psi_0^3(N)\\
\Gamma(N)\, \Delta^{(4)}(N) &= - \psi_3(N) + 4\psi_2(N)\psi_0(N) + 3\psi_1^2(N) -6\psi_1(N)\psi_0^2(N) + \psi_0^4(N)
\end{align}
\end{subequations}
and, when $N=1$, we have
\begin{subequations}
\begin{align}
\Delta^{(1)}(1) &= \gammae  \\
\Delta^{(2)}(1) &= \gammae^2 -\zeta_2 \\
\Delta^{(3)}(1) &= \gammae^3 -3\gammae \zeta_2 + 2\zeta_3 \\
\Delta^{(4)}(1) &= \gammae^4 -6\gammae^2\zeta_2 + 8\gammae \zeta_3 + 3\zeta_2^2 - 6\zeta_4,
\end{align}
\end{subequations}
having used
\begin{subequations}\label{eq:psi_k_in_1}
\begin{align}
\psi_0(1) &= -\gammae \\
\psi_1(1) &= \zeta_2 \\
\psi_2(1) &= -2\zeta_3 \\
\psi_3(1) &= 6\zeta_4.
\end{align}
\end{subequations}
Using instead the recursion
\beq
\Gamma^{(k+1)}(N) = \sum_{j=0}^{k} \binom{k}{j} \Gamma^{(j)}(N)\, \psi_{k-j}(N)
\eeq
and Eqs.~\eqref{eq:psi_k_in_1}, we get for the $\Gamma$ derivatives in $N=1$
\begin{subequations}
\begin{align}
\Gamma^{(1)}(1) &= -\gammae  \\
\Gamma^{(2)}(1) &= \gammae^2 +\zeta_2 \\
\Gamma^{(3)}(1) &= -\gammae^3 -3\gammae \zeta_2 - 2\zeta_3 \\
\Gamma^{(4)}(1) &= \gammae^4 +6\gammae^2\zeta_2 + 8\gammae \zeta_3 + 3\zeta_2^2 + 6\zeta_4.
\end{align}
\end{subequations}
Concerning Eq.~\eqref{eq:Mellin_log3} and Eq.~\eqref{eq:Mellin_log4},
we can find the recursion
\begin{align}
\Upsilon_{k+1}(N,0) &=
-\sum_{j=0}^k \binom{k}{j} \frac12 \[\frac{1}{2^j} + \frac{1}{(-2)^j} \]\, \Upsilon_{k-j}(N,0)\, \psi_j(N) \nonumber\\
&= -\sum_{j=0,\,\text{even}}^k \binom{k}{j} \frac{1}{2^j}\, \Upsilon_{k-j}(N,0)\, \psi_j(N),
\end{align}
from which we get
\begin{subequations}
\begin{align}
\Upsilon_1(N,0) &= -\psi_0(N) \\
\Upsilon_2(N,0) &= \psi_0^2(N) \\
\Upsilon_3(N,0) &= -\psi_0^3(N) - \frac{1}{4}\psi_2(N) \\
\Upsilon_4(N,0) &= \psi_0^4(N) + \psi_2(N)\psi_0(N).
\end{align}
\end{subequations}
Finally, the constant difference Eq.~\eqref{eq:Mellin_constant_difference}
can be easily computed from the recursion
\beq
C^{(k+1)}(0) = \sum_{j=0}^k \binom{k}{j} C^{(k-j)}(0) \[1-\frac{1+(-1)^j}{2^{j+1}}\] \psi_j(1)
\eeq
and can be expressed in terms of the Riemann $\zeta$ function according to Eq.~\eqref{eq:psi_n_Laurent}.
The first few terms are then given by
\begin{subequations}
\begin{align}
C^{(1)}(0) &= 0 \\
C^{(2)}(0) &= \zeta_2 \\
C^{(3)}(0) &= -\frac32 \zeta_3 \\
C^{(4)}(0) &= 3\zeta_2^2+6\zeta_4.
\end{align}
\end{subequations}

\chapter{Analytical expressions}

\minitoc

\noindent
We collect in this Appendix various explicit analytical expressions which complete
the discussions throughout the text.

\section{Drell-Yan process at fixed perturbative order}
\label{sec:DY_fixed-order}

To begin with, we present the LO and NLO expressions for the
rapidity distribution and inclusive cross-section;
the NNLO has been computed in Ref.~\cite{Hamberg:1990np} for the inclusive cross-section
and in Ref.~\cite{admp} for the rapidity distribution,
but we won't report their lengthy expressions here.

\subsection{Rapidity distribution}
Recalling the notation introduced in Sect.~\ref{sec:rap_resumm}, we have
\beq
\frac{1}{\tau}\frac{d\sigma}{dM^2 dY} =
\sum_{i,j} \int_\tau^1 \frac{dz}{z} \int_0^1 du\;
\Lrap_{ij}(z,u,\muf^2)\,
\bar C_{ij}\(z,u,\as(\mur^2),\frac{M^2}{\muf^2},\frac{M^2}{\mur^2}\)
\eeq
where $Y$ is the hadronic rapidity and we restored the
full dependence on renormalization and factorization scales
(the dependence of $\Lrap_{ij}$ on $\tau$ and $Y$, Eq.~\eqref{eq:Lrap},
is implicitly understood).
At next-to-leading order, the rapidity distribution receives contributions
from quark-antiquark and quark-gluon subprocesses:
\beq
\frac{d\sigma^{\rm NLO}}{dM^2\,dY} =
\frac{d\sigma^{\rm NLO}_{q\bar q}}{dM^2\,dY} +
\frac{d\sigma^{\rm NLO}_{qg+gq}}{dM^2\,dY}.
\eeq
The $q\bar q$ contribution is given by
\beq\label{eq:sigmaNLOqqbar}
\bar C_{q\bar q}\(z,u,\as,\frac{M^2}{\muf^2},\frac{M^2}{\mur^2}\)
=\delta(1-z) + \frac{\as}{\pi}\,C_F\, F_q\(z,u,\frac{M^2}{\muf^2}\) + \Ord\(\as^2\)
\eeq
where
\begin{align}
F_q\(z,u,\frac{M^2}{\muf^2}\) &= \frac{\delta(u)+\delta(1-u)}{2} \Bigg[ \delta(1-z) 
\(\frac{\pi^2}{3} -4\)
+2\,(1+z^2) \plus{\frac{\log(1-z)}{1-z}} \nonumber \\
&\qquad\qquad\qquad\qquad\quad
+\log\frac{M^2}{\muf^2}\plus{\frac{1+z^2}{1-z}}
-\frac{1+z^2}{1-z}\log z +1-z \Bigg] 
\nonumber \\
&\qquad
+\frac12 \,\frac{1+z^2}{1-z}\[\plus{\frac{1}{u}}+\plus{\frac{1}{1-u}}\]-(1-z).
\label{Fzu}
\end{align}
The $qg$ and $gq$ contribution are given by
\begin{align}
\bar C_{qg}\(z,u,\as,\frac{M^2}{\muf^2},\frac{M^2}{\mur^2}\)
&=\frac{\as}{2\pi}\,T_F\, F_g\(z,u,\frac{M^2}{\muf^2}\) + \Ord\(\as^2\)\\
\bar C_{gq}\(z,u,\as,\frac{M^2}{\muf^2},\frac{M^2}{\mur^2}\)
&=\frac{\as}{2\pi}\,T_F\, F_g\(z,1-u,\frac{M^2}{\muf^2}\) + \Ord\(\as^2\)
\end{align}
where 
\begin{multline}
F_g\(z,u,\frac{M^2}{\muf^2}\)
=\delta(u)\left[\(z^2+(1-z)^2\)\(\log\frac{(1-z)^2}{z} 
+ \log\frac{M^2}{\muf^2} \) +2z(1-z) \right] \\
+\(z^2+(1-z)^2\)\plus{\frac{1}{u}} +2z(1-z) +(1-z)^2 u.
\end{multline}

\subsection{Invariant mass distribution}
The inclusive coefficient functions are found integrating
over the hadronic rapidity $Y$, whose dependence is all contained in the definition
Eq.~\eqref{eq:Lrap} of $\Lrap_{ij}$. We get
\beq
\frac{d\sigma^{\rm NLO}}{dM^2} =
\frac{d\sigma^{\rm NLO}_{q\bar q}}{dM^2} +
\frac{d\sigma^{\rm NLO}_{qg+gq}}{dM^2}
\eeq
with
\begin{align}
\frac{1}{\tau}\frac{d\sigma^{\rm NLO}_{q\bar q}}{dM^2} &=
\int_\tau^1 \frac{dz}{z}\,\Lum_{q\bar q}\(\frac{\tau}{z}\) \left[
\delta(1-z) + \frac{\as}{\pi}\, C_F \, F_q^{\rm int}\(z,\frac{M^2}{\muf^2}\)
\right] \label{eq:NLO}\\
\frac{1}{\tau}\frac{d\sigma^{\rm NLO}_{qg+gq}}{dM^2} &=
\frac{\as}{2\pi}\,T_F
\int_\tau^1 \frac{dz}{z} \, 
\left[\Lum_{qg}\(\frac{\tau}{z}\) 
+\Lum_{gq}\(\frac{\tau}{z}\) \right]F_g^{\rm int}\(z,\frac{M^2}{\muf^2}\)
\end{align}
and
\begin{align}
F_q^{\rm int}\(z,\frac{M^2}{\muf^2}\) &=
2\,(1+z^2)\plus{\frac{\log\frac{1-z}{\sqrt{z}}}{1-z}}
-4\,\delta(1-z) + 2\log\frac{M^2}{\muf^2} \plus{\frac{1+z^2}{1-z}}\\
F_g^{\rm int}\(z,\frac{M^2}{\muf^2}\) &= \(z^2+(1-z)^2\) \( \log\frac{(1-z)^2}{z} 
+ \log\frac{M^2}{\muf^2} \) +\frac{1}{2} + 3z -\frac{7}{2}z^2.
\end{align}
It is interesting to rewrite the NLO $q\bar q$ coefficient function
in several forms ($\muf=M$):
\begin{subequations}\label{eq:C1qqDY}
\begin{align}
C_{q\bar q}^{(1)}(z,1)
&= \frac{2C_F}{\pi}\left\{
(1+z^2)\plus{\frac{\log\frac{1-z}{\sqrt{z}}}{1-z}}
+\(2\zeta_2-2\)\delta(1-z)
\right\}\\
&= \frac{2C_F}{\pi}\left\{
(1+z^2)\plus{\frac{\log\frac{1-z}{\sqrt{z}}}{1-z}}^{\prime}
+\(\zeta_2 -2\)\delta(1-z)
\right\}\\
&= \frac{2C_F}{\pi}\left\{
(1+z^2)\plus{\frac{\log(1-z)}{1-z}}
-\frac{1+z^2}{1-z}\log\sqrt{z}
+\(\zeta_2-2\)\delta(1-z)
\right\}\\
&= \frac{2C_F}{\pi}\left\{
2\plus{\frac{\log(1-z)}{1-z}}
-\frac{\log z}{1-z}-(1+z)\log\frac{1-z}{\sqrt{z}} +\(\zeta_2 -2\)\delta(1-z)
\right\}
\end{align}
\end{subequations}
where we have used the definition of the primed plus distribution, Eq.~\eqref{eq:plus_prime},
and the result Eq.~\eqref{eq:logz_zeta2}.

\subsection{$\xF$ distribution}
Especially in fixed-target experiments, distributions sometimes are
given in terms of the Feynman $\xF$ variable instead of rapidity. 
The variable $\xF$ is defined by
\beq
\xF=\frac{2\kl}{\sqrt{s}},
\label{xfdef}
\eeq
where $\kl$ is the longitudinal momentum of the Drell-Yan pair ($k=k_1+k_2$),
and it is related to the rapidity $Y$ and the transverse momentum $\kt$
by
\beq
Y=\frac{1}{2}\log\frac
{\sqrt{\xF^2+4\tau(1+\kth^2)}+\xF}
{\sqrt{\xF^2+4\tau(1+\kth^2)}-\xF}
\label{xfy}
\eeq
where $\kth^2=\kt^2/M^2$.
At leading order $\kt=0$, and at NLO $\kt$ is fixed uniquely in terms
of $\xF$ by the kinematics, so up to this order the  
$\xF$ and rapidity distributions are
simply proportional:
\beq
\left.\frac{d\sigma}{dM^2 dY}\right|_{\rm NLO}=
\sqrt{\xF^2 + 4\tau(1+\kth^2)}
\left.\frac{d\sigma}{dM^2 d\xF}\right|_{\rm NLO},
\eeq
though at higher orders they will be different.

\section{Higgs production at fixed perturbative order}
\label{sec:Higgs_fixed-order}

In the large-$m_t$ limit, i.e.\ when the Higgs mass $m_H$ is less than $2m_t$ enough
(tipically $m_H\lesssim 200$~GeV) we can use the effective lagrangian
\beq
\mathcal{L} = -\frac{1}{4v} W\, G_{\mu\nu}^a G^{a,\mu\nu}\,H, \qquad v= \(2G_F^2\)^{-1/4}
\eeq
for the computation of the Higgs production at hadron colliders.
The Wilson coefficient in the \MSbar\ scheme up to order $\as^3$~\cite{Chetyrkin:1997un,Harlander:2001is}
is given by
\beq
W = -\frac{\as}{3\pi} \[ 1 + \frac{11}{4}\frac{\as}{\pi} +
\[ \frac{2777}{288} + \frac{19}{16} \log\frac{\mu^2}{m_t^2} + n_f \( -\frac{67}{96} + \frac{1}{3} \log\frac{\mu^2}{m_t^2} \) \]
\(\frac{\as}{\pi}\)^2  +\Ord(\as^3) \].
\eeq
The total cross-section for the Higgs production at hadron colliders can be written as
\begin{align}
\sigma(\tau,M^2) &= \sum_{i,j}\int_0^1 dx_1\int_0^1 dx_2\int_0^1 dz\;
f_i(x_1)\, f_j(x_2)\, \hat\sigma_{ij}(z)\, \delta\(z-\frac{\tau}{x_1x_2}\)
\nonumber \\
&= \tau\, \sigma_0\, \as^2 \sum_{i,j}\int_\tau^1 \frac{dz}{z}\, \Lum_{ij}\(\frac{\tau}{z}\) C_{ij}(z)
\label{eq:Higgs_cs}
\end{align}
where
\begin{align}
&\tau = \frac{M^2}{s}\\
&\sigma_0 = \frac{G_F}{288\pi \sqrt{2}} \abs{\sum_q A(x_q)}^2, \qquad x_q=\frac{4m_q^2}{m_H^2}\\
&A(x) = \frac32 x \[ 1+(1-x)f(x) \]\\
&f(x) =
\begin{cases}
  \arcsin^2\frac{1}{\sqrt x} & x\geq 1\\
  -\frac14 \[\log\frac{1+\sqrt{1-x}}{1-\sqrt{1-x}} - i\pi\]^2 & x<1
\end{cases}\\
&\Lum_{ij}(z) = \int_z^1 \frac{dx}{x}\;f_i(x)\,f_j\(\frac{z}{x}\)
\end{align}
The partonic cross-section is
\beq
\hat\sigma_{ij}(z) = \sigma_0 \,z\, \as^2\, C_{ij}(z,\as)
\eeq
and the coefficient functions $C_{ij}(z)$ have an expansion in $\as$:
\beq
C_{ij} = C_{ij}^{(0)} + \as \, C_{ij}^{(1)} + \as^2 C_{ij}^{(2)} + \ldots
\eeq
At LO only $C_{gg}$ contributes
\beq
C_{ij}^{(0)}(z) = \delta_{ig} \, \delta_{jg} \, \delta(1-z).
\eeq
At NLO, in the large-$m_t$ limit, we have~\cite{Dawson:1990zj,Dawson:1993qf}\footnote{Actually
there is a missing term in these references: the correct result can be found for example in
Refs.~\cite{Anastasiou:2002yz,Catani:2003zt}.}
\begin{align}
  C_{gg}^{(1)}(z) &= \bar C_1(z)
  + 2\tilde P_{gg}^{(0)}(z)\, \log\frac{M^2}{\muf^2}
  +2\beta_0\delta(1-z)\log\frac{\mur^2}{\muf^2} \\
  C_{gq}^{(1)}(z) &= P_{gq}^{(0)}(z) \[\log\frac{(1-z)^2}{z} + \log\frac{M^2}{\muf^2} \]
  + \frac{C_F}{2\pi} z - \frac{3C_F}{4\pi}\frac{(1-z)^2}{z} \\
  &= P_{gq}^{(0)}(z) \[1+\log\frac{(1-z)^2}{z} + \log\frac{M^2}{\muf^2} \] - \frac{C_F}{4\pi} (1-z)(7-3z)\\
  C_{qg}^{(1)}(z) &= C_{gq}^{(1)}(z)\\
  C_{qq}^{(1)}(z) &= 0 \\
  C_{q\bar q}^{(1)}(z) &= \frac{32}{27\pi}\frac{(1-z)^3}{z}
\end{align}
where
\begin{align}
\bar C_1(z) &= \frac{2C_A}{\pi} \(2\plusq{\frac{\log(1-z)}{1-z}} - \frac{\log z}{1-z} + \zeta_2\delta(1-z)\)
+\frac{11}{2\pi}\delta(1-z)
\nonumber\\
&\quad +2P_{gg}^{{\rm reg}}(z)\,\log\frac{(1-z)^2}{z}  - \frac{11}{2\pi}\frac{(1-z)^3}{z}
\nonumber\\
&= \frac{4C_A}{\pi} \plus{\frac{\log\frac{1-z}{\sqrt{z}}}{1-z}} + \(\frac{4C_A}\pi\zeta_2 + \frac{11}{2\pi}\) \delta(1-z)
\nonumber\\
&\quad + 2 P_{gg}^{{\rm reg}}(z)\,\log\frac{(1-z)^2}{z} - \frac{11}{2\pi}\frac{(1-z)^3}{z}
\label{eq:Higgs_C1}
\end{align}
and
\beq
P_{gg}^{(0)}(z)=\tilde P_{gg}^{(0)}(z) + \beta_0\delta(1-z),\qquad
P_{gq}^{(0)}(z)=\frac{C_F}{2\pi}\,\frac{1+(1-z)^2}{z}
\eeq
with
\beq
\tilde P_{gg}^{(0)}(z) = \frac{C_A/\pi}{\plus{1-z}} + P_{gg}^{{\rm reg}}(z),\qquad
P_{gg}^{{\rm reg}}(z) = \frac{C_A}{\pi}\left[\frac{1}{z}-2+z(1-z)\right].
\eeq
With the help of the results of Sect.~\ref{sec:exact-Mellin-logs},
the Mellin transforms of each contribution to $\bar C_1(z)$ are
\begin{align}
  \frac{4C_A}{\pi}\,\Mell\[\plus{\frac{\log(1-z)}{1-z}}\] &=
  \frac{2C_A}{\pi} \[ \psi_0^2(N) - \psi_1(N) + 2\gammae\psi_0(N) + \zeta_2 + \gammae^2 \]
  \nonumber\\
  - \frac{2C_A}{\pi}\,\Mell\[\frac{\log z}{1-z}\] &=
  \frac{2C_A}{\pi}\, \psi_1(N)
  \nonumber\\
  \( \frac{11}{2\pi} + \frac{2C_A}{\pi} \zeta_2\)\,\Mell\[ \delta(1-z) \] &= \( \frac{11}{2\pi} + \frac{2C_A}{\pi} \zeta_2\)
  \nonumber\\
  2\, \Mell\[ P_{gg}^{{\rm reg}}(z)\,\log\frac{(1-z)^2}{z}\] &= \frac{2C_A}{\pi} \[ L(N-1)-2L(N) + L(N+1) - L(N+2) \]
  \nonumber\\
  - \frac{11}{2\pi}\,\Mell\[\frac{(1-z)^3}{z}\] &=
  - \frac{11}{2\pi} \[\frac{1}{N-1} - \frac{3}{N} + \frac{3}{N+1}-\frac{1}{N+2}\]
\end{align}
where we have defined
\beq
L(N) \equiv \Mell\[\log\frac{(1-z)^2}{z}\] = -\frac{2}{N}\[\psi_0(N+1)+\gammae\] +\frac{1}{N^2}.
\eeq
The full NNLO correction has been computed in~\cite{Anastasiou:2002yz} in the large-$m_t$ approximation.
The soft part is correctly reproduced in the large-$m_t$ approximation~\cite{Harlander:2001is}.

\subsection{Finite top mass}
\label{sec:Higgs_finite-mt}

The large-$m_t$ is good for the study of the threshold limit
but fails in the description of the small-$x$ region.
At NLO, the full $m_t$ dependence can be found in \cite{Bonciani:2007ex}.
In practice, getting rid correctly of the top mass amounts to the substitutions
\begin{align}
\frac{11}{2\pi}\,\delta(1-z) &\to \frac{{\cal G}}\pi \,\delta(1-z) \\
- \frac{11}{2\pi}\frac{(1-z)^3}{z} &\to \frac{C_A}{\pi} \,{\cal R}_{gg}(z)
\end{align}
in Eq.~\eqref{eq:Higgs_C1} for the $gg$ channel. The constant ${\cal G}$
can be found in Ref.~\cite{Aglietti:2006tp} and
the function $\mathcal{R}_{gg}$ is given by~\cite{Bonciani:2007ex}
\beq\label{eq:finite_mt_Rgg}
\mathcal{R}_{gg}(z) = \frac{1}{z(1-z)}\int_0^1 \frac{dv}{v(1-v)}
\left\{ \frac{z^4}{B}r_{gg}(s_t, t_t, u_t) - \frac14 \[1+z^2+(1-z)^2\]^2 \right\}
\eeq
where $B$ is proportional to the Born contribution and $r_{gg}$ is a function of
\beq
s_t = \frac{1}{yz}, \qquad
t_t = -\frac{1-z}{yz}(1-v), \qquad
u_t = -\frac{1-z}{yz}v; \qquad
y = \frac{m_t^2}{m_H^2}
\eeq
(see Ref.~\cite{Bonciani:2007ex} for details).
From this expression it is easy to see that the limits $m_t\to\infty$ and $z\to0$ do not commute.
Indeed, $m_t\to\infty$ means
\beq
y^{-1} \to 0,
\eeq
which conflicts with $z\to0$, because they appear in the combination $y^{-1}/z$, which can be
either $0$ (if the limit $m_t\to\infty$ is taken before) or $\infty$ (if the limit $z\to0$ is taken before).

\section{Soft-gluons resummation formulae}
\label{sec:app-resumm}

The general structure valid for both Drell-Yan pair and Higgs production
for the resummed coefficient function is given by~\cite{mvv}
\beq
C^{\rm res}\(N,\as(M^2)\) = g_0\(\as(M^2)\) \, \exp G(N,M^2)
\eeq
where $G$ is just the Sudakov exponent
\beq
G(N,M^2) = \Sud\(\ab \log\frac{1}{N}, \ab\)
\eeq
with a redefinition of the arguments.
Formally, $G$ is given by
\beq\label{eq:GNres}
G(N,M^2) = \int_0^1 dz \,\frac{z^{N-1}-1}{1-z}\[
2\int_{M^2}^{(1-z)^2M^2} \frac{d\mu^2}{\mu^2}\, A\(\as(\mu^2)\)
+ D\(\as([1-z]^2M^2)\)
\],
\eeq
where the functions $A(\as)$ and $D(\as)$ have the expansions
\begin{align}
  A(\as) &= A_1\,\as + A_2\,\as^2 + A_3\,\as^3 + \ldots\\
  D(\as) &= D_1\,\as + D_2\,\as^2 + \ldots
\end{align}
and, for Drell-Yan and Higgs, $D_1=0$.
The function $A(\as)$, sometimes called $\Gamma_{\rm cusp}(\as)$,
is given by the coefficient of the distributional part of
the GLAP splitting function:
\beq
P_{ii}(\as,x) = \frac{A(\as)}{\plus{1-x}} + \ldots
\eeq
with $i=q$ for Drell-Yan and $i=g$ for Higgs. The corresponding $A$ functions
are related by the colour-charge relation, Eq.~\eqref{eq:Pii_color-charge}.
To N$^k$LL accuracy, the function $A(\as)$ must be included up to order $\as^{k+1}$,
and the function $D(\as)$ up to order $\as^k$.

In the computation of $G(N,M^2)$ from Eq.~\eqref{eq:GNres}
the Mellin integral hits the Landau singularity of the strong coupling.
Therefore, Eq.~\eqref{eq:GNres} is only a formal definition and
a prescription is needed to compute the functions $g_i$.
This is done computing the Mellin transform in the large-$N$ limit
and keeping only terms up to the desired accuracy.
Using Eq.~\eqref{eq:Mellin_log1} we have in the large-$N$ limit~\cite{fr}
\begin{align}
\Mell\[\plus{\frac{\log^p(1-z)}{1-z}}\]
&= \frac1{p+1} \sum_{j=0}^{p+1}\binom{p+1}{j} \Gamma^{(j)}(1)\,\log^{p+1-j}\frac1N +\Ord\(\frac1N\)\nonumber\\
&=
-\sum_{j=0}^{p+1}
\frac{\Gamma^{(j)}(1)}{j!}
\frac{d^j}{d\log^j\frac1N}\int_0^{1-1/N}dz\,\frac{\log^p(1-z)}{1-z}
+\Ord\(\frac{1}{N}\);
\end{align}
a N$^k$LL accuracy is then obtained truncating the sum at $j=k$.
Hence, we finally have
\begin{multline}
G_k(N,M^2) =
-\sum_{j=0}^{k}
\frac{\Gamma^{(j)}(1)}{j!}
\frac{d^j}{d\log^j\frac1N}\\
\int_0^{1-1/N}dz
\,\frac{1}{1-z}\[
2\int_{M^2}^{(1-z)^2M^2} \frac{d\mu^2}{\mu^2}\, A\(\as(\mu^2)\)
+ D\(\as([1-z]^2M^2)\)
\].
\end{multline}
For technical reasons, the constant terms in $N$-space are usually
all included in $g_0$, and then the $(p+1)$-th term in the sum
must not contribute in the computation of $G(N,M^2)$.
Then, at N$^k$LL, for those terms with a power $p$ of $\log(1-z)$ smaller than $k$,
the sum must be truncated at $j=p$ in order not to include the constant term.

\subsection{Resummed cross-section for the Drell-Yan process}

In this Appendix we give the explicit expressions of the
functions $g_i$ which appear in the resummed Drell-Yan cross-section,
Eq.~\eqref{eq:S}.
We have~\cite{mvv}
\begin{subequations}\label{eq:g_i}
\begin{align}
g_0(\as) =& \, 1 + \as\, g_{01} + \as^2\, g_{02} + \Ord(\as^3)\\
g_1(\lambda) =&\, \frac{2A_1}{\beta_0} \left[ (1+\lambda)\log(1+\lambda) -\lambda \right] \\
g_2(\lambda) =&\, \frac{A_2}{\beta_0^2} \left[ \lambda - \log(1+\lambda) \right]
   + \frac{A_1}{\beta_0} \left[ \log(1+\lambda) \( \log\frac{M^2}{\mur^2} -2\gammae \) -\lambda \log\frac{\muf^2}{\mur^2} \right] \nonumber\\
       &\,+ \frac{A_1b_1}{\beta_0^2} \left[ \frac{1}{2}\log^2(1+\lambda) +\log(1+\lambda) -\lambda \right]\\
g_3(\lambda) =&\, \frac{1}{4\beta_0^3} \( A_3 - A_1  b_2 + A_1  b_1^2 - A_2  b_1 \) \frac{\lambda^2}{1+\lambda} \nonumber\\
&\, + \frac{A_1 b_1^2}{2\beta_0^3} \; \frac{\log(1+\lambda)}{1+\lambda} \left[ 1 + \frac{1}{2} \log(1+\lambda)\right]
      + \frac{ A_1  b_2 - A_1  b_1^2 }{2\beta_0^3} \; \log(1+\lambda) \nonumber\\
&\, +  \( \frac{A_1  b_1}{\beta_0^2}  \gammae 
          + \frac{A_2  b_1 }{2\beta_0^3}
          \) \left[ \frac{\lambda}{1+\lambda} 
- \frac{\log(1+\lambda)}{1+\lambda} \right]
\nonumber\\
&\, - \left(
            \frac{A_1  b_2 }{2\beta_0^3}
          + \frac{A_1}{\beta_0}   (\gammae^2 + \zeta_2) 
          + \frac{A_2}{\beta_0^2}  \gammae
          - \frac{D_2}{4\beta_0^2} 
         \right) 
           \frac{\lambda}{1+\lambda} 
\nonumber\\
&\,
       +   \left[
         \(
            \frac{A_1}{\beta_0}  \gammae
          + \frac{A_2 - A_1  b_1 }{2\beta_0^2}
          \) 
           \frac{\lambda}{1+\lambda} 
       +  \frac{A_1  b_1}{2\beta_0^2} \;
          \frac{\log(1+\lambda)}{1+\lambda}
          \right] \log\frac{M^2}{\mur^2}
\nonumber\\
&\,
       -  \frac{A_2}{2\beta_0^2} \, \lambda \, \log\frac{\muf^2}{\mur^2}
       + \frac{A_1}{4\beta_0} \left[
         \lambda\, \log^2\frac{\muf^2}{\mur^2}
          - \frac{\lambda}{1+\lambda} \, \log^2\frac{M^2}{\mur^2}
          \right].
\end{align}
\end{subequations}
The coefficients $g_{0k}$ can be found in~\cite{Moch:2005ky}, but without 
scale-dependent terms. Their full expression is given by
\begin{subequations}
\begin{align}
g_{01} =&\, \frac{C_F}{\pi}\left[ 4\zeta_2 -4 + 2\gammae^2 
+\(\frac{3}{2}-2\gammae\)\log\frac{M^2}{\muf^2} \right]\\
g_{02} =&\, \frac{C_F}{16\pi^2} \Bigg\{
          C_F \( \frac{511}{4} - 198\,\zeta_2 - 60\,\zeta_3 
          + \frac{552}{5}\,\zeta_2^2 
              - 128\,\gammae^2 + 128\,\gammae^2 \zeta_2 
              + 32\,\gammae^4 \) \nonumber\\
       &\qquad + C_A \bigg( - \frac{1535}{12} + \frac{376}{3}\,\zeta_2 
                  + {604 \over 9}\,\zeta_3 
                  - \frac{92}{5}\,\zeta_2^2 + \frac{1616}{27}\,\gammae
                  - 56\,\gammae \zeta_3\nonumber\\
       &\qquad\qquad\qquad+ {536 \over 9}\,\gammae^2
                  - 16\,\gammae^2 \zeta_2 + {176 \over 9}\,\gammae^3 \bigg)
\nonumber\\
&\qquad + n_f \( {127 \over 6} - {64 \over 3}\,\zeta_2 
+ {8 \over 9}\,\zeta_3 
                - {224 \over 27}\,\gammae - {80 \over 9}\,\gammae^2 
                - {32 \over 9}\,\gammae^3 \) \nonumber\\
&\qquad + \log^2\frac{M^2}{\muf^2} \left[ C_F \(32\gammae^2-48\gammae+18\)
 +C_A \(\frac{44}{3}\gammae-11\) +n_f\(2-\frac{8}{3}\gammae\) \right]
\nonumber\\
&\qquad + \log\frac{M^2}{\muf^2} 
\Bigg[ C_F \(48\zeta_3 +72\zeta_2 -93 -128\gammae \zeta_2 +128\gammae + 48\gammae^2 -64\gammae^3 \) \nonumber\\
       &\qquad\qquad\qquad +C_A \(\frac{193}{3} -24\zeta_3 -\frac{88}{3}\zeta_2 +16\gammae\zeta_2 -\frac{536}{9}\gammae -\frac{88}{3}\gammae^2 \) \nonumber\\ 
       &\qquad\qquad\qquad +n_f \(\frac{16}{3}\zeta_2 -\frac{34}{3} +\frac{80}{9}\gammae +\frac{16}{3}\gammae^2 \) \Bigg]
       \Bigg\} \nonumber\\
       &\,- \frac{\beta_0 C_F}{\pi}\left[ 4\zeta_2 -4 + 2\gammae^2 +\(\frac{3}{2}-2\gammae\)\log\frac{M^2}{\muf^2} \right] \log\frac{\muf^2}{\mur^2} \; .
\end{align}
\end{subequations}
The coefficients appearing in the previous functions are
\begin{align}
A_1 &= \frac{C_F}{\pi} = \frac{4}{3\pi}\\
A_2 &= \frac{C_F}{2\pi^2}
\left[ C_A \(\frac{67}{18} - \frac{\pi^2}{6}\) - \frac{10}{9}T_F\, n_f \right] 
= \frac{201-10n_f}{27\pi^2}-\frac{1}{3}\\
A_3 &= \frac{C_F}{4\pi^3} \left[ C_A^2 
\left( \frac{245}{24} - \frac{67}{9}\,\zeta_2
+ \frac{11}{6}\,\zeta_3 + \frac{11}{5}\,\zeta_2^2 \right) 
+ \left( -  \frac{55}{24}  + 2\,\zeta_3 \right)C_F \,n_f  \right. 
\nonumber\\ & 
\left. \mbox{} \qquad
+ \left( - \frac{209}{108} + \frac{10}{9}\,\zeta_2 -\frac{7}{3}\zeta_3 \right)
C_A \,n_f - \frac{1}{27} \, n_f^2 \right]
\\
D_2 &= \frac{C_F}{16\pi^2}
\left[ C_A\(-\frac{1616}{27}+\frac{88}{9}\pi^2+56\zeta_3\) 
+ \(\frac{224}{27}-\frac{16}{9}\pi^2\) n_f \right].
\end{align}

\subsection{Resummed cross-section for the Higgs production}

The Higgs resummation coefficients are simply related to those
of the Drell-Yan process. The coefficient for the Higgs are obtained as~\cite{mvv}
\beq
A_k^{\rm H} = \frac{C_A}{C_F} A_k^{\rm DY}
,\qquad
D_k^{\rm H} = \frac{C_A}{C_F} D_k^{\rm DY}.
\eeq
The function $g_0$ collecting constant terms is instead different;
the first two coefficients can be found in Ref.~\cite{Catani:2003zt}%
\footnote{These constants can be also found in Ref.~\cite{Moch:2005ky};
there, however, the constant contributions from the Wilson coefficient
in the large-$m_t$ approximation are not included.
However, also such constants must be there in order to reproduce correctly
the fixed-order result.
Therefore one could use the constants from Ref.~\cite{Moch:2005ky}
provided the Wilson coefficient is left in front of the resummed expression,
or (more consistently) of the whole matched result.
}
with full scale-dependence.
We report here for completeness the first:
\beq
g_{01}= \frac{C_A}\pi \[ 4\zeta_2 + 2\gammae^2 + \frac23\pi\beta_0\log\frac{\mur^2}{\muf^2} - 2\gammae \log\frac{M^2}{\muf^2}\]
+ \frac{11}{2\pi}.
\eeq

\subsection{Matching}\label{sec:matching}
Here we compute the terms to be subtracted in the resummed result in order
to avoid double counting.
The first step consists in expanding Eqs.~\eqref{eq:g_i} in powers of their argument $\lambda$:
\begin{subequations}
\begin{align}
g_1(\lambda) 
=&\, \frac{A_1}{\beta_0} \left[ \lambda^2-\frac{1}{3}\lambda^3 
+ \Ord(\lambda^4) \right] \\
g_2(\lambda) 
=&\, \frac{A_1}{\beta_0} \(\log\frac{M^2}{\muf^2}-2\gammae\) \lambda
+ \( \frac{A_2}{2\beta_0^2} + \frac{A_1}{2\beta_0} 
\(2\gammae - \log\frac{M^2}{\mur^2}\) \) \lambda^2
            + \Ord(\lambda^3) \\
g_3(\lambda) 
=&\, \(-\frac{A_1}{\beta_0}(\gammae^2+\zeta_2)
-\frac{A_2}{\beta_0^2}\gammae+\frac{D_2}{4\beta_0^2}\) \lambda 
\nonumber\\
&\,+\( \frac{A_1}{\beta_0}\gammae\log\frac{M^2}{\mur^2} 
+\frac{A_2}{2\beta_0^2}\log\frac{M^2}{\muf^2}
 - \frac{A_1}{4\beta_0}\(\log^2\frac{M^2}{\mur^2}-\log^2\frac{\muf^2}{\mur^2}\)
            \) \lambda + \Ord(\lambda^2).
\end{align}
\end{subequations}
The double counting term are found as the Taylor expansion
of $g_0(\as)\,\exp\Sud(\lambda,\ab)$ in powers of $\as$:
\begin{align}
g_0(\as)\,\exp\Sud(\lambda,\ab) &= (1 + \as\, g_{01} + \as^2\, g_{02} 
+ \ldots)e^{\as\,\Sud_1+\as^2\,\Sud_2+\ldots}
\nonumber \\
&=1 + (\Sud_1+g_{01})\as + \(\frac{\Sud_1^2}{2} + \Sud_2 + \Sud_1\,g_{01} + g_{02}\)
 \as^2 + \Ord(\as^3).
\end{align}
Using the expansions above, we get
\begin{align}
\as\, \Sud_1 
&=\left[\frac{A_1}{\beta_0} \(\frac{\lambda}{\ab} + \log\frac{M^2}{\muf^2}
 - 2\gammae \)\right] \lambda\\
\as^2\, \Sud_2 &= \Bigg[ - \frac{A_1}{3\beta_0} \frac{\lambda}{\ab}
+ \( \frac{A_2}{2\beta_0^2} + \frac{A_1}{2\beta_0}
\(2\gammae - \log\frac{M^2}{\mur^2}\) \)  
              +\bigg\{-\frac{A_1}{\beta_0}(\gammae^2+\zeta_2)
-\frac{A_2}{\beta_0^2}\gammae+\frac{D_2}{4\beta_0^2} \nonumber\\
&\qquad\qquad
            +\frac{A_1}{\beta_0}\gammae\log\frac{M^2}{\mur^2} 
+\frac{A_2}{2\beta_0^2}\log\frac{M^2}{\muf^2}
            - \frac{A_1}{4\beta_0}\(\log^2\frac{M^2}{\mur^2}-
\log^2\frac{\muf^2}{\mur^2}\)\bigg\} \frac{\ab}{\lambda} \Bigg] \lambda^2.
\end{align}
Note that, since $\lambda=\ab\log\frac{1}{N}$,
it can be seen as an expansion in $\lambda$ with $\lambda/\ab$ fixed.

\section{High-energy resummation at NLO}
\label{sec:app-SX-NLO}

In this Appendix we collect many details on the construction of the full resummed NLO
anomalous dimensions.

\subsection{BFKL kernel at NLO}
\label{sec:NLO_BFKL_kernel}

The NLO BFKL kernel $\chi_1^\sigma(M) = \chi_1^L(M) + \chi_1^R(M)$ in symmetric variables
and in the scheme used by Ref.~\cite{FL} is given by~\cite{abf742}%
\footnote{There are some misprints in Ref.~\cite{abf742} that we correct here.}
\begin{align}
\chi_1^\sigma(M) &= -\frac{N_c\beta_0}{2\pi}
\( \frac{\pi^2}{N_c^2}\chi_0^2(M) - \psi_1(M) - \psi_1(1-M) \)
\nonumber\\ &\quad
+\frac{N_c^2}{4\pi^2}
\Bigg\{
\( \frac{67}{9} - \frac{\pi ^2}{3} - \frac{10}9\frac{n_f}{N_c} \)
\[ \psi(1)-\psi(M) \]
+ \psi_2(M) + 3\zeta_3
- 4\phi^+(M)
\nonumber\\ &\quad\phantom{+\frac{N_c^2}{4\pi^2}\Bigg\{}
+\frac{\pi^2}{2} \[ \psi\(\frac{1+M}{2}\)-\psi\(\frac{M}{2}\) \]
\nonumber\\ &\quad\phantom{+\frac{N_c^2}{4\pi^2}\Bigg\{}
+\bigg[\frac{3}{4(1-2M)}
+\frac{2+3M(1-M)}{32}\( 1+\frac{n_f}{N_c^3}\)
\nonumber\\ &\qquad\qquad\qquad\qquad\qquad\qquad
\times\(
\frac{2}{1-2M}
+\frac{1}{1+2M}
-\frac{1}{3-2M}
\) \bigg]
\nonumber\\ &\quad\phantom{+\frac{N_c^2}{4\pi^2}\Bigg\{}\qquad\qquad
\times\[ \psi_1\(\frac{1+M}{2}\)-\psi_1\(\frac{M}{2}\) \]
\nonumber\\ &\quad\phantom{+\frac{N_c^2}{4\pi^2}\Bigg\{}
+(M\to 1-M)
\Bigg\}\label{eq:chi1_sigma}
\end{align}
where $\chi_0(M)$ is the LO BFKL kernel, Eq.~\eqref{eq:chi0},
$\beta_0$ is the first coefficient of the $\beta$-function, Eq.~\eqref{eq:beta0},
$\zeta_k$ is the Riemann Zeta function (see App.~\ref{sec:RiemannZeta}),
and $\phi^+(M)$ is given by
\beq\label{eq:phi_plus}
\phi^+(M) = -\int_0^1\frac{dx}{1+x}\, x^{M-1} \int_x^1\frac{dt}t\log (1-t);
\eeq
series representation of this function will be given in Sec.~\ref{sec:app-phi}.

Taking into account the reshuffling due to operator ordering in the running coupling,
the kernel acquires the non-symmetric term given in Eq.~\eqref{eq:chiNLO_ordering}.
Passing to DIS variables, Eq.~\eqref{eq:symm_to_DIS}, one then gets
\beq\label{eq:chi1}
\chi_1(M) = \frac{N_c^2}{4\pi^2} \delta(M) -\frac12 \chi_0(M)\chi_0'(M)
\eeq
where
\begin{align}
\delta(M) &= -\frac{2\pi\beta_0}{N_c}
\( \frac{\pi^2}{N_c^2}\chi_0^2(M) - \psi_1(M) + \psi_1(1-M) \)
+\( \frac{67}{9} - \frac{\pi ^2}{3} - \frac{10}9\frac{n_f}{N_c} \)
\bar\psi(M) \nonumber\\
&\quad
+\frac{\pi^3}{\sin(\pi M)} - \frac{\pi^2\cos(\pi M)}{\sin^2(\pi M)(1-2M)}
\left[ 3 + \( 1+\frac{n_f}{N_c^3}\) \frac{2+3M(1-M)}{(3-2M)(1+2M)}
\right] \nonumber\\
&\quad
- 4\phi(M) + \psi_2(M) + \psi_2(1-M)
+ 6\zeta_3, \label{eq:delta_term}
\end{align}
having used \eqref{eq:sin_of_psi} and
\beq\label{eq:cos_psi}
\frac{4\pi^2\cos(\pi M)}{\sin^2(\pi M)} = \psi_1\(\frac{M}{2}\) - \psi_1\(\frac{M+1}{2}\) - (M\to 1-M)
\eeq
and defining
\beq
\phi(M) = \phi^+(M) + \phi^-(M)
\eeq
where $\phi^-(M) = \phi^+(1-M)$, see Sect.~\ref{sec:app-phi}.

To change scheme to the \MSbar\ scheme, a further term must be added~\cite{Ball:1999sh}:
\beq\label{eq:chi1MSbar}
\chi_1^{\text{\MSbar}}(M) = \chi_1(M) + \frac{\beta_0 N_c}{2\pi} \[ \bar\psi^2(M) + 2\psi_1(1) - \psi_1(M) -\psi_1(1-M) \].
\eeq

\subsubsection{Series representation of $\phi(M)$}
\label{sec:app-phi}

The function $\phi^+(M)$, Eq.~\eqref{eq:phi_plus}, has a series representation
given by
\beq\label{eq:phiFL_series}
\phi^+(M) = \sum_{n=0}^\infty (-)^n \frac{\psi (n+1+M )-\psi (1)}{(n+M)^2}.
\eeq
It can be obtained expanding
\beq
\frac{1}{1+x} = \sum_{n=0}^\infty (-x)^n
\eeq
and computing the integrals in reverse order
\begin{align}
\phi^+(M) &= -\sum_{n=0}^\infty (-)^n \int_0^1 \frac{dt}{t} \log(1-t) \int_0^t dx \,x^n \,x^{M -1} \nonumber\\
&= -\sum_{n=0}^\infty (-)^n \int_0^1 \frac{dt}{t} \log(1-t) \,\frac{t^{n+M}}{n+M} .
\end{align}
The computation of the second integral can be done with the trick
\begin{align}
\int_0^1 dt\; t^{c-1} \log(1-t)
&= \left.\frac{d}{d\epsilon}\int_0^1 dt\; t^{c-1} (1-t)^\epsilon \right|_{\epsilon=0} \nonumber\\
&= \Gamma(c) \left.\frac{d}{d\epsilon} \frac{\Gamma(1+\epsilon)}{\Gamma(c+1+\epsilon)} \right|_{\epsilon=0}\nonumber\\
&= \frac{\psi(1) - \psi(c+1)}{c},
\end{align}
bringing directly to Eq.~\eqref{eq:phiFL_series}
(remember that $\psi(1) = -\gammae$, see App.~\ref{sec:Gamma}).
From Eq.~\eqref{eq:phiFL_series} it is clear that $\phi^+(M)$ has poles in
$M=0, -1,-2,\ldots$ coming from the collinear region;
the function $\phi^-(M) = -\phi^+(1-M)$ has instead poles in $M=1,2,3,\ldots$
coming from the anti-collinear region.

An alternative expansion can be obtained noting that
\beq
\int_x^1\frac{dt}t\log(1-t) = \Li_2(x) - \zeta(2)
\eeq
as one can verify expanding the logarithm
\beq
\log(1-t) = -\sum_{k=1}^\infty \frac{t^k}{k}
\eeq
and recalling the definitions
\beq
\Li_s(x) = \sum_{k=1}^\infty \frac{x^k}{k^s},\qquad
\zeta(s) = \sum_{k=1}^\infty \frac{1}{k^s} = \Li_s(1).
\eeq
Then we can write
\beq
\phi(M) = \frac{\pi \zeta(2)}{\sin(\pi M)} - \phi_L^+(M) - \phi_L^-(M)
\eeq
where
\begin{align}
  \phi_L^+(M) &= \int_0^1\frac{dx}{1+x}\, x^{M-1} \Li_2(x)
  = \sum_{k=1}^\infty \frac{1}{2k^2} \left[ \psi\(\frac{M+1+k}{2}\) - \psi\(\frac{M+k}{2}\) \right] \\
  \phi_L^-(M) &= \phi_L^+(1-M).
\end{align}
Using the relation
\beq\label{eq:sin_of_psi}
\frac{2\pi}{\sin(\pi M)} = 2\int_0^1\frac{dx}{1+x} \(x^{M-1} + x^{-M}\) = \psi\(\frac{M+1}{2}\) - \psi\(\frac{M}{2}\) + (M\to 1-M)
\eeq
we have the identification
\beq
\phi^+(M) = \frac{\zeta(2)}{2}\[\psi\(\frac{M+1}{2}\) - \psi\(\frac{M}{2}\)\] - \phi_L^+(M).
\eeq
Also in this case, even if in a less transparent way, the collinear and anticollinear poles
are separated.

Practically, it seems that the convergence of expansion~\eqref{eq:phiFL_series} is
faster, and hence we use it for numerical applications.

\subsubsection{Expansion around $M=0$ and relation with the singular expansion}

By using Eq.~\eqref{eq:psi_n_Laurent} and the series representation of
$\phi^+$ Eq.~\eqref{eq:phiFL_series}, the behaviour of the kernel $\chi_1$
can be found.
By writing the Laurent expansion as\footnote{We know that the $\Ord(\as^{k+1})$ BFKL
kernel $\chi_k(M)$ has a $(k+1)$-th order pole in $M=0$.}
\beq
\chi_k(M) = \sum_{j=-\infty}^{k+1} \frac{\chi_{k,j}}{M^j}
\eeq
we have
\begin{subequations}\label{eq:chi_kj}
\begin{align}
\chi_{0,1} &= \frac{N_c}{\pi} \\
\chi_{0,0} &= 0 \\
\chi_{1,2} &= -\frac{11N_c^2+2n_f/N_c}{12\pi^2} \\
\chi_{1,1} &= -\frac{n_f}{36\pi^2 N_c}\(10N_c^2+13\)
.
\end{align}
\end{subequations}
On the other hand, the functions $\chi_s$ and $\chi_{ss}$ have a series expansion%
\footnote{Note that from a formal point of view the function $\chi_s(\as/M)$ and $\chi_{ss}(\as/M)$
could be expanded in positive powers of $M/\as$; however, for our purposes we are interested in a
perturbative expansion in $\as$.}
\begin{align}
\chi_s\(\frac{\as}{M}\) &= \sum_{k=0}^\infty \chi_{s,k} \,\frac{\as^k}{M^k} \label{eq:chi_s_exp}\\
\chi_{ss}\(\frac{\as}{M}\) &= \sum_{k=0}^\infty \chi_{ss,k} \,\frac{\as^k}{M^k}.
\end{align}
To obtain the coefficients $\chi_{s,k}$,
we put the series \eqref{eq:chi_s_exp} as argument of $\gamma_0$
and match the coefficients order by order in $\as/M$%
(a general procedure is described in App.~\ref{sec:inverse_chi0})%
.
We obtain for the first three coefficient
\beq
\chi_{s,0} = 0,\qquad
\chi_{s,1} = \frac{C_A}{\pi},\qquad
\chi_{s,2} = -\frac{11 C_A^2}{12\pi^2} + \frac{n_f}{6\pi^2}(2C_F-C_A)
\eeq
or, using the definitions $C_F=(N_c^2-1)/2N_c$, $C_A=N_c$,
\beq
\chi_s(\as/M) = \frac{N_c}{\pi}\frac{\as}{M} - \frac{11 N_c^2 + 2n_f/N_c}{12\pi^2} \frac{\as^2}{M^2} +\ldots
\eeq
Note that this result is the same if the largest eigenvalue is taken
to be the largest at each $N$ or just in $N=0$ (see discussion at the end of Sect.~\ref{sec:glap:projectors});
higher orders will depend on the choice used.
For the $n_f=0$ case there are no ambiguities and these results can be estabilished
just by looking at $\gamma_{gg}$.
The coefficients $\chi_{ss,k}$ can be simply obtained from the knowledge of the expansion of $\chi_s$
and the definition \eqref{eq:chiss_def}; using the general form
\beq
\gamma_1(N) = \frac{a_0}{N^2} + \frac{a_1}{N} + a_2 + \Ord(N)
\eeq
we get
\beq
\chi_{ss,0} = \frac{a_0 \pi}{C_A},\qquad
\chi_{ss,1} = a_1.
\eeq
From Eq.~\eqref{eq:gamma1expansion} we have
\beq
a_0 = 0,\qquad
a_1 = -\frac{n_f(23C_A-26C_F)}{36\pi^2}
\eeq
and hence (using again $C_F=(N_c^2-1)/2N_c$, $C_A=N_c$)
\beq
\chi_{ss,0} = 0,\qquad
\chi_{ss,1} = -\frac{n_f}{36\pi^2 N_c}\(10N_c^2+13\).
\eeq
With the help of Fig.~\ref{fig:DL_structure} it is simple to match these coefficients
to those of $\chi_0$ and $\chi_1$:
\begin{align}
  \chi_{0,1} &= \chi_{s,1} & \chi_{0,0} &= \chi_{ss,0} \\
  \chi_{1,2} &= \chi_{s,2} & \chi_{1,1} &= \chi_{ss,1}.
\end{align}
Of course, they coincide.

\subsubsection{Computing the inverse function of $\chi_0$ or $\gamma_0$}
\label{sec:inverse_chi0}
We now want to solve the duality relation
\beq
\gamma_0\(\chi_s(z)\) = \frac1z
\eeq
in an algorithmic way. We concentrate on $\chi_s$, but the procedure applies to $\gamma_s$ as well.
Using the expansion
\beq
\gamma_0(N) = \sum_{k=-1}^\infty \gamma_k N^k
\eeq
we have ($\chi_{s,0}=0$)
\beq
\sum_{k=-1}^\infty \gamma_k \(\sum_{j=1}^\infty \chi_{s,j} z^j\)^k = \frac1z
\eeq
or, rearranging terms,
\beq
\sum_{k=0}^\infty \gamma_{k-1} z^k \(\sum_{j=0}^\infty \chi_{s,j+1} z^j\)^k = \sum_{j=0}^\infty \chi_{s,j+1} z^j
\eeq
Using the result Eq.~\eqref{eq:power_of_a_series} with $a_j = \chi_{s,j+1}$ we get
\beq
\sum_{n=0}^\infty \[\chi_{s,n+1} - \sum_{k=0}^n \gamma_{k-1} c_{k,n-k}\] z^n = 0
\eeq
where
\beq
c_{k,0} = \chi_{s,1}^k, \qquad
c_{k,p} = \frac{1}{\chi_{s,1} p} \sum_{j=1}^p (jk+j-p)\,\chi_{s,j+1}\,c_{k,p-j}.
\eeq
Then, for $n=0$ we get
\beq
\chi_{s,1} = \gamma_{-1}
\eeq
while for $n>0$ we can compute recursively the coefficients
\beq
\chi_{s,n+1} = \sum_{k=1}^n \gamma_{k-1} c_{k,n-k}.
\eeq

\subsubsection{Subtract double counting}
From the computational point of view, it is better to subtract double counting
between the fixed-order kernel and the singular expansion
directly from the $\psi_k$ functions in the fixed-order kernels,
in order to avoid differences of huge numbers.
Using the results
\begin{align}
\chi_0^2(M)-\text{d.c.} &= \tilde\chi_0^2(M) + 2\frac{N_c}{\pi}\,\frac{\tilde\chi_0(M)+N_c/\pi}{M(1-M)}\\
\phi^+(M)-\text{d.c.} &= \tilde\phi^+(M) = \phi^+(M)-\frac{\zeta(2)}{M}\\
\frac{1}{1-2M}\psi_1(M/2)-\text{d.c.} &= \frac{1}{1-2M} \[\psi_1(1+M/2)+16\]\label{eq:poles_sub_1}\\
\frac{2+3M(1-M)}{1-2M}\psi_1(M/2)-\text{d.c.} &= \frac{2+3M(1-M)}{1-2M} \psi_1(1+M/2) + \frac{44}{1-2M}\\
\frac{2+3M(1-M)}{1+2M}\psi_1(M/2)-\text{d.c.} &= \frac{2+3M(1-M)}{1+2M} \psi_1(1+M/2) - \frac{4}{1+2M}\\
\frac{2+3M(1-M)}{3-2M}\psi_1(M/2)-\text{d.c.} &= \frac{2+3M(1-M)}{3-2M} \psi_1(1+M/2) - \frac{4/9}{3-2M}\label{eq:poles_sub_4}
\end{align}
we get\footnote{In this equation there are some terms which cancel with the analogous in the $M\to1-M$ part;
however we are not going to use this expression for numerical computations, but we will use
Eq.~\eqref{eq:chi1_sigma_minus_dc_final}, which is optimized.}
\begin{align}
\tilde\chi_1^\sigma(M) &= \chi_1^\sigma(M) -\text{d.c.} \nonumber\\
&= -\frac{N_c\beta_0}{2\pi}
\( \frac{\pi^2}{N_c^2}\tilde\chi_0^2(M) +
2\,\frac{\pi\tilde\chi_0(M)/N_c+1}{M(1-M)} - \psi_1(1+M) - \psi_1(2-M)\)
\nonumber\\ &\qquad
+\frac{N_c^2}{4\pi^2}
\Bigg\{
\( \frac{67}{9} - \frac{\pi ^2}{3} - \frac{10}9\frac{n_f}{N_c} \)
\[ \psi(1)-\psi(1+M) \]
\nonumber\\ &\qquad\phantom{+\frac{N_c^2}{4\pi^2}\Bigg\{}
+ \psi_2(1+M) + 3\zeta(3)
- 4\tilde\phi^+(M)
\nonumber\\ &\qquad\phantom{+\frac{N_c^2}{4\pi^2}\Bigg\{}
+\frac{\pi^2}{2} \[ \psi\(\frac{1+M}{2}\)-\psi\(1+\frac{M}{2}\) \]
\nonumber\\ &\qquad\phantom{+\frac{N_c^2}{4\pi^2}\Bigg\{}
+\frac{3}{4(1-2M)}\[ \psi_1\(\frac{1+M}{2}\)-\psi_1\(1+\frac{M}{2}\)-16 \]
\nonumber\\ &\qquad\phantom{+\frac{N_c^2}{4\pi^2}\Bigg\{}
+\frac{2+3M(1-M)}{32}\( 1+\frac{n_f}{N_c^3}\)
\nonumber\\ &\hskip3cm
\times\Bigg(
\frac{2}{1-2M}
\[ \psi_1\(\frac{1+M}{2}\)-\psi_1\(1+\frac{M}{2}\) \]
\nonumber\\ &\hskip3.7cm
+\frac{1}{1+2M}
\[ \psi_1\(\frac{1+M}{2}\)-\psi_1\(1+\frac{M}{2}\) \]
\nonumber\\ &\hskip3.7cm
-\frac{1}{3-2M}
\[ \psi_1\(\frac{1+M}{2}\)-\psi_1\(1+\frac{M}{2}\) \]
\Bigg)
\nonumber\\ &\qquad\phantom{+\frac{N_c^2}{4\pi^2}\Bigg\{}
+\frac{1}{32}\( 1+\frac{n_f}{N_c^3}\)
\( -\frac{88}{1-2M} + \frac{4}{1+2M} - \frac{4/9}{3-2M} \)
\nonumber\\ &\qquad\phantom{+\frac{N_c^2}{4\pi^2}\Bigg\{}
+(M\to 1-M)
\Bigg\} \label{eq:chi1_sigma_minus_dc}
\end{align}
where
\beq
\tilde\chi_0(M) = \chi_0(M)-\text{d.c.}
= \frac{N_c}{\pi}\[ 2\psi(1) - \psi(1+M) - \psi(2-M) \].
\eeq
Eq.~\eqref{eq:chi1_sigma_minus_dc} has no poles in the range $-1<M<2$;
however, there are simple poles in the expression for $M=-\frac12,\frac32$
that cancel adding the $M\to1-M$ contribution. For numerical convenience,
it is better to subtract these poles directly in each collinear and anticollinear contributions.
The resulting expression is (see Sect.~\ref{sec:spurious_pole_subtr} for details)
\begin{align}
\tilde\chi_1^\sigma(M) &= -\frac{N_c\beta_0}{2\pi}
\( \frac{\pi^2}{N_c^2}\tilde\chi_0^2(M) +
2\,\frac{\pi\tilde\chi_0(M)/N_c+1}{M(1-M)} - \psi_1(1+M) - \psi_1(2-M)\)
\nonumber\\ &\qquad
+\frac{N_c^2}{4\pi^2}
\Bigg\{
\( \frac{67}{9} - \frac{\pi ^2}{3} - \frac{10}9\frac{n_f}{N_c} \)
\[ \psi(1)-\psi(1+M) \]
\nonumber\\ &\qquad\phantom{+\frac{N_c^2}{4\pi^2}\Bigg\{}
+ \psi_2(1+M) + 3\zeta(3)
- 4\tilde\phi^+(M)
\nonumber\\ &\qquad\phantom{+\frac{N_c^2}{4\pi^2}\Bigg\{}
+\frac{\pi^2}{2} \[ \psi\(\frac{1+M}{2}\)-\psi\(1+\frac{M}{2}\) \]
\nonumber\\ &\qquad\phantom{+\frac{N_c^2}{4\pi^2}\Bigg\{}
+\frac{3}{4(1-2M)}\[ \psi_1\(\frac{1+M}{2}\)-\psi_1\(1+\frac{M}{2}\) - \psi_1\(\frac34\) + \psi_1\(\frac54\) \]
\nonumber\\ &\qquad\phantom{+\frac{N_c^2}{4\pi^2}\Bigg\{}
+\frac{2+3M(1-M)}{32}\( 1+\frac{n_f}{N_c^3}\)
\nonumber\\ &\hskip3cm
\times\Bigg(
\frac{2}{1-2M}
\[ \psi_1\(\frac{1+M}{2}\)-\psi_1\(1+\frac{M}{2}\) - \psi_1\(\frac34\) + \psi_1\(\frac54\) \]
\nonumber\\ &\hskip3.7cm
+\frac{1}{1+2M}
\[ \psi_1\(\frac{1+M}{2}\)-\psi_1\(1+\frac{M}{2}\) - \psi_1\(\frac14\) + \psi_1\(\frac34\) \]
\nonumber\\ &\hskip3.7cm
-\frac{1}{3-2M}
\[ \psi_1\(\frac{1+M}{2}\)-\psi_1\(1+\frac{M}{2}\) - \psi_1\(\frac54\) + \psi_1\(\frac74\) \]
\Bigg)
\nonumber\\ &\qquad\phantom{+\frac{N_c^2}{4\pi^2}\Bigg\{}
+ \frac23 \( 1+\frac{n_f}{N_c^3}\)
\nonumber\\ &\qquad\phantom{+\frac{N_c^2}{4\pi^2}\Bigg\{}
+(M\to 1-M)
\Bigg\} \label{eq:chi1_sigma_minus_dc_final}
\end{align}
where the constant term in the second-last line arises from the asymmetric
contributions of the previous two lines.

Following Ref.~\cite{abf742}, we define
\beq\label{eq:breve_chi1}
\breve\chi_1(M) = \chi^\sigma_1(M) - \frac{N_c}{2\pi}\chi_0(M)\[ 2\psi_1(1) - \psi_1(M) -\psi_1(1-M) \];
\eeq
after subtracting double counting we have
\beq
\tilde\chi_1(M)
= \breve\chi_1(M)
- \chi_{1,2}\(\frac{1}{M^2}+\frac1{(1-M)^2}\)
- \chi_{1,1}\(\frac{1}{M}+\frac1{1-M}\)
\eeq
with $\chi_{1,1},\chi_{1,2}$ defined in \eqref{eq:chi_kj}.
Using for numerical convenience the previous results, we can write
\begin{multline}
\tilde\chi_1(M) = \tilde\chi_1^\sigma(M)
-\frac{N_c}{2\pi}\tilde\chi_0(M)\[ 2\psi_1(1) - \psi_1(1+M) -\psi_1(2-M) \]\\
+\frac{N_c}{2\pi}\bigg[ \frac{\tilde\chi_0(M)+N_c(1+M)/\pi}{M^2} + \frac{\tilde\chi_0(M)+N_c(2-M)/\pi}{(1-M)^2}\\
-\frac{N_c}{\pi}\frac{2\psi_1(1) -\psi_1(1+M) -\psi_1(2-M) -1}{M(1-M)} \bigg].
\label{eq:tilde_chi1}
\end{multline}

\subsubsection{Spurious poles subtraction in Eq.~\eqref{eq:chi1_sigma_minus_dc_final}}
\label{sec:spurious_pole_subtr}

In Eqs.~\eqref{eq:chi1_sigma},~\eqref{eq:chi1_sigma_minus_dc}
there are explicit simple poles in $M=-\frac12,\frac32$
that cancel adding the $M\to1-M$ contribution.
When extending off-shell, the cancellation no longer takes place;
for this it is useful to add some terms in order for the cancellation to take place directly
in each collinear and anticollinear term.
Let us recall Eq.~\eqref{eq:cos_psi} and define
\beq
g(M) = \psi_1\(\frac{M}{2}\) - \psi_1\(\frac{1+M}{2}\)
\eeq
such that
\beq
\frac{4\pi^2\cos(\pi M)}{\sin^2(\pi M)} = g(M) - g(1-M).
\eeq
The structure of the terms in Eq.~\eqref{eq:chi1_sigma} containing this function is
\beq
h(M) =
f(M)\frac{4\pi^2\cos(\pi M)}{\sin^2(\pi M)} = f(M) g(M) + (M\to1-M)
\eeq
with
\beq\label{eq:aux_f_antisymm}
f(1-M) = -f(M).
\eeq

Let us consider first the case in which
\beq\label{eq:example_f1}
f_1(M) = \frac{A(M)}{1-2M}
\eeq
where $A(M)=A(1-M)$ can be identified either with a constant or with $2+3M(1-M)$,
but it is irrelevant for this discussion.
Both $f_1(M)g(M)$ and $f_1(1-M)g(1-M)$ have a simple pole in $M=\frac12$,
but this pole cancels in the sum, because $g(M)-g(1-M)$ has a zero in $M=\frac12$.
We can subtract the pole in each term
\beq\label{eq:example_h1}
h_1(M) = f_1(M)\[ g(M)-g\(\frac12\)\] + (M\to1-M)
\eeq
without adding any other term (the $g(1/2)$ term cancels with the analogous in the $M\to1-M$ part).

Now let us move to
\beq\label{eq:example_f2}
f_2(M) = \big(2+3M(1-M)\big)\[\frac{1}{1+2M}-\frac{1}{3-2M}\].
\eeq
This case is a bit more complicated, because in order to preserve
the anti-symmetry \eqref{eq:aux_f_antisymm} we cannot separate $f_2(M)$
into the sum of two functions with just one pole each.
Indeed, for the symmetry $M\to1-M$, the first term in the square brackets goes into (minus) the second
and vice-versa.
For the same reason as before, also in this case the sum has no poles,
since $g(M)-g(1-M)$ has zeros in $M=-\frac12$ and $M=\frac32$, where $f_2(M)$ has poles.
The subtraction must be done separately for the two terms in the square brackets
\beq\label{eq:example_h2}
h_2(M) = \big(2+3M(1-M)\big)\[ \frac{g(M)-g(-1/2)}{1+2M} - \frac{g(M)-g(3/2)}{3-2M} \] + (M\to1-M).
\eeq
This expression is exact again, but now the cancellation of the
auxiliary terms takes place thanks to the relation
\beq
g\(\frac32\)-g\(-\frac12\) = 0
\eeq
peculiar of the actual form of $g$, as one can verify using
\beq
\psi_1\(\frac{3}{4}\) = \psi_1\(-\frac{1}{4}\) -16
, \qquad
\psi_1\(\frac{5}{4}\) = \psi_1\(\frac{1}{4}\) -16
\eeq
coming from the recursion formula
\beq
\psi_k(x+1) = \psi_k(x) + \frac{(-)^k k!}{x^{k+1}}.
\eeq
Also in this case, the numerator hasn't played any role.

Now turn to the case in wich the double counting poles in
$M=0,1$ are removed from $h(M)$, leading to Eq.~\eqref{eq:chi1_sigma_minus_dc}.
In this case we can define
\beq
\tilde g(M) = g(M) + \frac4{M^2}
= \psi_1\(1+\frac{M}{2}\) - \psi_1\(\frac{1+M}{2}\)
\eeq
which is free of poles.
The poles of $h(M)$ can be found by considering its divergent behaviour
\beq
h(M) \sim -\frac{4f(M)}{M^2} + (M\to1-M) \sim -\frac{4f(0)}{M^2} - \frac{4f'(0)}{M} + (M\to1-M)
\eeq
and then after subtracting these poles we get
\begin{align}
  \tilde h(M) &= h(M) - \text{d.c.} \nonumber \\
  &= \tilde g(M) f(M) -4\,\frac{f(M)-f(0)-f'(0) M}{M^2} + (M\to1-M) \label{eq:example_tildeh1}
\end{align}
(with this formula we have obtained Eqs.~\eqref{eq:poles_sub_1}$\div$\eqref{eq:poles_sub_4}).
Considering now again $f_1$, Eq.~\eqref{eq:example_f1}, we have
\beq
\frac{f_1(M)-f_1(0)-f_1'(0) M}{M^2} \propto f_1(M)
\eeq
in both cases for $A(M)$; in fact, this proportionality holds
as long as $A(M)$ is a polynomial of order $2$ at most.
This means that the second term in Eq.~\eqref{eq:example_tildeh1}
cancels with the analogous in the $M\to1-M$ part,
since $f(1-M)=-f(M)$.
Going to $\tilde h_1$ we have then, similarly to Eq.~\eqref{eq:example_h1},
\beq
\tilde h_1(M) = f_1(M)\[ \tilde g(M)- \tilde g\(\frac12\)\] + (M\to1-M).
\eeq
In this case, as before, the $\tilde g(1/2)$ term cancels with the analogous in
the $M\to1-M$ term.
Consider instead $f_2$, Eq.~\eqref{eq:example_f2}; this case is now non-trivial,
for two facts:
\begin{itemize}
\item the second term in Eq.~\eqref{eq:example_tildeh1} doesn't cancel
\item $\tilde g\(\frac32\)-\tilde g\(-\frac12\) \neq 0$.
\end{itemize}
Defining for convenience $\tilde h_2^{\rm res}$ as
\begin{multline}
\tilde h_2(M) = \bigg\{
\big(2+3M(1-M)\big)\[ \frac{\tilde g(M) - \tilde g(-1/2)}{1+2M} - \frac{\tilde g(M) - \tilde g(3/2)}{3-2M} \]\\
+ (M\to1-M) \bigg\}
+ \tilde h_2^{\rm res}(M)
\end{multline}
we find, after a tedious but straightforward computation,
\beq
\tilde h_2^{\rm res}(M) = \frac{128}{3}.
\eeq
We can then more conveniently write
\begin{multline}
\tilde h_2(M) = \left\{
\big(2+3M(1-M)\big)\[ \frac{\tilde g(M) - \tilde g(-1/2)}{1+2M} - \frac{\tilde g(M) - \tilde g(3/2)}{3-2M} \]
+\frac{64}{3}
\right\}\\
+ (M\to1-M).
\end{multline}
All together, this brings to Eq.~\eqref{eq:chi1_sigma_minus_dc_final}.

\subsubsection{Off-shell extension of $\chi_1(M)$}

The off shell extension (in symmetric variables) of Eq.~\eqref{eq:chi1_sigma} can be simply obtained
by substituting in the first line
\begin{multline}
\frac{\pi^2}{N_c^2}\chi_0^2(M) - \psi_1(M) - \psi_1(1-M) \to\\
\frac{\pi^2}{N_c^2}\bar\chi_0^2(M,N) - \psi_1\(M+\frac{N}{2}\) - \psi_1\(1-M+\frac{N}{2}\)
\end{multline}
and in the following
\beq
M\to M+\frac{N}{2}
\eeq
but in the last line, which should remain $(M\to1-M)$.
The off-shell extension of Eq.~\eqref{eq:chi1_sigma_minus_dc_final}
can be done in the same way, but the first line transforms as
\begin{multline}
\frac{\pi^2}{N_c^2}\tilde\chi_0^2(M) +
\[ 2\,\frac{\pi\tilde\chi_0(M)/N_c+1}{M} - \psi_1(1+M) + (M\to1-M) \]
\to\\
\frac{\pi^2}{N_c^2}\tilde{\bar\chi}_0^2(M,N) +
2\,\frac{(1+N)\pi\tilde{\bar\chi}_0(M,N)/N_c+1}{(M+N/2)(1-M+N/2)} - \psi_1\(1+M+\frac{N}{2}\) - \psi_1\(2-M+\frac{N}{2}\)
\end{multline}
The off-shell extension of $\breve\chi_1$, Eq.~\eqref{eq:breve_chi1}, is
\beq
\breve{\bar\chi}_1(M,N) = \bar\chi_1^\sigma(M,N)
-\frac{N_c}{\pi} \bar\chi_0(M,N) \[2\psi_1(1+N) - \psi_1\(M+\frac{N}{2}\) - \psi_1\(1-M-\frac{N}{2}\)\]
\eeq
which do not to introduce spurious singularities.
After subtracing double counting, the off-shell extension of Eq.~\eqref{eq:tilde_chi1} is
\begin{multline}
\tilde{\bar\chi}_1(M,N) = \tilde{\bar\chi}_1^\sigma(M,N)
-\frac{1}{2}\tilde{\bar\chi}_0(M,N)\tilde\psi(M,N)
\\
+\frac{N_c}{2\pi}\Bigg\{ \frac{1}{(M+N/2)^2} \[\tilde{\bar\chi}_0(M,N) - \tilde{\bar\chi}_0\(-\frac N2,N\) - \(M+\frac N2\)\tilde{\bar\chi}_0'\(-\frac N2,N\) \]
\\
-\frac{1}{M+N/2} \[ \tilde\psi(M,N) - \tilde\psi\(-\frac N2,N\) \]
\\
+ (M\to1-M)\Bigg\}
\end{multline}
where we have defined for convenience
\beq
\tilde\psi(M,N) = \frac{N_c}{\pi}\[ 2\psi_1(1+N) - \psi_1\(1+M+\frac N2\) -\psi_1\(2-M+\frac N2\) \]
\eeq
and the prime denotes derivatives with respect to the first argument:
\beq
\tilde{\bar\chi}_0'\(-\frac{N}{2},N\) = \frac{N_c}{\pi} \[\psi_1(2+N) - \psi_1(1)\]
\eeq
It is very important, from a numerical point of view, to provide
an analytic evaluation of some critical points, i.e.
\beq
M+\frac N2 = 0,\frac12,-\frac12,\frac32,
\eeq
and the same with $M\to 1-M$.

\subsection{Resummation at NLO}

We now present the construction of the NLO resummed anomalous dimension.
We do not pretend to explain the construction in detail, but just to show the result:
full details can be found on Ref.~\cite{abf742}.

The resummed anomalous dimension at NLO is given by
\begin{align}
\gamma_{\rm res}^{\rm NLO}(\as,N) &= \gamma_B(\as,N) - \gamma_{B,s}(\as,N) - \gamma_{B,ss}(\as,N) \nonumber\\
&\quad+ \gamma_{\rm DL}^\sigma(\as,N) - \gamma_{ss}^{\rm rc}(\as,N) + \frac{N}{2} + \gamma_{\rm match}(N) + \gamma_{\rm mom}(N),
\label{eq:gamma_resNLO_final}
\end{align}
which has the same structure of the LO result Eq.~\eqref{eq:gamma_res_final},
with some modifications that we are going to discuss.

First, the coefficients $c(\as)$ and $\kappa(\as)$ in the Bateman anomalous dimension
and its asymptotic subtracted terms are obtained from a NLO off-shell kernel given by
\begin{align}
\bar \chi^\sigma_B(\as,M,N) &= \bar\chi_s^\sigma(\as,M,N) + \as \bar\chi_{ss}^\sigma(\as,M,N)\nonumber\\
&\quad+\as \tilde{\bar\chi}_0(M,N) + \as^2\[\tilde{\bar\chi}_1(M,N) - \tilde{\bar\chi}_1\(-\frac{N}{2},N\) + \tilde{\bar\chi}_1(0,0)\]\nonumber\\
&\quad+\bar\chi_s^{\beta_0}(\as,M,N) + \bar\chi_0^{\beta_0}(\as,M,N) + \bar\chi_i^{\beta_0}(\as,M,N),
\end{align}
where the functions in the last line are due to running coupling commutators at LO
and are given by
\begin{align}
\bar\chi_s^{\beta_0}(\as,M,N) &= \frac{\beta_0}2\Bigg[
\(\frac{\as}{M+\frac N2}\)^3 \chi_s''\(\frac{\as}{M+\frac N2}\) -
\(\frac{\as}{1-M+\frac N2}\)^3 \chi_s''\(\frac{\as}{1-M+\frac N2}\)\nonumber\\
&\qquad- 2\(\frac{\as}{1-M+\frac N2}\)^2 \chi_s'\(\frac{\as}{1-M+\frac N2}\)
\Bigg]\\
\bar\chi_0^{\beta_0}(\as,M,N) &= - \as^2 \beta_0 \frac{C_A}{\pi}\psi_1\(2-M+\frac N2\)\\
\bar\chi_i^{\beta_0}(\as,M,N) &= \frac1{M+\frac N2} \as^2 \beta_0 \[c_{\rm LO}'(\as) + \frac12\kappa_{\rm LO}'(\as)\(M-\frac12\)^2\].
\end{align}
Note that the minimum is no longer in $M=\frac12$.

The function $\gamma_{\rm DL}^\sigma(\as,N)$ is obtained instead from the following NLO
off-shell kernel (remember that for $n_f\neq0$ the approximation described in Sect.~\ref{sec:rat_approx_improved}
is used)
\begin{align}
\bar \chi^\sigma_B(\as,M,N) &= \bar\chi_s^\sigma(\as,M,N) + \as \bar\chi_{ss}^\sigma(\as,M,N)\nonumber\\
&\quad+\as \tilde{\bar\chi}_0(M,N) + \as^2\[\tilde{\bar\chi}_1(M,N) - \tilde{\bar\chi}_1\(-\frac{N}{2},N\) + \tilde{\bar\chi}_1(0,0)\]\nonumber\\
&\quad+\bar\chi_0^{\beta_0}(\as,M,N) + \bar\chi_{\rm int}^{\beta_0}(\as,M,N),
\end{align}
with
\beq
\bar\chi_{\rm int}^{\beta_0}(\as,M,N) = \frac{\as^2 \beta_0}{M+\frac N2} \frac{C_A}{\pi}
\[\psi(1) +\psi(1+N) - \psi\(1+M+\frac N2\) - \psi\(1-M+\frac N2\)\].
\eeq
The function $\gamma_{ss}^{\rm rc}$ is given by
\beq
\gamma_{ss}^{\rm rc}(\as,N) = \beta_0\[ \frac{N\,\chi_0''(\gamma_s)}{2\,[\chi_0'(\gamma_s)]^2} - \as \].
\eeq

Finally, since the parameters $c(\as)$ and $\kappa(\as)$ are different between the
Bateman and the DL parts of the result, in order to allow the cancellation of spurious
square-root branch-cut the function $\gamma_{\rm match}$ is added. It is given by
\beq
\gamma_{\rm match}(N) = \sqrt{\frac{N-c_1}{\kappa_1/2}} - \sqrt{\frac{N+1}{\kappa_1/2}} + \frac{1+c_1}{\sqrt{2\kappa_1(1+N)}} - (1\to 2)
\eeq
where $c_1$ and $\kappa_1$ are the parameters from $\gamma_{\rm DL}^\sigma$ and $c_2$ and $\kappa_2$
from $\gamma_B$.

\subsection{Resummation of quark anomalous dimension with running coupling}
\label{sec:resummation_quark_AD}

The $qg$ anomalous dimension is resummed as in Eq.~\eqref{eq:gamma_qg_ss_rc},
which needs the computation of the series
\beq\label{eq:series_h_rc}
h\([\gamma]\) = \sum_{k=0}^{\infty} h_k\, \[\gamma^k\]
\eeq
where $[\gamma^k]$ are defined by
\beq
\[\gamma^k\] = \(\frac{\dot\gamma}{\gamma}\)^k \frac{\Gamma\(\gamma^2/\dot\gamma+k\)}{\Gamma\(\gamma^2/\dot\gamma\)}
= \(\frac{\dot\gamma}{\gamma}\)^k \(\frac{\gamma^2}{\dot\gamma}\)_k,
\eeq
where $(a)_k = \Gamma(a+k)/\Gamma(a)$ is the Pochammer symbol, or, recursively,
\beq
\[\gamma^{k+1}\] = \gamma \(1+k\frac{\dot\gamma}{\gamma^2}\)\[\gamma^k\]
= \frac{\dot\gamma}{\gamma} \(\frac{\gamma^2}{\dot\gamma} + k\)\[\gamma^k\], \qquad
\[\gamma\] = \gamma.
\eeq
Note that, if all $h_k=1$, we get a divergent series
\beq
\sum_{k=0}^\infty \[\gamma^k\] = t^a e^{t} \Gamma\(1-a,t\),\qquad
t=-\frac{\gamma}{\dot\gamma}, \qquad a=\frac{\gamma^2}{\dot\gamma}=-t\gamma,
\eeq
which is the asymptotic expansion of the incomplete Gamma function.
Note also that, at fixed coupling, $\dot\gamma=0$, which corresponds to the limit $\abs{t}\to\infty$;
taking the limit on both sides we get
\beq
\sum_{k=0}^\infty \gamma^k = \frac{1}{1-\gamma},
\eeq
which is of course correct (geometric series). This series convergences for $\abs{\gamma}<1$,
and in particular a pole in $\gamma=1$ is present. The running coupling result is instead well
defined for all $\gamma$ ($t>0$), even if the series has zero radius of convergence:
\beq
\lim_{k\to\infty}\abs{\frac{\[\gamma^{k+1}\]}{\[\gamma^k\]}} = \lim_{k\to\infty}\abs{\gamma + k\frac{\dot\gamma}{\gamma}} = \infty.
\eeq
Hence, the running of the coupling constant $\as$ reduces the convergence radius of the series
by a factorial term.

This computation suggests that the series with $h_k=1$ could be
$\Bor_1$-summable and $\Bor_2^*$-summable (see App.~\ref{sec:Borel} for notations).
The order-$1$ Borel transform is
\beq\label{eq:toy_rc_borel1}
\sum_{k=0}^\infty \frac1{k!} \[\gamma^k\] w^k
= \sum_{k=0}^\infty \frac1{k!} \(w\frac{\dot\gamma}{\gamma}\)^k \(\frac{\gamma^2}{\dot\gamma}\)_k
= {}_1F_0\(\frac{\gamma^2}{\dot\gamma}; ; w\frac{\dot\gamma}{\gamma}\)
= \(1- w\frac{\dot\gamma}{\gamma}\)^{-\gamma^2/\dot\gamma}
\eeq
with radius of convergence $\abs{w}<\gamma/\dot\gamma$.
The order-$2$ Borel transform is
\beq
\sum_{k=0}^\infty \frac1{(k!)^2} \[\gamma^k\] w^k
= \sum_{k=0}^\infty \frac1{k!} \(w\frac{\dot\gamma}{\gamma}\)^k \frac{(\gamma^2/\dot\gamma)_k}{(1)_k}
= {}_1F_1\(\frac{\gamma^2}{\dot\gamma}; 1; w\frac{\dot\gamma}{\gamma}\)
\eeq
with infinite radius of convergence.
In the first case ($n_{\rm rc}=1$), the Borel inversion integral converges
provided $\gamma$ and $\dot\gamma$ are not both real and positive, otherwise
the branch-cut of Eq.~\eqref{eq:toy_rc_borel1} lies on the integration path.
In the second case ($n_{\rm rc}=2$), the Borel inversion integral converges only for $\Re\!\[\gamma^2/\dot\gamma\]>0$.

The series $h(M)$ has a finite radius of convergence~\cite{Catani:1994sq,abf799}:
then one would expect that even the actual series Eq.~\eqref{eq:series_h_rc} is 
$\Bor_1$- and $\Bor^*_2$-summable.
However, the coefficients $h_k$ are only known recursively, and it's pretty hard
to work many of them out. Then, only a truncated method can be adopted.
Originally~\cite{abf799}, the truncated Borel method described in App.~\ref{sec:Borel_truncated}
was adopted:
\beq
h\([\gamma]\) \simeq \int_0^\Lambda dw\, B_2(w) \sum_k^K h_k \[\gamma^k\] \frac{w^k}{(k!)^2}.
\eeq
However, it turns out that this is not enough for the result to be stable.
For some values of $\gamma$ and $\dot\gamma$, the integral is not convergent at infinity,
and its truncation up to $w=\Lambda$ is not stable for variations of $\Lambda$.
One could try to improve the approximation if the behaviour at large $w$
of the Borel transform is known analytically somehow.
In this case we could consider the step-defined Borel transform
\beq
\hat s_2(w) =
\begin{cases}
  \sum_k^K h_k\[\gamma^k\] \frac{w^k}{(k!)^2} & \text{for }w\leq\Lambda \\
  \hat s_2^{\rm asympt}(w) & \text{for }w>\Lambda
\end{cases}
\eeq
and extend the integration to infinity,
\beq
h(M) \simeq \int_0^\infty dw\, B_2(w) \, \hat s_2(w).
\eeq
However, it is not clear whether this approximation is good or not.

A better method consists in using the Borel-Pad\'e method described
in App.~\ref{sec:Borel-Pade}. This method works also with an order-$1$ Borel,
since the Borel transform is approximated with a Pad\'e approximant
which is analytically defined.
Then one can consider the order-$n$ Borel transform
\beq
\hat s_n = \sum_k^\infty h_k\[\gamma^k\] \frac{w^k}{(k!)^n}
\eeq
and approximate it with a $[p/p]$ Pad\'e (obtained using the terms up to $K=2p$
of the series), obtaining
\beq
h\(\gamma\) \simeq \int_0^\infty dw\, B_n(w) \, \[p/p\]_{\hat s_n}(w).
\eeq
As already said, this method works for both $n=1,2$.

\subsection{Resummation of coefficient functions}
\label{sec:resummation_coeff_funct}

The resummation of coefficient functions has been performed in Ref.~\cite{Catani:1994sq}
and extended to include running coupling resummation in Ref.~\cite{abf799}.
In the \QMSbar\ scheme the coefficients functions at fixed coupling are given at NLL by
\begin{subequations}\label{eq:resumm_coeff_func}
\begin{align}
  C_{Lg}(\as,N) &= \frac{\as}{2\pi} \,\frac{2n_f}3 \,\left.\tilde h_L\(M\)\right|_{M=\gamma_s(\as/N)}\\
  C_{2g}(\as,N) &= \frac{\as}{2\pi} \,\frac{2n_f}3\,\left.\frac{\tilde h_2^{\rm CH}(M) - \tilde h_{qg}(M)}{M}\right|_{M=\gamma_s(\as/N)}\\
  &= \frac{\as}{2\pi} \,\frac{n_f}3 \,\left.\tilde h_2\(M\)\right|_{M=\gamma_s(\as/N)}\\
\end{align}
\end{subequations}
where (the superscript CH stands for Catani-Hautmann, Ref.~\cite{Catani:1994sq})
\begin{align}
  \tilde h_L(M) &= \frac{3(1-M)}{3-2M}\,\frac{\Gamma^3(1-M)\,\Gamma^3(1+2M)}{\Gamma(2-2M)\,\Gamma(2+2M)}\\
  \tilde h_2^{\rm CH}(M) &= \frac{3(2+3M-3M^2)}{2(3-2M)}\,\frac{\Gamma^3(1-M)\,\Gamma^3(1+2M)}{\Gamma(2-2M)\,\Gamma(2+2M)}\\
  h_{qg}(M) &= \frac{\as}{2\pi} \,\frac{2n_f}{3} \, \tilde h_{qg}(M)\\
  \tilde h_2(M) &= 2\,\frac{\tilde h_2^{\rm CH}(M) - \tilde h_{qg}(M)}{M}.
\end{align}
With these definitions, all the functions $\tilde h_i$ have a series expansion starting
with $1$, i.e.\ $\tilde h_i(0)=1$.
The coefficients of the series expansion of $\tilde h_{qg}$ can be worked out as described
in Ref.~\cite{Catani:1994sq}: with these coefficients at the hand, it is straightforward
to obtain also the coefficient of the series expansion of $\tilde h_2$.
Then, one can use the procedure described in Sect.~\ref{sec:resummation_quark_AD}
for the resummation of the quark anomalous dimensions
to resum the coefficient functions as well.

The singular part of the quark-singlet coefficient functions is obtained from Eqs.~\eqref{eq:resumm_coeff_func}
by colour-charge relation~\cite{Catani:1994sq}
\begin{align}
  C_{Lq}(\as,N) &= \frac{C_F}{C_A}\[C_{Lg}(\as,N) - \frac{\as}{2\pi} \,\frac{2n_f}3 \] \\
  C_{2q}(\as,N) &= \frac{C_F}{C_A}\[C_{2g}(\as,N) - \frac{\as}{2\pi} \,\frac{n_f}3 \].
\end{align}

\chapter{Series and divergent series}

\minitoc

\noindent
In this Appendix we will mainly talk about divergent series, which recur
several times throughout the text.
A complete reference on the subject is \cite{hardy}.

\section{Series}

\begin{definition}\label{def:convergence}
The series
\beq
\sum_k c_k
\eeq
is said to be \emph{convergent} if, given the partial sums
\beq
s_n = \sum_k^n c_k,
\eeq
the limit
\beq
s = \lim_{n\to\infty} s_n
\eeq
exists and is finite; in this case such limit $s$ is called the \emph{sum} of the series.
Otherwise, the series is said to be \emph{divergent}.
\end{definition}
\begin{definition}\label{def:abs_convergence}
The series
\beq
\sum_k c_k
\eeq
is said to be \emph{absolutely convergent} if the series
\beq
\sum_k |c_k|
\eeq
is convergent.
\end{definition}
The sum of a series which is convergent but not absolutely convergent can be any number.
Hence, we could recast the definition of absolute convergence by the requirement that
the limit of partial sums exists, is finite and is \emph{unique}.

There are several convergence test. We are not going to review all of them;
we will just show one of them, which is quite common and useful: the \emph{ratio test}.
Defining
\beq
\rho = \lim_{k\to\infty} \abs{\frac{c_{k+1}}{c_k}}
\eeq
whe have that
\begin{itemize}
\item if $\rho<1$ the series converges absolutely
\item if $\rho>1$ the series diverges
\item if $\rho=1$ the test is inconclusive, and we need to use another test.
\end{itemize}
A typical example in which the test is inconclusive is the case of $c_k=k^\sigma$:
indeed $\rho=1$ for any $\sigma$, and we know from other tests (for instance, the root test)
that for $\sigma<-1$ the series of $c_k$ converges.

\subsection{Power series}

Sometimes it is useful to talk about power series, i.e.\ series in which the $k$-th coefficient
$c_k$ is splitted into a numeric coefficient (we will call it $c_k$ again) times the $k$-th power
of a generic complex variable $z$:
\beq
s(z) = \sum_k c_k z^k.
\eeq
Of course, this definition doesn't add anything to the discussion above: indeed, for any fixed $z$
we have again a series with $k$-th coefficient $c_k z^k$ and everything is like before.

However, power series are very useful, since they arise in functional analysis and in perturbation theory.
In particular, we can introduce the concept of \emph{radius of convergence}.
Indeed, using the ratio test, we get
\beq
\rho = \lim_{k\to\infty} \abs{\frac{c_{k+1}z^{k+1}}{c_k z^k}} = \abs{z} \lim_{k\to\infty} \abs{\frac{c_{k+1}}{c_k}} = \frac{\abs{z}}{r}
\eeq
where we have defined
\beq
r = \lim_{k\to\infty} \abs{\frac{c_k}{c_{k+1}}};
\eeq
then, the convergence condition $\rho<1$ gives
\beq
\abs{z} < r,
\eeq
i.e.\ the power series $s(z)$ converges in a circle of radius $r$.
Since the series converges inside such a circle, the sum $s(z)$ is analytic
in the same region. Then, reversing the argument, is we are going to expand a function
$f(z)$ around some point $z_0$ and this function has poles in the complex plane,
the radius of convergence of the expansion can be at most the distance between
$z_0$ and the closest pole to it.

\subsection{Asymptotic expansions}

It happens in some cases that we are able to find a series expansion
of a function $f(z)$ around some values of $z$ which is actually a divergent
series. In these cases, we will call it an \emph{asymptotic expansion}.
Formally (supposing for simplicity to consider an expansion around $z=0$)
we have the following
\begin{definition}
A series expansion
\beq
s(z) = \sum_k c_k z^k
\eeq
is said to be \emph{asymptotic} to $f(z)$ in the sense of Poincar\'e if
\beq
\lim_{z\to0} z^{-n}\[f(z) - s_n(z) \] = 0
\eeq
for all $n>0$ ($s_n$ is the $n$-th partial sum).
\end{definition}
A stronger definition of asymptotic series can be obtained by requiring that
there exists a constant $C$ such that
\beq
\abs{f(z) - s_n(z)} \leq C\, c_{k+1} \abs{z}^{k+1}
\eeq
for all $n$.

\subsection{Operation with series}

Raising a series to a power gives
\beq\label{eq:power_of_a_series}
\(\sum_{j=0}^\infty a_j z^j\)^n = \sum_{k=0}^\infty c_{n,k} z^k
\eeq
with
\beq
c_{n,0} = a_0^n, \qquad
c_{n,k} = \frac{1}{a_0 k} \sum_{j=1}^k (jn+j-k)\,a_j\,c_{n,k-j}.
\eeq

\section{Divergent series and their sum}
\label{sec:divergent_series}

When the limit of the partial sums of a series does not exist or is not finite
the series is said to be \emph{divergent} (Def.~\ref{def:convergence}).
This, however, doesn't mean that the sum of the series is infinite;
indeed, it simply means that the definition of the sum of the series as the
limit of the partial sums is not a good definition.
Then, we need a definition for the sum of a divergent series.

As an example, let's consider a classical divergent series
\beq\label{eq:111}
\sum_{k=0}^\infty (-1)^k = 1-1+1-1+\ldots;
\eeq
it diverges because the partial sums
\beq
s_n = \sum_{k=0}^n (-1)^k =
\begin{cases}
0 & \text{for $n$ even}\\
1 & \text{for $n$ odd}
\end{cases}
\eeq
oscillate between $0$ and $1$, and the limit for $n\to\infty$ of $s_n$ does not exist.
However there are some simple arguments to believe that the sum \emph{should} be $1/2$.
For example, if we call $s$ the sum of the series, we can manipulate it to obtain an equation for $s$:
\begin{align}
  s &= \sum_{k=0}^\infty (-1)^k \\
  &= 1 + \sum_{k=1}^\infty (-1)^k \\
  &= 1 + \sum_{k=0}^\infty (-1)^{k+1} \\
  &= 1 - \sum_{k=0}^\infty (-1)^k \\
  &= 1-s
\end{align}
from which we get immediately $s=1/2$.
Of course, such manipulations are allowed if the series is convergent;
here they might be not allowed.
However there are other ways to obtain $s=1/2$. For example, consider the power series
\beq\label{eq:111_z}
s(z) = \sum_{k=0}^\infty (-z)^k
\eeq
which converges in the complex circle $|z|<1$, where the sum is
\beq
s(z) = \frac{1}{1+z};
\eeq
this result can be analytically continued to the whole complex plane but $z=-1$.
The series \eqref{eq:111} is recovered when $z=1$, which is outside the convergence circle,
but using the analytical extension of the sum we get again $s=s(1)=1/2$.

In the following we will provide the necessary mathematical theorems and tools
to \emph{define} the sum of a divergent series in such a way that we can really \emph{prove}
that the sum of the series \eqref{eq:111} is $1/2$.

\subsection{Linear transformation and regularity}

The way forward the definition of the sum of a divergent series
starts by considering a linear transformation $T$ of the sequence
$\left\{ s_n \right\}_{n\geq 0}$ of the partial sums into the sequence $\left\{ t_m \right\}_{m\geq 0}$:
\beq\label{eq:series_LT}
t_m = \sum_n a_{mn} s_n,
\eeq
where $a_{mn}$ are complex coefficients. This transformation can be generalized to the case
in which $m$ is a continuous index:
\beq
t(x) = \sum_n a_n(x) s_n.
\eeq
We have the following
\begin{definition}\label{def:regularity}
The linear transformation \eqref{eq:series_LT}
is said to be \emph{regular} if, whenever
\beq
\lim_{n\to\infty} s_n = s
\eeq
(that is, the original series converges), we have
\beq
\lim_{m\to\infty} t_m = s.
\eeq
\end{definition}
A regular linear transformation is then a transformation which doesn't change the
sum for convergent series.
Applying a regular linear transformation to a divergent series may
lead to a finite limit
\beq
t = \lim_{m\to\infty} t_m.
\eeq
When this is the case, we are then tempted to consider $t$ as the sum of the divergent series.
Of course, if different linear transformation gave different sums, this would not be a good definition.

Let's call the entire set of sequences $S$, and $C\subset S$ the subset of convergence sequences.
A regular linear transformation $T$ gives a finite result at least in $C$, but possibly
in a wider subset $S_T\subseteq S$.
Now consider two regular linear transformation $T_1$ and $T_2$: when the results
obtained with both transformations coincide for all the sequences in $S_{T_1}\cap S_{T_2}$
the two transformations are said to be \emph{consistent}.
Consistent transformations are good candidates for extending the definition
of the sum of a series to divergent series.
In the literature, several regular linear transformations are known to be consistent.

\subsection{Summation methods}

Several summation methods have been proposed for long;
the most well known are by Cesaro, Abel, Euler, etc.
For a complete review, we refer the Reader to Ref.~\cite{hardy}.
All these methods are based on regular linear transformations.
There exists other summation methods which are not linear, but they are not
supported by strong theorems, and therefore we do not discuss them here:
some details can be found in Ref.~\cite{Weniger}.

In the following, we will concentrate on one method, by Borel, which turns out to be
very flexible especially for physical application and for numerical implementations.

\section{Borel summation}
\label{sec:Borel}

The Borel method for summing divergent series is based on a continuous 
linear transformation.
In its original form it can be formulated as
\beq\label{eq:Borel_method}
\sum_{k} c_k \overset{\Bor}{=} \int_0^\infty dw \, e^{-w} \sum_k \frac{c_k}{k!}w^k.
\eeq
Its regularity can be easily proven by noting that, for absolutely convergent series,
the integral and the sum can be exchanged and
\beq
\frac{1}{k!} \int_0^\infty dw\,e^{-w}\,w^k = 1;
\eeq
pictorially, a nice way to wiew the Borel method is to start from the divergent
series, multiply each term by $1$, write $1$ as above with the appropriate $k$ in each term,
and finally exchange the sum and the integral.

It can be useful to introduce the following
\begin{definition}
  The inner power series in the right-hand-side of Eq.~\eqref{eq:Borel_method},
\beq
\hat s(w) = \sum_k \frac{c_k}{k!}w^k,
\eeq
is called the \emph{Borel transform} of the series $\sum_k c_k$.
\end{definition}
Since in the Borel transform there is a $k!$ factorial in the denominator, this sum
has more chances to converge, even if the original sum is divergent.
Hence, we have the following
\begin{definition}\label{def:Borel}
Given a divergent series,
\begin{itemize}
\item if its Borel transform converges
\item if its sum $\hat s(w)$ is defined on (or can be analitically extended to) $0\leq w \leq \infty$
\item if the integral converges
\end{itemize}
the series is said to be \emph{Borel-summable} or \emph{$\Bor$-summable}.
If the second requirement is made stronger by requiring that the Borel transform actually
has infinite radius of convergence we will call the series \emph{$\Bor^*$-summable}.
\end{definition}

The Borel method can be generalized for ``more divergent'' series, by iterating the
standard method:
\beq\label{eq:generalized_Borel_1}
\sum_{k} c_k \overset{\Bor_n}{=} \int_0^\infty dw_1 \cdots \int_0^\infty dw_n \, e^{-w_1 \ldots -w_n} \sum_k \frac{c_k}{(k!)^n}(w_1\cdots w_n)^k.
\eeq
In this way, because at the denominator there is a $(k!)^n$ the inner sum
(which we'll call generalized Borel transform $\hat s_n(w_1\cdots w_n)$)
has many more chances to converge, provided $n$ is large enough.
A divergent series which can be summed with this generalized Borel method of order $n$
is said $\Bor_n$-summable (or $\Bor^*_n$-summable in the stronger case).
The standard method Eq.~\eqref{eq:Borel_method} is recovered for $n=1$.

We can recast Eq.~\eqref{eq:generalized_Borel_1} in a form more similar to the original one:
by changing integration variables $w_n = w/(w_1\cdots w_{n-1})$ we get
\beq\label{eq:generalized_Borel_2}
\sum_{k} c_k \overset{\Bor_n}{=} \int_0^\infty dw \, B_n(w) \sum_k \frac{c_k}{(k!)^n}w^k.
\eeq
where
\begin{align}
\label{eq:Borel_Bn}
B_n(w) &= \int_0^\infty \frac{dw_1}{w_1} \cdots \int_0^\infty \frac{dw_{n-1}}{w_{n-1}} \, \exp\[-w_1 \ldots -w_{n-1} - \frac{w}{w_1\cdots w_{n-1}}\]\\
&= \int_0^\infty \frac{dw_1}{w_1} \cdots \int_0^\infty \frac{dw_{n-k}}{w_{n-k}} \, \exp\[-w_1 \ldots -w_{n-k}\] B_k\(\frac{w}{w_1\cdots w_{n-k}}\)
\end{align}
which satisfy, for regularity,
\beq
(k!)^n = \int_0^\infty dw\,B_n(w)\,w^k.
\eeq
Explicit expressions of $B_n(w)$ for the first few $n=1,2$ are
\begin{align}
B_1(w) &= e^{-w}\\
B_2(w) &= 2K_0(2\sqrt{w})
\end{align}
but no explicit expressions in terms of common functions are available for higher orders.
Anyway, we can recast the integrals for a generic $n$ into a single integral
\beq\label{eq:Borel_Bn_Gamma}
B_n(w) = \frac{1}{2\pi i} \int ds\,w^{-s}\,\Gamma^n(s) = G^{n0}_{0n}(w;;;\underbrace{0,\ldots,0}_{n};)
\eeq
where the integration contour is the same as for a Mellin inversion
(passing to the right of $s=0$), and the function
is called a Meijer $G$ function.
The proof of Eq.~\eqref{eq:Borel_Bn_Gamma} can be easily obtained from Eq.~\eqref{eq:Borel_Bn}
by using the formula
\beq
e^{-x} = \frac{1}{2\pi i} \int ds\,x^{-s}\,\Gamma(s)
\eeq
to rewrite $\exp\[- \frac{w}{w_1\cdots w_{n-1}}\]$ and computing the $w_i$ integrals.
We may then recast the order $n$ Borel method in a third form
\beq\label{eq:generalized_Borel_3}
\sum_{k} c_k \overset{\Bor_n}{=} \frac{1}{2\pi i} \int ds\,\Gamma^n(s) \int_0^\infty dw \, w^{-s} \sum_k \frac{c_k}{(k!)^n}w^k
\eeq
which may be more convenient for some applications.
Indeed, whence Eqs.~\eqref{eq:generalized_Borel_1} and \eqref{eq:generalized_Borel_2}
are completely equivalent, this equation is somehow different because we have
swapped the two integrals. In particular, we have to specify the integration path:
after the $w$ integral some other $s$-singularities may arise, and we need a prescription
how the contour has to pass through them.\footnote{%
The correct choice seems to be that the path has to cross the real axis
to the right of the rightmost singularity of $\Gamma(s)$, i.e.\ $s=0$, and to the left
of any other singularity wich may appear out of the $w$ integral.}

Note that we may also consider the case in which $n=0$: in practice, it means that
we add to the original series a $w^k$ in each term, obtaining then a power series;
if this power series can be summed, the Borel sum consists in computing this $0$-order
Borel transform in $w=1$.
Formally, this is obtained with $B_0(w)=\delta(1-w)$, which can be also obtained
from the general expression Eq.~\eqref{eq:Borel_Bn_Gamma}.
This is, for example, the second way we used to find a sum of
the series Eq.~\eqref{eq:111}, see Eq.~\eqref{eq:111_z} and discussion there.

If a series is $\Bor_n$-summable it is also $\Bor_k$-summable, $\forall k>n$ (the same for $\Bor^*_n$).
Indeed, in order for a series to be $\Bor_n$-summable, the Borel transform $\hat s_k(w)$
for any $k>n$ must have an infinite radius of convergence (since $\hat s_n(w)$ has, at least,
a finite radius of convergence). Then, in the $\Bor_k$ expression for the sum,
one can swap back one of the integrals with the sum without problems
(because the series converges everywhere), obtaining a completely equivalent
expression: the integral can now be computed explicitly, giving the formal expression
for the $\Bor_{k-1}$ sum of the series. Since these manipulations are all
completely legal, the result will be the same: this completes the proof.

Therefore, it is useful to consider the minimal value of $n$ for which a series is $\Bor_n$-summable
(or $\Bor_n^*$-summable).
For a wide class of divergent series, it happens that if a series is minimally
$\Bor_n$-summable, then it is also minimally $\Bor^*_{n+1}$-summable.
This comes simply from the fact that, often, the Borel transform
$\hat s_n(w)$ has a finite radius of convergence, and then
straightforwardly $\hat s_n(w)$ has infinite radius of convergence.
However there are special cases in which the same $n$ is minimal
for both $\Bor$- and $\Bor^*$-summability, but they are not so common.

Note that this is valid for a divergent series; conversely, if a series is convergent,
it is for sure $\Bor_0$-summable, but it can be or not $\Bor^*_0$-summable, depending on
the radius of convergence of its $0$-order Borel transform $\hat s_0(w)$.
We now see some examples.

\subsection{Applications of the Borel method: some examples}
\label{sec:borel_examples}
\subsubsection{Example 0}

Consider as first example the series Eq.~\eqref{eq:111}.
With standard Borel method ($n=1$), we get
\beq
\hat s_1(w) = \sum_k\frac{(-w)^k}{k!} = e^{-w}
\eeq
with infinite radius of convergence, and then
\beq
s = \int_0^\infty dw\, e^{-w} e^{-w} = \frac12,
\eeq
again.
Higher order methods give, of course, the same result.

\subsubsection{Example 1a}

Consider the divergent series
\beq
s=\sum_{k=0}^\infty (-1)^k k!.
\eeq
Its Borel transform for $n=1,2,3$ is
\beq\label{eq:borel_transforms_k!}
\hat s_1(w) = \frac{1}{1+w}, \qquad
\hat s_2(w) = e^{-w}, \qquad
\hat s_3(w) = J_0\(2\sqrt{w}\),
\eeq
where the first transform has convergence radius $|w|<1$,
while the others have infinite convergence radius
($J_0(x)$ is a cylindrical Bessel function).
The Borel sums with $n=1,2$ are given by
\beq
s \overset{\Bor_1}{=} \int_0^\infty dw\,e^{-w} \frac{1}{1+w}, \qquad
s \overset{\Bor_2}{=} \int_0^\infty dw\,e^{-w} 2K_0(2\sqrt{w});
\eeq
as expected, the result is the same
\beq\label{eq:borel_example1a_res}
s = -e\,\Ei(-1) = 0.596347.
\eeq

\subsubsection{Example 1b}

Consider now the divergent series, closely related to that of the previous example,
\beq
s=\sum_{k=0}^\infty k!.
\eeq
Its Borel transform for $n=1,2,3$ is
\beq
\hat s_1(w) = \frac{1}{1-w}, \qquad
\hat s_2(w) = e^w, \qquad
\hat s_3(w) = I_0\(2\sqrt{w}\),
\eeq
where the first transform has again convergence radius $|w|<1$,
while the others have infinite convergence radius
($I_0(x)$ is a modified cylindrical Bessel function).
However, in this case, for every $n$ the Borel inversion intergal does not converge:
for $n=1$, because of a pole in the integration path, $w=1$, the others
for the bad behaviour at $w\to\infty$.

However, the standard method $n=1$
\beq
s \overset{\Bor_1}{=} \int_0^\infty dw\,e^{-w} \frac{1}{1-w}
\eeq
can still give a result, by deforming
the integration contour in the complex $w$-plane to avoid the pole in $w=1$;
nevertheless, the result has an ambiguity, given by the two possible
way of avoiding the pole (above or below).
The result is (a simple way to compute the integral is to deform the contour into
the two straight lines $0\to i\epsilon\to i\epsilon+\infty$)
\beq\label{eq:borel_sum_series_k!_1}
s \overset{\Bor_1}{=} e^{-1}\[\Ei(1) \pm i\pi \].
\eeq
Note that here, even if with $n=1$ the convergence radius of the Borel transform is finite,
we are nevertheless able to find a (even ambiguous) sum.

We could try to interpret the result of the $n\geq2$ method imagining that
in those case the pole appearing on the first approach is somehow pushed at infinity.
Indeed, for $n=2$, $e^w$ has an essential singularity at the complex infinity, so we might
try to do the same job as in the $n=1$ case by going to infinity in a different direction.
For example, we choose a contour on the positive imaginary axis, giving as result
\beq
-e\,\Ei(-1) + 1.32387\cdot10^{-8} i.
\eeq
Using instead the negative imaginary axis, we get the opposite sign in the imaginary part.
We may then quote as result
\beq
s \overset{\Bor_2}{=} -e\,\Ei(-1) \pm 1.32387\cdot10^{-8} i,
\eeq
which however differs a lot from the $n=1$: this may mean that the series is seriously
too divergent, and hence any attempt to give it a meaning is going to fail, or to give
meaningless results. We see in the next example which is the correct interpretation of these results.

\subsubsection{Example 2}

Of course, examples 1a and 1b are related, and we want to consider here their relation.
Then we concentrate our attention on the power series
\beq\label{eq:series_k!}
s(z)=\sum_{k=0}^\infty (-z)^k k!,
\eeq
which gives back the first two examples when $z=1$ and $z=-1$, respectively.
We can working on this series as in the examples above; for instance, the Borel transforms
are essentially the same as for example 1a, but with $w\to wz$:
\beq
\hat s_1(w) = \frac{1}{1+wz}, \qquad
\hat s_2(w) = e^{-wz}, \qquad
\hat s_3(w) = J_0\(2\sqrt{wz}\).
\eeq
Of course, the radius of convergence of such series depends on $z$.
Let's for a while forget about this radius and compute the integrals;
each method gives
\beq
s(z) = \frac1z \, \Gamma\(0,\frac1z\)\, \exp\frac1z 
\eeq
which reproduces the result Eq.~\eqref{eq:borel_example1a_res} for $z=1$.

\begin{figure}[t]
  \centering
  \includegraphics[width=0.49\textwidth]{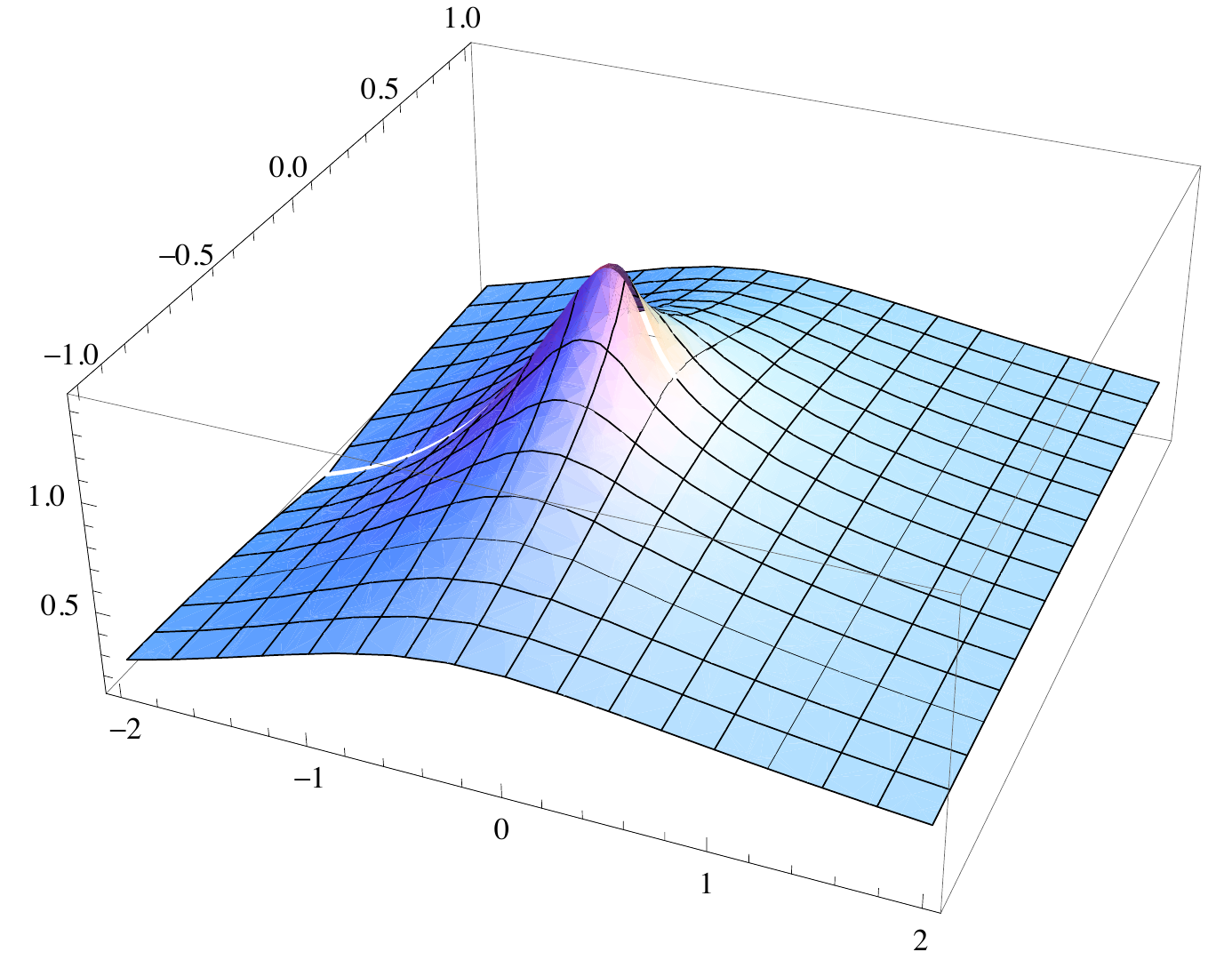}
  \includegraphics[width=0.49\textwidth]{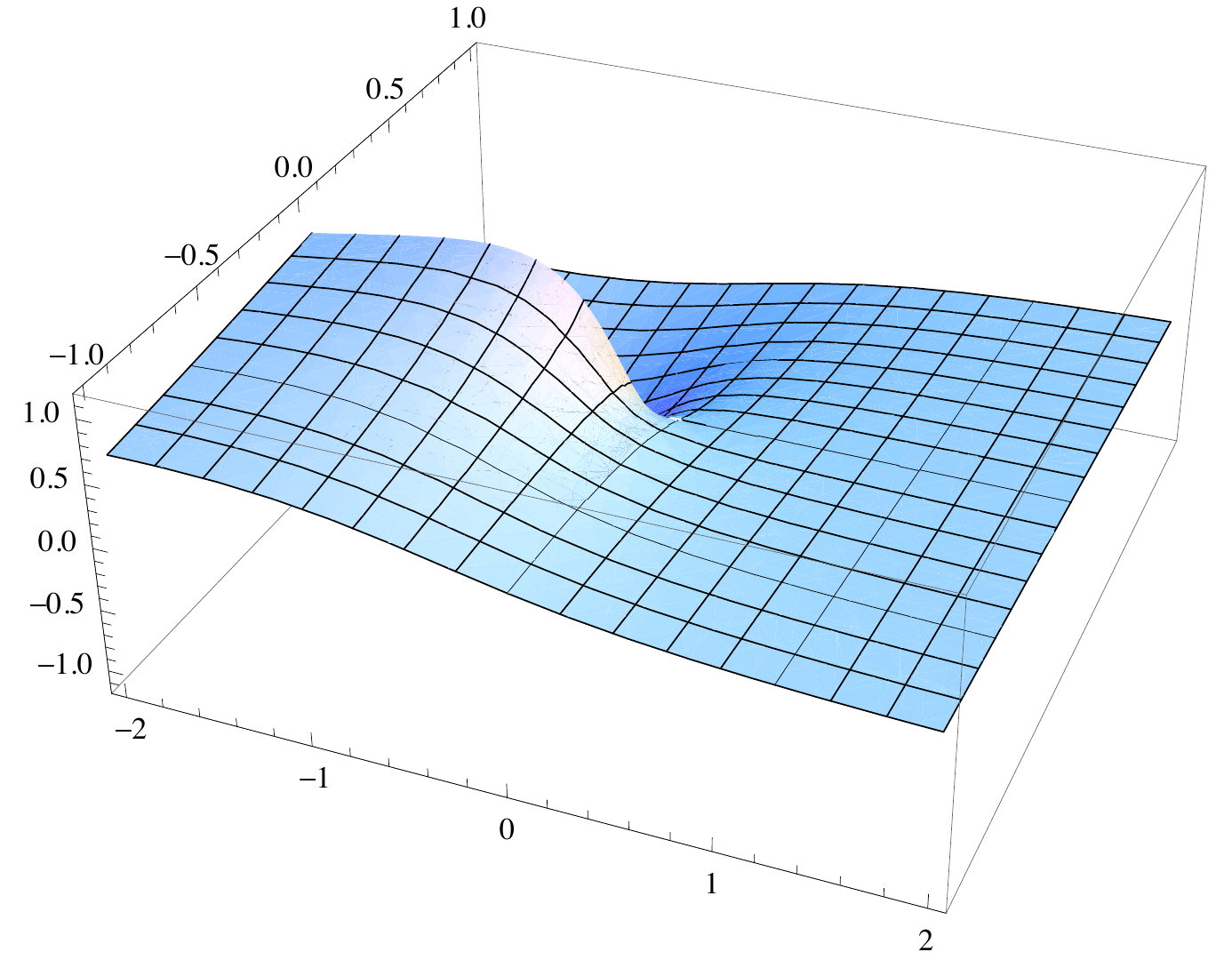}
  \caption{Real and imaginary part of the function $s(z)$, Eq.~\eqref{eq:borel_sum_series_k!}}
  \label{fig:borel_sum_series_k!}
\end{figure}

As a function of $z$ in the complex plane, $s(z)$ has a branch-cut along the negative real axis,
where the real part could be analytically continued but the imaginary part has a discontinuity,
see Fig.~\ref{fig:borel_sum_series_k!}.
In particular, the imaginary part along the negative real axis is given by
\beq\label{eq:series_k!_disc}
\Im s(-x+i0^+) = -\frac{\pi}{x} \,e^{-1/x},
\eeq
as one can easily find knowing that $\Gamma(0,-1/x)$ has a discontinuity of $2\pi$.

If one consider now the limit $z\to-1$, the result depends (because of the discontinuity)
if the point $z=-1$ is approached from imaginary part of $z$ positive or negative,
and hence it has an ambiguous imaginary part. The result is
\beq
s(-1) = e^{-1}\[\Ei(1) \pm i\pi \],
\eeq
which is exactly what we found in Eq.~\eqref{eq:borel_sum_series_k!_1} with the standard Borel method $n=1$.
This means that, in this case, the $n=1$ method gives the \emph{correct} result also if the
series is not $\Bor_1$-summable in the sense of Def.~\ref{def:Borel}.
Of course, this is not a case: indeed, for instance, the upper sign of the imaginary part is obtained
when approaching $z=-1$ from the lower half-plane, i.e.\ we have
\beq
s(-1-i0^+) = e^{-1}\[\Ei(1) + i\pi \].
\eeq
If one builds the divergent series for $z=-1-i0^+$, the Borel transform for $n=1$ becomes
\beq
\hat s_1(w) = \frac{1}{1-w(1+i0^+)} = \frac{1}{1-i0^+-w}
\eeq
which automatically include a prescription to circumvent the pole.
When $n=2$, the Borel transform is
\beq
\hat s_2(w) = e^{w(1+i0^+)},
\eeq
and again we automatically have a prescription to compute the inversion integral,
different from the one we tried to use before.
However, in this case, the integral cannot be computed even with the prescription,
remarking again that higher orders in this case don't work.

Note, however, that there is no hope to find a numerical method to automatically extract
the result for this sum: indeed, here we use analytic continuation, and before we deformed the contour
of Borel inversion, having already analytically computed the Borel transform.

Note that the function can be continued to $z=0$ giving as result
\beq\label{eq:borel_sum_series_k!}
s(0) = 1,
\eeq
as one would naively expect from the series (only the first term remains for $z=0$).
However, the function is not analytical in $z$, because of the presence of the branch-cut
from $z=0$ down to the negative real axis: indeed, the series \eqref{eq:series_k!}
has the form of the Taylor expansion of $s(z)$ around $z=0$, and the fact that the series
has zero radius of convergence (is divergent for all $z$) is related to the non-analyticity of $s(z)$ in $z=0$.
To exploit this relation, we have to show that because of the branch-cut the derivatives
of $s(z)$ in $z=0$ give the coefficients of the divergent series.
However, since $s(z)$ is not analytical in $z=0$, we cannot compute such derivatives
in a straightforward way.
The idea here is to use the Cauchy formula for derivatives
\beq
s^{(k)}(z) = \frac{k!}{2\pi i} \oint dx\, \frac{s(x)}{(x-z)^{k+1}}
\eeq
where the integration contour encircles the point $z$.
Then we can modify the contour to a Pac-Man contour eating the negative real axis,
and, if
\begin{itemize}
\item $z s(z)\to 0$ as $z\to0$ and
\item $s(z) \to 0$ as $z\to\infty$,
\end{itemize}
(as it is in the present case), the contribution from the circles
at infinity and around $z=0$ vanishes, leaving
\beq
s^{(k)}(z) = \frac{k!}{\pi} \int_{-\infty}^0 dx\, \frac{\Im s(x+i0^+)}{(x-z)^{k+1}}.
\eeq
We can then push this expression to be valid also in the limit $z=0$,
from which we can obtain the coefficients of the Taylor expansion:
\beq
c_k = \frac1{k!}s^{(k)}(0) = \frac{1}{\pi} \int_{-\infty}^0 dx\, \frac{\Im s(x+i0^+)}{x^{k+1}}.
\eeq
Using Eq.~\eqref{eq:series_k!_disc}, with a straightforward calculation we find
\beq
c_k = (-1)^k k!,
\eeq
which are exactly the coefficient of the power series Eq.~\eqref{eq:series_k!},
as expected.
Summarizing, this computation shows a deep connection between the non-analyticity
of a function and a divergent power series, and in particular it relates
the discontinuity along a cut of the non-analytic function to the large-$k$
behaviour of the terms of its power series.

\subsubsection{Example 3}

There are many other examples that may be built out of example 2 above, by choosing
different values of the variable $z$.
For example, let us choose $z=-i$ to obtain
\beq
s=\sum_{k=0}^\infty (-i)^k k! = \sum_{k=0}^\infty (-1)^k (2k)! + i \sum_{k=0}^\infty (-1)^k (2k+1)!.
\eeq
We know that the sums of the two series are, respectively, the real and the imaginary parts of
\beq
i \, \Gamma\(0,i\)\, e^i.
\eeq
However, separately, the two series are not $B_1$-summable, because the terms of the series
diverge more than $k!$. Of course, we know that this is due to the fact that we have separated
the two sums: the original sum is a special case of the one we have already
resummed before, Eq.~\eqref{eq:series_k!}.
As a toy exercise, we can then try to use $n=2$ Borel method. Concentrating on
the real part (the other series has the same properties), we have
\beq
\hat s_2(w) = \sum_{k=0}^\infty \frac{(2k)!}{(k!)^2} (-w)^k = \frac{1}{\sqrt{1+4w}}
\eeq
with convergence radius $\abs{w}<1/4$, and hence
\beq
s \overset{\Bor_2}{=} \int_0^\infty dw\, \frac{2K_0(2\sqrt{w})}{\sqrt{1+4w}} = 0.62145
\eeq
which is numerically equal to what expected.

\subsubsection{Example 4}

Now let's consider a more divergent series like
\beq
\sum (-1)^k (k!)^p,\qquad p>1.
\eeq
From the last examples, we immediately understand that the series is
$\Bor_p$-summable and $\Bor^*_{p+1}$-summable.
Its order $p$ and $p+1$ Borel transforms are trivially
\beq
\hat s_p(w) = \frac{1}{1+w},\qquad
\hat s_{p+1}(w) = e^{-w}.
\eeq
However, here the standard formulae Eqs.~\eqref{eq:generalized_Borel_1}
and \eqref{eq:generalized_Borel_2} are complicated because of the many integrals
to be computed.
Indeed, for general $p$, it's better to use Eq.~\eqref{eq:generalized_Borel_3}:
the $w$ integral gives, in the $p$ and $p+1$ cases respectively,
\beq
\frac{\pi}{\sin(\pi s)}, \qquad\text{and} \qquad
\Gamma(1-s),
\eeq
and the sum of the series is
\begin{align}
\sum (-1)^k (k!)^p
&\overset{\Bor_p}{=} \frac1{2i}\int ds\, \frac{\Gamma^p(s)}{\sin(\pi s)}\\
&\overset{\Bor_{p+1}}{=} \frac1{2\pi i}\int ds\, \Gamma^{p+1}(s) \Gamma(1-s),
\end{align}
and the two results are trivially related by the Euler reflection formula.
Note the presence of the poles for positive integer values of $s$: the integration path
has to cross the real axis in the region $0<s<1$.
Even if an analytical expression of the integrals can't be found, the numerical
result for any given $p$ is very easy to compute.

\subsection{Truncated Borel sum}
\label{sec:Borel_truncated}

Sometimes, even if a series is $\Bor_n$-summable, it is not possible to analytically
compute the sum of the Borel transform.
In these cases, it's impossible to use the (generalized) Borel method in a straightforward way,
since any truncation of the Borel transform series would reproduce a truncation of
the original divergent series.

\begin{figure}[th]
  \centering
  \includegraphics[width=0.49\textwidth,page=1]{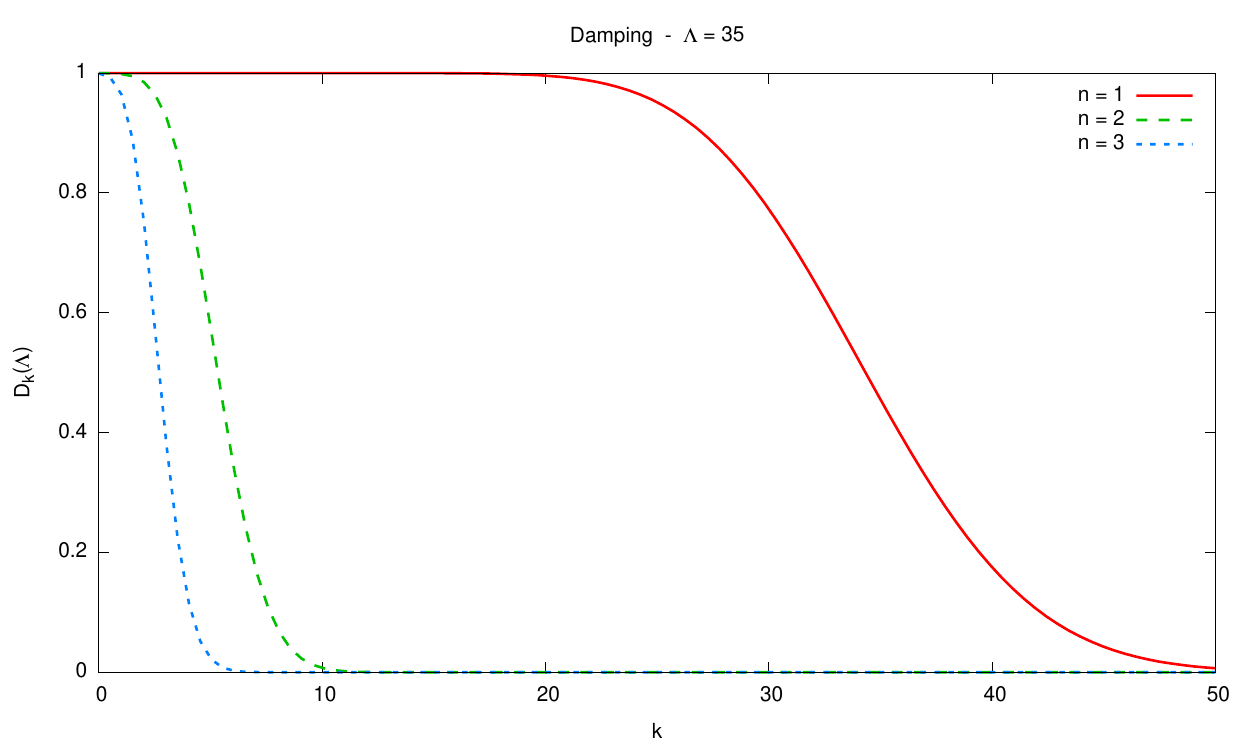}
  \includegraphics[width=0.49\textwidth,page=2]{plot_borel_damping.pdf}
  \caption{The function $D_k^{(n)}(\Lambda)$ defined in Eq.~\eqref{eq:borel_damping} for $n=1,2,3$.
    In the left plot $\Lambda=35$ for all curves, while in the right plot $\Lambda$ is chosen in such a way
    to have the medium damping point at the same value $k=10$.}
  \label{fig:borel_damping}
\end{figure}

A way to use the Borel method in such cases is to truncate the $w$ integral up to some
cutoff $\Lambda$: if the series is $\Bor_n$-summable, the integral is convergent,
and it means that the truncated integral can give an arbitrarily precise result
provided $\Lambda$ is large enough.
With the cutoff $\Lambda$, we have then
\begin{align}
  \sum_k c_k &\overset{{\rm B}_n}{=} \int_0^\infty dw\, B_n(w) \sum_k c_k \frac{w^k}{(k!)^n} \nonumber\\
  &\simeq  \int_0^\Lambda dw\, B_n(w) \sum_k c_k \frac{w^k}{(k!)^n} \nonumber\\
  &= \sum_k c_k\, D_k^{(n)}(\Lambda)\label{eq:Borel_cutoff}
\end{align}
where
\begin{align}
D_k^{(n)}(\Lambda) &= \frac{1}{(k!)^n} \int_0^\Lambda dw\, B_n(w)\, w^k \nonumber\\
&= \frac{1}{2\pi i} \int ds\, \frac{\Lambda^{k+1-s}}{k+1-s}\(\frac{\Gamma(s)}{k!}\)^n
\label{eq:borel_damping}
\end{align}
is a damping factor for $k$, showed in Fig.~\ref{fig:borel_damping};
the second row is obtained using Eq.~\eqref{eq:generalized_Borel_3}.
If this damping is enough to suppress the divergent terms in the series,
we can truncate the series at some $k=K$, and the subsequent terms would give
smaller and smaller contributions. Then we are again able to have a result as accurate
as needed by chosing $K$ large enough.
We have then
\begin{align}\label{eq:Borel_truncated}
\sum_k c_k &\simeq \sum_k^K c_k\, D_k^{(n)}(\Lambda)\nonumber\\
&= \int_0^\Lambda dw\, B_n(w) \sum_k^K c_k \frac{w^k}{(k!)^n}\nonumber\\
&= \frac{1}{2\pi i} \int ds\, \Gamma^n(s) \sum_k^K c_k \frac{\Lambda^{k+1-s}}{(k+1-s) (k!)^n}
\end{align}
which can be used to evaluate the Borel sum of a divergent series
to arbitrary precision by chosing appropriate values for $\Lambda$ and $K$.
In particular, using the last row, always just one single integral has to be computed,
independently on the order $n$.

We emphasize that for the applicability of Eq.~\eqref{eq:Borel_truncated}
there is the strong requirement that the damping be enough to cure the divergence of the series,
that is, the series in the last line of Eq.~\eqref{eq:Borel_cutoff} has to be convergent.
This is related to the radius of convergence of the Borel transform: indeed,
between the second an third lines of Eq.~\eqref{eq:Borel_cutoff} we exchanged the integral
and the sum, and this can be done provided the sum is absolutely convergent for all
values of $w$ in the integration range $0\leq w\leq \infty$, that is to say that the Borel transform
$\hat s_n(w)$ has infinite radius of convergence.
We conclude that the truncated Borel method Eq.~\eqref{eq:Borel_truncated}
applies only to those divergent series which are, according to Def.~\ref{def:Borel},
$\Bor^*_n$-summable.
Fortunately, if a series is not $\Bor^*_n$-summable for a given $n$,
there are chances that it is for a larger $n$.

\subsubsection{Damping factor}

The damping factor is, for $n=1$,
\beq\label{eq:damping1}
D_k^{(1)}(\Lambda) = 1 - \frac{\Gamma(k+1,\Lambda)}{\Gamma(k+1)}
\eeq
where $\Gamma(k+1,\Lambda)$ is the incomplete Gamma function.
It can be computed directly (and easily) from the integral in the first row of Eq.~\eqref{eq:borel_damping};
otherwise, we can use the second row formulation, deform the contour to encircle
the poles of $\Gamma(s)$, obtaining
\beq
D_k^{(1)}(\Lambda) = \frac{1}{k!} \sum_{j=0}^\infty \frac{(-1)^j \Lambda^{k+1+j}}{(k+1+j) j!}
\eeq
which gives the same result.

The same kind of double computation may be done for $n=0$: indeed,
recalling that $B_0(w)=\delta(1-w)$, from the integral we get
\beq
D_k^{(0)}(\Lambda) = \thetaH(\Lambda-1).
\eeq
This is not a damping, but this is expected, since $n=0$ does not manipulate the original series:
it simply means that, provided $\Lambda>1$, the cutoff is ineffective,
while $\Lambda<1$ is meaningless (too small).
With the second formulation, the same result is obtained by noting that the only pole
is at $s=k+1$, with residue $1$: then, when $\Lambda>1$, we can close the path on the right,
where the contribution is given only by the pole, and we get $1$,
while for $\Lambda<1$ we can deform the path to the left where there are no singularities,
and we get $0$.

In the case of $n=2$ we have instead a complicated analytical expression
in terms of hypergeometric functions; for practical purposes it's easier to
numerically compute the integral. For $n=3$ or higher no analytic results
are available.

\begin{figure}[th]
  \centering
  \includegraphics[width=0.6\textwidth,page=3]{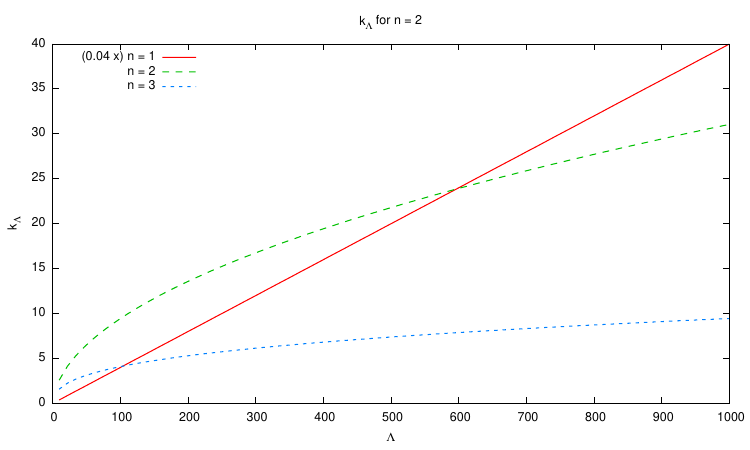}
  \caption{$k_\Lambda^{(n)}$ as defined in Eq.~\eqref{eq:k_Lambda} for $n=1,2,3$.}
  \label{fig:k_Lambda}
\end{figure}

The approximation Eq.~\eqref{eq:Borel_truncated} is good provided $K$
is larger than the point where the damping starts to suppress 
the terms in the series. Let's call such a point $k_\Lambda^{(n)}$.
Of course it may depend also on the series, hence it's impossible to have a general expression for it.
Nevertheless, as one can see from Fig.~\ref{fig:borel_damping}, there is a relatively small
region in which the damping factor goes from being approximately $1$ to be approximately $0$
(the larger $n$, the smaller this region).
Hence, $k_\Lambda^{(n)}$ cannot be too far from the medium point of this damping:
then, to have a precise definition of $k_\Lambda^{(n)}$, let's define it as the value
\beq\label{eq:k_Lambda}
D_{k_\Lambda^{(n)}}^{(n)}(\Lambda) = \frac12.
\eeq
For $n=1$, we have (numerically)
\beq
k_\Lambda^{(1)} \simeq \Lambda
\eeq
For $n=2,3$ see Fig.~\ref{fig:k_Lambda}.

\subsubsection{Example}

Let's consider again the divergent series discussed in the example 1a,
\beq
s=\sum_{k=0}^\infty (-1)^k k!.
\eeq
We have seen that this series is minimally $\Bor_1$-summable and $\Bor^*_2$-summable.
Hence, we expect that the truncated Borel method can be used only for $n\geq2$.
\begin{figure}[th]
  \centering
  \includegraphics[width=0.9\textwidth,page=4]{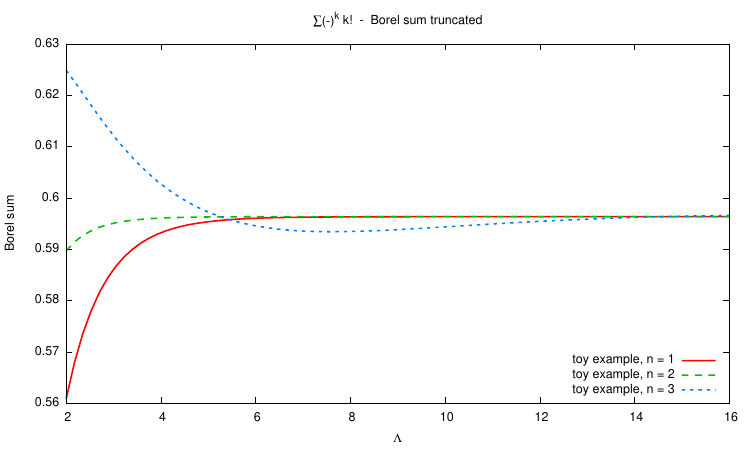}
  \caption{Dependence of Eq.~\eqref{eq:borel_series_k!_cutoff} on the cutoff $\Lambda$.}
  \label{fig:borel_series_k!_Lambda}
\end{figure}
To begin with, we show in Fig.~\ref{fig:borel_series_k!_Lambda} the dependence
on the cutoff $\Lambda$ for $n=1,2,3$ of
\beq\label{eq:borel_series_k!_cutoff}
\int_0^\Lambda dw\, B_n(w) \,\hat s_n(w)
\eeq
in the case in which we still use the exact (non-truncated) Borel transforms Eq.~\eqref{eq:borel_transforms_k!}.
The faster convergence is obtained for $n=2$, which is the minimal value
for $\Bor^*_n$-summability. We will consider this fact as an hint that the minimal
$n$ for which a series is $\Bor^*_n$-summable is the best one for numerical applications.
Nevertheless, at this level also the other values of $n$ can be used as well.
In particular, for $n=3$, we need to go to much larger values of $\Lambda$ (outside the plot range)
to obtain the same accuracy as with $n=1,2$;
this can be easily understood by looking at Fig.~\ref{fig:borel_damping} or Fig.~\ref{fig:k_Lambda},
where we see that the damping for $n=3$ is much stronger and we need larger $\Lambda$ to
effectively include the first few terms in the sum (if the included terms are too few,
the asymptotic behaviour of the coefficients is not probed and
the method cannot accurately guess the correct sum).

\begin{figure}[th]
  \centering
  \includegraphics[width=0.9\textwidth,page=6]{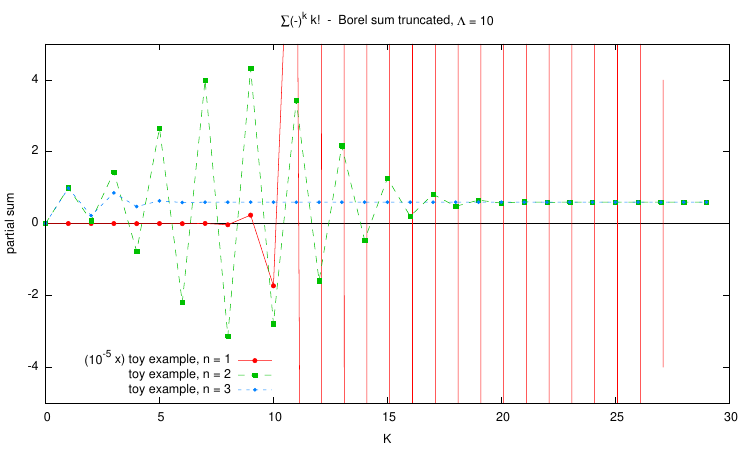}
  \caption{Behaviour of the partial sums Eq.~\eqref{eq:borel_series_k!_cutoff_exchanged} as a function of $K$ for $\Lambda=10$.}
  \label{fig:borel_series_k!_cutoff_terms}
\end{figure}
Now, let's fix $\Lambda=10$, and look at the partial sums in the series obtained
exchanging sum and integral in Eq.~\eqref{eq:borel_series_k!_cutoff}:
\beq\label{eq:borel_series_k!_cutoff_exchanged}
\sum_k^K (-1)^k  k!\, D_k^{(n)}(\Lambda).
\eeq
These are shown in Fig.~\ref{fig:borel_series_k!_cutoff_terms} for $n=1,2,3$.
We see immediately from this plot that method $n=1$ has no chances to work
at the truncated level, as expected.
The other two converge rapidly to the appropriate value given $\Lambda=10$,
which can be read out of the plot in Fig.~\ref{fig:borel_series_k!_Lambda}.

\subsubsection{On the domain of analyticity}

Borel summation extends the convergence radius of a power series:
for divergent series (radius zero) it may find a sum for values of the variable $z$
in some domain, but also if the series is convergent with finite radius
it may extend this convergence domain.

Let's see this with an example. Consider the geometric series
\beq
\sum_k (-z)^k = \frac{1}{1+z},\qquad \abs{z}<1;
\eeq
the convergence radius $1$ means that any truncated sum would
converge to the sum if $z$ is within the convergence circle, and would
have no limit if $z$ is outside. Of course, having an analytical result
for the sum, it can be extended analytically to the whole complex plane
but $z=-1$.

Using a Borel of order $n=1$, we get
\begin{align}
\sum_k (-z)^k
&= \int_0^\infty dw \, e^{-w} \sum \frac{(-zw)^k}{k!}\nonumber\\
&= \int_0^\infty dw \, e^{-w(1+z)}\nonumber\\
&= \frac{1}{1+z},\qquad \Re z>-1,\label{eq:trunc_ex_anal_ext}
\end{align}
which converges in the whole half-plane $\Re z>-1$.
This means that, provided $\Re z>-1$,
the integral can be computed numerically, or equivalently that
cutoffing the integral to an upper value $\Lambda$ the limit
$\Lambda\to\infty$ can be safely taken only for $\Re z>-1$.
Also in this case, of course, being the sum the same as before,
the result can be analytically extended everywhere but in $z=-1$.

We are now interested in what happens when considering a truncated Borel sum.
We then consider
\beq
\sum_k (-z)^k \simeq \sum_k^K (-z)^k D_k^{(1)}(\Lambda);
\eeq
strictly speaking, this series has still convergence radius $1$, as one
can see immediately with the square ratio test:
\beq 
\frac1\rho = \lim_{k\to\infty}\frac{D_{k+1}^{(1)}(\Lambda)}{D_k^{(1)}(\Lambda)}
= \lim_{k\to\infty}\frac{\gamma(k+2,\Lambda)}{(k+1)\,\gamma(k+1,\Lambda)} = 1.
\eeq
This is a consequence of the fact that, after truncating the series, the
integral and the sum has been exchanged back to their original position;
otherwise stated, the extension in the analyticity domain in Eq.~\eqref{eq:trunc_ex_anal_ext}
is achieved thanks to the fact that the Borel transform is explicitly summed
and analytically continued.
However, for any given finite $\Lambda$, when $\abs{z}>1$ one could treat the
sum as an asymptotic expansion, and find the optimal value of $K$
which stabilizes the sum.
Then, at least for $\Lambda$ reasonably small, the truncated Borel sum
could give an approximation of the result even outside the convergence region.

\subsection{Pad\'e approximants and the Borel-Pad\'e method}
\label{sec:Borel-Pade}

A very powerful method to compute numerically the sum of a divergent series
consists in mixing the use of a minimal Borel method with a Pad\'e approximant \cite{borel1}.

Indeed, as discussed above, when a series is minimally $\Bor_n$-summable
the Borel transform has a finite radius of convergence. Hence, numerically,
any truncation of the series cannot reproduce the Borel transform outside
the convergence radius, while it would be required for the Borel integral.
Here comes the idea to use a Pad\'e approximant to the Borel transform.

A Pad\'e approximant to a function $f(z)$ is a ratio of two polynomials
\beq
\[p/m\]_f(z) = \frac{a_0+a_1z+\ldots+a_p z^p}{1+b_1z+\ldots +b_m z^m}
\eeq
of order $p$ and $m$, respectively, such that its Taylor expansion
corresponds up to the $(p+m)$-th order to that of $f(z)$:
\beq
\[p/m\]_f(z) - f(z) = \Ord(z^{p+m+1}).
\eeq
This equation can be easily solved multiplying $f(z)$ by the denominator
\beq
a_0+a_1z+\ldots+a_p z^p = \(1+b_1z+\ldots +b_m z^m\) f(z) + \Ord(z^{p+m+1})
\eeq
and equating term by term; one gets a system of equations to express the $p+m+1$ coefficients
$a_k,b_k$ in terms of the first $p+m+1$ coefficients $c_k$ of the Taylor expansion
of $f(z)$:
\begin{align*}
  a_0&= c_0\\
  a_1&= c_1 + c_0 b_1\\
  a_2&= c_2 + c_1 b_1 + c_0 b_2\\
  &\vdots\\
  a_p&= c_p + c_{p-1} b_1 + \ldots + c_0q_p\\
  0  &= c_{p+1} + c_p b_1 + \ldots + c_{p-m+1}q_m\\
  &\vdots\\
  0  &= c_{p+m} + c_{p+m-1} b_1 + \ldots + c_{p}q_m
\end{align*}
(we intend that if the index of a coefficient is less than zero the coefficient is zero).
Actually, the system splits into two pieces: the second set of equations are
$m$ equations for the $m$ unknowns $b_1,\ldots,b_m$, and we can solve it first.
Then, with the $b_k$, we can compute the $a_k$ directly using the first set of $p+1$ equations.

The Borel-Pad\'e method can then be formulated in this way:
\beq
s\overset{\Bor_n}{=}\int_0^\infty dw\,B_n(w)\,\[p/m\]_{\hat s_n}(w)
\eeq
where $\[p/m\]_{\hat s_n}(w)$ is the Pad\'e approximant of the Borel transform $\hat s_n(w)$.
Since the Pad\'e approximant has a polynomial at the denominator, it will produce poles in the
complex plane. It has been observed \cite{borel1} that for diagonal ($p=m$) or quasi-diagonal ($p+1=m$)
Pad\'e's such poles reproduce quite well the poles in the Borel transform, and, if the Borel transform has
a branch-cut, its Pad\'e approximant mimics the cut with a sequence of poles along it.
This would mean that, using a Pad\'e approximant to the Borel transform, even few terms in
the series (which, remember, has finite radius of convergence) can actually provide
a good approximation even outside the convergence radius.
Moreover, where some poles lie on the real axis, by deforming the integral path to
avoid the singularities one can obtain \cite{borel2} the correct (ambiguous) sum
(see Example 1b of Sect.~\ref{sec:borel_examples});
note that increasing the order doesn't help in this case (see the same example),
meaning that this method is more powerful than the truncated method described in
Sect.~\ref{sec:Borel_truncated}.

Numerically, a very efficient and fast way for computing a diagonal or quasi-diagonal
Pad\'e approximant is the Wynn epsilon algorithm (for a review, see \cite{Weniger}).
However, it computes the full approximant at a given $z$, and not the coefficients,
so it is not convenient to use it here, since the result must be integrated (Borel integral).
Hence, for practical purposes of computing Borel sums using few coefficients of the series,
it is better to solve the system once for all in order to have all the coefficients available,
and then we compute the Borel integral.

\chapter{Special functions}

\minitoc

\noindent
In this Appendix we collect properties and relation of some special functions which appear
in the text. We will mainly concentrate on those properties which are useful
in the context of resummation. This Appendix is \emph{not} intended as a complete overview on the
subject.

\section{Euler Gamma and related functions}
\label{sec:Gamma}

The Euler Gamma function is defined as
\beq\label{eq:Gdef}
\Gamma(z) \equiv \int_0^\infty dt \, e^{-t} \, t^{z-1};
\eeq
the integral converges in the half-plane $\Re z>0$.
The Gamma function is a real function, i.e.\ it satisfies
\beq
\Gamma(\bar z) = \overline{\Gamma(z)};
\eeq
in particular, $\Im \Gamma(x) = 0$ for $x\in \mathbb{R}$.
Integrating by parts, it is easy to show that $\Gamma(z)$ satisfy the recursion relation
\beq\label{eq:Grec}
\Gamma(z+1) = z\, \Gamma(z).
\eeq
Then, because additionally $\Gamma(1)=1$, for $z=n\in\mathbb{N}$ we have
\beq
\Gamma(n+1) = n!
\eeq
extending the factorial to complex values.
Eq.~\eqref{eq:Grec} in reverse allows to analytically extend the Gamma function to the whole
complex plane, apart from some singular points.
Indeed, the Gamma function has poles for non-positive integer values of its argument:
more precisely, for $n\in\mathbb{N}$ $\Gamma(-n)$ has a simple pole with residue
\beq
\Res{z=-n}\Gamma(z) = \frac{(-1)^n}{n!}.
\eeq
Around one of such poles, the Gamma function satisfies the expansion
\beq
\Gamma(z-n) = \frac{(-1)^n}{n!} \left[ \frac{1}{z} + \psi(n+1) + \Ord(z) \right]
\eeq
where $\psi$ is the DiGamma function, defined in Eq.~\eqref{eq:Gpsi}.

The Gamma function satisfies the so called \emph{Euler reflection formula}
\beq
\Gamma(z)\, \Gamma(1-z) = \frac{\pi}{\sin(\pi z)},
\eeq
which is useful to relate the region of convergence of the integral in Eq.~\eqref{eq:Gdef}
to the other half-plane; also the pole structure is evident from such formula.

At large $\abs{z}$, for $\abs{\arg z}<\pi$, the Gamma function has the asymptotic expansion
\beq\label{eq:GammaStirling}
\Gamma(z) = e^{-z} z^{z-\frac12} \sqrt{2\pi} \[1+\frac1{12z}+\frac1{288z^2} + \ldots\]
\eeq
which reduces to the well known Stirling approximation keeping only the first term.
It has to be noticed that even the Stirling approximation is very precise also for small values of $z$:
for example, in the Stirling approximation we have
\beq
\Gamma(2) \simeq 0.96\ldots
\eeq
which is very close to the exact value $\Gamma(2)=1$.

The inverse Gamma function
\beq\label{eq:delta}
\Delta(z) \equiv \frac{1}{\Gamma(z)}
\eeq
is an entire function, which satisfies the recursion relation
\beq
\Delta(z) = z\, \Delta(z+1).
\eeq
An integral representation is given by
\beq
\Delta(z) = \frac{1}{2\pi i}\int_{c-i\infty}^{c+i\infty} ds\, e^{s} \, s^{-z}.
\eeq
From the recursion relation we can derive a relation for its derivatives,
\beq
\Delta^{(k)}(z) = z \, \Delta^{(k)}(z+1) + k\, \Delta^{(k-1)}(z+1)
\eeq
from which it follows, in particular, for $z=0$
\beq\label{eq:Delta_rec_rel_0}
\Delta^{(k)}(0) = k\, \Delta^{(k-1)}(1).
\eeq

\subsection{PolyGamma functions}

The logarithmic derivative of the Gamma function is usually called DiGamma function:
\beq\label{eq:Gpsi}
\psi(z) \equiv \frac{d}{dz}\log\Gamma(z).
\eeq
Form the recursion relation \eqref{eq:Grec} it follows that
\beq\label{eq:Gpsi_rec}
\psi(z+1) = \psi(z) + \frac{1}{z};
\eeq
for a positive integer $n$, we can iterate the recursion to obtain
\beq\label{eq:Gpsi_n+1}
\psi(n+1) = \psi(1) + 1+\frac{1}{2}+\frac{1}{3}+\ldots+\frac{1}{n}.
\eeq
The value of $-\psi(1)$ is called the Euler-Mascheroni constant and is
\beq
\gammae \equiv -\psi(1) = 0.577216\ldots
\eeq
Higher order derivatives, the PolyGamma functions, are usually indicated as
\beq
\psi_n(z) \equiv \psi^{(n)}(z) = \frac{d^{n+1}}{dz^{n+1}}\log\Gamma(z);
\eeq
with this notation, the DiGamma function is also indicated as $\psi_0$.
Derivating Eq.~\eqref{eq:Gpsi_rec} we obtain the recursion relation
\beq
\psi_n(z+1) = \psi_n(z) + (-1)^n\, n!\,\frac{1}{z^{n+1}}.
\eeq
From this relation one can write a Laurent expansion of $\psi_n(z)$
around $z=0$,
\beq\label{eq:psi_n_Laurent}
\psi_n(z) = \frac{(-)^{n+1}n!}{z^{n+1}} + \sum_{k=0}^\infty \frac{\psi_{n+k}(1)}{k!} z^k,
\qquad \psi_k(1) = (-)^{k+1} k! \zeta(k+1)
\eeq
where $\zeta$ is the Riemann Zeta function defined in Sect.~\ref{sec:RiemannZeta}.

An asymptotic expansion at large $\abs{z}$, for $\abs{\arg z}<\pi$, is given by
\begin{subequations}\label{eq:psi0_asympt}
\begin{align}
\psi(1+z)
&=\log z + \frac{1}{2z} + \sum_{j=1}^{\infty}\frac{\zeta(1-2j)}{z^{2j}}\\
&=\log z + \frac{1}{2z} + 2\sum_{j=1}^{\infty}\frac{(-1)^k (2k-1)! \zeta(2j)}{(2\pi z)^{2j}},
\end{align}
\end{subequations}
where in the second equality we have used Eq.~\eqref{eq:zeta_reflection_integer}.
By shifting the argument of the logarithm, we can mimic the first subleading term
in the expansion, obtaining the approximation
\beq\label{eq:digamma_log_approx}
\psi(1+z) = \log\(\frac12 + z\) + \Ord\(\frac{1}{z^2}\),
\eeq
which is surprisingly good even for small $z$ (apart along the negative real axis,
where the log has a cut and the DiGamma a sequence of poles).
All the PolyGamma $\psi_k$ with $k\geq1$ vanish as $1/z^k$ at large $\abs{z}$, $\abs{\arg z}<\pi$.

\subsection{Generalized Gamma functions}
The incomplete Gamma function (or plica function) is defined as
\beq
\Gamma(z,a) \equiv \int_a^\infty dt \, e^{-t} \, t^{z-1};
\eeq
it ha a branch-cut on the negative real axis in the $a$ complex plane.
Its complement to the Gamma function is sometimes called the truncated Gamma function,
\beq
\gamma(z,a) \equiv \int_0^a dt \, e^{-t} \, t^{z-1},
\eeq
and of course
\beq
\Gamma(z) = \Gamma(z,a) + \gamma(z,a).
\eeq
For $z=k+1$ integer, integrating repeatedly by parts we get
\beq
\frac{\Gamma(k+1,a)}{\Gamma(k+1)} = e^{-a} \sum_{n=0}^{k}\frac{a^n}{n!},
\eeq
where $\Gamma(k+1)=k!$.

Another useful function is the so called Beta function, defined by
\beq
B(a,b) \equiv \int_0^1 dx \;x^{a-1} (1-x)^{b-1},
\eeq
which can be written in terms of Gamma functions as
\beq
B(a,b) = \frac{\Gamma(a)\,\Gamma(b)}{\Gamma(a+b)}.
\eeq

\section{Riemann Zeta function}
\label{sec:RiemannZeta}

A commonly appearing function is the so called Riemann Zeta function, defined bi the series
\beq\label{eq:zeta_series}
\zeta(s) = \sum_{k=1}^\infty k^{-s}.
\eeq
The series converges for $\Re s>1$, but the function can be analytically extended to the whole complex plane,
apart from $s=1$, where it reduces to the divergent harmonic series.
The analytic continuation is based on the reflection formula
\beq
\zeta(1-s) = 2(2\pi)^{-s}\cos\(\frac{\pi s}{2}\)\Gamma(s)\zeta(s);
\eeq
in particular we can relate the Zeta function for values of its argument less than $1$
to the Zeta function computed values where the series \eqref{eq:zeta_series} converges.
For positive integers $s=k>0$, we have two cases:
when $k=2j+1$ is odd, the cosine vanishes and we get
\beq
\zeta(-2j) = 0, \qquad j\in\mathbb{N}, j>0,
\eeq
while when $k=2j$ is even, we have
\beq\label{eq:zeta_reflection_integer}
\zeta(1-2j) = \frac{2(2j-1)!}{(2\pi)^{2j}} \zeta(2j), \qquad j\in\mathbb{N}, j>0.
\eeq
An integral expression for the Zeta function valid for real $s>1$ is
\beq
\zeta(s) = \frac{1}{\Gamma(s)} \int_0^\infty dt\, \frac{t^{s-1}}{e^t-1};
\eeq
by expanding the denominator as a geometric series and exchanging the integral and the sum
we get back the series definition Eq.~\eqref{eq:zeta_series}.

Special values of the Zeta function are
\begin{align}
  \zeta(2) &= \frac{\pi^2}{6} \\
  \zeta(3) &= 1.2020569\ldots \\
  \zeta(4) &= \frac{\pi^4}{90}.
\end{align}
In the text, the values of the Zeta function for integer arguments are usually
indicated as
\beq
\zeta_k \equiv \zeta(k).
\eeq

\section{Hypergeometric and related functions}
\label{sec:hypergeometric}

The hypergeometric functions are a class of very general functions, which reduces
to simpler special functions in various cases.
The most general definition can be given in term of the power series
\beq
{}_pF_q(a_1,\ldots,a_p; b_1,\ldots,b_q; z) =
\sum_{k=0}^\infty \frac{{(a_1)}_k \cdots {(a_p)}_k}{{(b_1)}_k \cdots {(b_q)}_k} \, \frac{z^k}{k!}
\eeq
where ${(a)}_k$ is the Pochhammer symbol
\beq
{(a)}_k = \frac{\Gamma(a+k)}{\Gamma(a)} = a(a+1)\cdots (a+k-1).
\eeq
When $p>q+1$, the series has zero radius of convergence, but with some
analytical relations a meaning can be given also to these functions,
and the power series serves as an asymptotic expansion.

To see some examples, the easiest case is given by $p=q=0$,
for which
\beq
{}_0F_0(;;z) = e^z.
\eeq
We discuss in the following other special cases which appear in the text.

\subsection{Airy functions}
\label{sec:Airy}
The (regular) Airy function can be written in terms of hypergeometric functions as
\beq
\Ai(z) = \frac{1}{3^{2/3}\Gamma(2/3)}\, {}_0F_1\(;\frac23;\frac19 z^3\) - \frac{z}{3^{1/3}\Gamma(1/3)}\, {}_0F_1\(;\frac43;\frac19 z^3\)
\eeq
and is one of the two linearly independent solutions of the Airy equation
\beq
f''(z) - z\,f(z) = 0.
\eeq
It is an entire function, with infinitely zeros on the negative real axis. The first zero (closest to the origin)
is located at $z=-2.33811$.

\subsection{Bateman functions}
\label{sec:Bateman}
The Bateman function $K_\nu(z)$ is a solution of the Bateman equation
\beq
-K_\nu''(z) + \(1-\frac{\nu}{z}\)K_\nu(z) =0.
\eeq
It can be expressed in integral form and written in terms of hypergeometric functions as
\begin{align}
K_\nu(z) &= \frac{2}{\pi} \int_0^{\pi/2} d\theta\;\cos\(z\tan\theta -\nu\theta\) \\
&= \frac{e^{-z}}{\Gamma\(1+\frac\nu2\)} U\(-\frac{\nu}{2},0,2z\)
\end{align}
where $U(a,b,z)$ is the \emph{confluent hypergeometric function of the second kind}
\begin{align}
U(a,b,z) &= z^{-a} \,{}_2F_0(a,1+a-b;;-z^{-1})\\
&= \frac{\pi}{\sin(\pi b)}\[ \frac{{}_1F_1(a;b;z)}{\Gamma(a-b+1) \Gamma(b)} - \frac{z^{1-b}\, {}_1F_1(a-b+1;2-b;z)}{\Gamma(a) \Gamma(2-b)} \]
\end{align}
and satisfies
\beq
U(a,b,z) = z^{1-b} \,U(a+1-b,2-b,z).
\eeq
The first form for exists only as a formal power series, but can be used
as an asymptotic expansion.

The $z$-derivative is related to the function itself:
\beq
\frac{d}{dz}U(a,b,z) = -a \,U(a+1,b+1,z).
\eeq
From this, the logarithmic derivative of the Bateman function can be written as
\beq
\frac{d}{dz}\log K_\nu(z) = -1 + \nu \frac{U\(1-\frac{\nu}{2},1,2z\)}{U\(-\frac{\nu}{2},0,2z\)}.
\eeq

\chapter{Numerical methods}

\minitoc

\noindent
In this Appendix we collect some numerical method used in developing
the codes used in the text.

\section{Numerical derivatives}
Numerical derivatives are usually approximated to the finite differences
\begin{align}
f'(x) &\sim \frac{f(x+h) - f(x-h)}{2h}\\
f''(x) &\sim \frac{f(x+h) + f(x-h) - 2f(x)}{h^2}
\end{align}
where $h$ is a ``small'' parameter: the smaller $h$ is,
the better should be the approximation. However, if $h$ is too small,
a roundoff error probably takes place.
Then, a better way to compute the derivarive is to compute the finite differences
for different values of $h$ not too small, and then use a polynomial interpolation to $h=0$.
Using 3 values of $h$ we can use a quadratic interpolation, and get as a best estimate
\beq
f'(x) \sim \frac{1}{h_2-h_1} \[ h_2\frac{h_1y_0 - h_0y_1}{h_1-h_0} - h_1\frac{h_2y_0 - h_0y_2}{h_2-h_0} \]
\eeq
where
\beq
y_i = \frac{f(x+h_i) - f(x-h_i)}{2h_i},\qquad i=0,1,2.
\eeq
If $h_i$ are chosen as
\beq
h_i = \frac{h_0}{2^i}
\eeq
we get the easier expression
\beq
f'(x) \sim \frac{y_0-6y_1+8y_2}{3}.
\eeq
The same technique also applies for the second derivative.

\section{Root finding}
For the computation of the inverse function which defines $\chi_s$,
Eq.~\eqref{eq:chis_def}, and for putting on-shell the off-shell kernels,
we use a root finding algorithm.

If we know the analytic derivative, the Newton algorithm is quite good (even if not so fast),
and moreover it is applicable also in the complex plane.

However, typically the analytic derivative is not known (or hard to work out)
and we use the secant method, which uses finite differences instead of the
actual derivative. The secant method is also valid in the complex plane,
but it is not much robust, and quite accurate initial guesses are needed
in order for the algorithm not to fail.

In the (rare) cases in which we just need tho find a root in the real axis
the falsi regula method works pretty well, since it is fast and robust.
However, the two initial guesses must surround the root, and this may
be not simple to realize.

We do not discuss these algorithms here, since the Reader can find
very accurate description of them throughout the web.

\section{Chebyshev polynomials}
\label{app:cheb}

In this Appendix we recall the definition and the main properties
of Chebyshev polynomials. The Chebyshev polynomials
\beq 
T_i(z) = \sum_{k=0}^i T_{ik}\, z^k,
\eeq
are defined in the range $z\in[-1,1]$ recursively by
\begin{subequations}\label{eq:cheb}
\begin{align}
T_0(z) &= 1 \\
T_1(z) &= z \\
T_2(z) &= 2z^2-1 \\
T_i(z) &= 2z\,T_{i-1}(z) -T_{i-2}. \label{eq:recursive}
\end{align}
\end{subequations}
A generic function $G(u)$ can be approximated
in the range $u\in[u_{\rm min}, u_{\rm max}]$ by its
expansion on the basis of Chebyshev
polynomials~\eqref{eq:cheb}, truncated at some order $n$:
\beq
G(u) \simeq -\frac{c_0}{2} +\sum_{i=0}^n c_i\, T_i(Au+B),
\eeq
where
\beq\label{eq:AB}
A=\frac{2}{u_{\rm max}-u_{\rm min}}, 
\qquad B=-\frac{u_{\rm max}+u_{\rm min}}{u_{\rm max}-u_{\rm min}}.
\eeq
Simple numerical algorithms for the calculation of the coefficients $c_i$
are available (we have used the routines of the \texttt{gsl}).

Simple algebra leads to
\beq
G(u)\simeq \sum_{k=0}^n \tilde c_k \, (Au+B)^k
\eeq
where
\beq
\tilde c_k = -\frac{c_0}{2}\delta_{k0} + \sum_{i=k}^n c_i\, T_{ik}.
\eeq

\subsection{Minimal Prescription}
\label{sec:cheb_minimal}

We have shown in Sect.~\ref{sec:resumm_numerical} that the minimal prescription
can be conveniently implemented by means of an analytic expression
for the Mellin transform of the luminosity 
$\Lum(z)$ (which can be either $\Lum_{q\bar q}(z)$ in the case of
invariant mass distributions or $\Lrap(z,1/2)$ in the case
of rapidity distributions). This can be obtained by expanding
$\Lum(z)$ on the basis of Chebyshev polynomials, truncated at
some finite order $n$, and then taking its analytical Mellin transform.
The luminosity itself, however, is very badly behaved in the range
$(0,1)$: it is singular at $z=0$, and varies by orders of magnitude in the
range
\beq
0\leq z\leq z_{\rm max}
;\qquad z_{\rm max}
=
\begin{cases}
1 &\text{for the rapidity-integrated cross-section}\\
e^{-2\abs{Y}} & \text{for the rapidity distribution}
\end{cases}
\eeq
It is therefore convenient to expand a regularized function
\beq
F(z) = \frac{z^\beta}{(z_{\rm max}-z)^\delta} \, \Lum(z).
\eeq 
Values of $\beta$ and
$\delta$ in the range $3\div 7$ are normally suited to
make $F(z)$ smooth enough to be approximated by a reasonably
small number of Chebyshev polynomials. Equation~\eqref{eq:AB} gives
\beq 
A=\frac{2}{z_{\rm max}},\qquad B=-1.
\eeq
and the approximation is
\beq
F(z)= \sum_{k=0}^n \tilde c_k\, \(2\frac{z}{z_{\rm max}}-1\)^k
= \sum_{p=0}^n \hat c_p \, z^p,
\eeq
where
\beq
\hat c_p = \(\frac{2}{z_{\rm max}}\)^p \sum_{k=p}^n \binom{k}{p} 
(-1)^{k-p}\, \tilde c_k.
\eeq
The luminosity is easily recovered:
\beq\label{eq:lum_cheb_app_1}
\Lum(z)
=(z_{\rm max}-z)^\delta \sum_{p=0}^n \hat c_p \, z^{p-\beta}
= \sum_{j=0}^\delta \binom{\delta}{j}\,z_{\rm max}^{\delta-j}\,(-1)^j
\sum_{p=0}^n \hat c_p \, z^{p+j-\beta}\;,
\eeq
where the last equality holds for $\delta$ integer.
It is now immediate to obtain the Mellin transform:
\beq
\Lum(N) = \int_0^{z_{\rm max}} dz\, z^{N-1} \, \Lum(z) =
\sum_{p=0}^n\sum_{j=0}^\delta \hat c_p \, \binom{\delta}{j}\,(-1)^j \,
\frac{z_{\rm max}^{N+p+\delta-\beta}}{N+p+j-\beta}.
\eeq

Alternatively, one may introduce the variable $u$
\beq
z= z_{\rm max}e^u
\eeq
and appoximate the function
\beq
F(u) = z\Lum(z)=z_{\rm max}\, e^u \Lum
\( z_{\rm max} e^u \)
\eeq
by an expansion on the basis of Chebyshev polynomials.
The variable $u$ ranges from $-\infty$ to 0 when $0\leq z\leq z_{\rm max}$;
however,
for practical purposes one only needs the luminosity 
for $z\geq z_{\rm  min}=z_{\rm max}e^{u_{\rm min}}$. We have therefore
\beq
A=-\frac{2}{u_{\rm min}}, \qquad B=1,
\eeq
and
\beq
F(u) = \sum_{k=0}^n \tilde c_k \(1-2 \frac{u}{u_{\rm min}}\)^k.
\eeq
We can now reconstruct $\Lum(z)$ through the replacement
$u=\log\frac{z}{z_{\rm max}}$. We get
\beq
\Lum(z) = \frac{1}{z} \sum_{p=0}^n (-2)^p\, u_{\rm min}^{-p}\,
\log^p\frac{z}{z_{\rm max}}\sum_{k=p}^n \binom{k}{p} \tilde c_k.
\eeq
The Mellin transform is computed using the result
\beq
\int_0^{z_{\rm max}} dz \, z^{N-2} \log^p\frac{z}{z_{\rm max}}
= z_{\rm max}^{N-1} \frac{(-1)^p\, p!}{(N-1)^{p+1}}.
\eeq
We obtain
\beq
\Lum(N) = \int_0^{z_{\rm max}} dz \, z^{N-1}\Lum(z) 
= z_{\rm max}^{N-1} \sum_{p=0}^n \frac{\bar c_p}{(N-1)^{p+1}},
\eeq
where
\beq
\bar c_p= \frac{2^p}{u_{\rm min}^p} \sum_{k=p}^n \frac{k!}{(k-p)!}\,\tilde c_k.
\eeq
In practice, we have found that the second method is to be preferred
for small values of $\tau$, $\tau\lesssim 0.1$, while the previous one
works better for $\tau\gtrsim 0.1$.

\subsection{Borel Prescription}
\label{sec:cheb_borel}

In this case, we look for an approximation of the function
$g(z,\tau)$, Eq.~\eqref{eq:g_zx}, as a function of $z\in[\tau,1]$. We have
\beq
g(z,\tau) = \sum_{k=0}^n \tilde c_k\, (Az+B)^k 
= \sum_{p=0}^n b_p\, (1-z)^p
\eeq
where
\beq
b_p = (-A)^p \sum_{k=p}^n \binom{k}{p} (A+B)^{k-p}\, \tilde c_k
\eeq
and
\beq
A=\frac{2}{1-\tau}, \qquad B= -\frac{1+\tau}{1-\tau}.
\eeq
Note that $A+B=1$ in this case. Therefore
\beq
b_p = \(\frac{-2}{1-\tau}\)^p \sum_{k=p}^n \binom{k}{p}\, \tilde c_k .
\eeq
In the case of the rapidity distributions, the variable $z$
is in the range $z\in[\tau e^{2\abs{Y}},1]$; therefore
\beq
b_p = \(\frac{-2}{1-\tau e^{2\abs{Y}}}\)^p \sum_{k=p}^n
\binom{k}{p}\, \tilde c_k.  
\eeq

For the original Borel prescription, Eq.~\eqref{eq:Borel_prescription},
it proves useful to approximate the function $\tilde g(z,\tau)$, Eq.~\eqref{eq:gztau_cheb_log1z},
in powers of $\log\frac1z$. Defining
\beq
t=\log\frac1z
\eeq
we get
\beq
\tilde g(z,\tau) = \sum_{k=0}^n \tilde c_k\, (At+B)^k =\sum_{p=0}^n \tilde b_p\, t^p
\eeq
where
\beq
\tilde b_p = \sum_{k=p}^n \binom{k}{p}\, \tilde c_k \, A^p B^{k-p}.
\eeq
Approximating the function in the range $0<t<t_{\rm max}$, corresponding to
$z_{\rm min}<z<1$, with $z_{\rm min}=e^{-t_{\rm max}}$, we have
\beq
A= \frac2{t_{\rm max}}, \qquad B=-1,
\eeq
and hence
\beq
\tilde b_p = \(-\frac{2}{t_{\rm max}}\)^p \sum_{k=p}^n \binom{k}{p}\, (-)^k\, \tilde c_k.
\eeq

\cleardoublepage
\phantomsection
\addstarredchapter{Acknowledgments}
\chapter*{Acknowledgments}
\markboth{Acknowledgments}{}

Here I am, at the end of my Ph.D.\ thesis, my last opportunity
to informally thank on a formal document all the people around me,
who helped me somehow in preparing this work.
I will try to do all my best to use this opportunity properly.

First, as usual, I would like to thank my supervisors:
of course, without them this thesis would not exist at all.
In particular, I am really grateful to Giovanni, who I worked with
for four years, and who has been a great teacher and a friend:
most things I know on the subject of this thesis I have learnt from him,
but most importantly he introduced me to the research world.
I also want to thank Stefano a lot: unfortunately we didn't
spend so much time working together face-to-face, but I'm always surprised
by the amount of things I can learn from almost every e-mail
he sends me (even if often I need some time to decrypt them...).

Then, I move to the other collaborators, or simply people
who I interacted with for the preparation of this thesis.
A first special thank goes to Tiziano Peraro, who wrote a small-$x$ code
complementary to mine, which made possible several checks
and allowed us to optimize the codes as much as possible;
moreover, I'm grateful to him 'cause I felt (and still, I feel) free to complain
with him whenever something is puzzling me.
The second thank is definitely for Richard Ball, who gave us
his first version of a small-$x$ code, and who helped me in understanding
how he practically realized several of the dark steps of small-$x$ resummation.
Then, I want to thank Guido Altarelli, who underlined me the relevance
of high-energy resummation and the importance of completing
the resummation task.
A thank goes to Alessandro Vicini, who gave me his code on
Higgs at NLO with full $m_t$ dependence, explaining it to me in detail.
Several thanks are for Juan Rojo, who I discussed with several times
and who is always interested in my codes, for Maria Ubiali
for useful discussions on the Talbot algorithm, and for Fabrizio Caola,
who discussed with me small-$x$ resummation at the beginning.
I thank Robert Harlander for giving me a piece of his code on Higgs at NNLO
with full $m_t$ dependence.

Finally, I have some formal thanks:
to the GGI, where I spent almost two months learning many stringy stuff and having much fun;
to the CERN Theory division, where I spent nine months last year, having the
opportunity to meet people, to participate to seminars and conferences, and
to enjoy staying in a very friendly and pleasing place;
and to DESY, where I'm going to stay a couple of years, which is giving me
the opportunity to complete my thesis.

Now I move to friends. I don't exactly know where to start from,
then I will choose the easiest way.
So, the first thank is for Ludovica Aperio Bella and Davide Caffari, aka \emph{the Conquis}:
I don't have to specify anything more, they were my family, and it was really funny.
Then, of course, I have to thank my band, \emph{the Suspenders}, i.e.\
Alexandros, Matias, Samir, Valerio, Lillian and Georgios:
I really miss them all, Friday's practices were the best moment of the week.
Maybe I should also thank the CERN Musiclub, which made all this possible.
Staying at CERN, I have many many other people to thank as well.
Riccardo Torre, for many things, often not related to physics (he knows what I mean);
Ennio Salvioni, even if he uses Windows sometimes;
Andrea Coccaro, because he always laughs;
Roberto Di Nardo, because he did it;
Manuela Venturi, for the jingle; all the others...

Moving to the University, I have several persons to thank.
A special thank goes to Guido Fernandes, who is really a good friend,
and who knows everything about almost everything, including product codes,
and can always help me because he already had that problem, or he
possesses that tool.
I want to thank Francesco Sanfilippo for his infinite knowledge on
\verb-C++- and \verb-BASH- tricks, and for having studied QFT with me.
A thank goes to Stefano Davini, for many useful discussions
about elegant coding, and for having shared with me the teaching experience.
I thank Daniele Musso, mainly for \emph{Blues in D-branes}, but also
because he made my exam on instantons much more interesting with his questions.
I'm also grateful to Giulia Piovano, who'll do all the Ph.D. exam paperwork for me.
Of course I want to thank also my lunch friends, my office mates,
and all the people with which I shared these years.

I don't have time to thank all my local friends from Sant'Olcese,
and more widely from Genova. I'm actually really sorry that in the last years
we didn't have many opportunities to meet, because it was really nice
when we were young and we went out until the wee hours of the morning.

I don't know exactly which category he should belong to, so I thank
Matteo Casu in a separate paragraph: for the endless conversations,
and because we always end up with very nerdy strange ideas.

Of course, a great thank goes to my parents and my relatives:
even if they stopped understanding what I'm doing long ago,
they have always supported me, and I'm pretty sure they are proud of me.

Last, but definitely not least, comes Sara.
To thank her is not enough, since what she gave and she's giving me is priceless.
I could spend an undefined amount of words to describe why she
occupies a special place in this page, but I don't want to be rhetorical.
She perfectly knows why, and not only her, and that's fine.
But just to satisfy her, let me just thank her for a specific topic.
Then, I want to thank Sara because she allowed me to cut and paste
from her thesis the description of the Drell-Yan process that
\emph{I~wrote} for her.

\vspace{2cm}
\flushright{\calligra\LARGE Marco Bonvini}

\cleardoublepage
\phantomsection
\addcontentsline{toc}{chapter}{\bibname}
\bibliographystyle{hep_bib_style}
\bibliography{/Users/cesium/Documents/university/tesi/notes/QCD.bib}

\providecommand{\href}[2]{#2}\begingroup\raggedright\begin{thebibliography}{100}

\bibitem{top_cdf}
{CDF Collaboration}, F.~Abe et al., {\em {Observation of top quark production
  in $\bar{p}p$ collisions}\/},
  \href{http://dx.doi.org/10.1103/PhysRevLett.74.2626}{Phys.Rev.Lett. {\bf 74}
  (1995)  2626--2631}, \href{http://arxiv.org/abs/hep-ex/9503002}{{\tt
  arXiv:hep-ex/9503002 [hep-ex]}}.

\bibitem{top_d0}
{D0 Collaboration}, S.~Abachi et al., {\em {Search for high mass top quark
  production in $p\bar{p}$ collisions at $\sqrt{s} = 1.8$ TeV}\/},
  \href{http://dx.doi.org/10.1103/PhysRevLett.74.2422}{Phys.Rev.Lett. {\bf 74}
  (1995)  2422--2426}, \href{http://arxiv.org/abs/hep-ex/9411001}{{\tt
  arXiv:hep-ex/9411001 [hep-ex]}}.

\bibitem{PQ}
R.~Peccei and H.~R. Quinn, {\em {CP Conservation in the Presence of
  Instantons}\/},
  \href{http://dx.doi.org/10.1103/PhysRevLett.38.1440}{Phys.Rev.Lett. {\bf 38}
  (1977)  1440--1443}.

\bibitem{esw}
R.~Ellis, W.~Stirling, and B.~Webber, {\em {QCD and collider physics}}, vol.~8.
\newblock Camb.Monogr.Part.Phys.Nucl.Phys.Cosmol., 1996.

\bibitem{KLN1}
T.~Kinoshita, {\em {Mass singularities of Feynman amplitudes}\/},  J.Math.Phys.
  {\bf 3} (1962)  650--677.

\bibitem{KLN2}
T.~Lee and M.~Nauenberg, {\em {Degenerate Systems and Mass Singularities}\/},
  \href{http://dx.doi.org/10.1103/PhysRev.133.B1549}{Phys.Rev. {\bf 133} (1964)
   B1549--B1562}.

\bibitem{mvv_ns}
S.~Moch, J.~Vermaseren, and A.~Vogt, {\em {The Three loop splitting functions
  in QCD: The Nonsinglet case}\/},
  \href{http://dx.doi.org/10.1016/j.nuclphysb.2004.03.030}{Nucl.Phys. {\bf
  B688} (2004)  101--134}, \href{http://arxiv.org/abs/hep-ph/0403192}{{\tt
  arXiv:hep-ph/0403192 [hep-ph]}}.

\bibitem{vmv}
A.~Vogt, S.~Moch, and J.~Vermaseren, {\em {The Three-loop splitting functions
  in QCD: The Singlet case}\/},
  \href{http://dx.doi.org/10.1016/j.nuclphysb.2004.04.024}{Nucl.Phys. {\bf
  B691} (2004)  129--181}, \href{http://arxiv.org/abs/hep-ph/0404111}{{\tt
  arXiv:hep-ph/0404111 [hep-ph]}}.

\bibitem{abf599}
G.~Altarelli, R.~D. Ball, and S.~Forte, {\em {Small x resummation and HERA
  structure function data}\/},
  \href{http://dx.doi.org/10.1016/S0550-3213(01)00023-2}{Nucl.Phys. {\bf B599}
  (2001)  383--423}, \href{http://arxiv.org/abs/hep-ph/0011270}{{\tt
  arXiv:hep-ph/0011270 [hep-ph]}}.

\bibitem{abf799}
G.~Altarelli, R.~D. Ball, and S.~Forte, {\em {Small x Resummation with Quarks:
  Deep-Inelastic Scattering}\/},
  \href{http://dx.doi.org/10.1016/j.nuclphysb.2008.03.003}{Nucl.Phys. {\bf
  B799} (2008)  199--240}, \href{http://arxiv.org/abs/0802.0032}{{\tt
  arXiv:0802.0032 [hep-ph]}}.

\bibitem{Korchemsky:1988si}
G.~Korchemsky, {\em {Asymptotics of the Altarelli-Parisi-Lipatov Evolution
  Kernels of Parton Distributions}\/},
\href{http://dx.doi.org/10.1142/S0217732389001453}{Mod.Phys.Lett. {\bf A4}
  (1989)  1257--1276}.

\bibitem{NNPDF1}
{NNPDF Collaboration}, R.~D. Ball et al., {\em {A Determination of parton
  distributions with faithful uncertainty estimation}\/},
  \href{http://dx.doi.org/10.1016/j.nuclphysb.2008.09.037,
  10.1016/j.nuclphysb.2009.02.027, 10.1016/j.nuclphysb.2008.09.037,
  10.1016/j.nuclphysb.2009.02.027}{Nucl.Phys. {\bf B809} (2009)  1--63},
  \href{http://arxiv.org/abs/0808.1231}{{\tt arXiv:0808.1231 [hep-ph]}}.

\bibitem{DDT}
Y.~Dokshitzer, D.~D'yakonov, and S.~Troyan, {\em On the transverse momentum
  distribution of massive lepton pairs\/},
  \href{http://dx.doi.org/10.1016/0370-2693(78)90240-X}{Physics Letters B {\bf
  79} (1978) no.~3, 269 -- 272}.

\bibitem{ParisiPetronzio}
G.~Parisi and R.~Petronzio, {\em {Small Transverse Momentum Distributions in
  Hard Processes}\/},
  \href{http://dx.doi.org/10.1016/0550-3213(79)90040-3}{Nucl.Phys. {\bf B154}
  (1979)  427}.

\bibitem{Curci:1979sk}
G.~Curci, M.~Greco, and Y.~Srivastava, {\em {COHERENT QUARK - GLUON JETS}\/},
  \href{http://dx.doi.org/10.1103/PhysRevLett.43.834}{Phys.Rev.Lett. {\bf 43}
  (1979)  834--837}.

\bibitem{CSS}
J.~C. Collins, D.~E. Soper, and G.~F. Sterman, {\em {Transverse Momentum
  Distribution in Drell-Yan Pair and W and Z Boson Production}\/},
  \href{http://dx.doi.org/10.1016/0550-3213(85)90479-1}{Nucl.Phys. {\bf B250}
  (1985)  199}.

\bibitem{bfr1}
M.~Bonvini, S.~Forte, and G.~Ridolfi, {\em {Borel resummation of transverse
  momentum distributions}\/},
  \href{http://dx.doi.org/10.1016/j.nuclphysb.2008.09.041}{Nucl.Phys. {\bf
  B808} (2009)  347--363}, \href{http://arxiv.org/abs/0807.3830}{{\tt
  arXiv:0807.3830 [hep-ph]}}.

\bibitem{Sterman:1986aj}
G.~F. Sterman, {\em {Summation of Large Corrections to Short Distance Hadronic
  Cross-Sections}\/},
  \href{http://dx.doi.org/10.1016/0550-3213(87)90258-6}{Nucl.Phys. {\bf B281}
  (1987)  310}.

\bibitem{Catani:1989ne}
S.~Catani and L.~Trentadue, {\em {Resummation of the QCD Perturbative Series
  for Hard Processes}\/},
  \href{http://dx.doi.org/10.1016/0550-3213(89)90273-3}{Nucl.Phys. {\bf B327}
  (1989)  323}.

\bibitem{fr}
S.~Forte and G.~Ridolfi, {\em {Renormalization group approach to soft gluon
  resummation}\/},
  \href{http://dx.doi.org/10.1016/S0550-3213(02)01034-9}{Nucl.Phys. {\bf B650}
  (2003)  229--270}, \href{http://arxiv.org/abs/hep-ph/0209154}{{\tt
  arXiv:hep-ph/0209154 [hep-ph]}}.

\bibitem{Catani}
S.~Catani, {\em {Higher order QCD corrections in hadron collisions: Soft gluon
  resummation and exponentiation}\/},  Nucl.Phys.Proc.Suppl. {\bf 54A} (1997)
  107--113, \href{http://arxiv.org/abs/hep-ph/9610413}{{\tt
  arXiv:hep-ph/9610413 [hep-ph]}}.

\bibitem{Weinberg1}
S.~Weinberg, {\em {The Quantum theory of fields. Vol. 1: Foundations}\/}, .

\bibitem{Sudakov}
V.~Sudakov, {\em {Vertex parts at very high-energies in quantum
  electrodynamics}\/},  Sov.Phys.JETP {\bf 3} (1956)  65--71.

\bibitem{Contopanagos:1996nh}
H.~Contopanagos, E.~Laenen, and G.~F. Sterman, {\em {Sudakov factorization and
  resummation}\/},
  \href{http://dx.doi.org/10.1016/S0550-3213(96)00567-6}{Nucl.Phys. {\bf B484}
  (1997)  303--330}, \href{http://arxiv.org/abs/hep-ph/9604313}{{\tt
  arXiv:hep-ph/9604313 [hep-ph]}}.

\bibitem{bfr2}
M.~Bonvini, S.~Forte, and G.~Ridolfi, {\em {Soft gluon resummation of Drell-Yan
  rapidity distributions: Theory and phenomenology}\/},
  \href{http://dx.doi.org/10.1016/j.nuclphysb.2011.01.023}{Nucl.Phys. {\bf
  B847} (2011)  93--159}, \href{http://arxiv.org/abs/1009.5691}{{\tt
  arXiv:1009.5691 [hep-ph]}}.

\bibitem{scetN}
A.~V. Manohar, {\em {Deep inelastic scattering as $x \to 1$ using soft
  collinear effective theory}\/},
  \href{http://dx.doi.org/10.1103/PhysRevD.68.114019}{Phys.Rev. {\bf D68}
  (2003)  114019}, \href{http://arxiv.org/abs/hep-ph/0309176}{{\tt
  arXiv:hep-ph/0309176 [hep-ph]}}.

\bibitem{Becher:2006nr}
T.~Becher and M.~Neubert, {\em {Threshold resummation in momentum space from
  effective field theory}\/},
  \href{http://dx.doi.org/10.1103/PhysRevLett.97.082001}{Phys.Rev.Lett. {\bf
  97} (2006)  082001}, \href{http://arxiv.org/abs/hep-ph/0605050}{{\tt
  arXiv:hep-ph/0605050 [hep-ph]}}.

\bibitem{bnx}
T.~Becher, M.~Neubert, and G.~Xu, {\em {Dynamical Threshold Enhancement and
  Resummation in Drell-Yan Production}\/},
  \href{http://dx.doi.org/10.1088/1126-6708/2008/07/030}{JHEP {\bf 0807} (2008)
   030}, \href{http://arxiv.org/abs/0710.0680}{{\tt arXiv:0710.0680 [hep-ph]}}.

\bibitem{bfgr}
M.~Bonvini, S.~Forte, M.~Ghezzi, and G.~Ridolfi, {\em {Threshold resummation in
  SCET vs. perturbative QCD: an analytic comparison}\/},
\href{http://arxiv.org/abs/1201.6364}{{\tt arXiv:1201.6364 [hep-ph]}}.

\bibitem{frru}
S.~Forte, G.~Ridolfi, J.~Rojo, and M.~Ubiali, {\em {Borel resummation of soft
  gluon radiation and higher twists}\/},
  \href{http://dx.doi.org/10.1016/j.physletb.2006.02.058}{Phys.Lett. {\bf B635}
  (2006)  313--319}, \href{http://arxiv.org/abs/hep-ph/0601048}{{\tt
  arXiv:hep-ph/0601048 [hep-ph]}}.

\bibitem{cmnt}
S.~Catani, M.~L. Mangano, P.~Nason, and L.~Trentadue, {\em {The Resummation of
  soft gluons in hadronic collisions}\/},
  \href{http://dx.doi.org/10.1016/0550-3213(96)00399-9}{Nucl.Phys. {\bf B478}
  (1996)  273--310}, \href{http://arxiv.org/abs/hep-ph/9604351}{{\tt
  arXiv:hep-ph/9604351 [hep-ph]}}.

\bibitem{afr}
R.~Abbate, S.~Forte, and G.~Ridolfi, {\em {A New prescription for soft gluon
  resummation}\/},
  \href{http://dx.doi.org/10.1016/j.physletb.2007.09.060}{Phys.Lett. {\bf B657}
  (2007)  55--63}, \href{http://arxiv.org/abs/0707.2452}{{\tt arXiv:0707.2452
  [hep-ph]}}.

\bibitem{Bonciani}
R.~Bonciani, S.~Catani, M.~L. Mangano, and P.~Nason, {\em {NLL resummation of
  the heavy quark hadroproduction cross-section}\/},
  \href{http://dx.doi.org/10.1016/S0550-3213(98)00335-6,
  10.1016/j.nuclphysb.2008.06.006}{Nucl.Phys. {\bf B529} (1998)  424--450},
  \href{http://arxiv.org/abs/hep-ph/9801375}{{\tt arXiv:hep-ph/9801375
  [hep-ph]}}.

\bibitem{Mukherjee:2006uu}
A.~Mukherjee and W.~Vogelsang, {\em {Threshold resummation for W-boson
  production at RHIC}\/},
  \href{http://dx.doi.org/10.1103/PhysRevD.73.074005}{Phys.Rev. {\bf D73}
  (2006)  074005}, \href{http://arxiv.org/abs/hep-ph/0601162}{{\tt
  arXiv:hep-ph/0601162 [hep-ph]}}.

\bibitem{bolz}
P.~Bolzoni, {\em {Threshold resummation of Drell-Yan rapidity
  distributions}\/},
  \href{http://dx.doi.org/10.1016/j.physletb.2006.10.064}{Phys.Lett. {\bf B643}
  (2006)  325--330}, \href{http://arxiv.org/abs/hep-ph/0609073}{{\tt
  arXiv:hep-ph/0609073 [hep-ph]}}.

\bibitem{Laenen:1992ey}
E.~Laenen and G.~F. Sterman, {\em {Resummation for Drell-Yan differential
  distributions}\/}, .

\bibitem{Bourilkov:2006cj}
D.~Bourilkov, R.~C. Group, and M.~R. Whalley, {\em {LHAPDF: PDF use from the
  Tevatron to the LHC}\/},  \href{http://arxiv.org/abs/hep-ph/0605240}{{\tt
  arXiv:hep-ph/0605240 [hep-ph]}}.

\bibitem{Furmanski:1981ja}
W.~Furmanski and R.~Petronzio, {\em {A METHOD OF ANALYZING THE SCALING
  VIOLATION OF INCLUSIVE SPECTRA IN HARD PROCESSES}\/},
  \href{http://dx.doi.org/10.1016/0550-3213(82)90398-4}{Nucl.Phys. {\bf B195}
  (1982)  237}.

\bibitem{Ellis:1978sf}
R.~Ellis, H.~Georgi, M.~Machacek, H.~Politzer, and G.~G. Ross, {\em
  {Factorization and the Parton Model in QCD}\/},
  \href{http://dx.doi.org/10.1016/0370-2693(78)90023-0}{Phys.Lett. {\bf B78}
  (1978)  281}.

\bibitem{Ellis:1978ty}
R.~Ellis, H.~Georgi, M.~Machacek, H.~Politzer, and G.~G. Ross, {\em
  {Perturbation Theory and the Parton Model in QCD}\/},
  \href{http://dx.doi.org/10.1016/0550-3213(79)90105-6}{Nucl.Phys. {\bf B152}
  (1979)  285}.

\bibitem{Catani:1990xk}
S.~Catani, M.~Ciafaloni, and F.~Hautmann, {\em {Gluon contributions to small x
  heavy flavor production}\/},
  \href{http://dx.doi.org/10.1016/0370-2693(90)91601-7}{Phys.Lett. {\bf B242}
  (1990)  97}.

\bibitem{Catani:1990eg}
S.~Catani, M.~Ciafaloni, and F.~Hautmann, {\em {High-energy factorization and
  small x heavy flavor production}\/},
  \href{http://dx.doi.org/10.1016/0550-3213(91)90055-3}{Nucl.Phys. {\bf B366}
  (1991)  135--188}.

\bibitem{Catani:1993ww}
S.~Catani, M.~Ciafaloni, and F.~Hautmann, {\em {High-energy factorization in
  QCD and minimal subtraction scheme}\/},
  \href{http://dx.doi.org/10.1016/0370-2693(93)90204-U}{Phys.Lett. {\bf B307}
  (1993)  147--153}.

\bibitem{Catani:1993rn}
S.~Catani and F.~Hautmann, {\em {Quark anomalous dimensions at small x}\/},
  \href{http://dx.doi.org/10.1016/0370-2693(93)90174-G}{Phys.Lett. {\bf B315}
  (1993)  157--163}.

\bibitem{Catani:1994sq}
S.~Catani and F.~Hautmann, {\em {High-energy factorization and small x deep
  inelastic scattering beyond leading order}\/},
  \href{http://dx.doi.org/10.1016/0550-3213(94)90636-X}{Nucl.Phys. {\bf B427}
  (1994)  475--524}, \href{http://arxiv.org/abs/hep-ph/9405388}{{\tt
  arXiv:hep-ph/9405388 [hep-ph]}}.

\bibitem{Colferai:1999em}
D.~Colferai, {\em {Small x processes in perturbative quantum
  chromodynamics}\/},  \href{http://arxiv.org/abs/hep-ph/0008309}{{\tt
  arXiv:hep-ph/0008309 [hep-ph]}}.
Ph.D thesis (Advisor: M. Ciafaloni).

\bibitem{bf405}
R.~D. Ball and S.~Forte, {\em {Asymptotically free partons at high-energy}\/},
  \href{http://dx.doi.org/10.1016/S0370-2693(97)00625-4}{Phys.Lett. {\bf B405}
  (1997)  317--326}, \href{http://arxiv.org/abs/hep-ph/9703417}{{\tt
  arXiv:hep-ph/9703417 [hep-ph]}}.

\bibitem{Lipatov:1976zz}
L.~Lipatov, {\em {Reggeization of the Vector Meson and the Vacuum Singularity
  in Nonabelian Gauge Theories}\/},
Sov.J.Nucl.Phys. {\bf 23} (1976)  338--345.

\bibitem{Kuraev:1976ge}
E.~Kuraev, L.~Lipatov, and V.~S. Fadin, {\em {Multi - Reggeon Processes in the
  Yang-Mills Theory}\/},
Sov.Phys.JETP {\bf 44} (1976)  443--450.

\bibitem{Kuraev:1977fs}
E.~Kuraev, L.~Lipatov, and V.~S. Fadin, {\em {The Pomeranchuk Singularity in
  Nonabelian Gauge Theories}\/},
Sov.Phys.JETP {\bf 45} (1977)  199--204.

\bibitem{Balitsky:1978ic}
I.~Balitsky and L.~Lipatov, {\em {The Pomeranchuk Singularity in Quantum
  Chromodynamics}\/},
Sov.J.Nucl.Phys. {\bf 28} (1978)  822--829.

\bibitem{FL}
V.~S. Fadin and L.~Lipatov, {\em {BFKL pomeron in the next-to-leading
  approximation}\/},
  \href{http://dx.doi.org/10.1016/S0370-2693(98)00473-0}{Phys.Lett. {\bf B429}
  (1998)  127--134}, \href{http://arxiv.org/abs/hep-ph/9802290}{{\tt
  arXiv:hep-ph/9802290 [hep-ph]}}.

\bibitem{Altarelli:1999vw}
G.~Altarelli, R.~D. Ball, and S.~Forte, {\em {Resummation of singlet parton
  evolution at small x}\/},
  \href{http://dx.doi.org/10.1016/S0550-3213(00)00032-8}{Nucl.Phys. {\bf B575}
  (2000)  313--329},
\href{http://arxiv.org/abs/hep-ph/9911273}{{\tt arXiv:hep-ph/9911273
  [hep-ph]}}.

\bibitem{abf742}
G.~Altarelli, R.~D. Ball, and S.~Forte, {\em {Perturbatively stable resummed
  small x evolution kernels}\/},
  \href{http://dx.doi.org/10.1016/j.nuclphysb.2006.01.046}{Nucl.Phys. {\bf
  B742} (2006)  1--40}, \href{http://arxiv.org/abs/hep-ph/0512237}{{\tt
  arXiv:hep-ph/0512237 [hep-ph]}}.

\bibitem{Salam:1998tj}
G.~Salam, {\em {A Resummation of large subleading corrections at small x}\/},
  JHEP {\bf 9807} (1998)  019,
\href{http://arxiv.org/abs/hep-ph/9806482}{{\tt arXiv:hep-ph/9806482
  [hep-ph]}}.

\bibitem{abf621}
G.~Altarelli, R.~D. Ball, and S.~Forte, {\em {Factorization and resummation of
  small x scaling violations with running coupling}\/},
  \href{http://dx.doi.org/10.1016/S0550-3213(01)00563-6}{Nucl.Phys. {\bf B621}
  (2002)  359--387}, \href{http://arxiv.org/abs/hep-ph/0109178}{{\tt
  arXiv:hep-ph/0109178 [hep-ph]}}.

\bibitem{Camici:1997ta}
G.~Camici and M.~Ciafaloni, {\em {k factorization and small x anomalous
  dimensions}\/},  \href{http://dx.doi.org/10.1016/S0550-3213(97)00261-7,
  10.1016/S0550-3213(97)00261-7}{Nucl.Phys. {\bf B496} (1997)  305--336},
\href{http://arxiv.org/abs/hep-ph/9701303}{{\tt arXiv:hep-ph/9701303
  [hep-ph]}}.

\bibitem{Lipatov:1985uk}
L.~Lipatov, {\em {The Bare Pomeron in Quantum Chromodynamics}\/},
Sov.Phys.JETP {\bf 63} (1986)  904--912.

\bibitem{Ball:2005mj}
R.~D. Ball and S.~Forte, {\em {All order running coupling BFKL evolution from
  GLAP (and vice-versa)}\/},
  \href{http://dx.doi.org/10.1016/j.nuclphysb.2006.02.020}{Nucl.Phys. {\bf
  B742} (2006)  158--175},
\href{http://arxiv.org/abs/hep-ph/0601049}{{\tt arXiv:hep-ph/0601049
  [hep-ph]}}.

\bibitem{Ciafaloni:1995bn}
M.~Ciafaloni, {\em {k(T) factorization versus renormalization group: A Small x
  consistency argument}\/},
  \href{http://dx.doi.org/10.1016/0370-2693(95)00801-Q}{Phys.Lett. {\bf B356}
  (1995)  74--78},
\href{http://arxiv.org/abs/hep-ph/9507307}{{\tt arXiv:hep-ph/9507307
  [hep-ph]}}.

\bibitem{Ciafaloni:2005cg}
M.~Ciafaloni and D.~Colferai, {\em {Dimensional regularisation and
  factorisation schemes in the BFKL equation at subleading level}\/},
  \href{http://dx.doi.org/10.1088/1126-6708/2005/09/069}{JHEP {\bf 0509} (2005)
   069},
\href{http://arxiv.org/abs/hep-ph/0507106}{{\tt arXiv:hep-ph/0507106
  [hep-ph]}}.

\bibitem{Ball:1995tn}
R.~D. Ball and S.~Forte, {\em {Momentum conservation at small x}\/},
  \href{http://dx.doi.org/10.1016/0370-2693(95)01090-D}{Phys.Lett. {\bf B359}
  (1995)  362--368},
\href{http://arxiv.org/abs/hep-ph/9507321}{{\tt arXiv:hep-ph/9507321
  [hep-ph]}}.

\bibitem{Appell:1988ie}
D.~Appell, G.~F. Sterman, and P.~B. Mackenzie, {\em {SOFT GLUONS AND THE
  NORMALIZATION OF THE DRELL-YAN CROSS-SECTION}\/},
  \href{http://dx.doi.org/10.1016/0550-3213(88)90082-X}{Nucl.Phys. {\bf B309}
  (1988)  259}.

\bibitem{Ball:2010de}
R.~D. Ball, L.~Del~Debbio, S.~Forte, A.~Guffanti, J.~I. Latorre, et al., {\em
  {A first unbiased global NLO determination of parton distributions and their
  uncertainties}\/},
  \href{http://dx.doi.org/10.1016/j.nuclphysb.2010.05.008}{Nucl.Phys. {\bf
  B838} (2010)  136--206}, \href{http://arxiv.org/abs/1002.4407}{{\tt
  arXiv:1002.4407 [hep-ph]}}.

\bibitem{Grazzini:2010zc}
M.~Grazzini, {\em {QCD Effects in Higgs Boson Production at Hadron
  Colliders}\/},  PoS {\bf RADCOR2009} (2010)  047,
  \href{http://arxiv.org/abs/1001.3766}{{\tt arXiv:1001.3766 [hep-ph]}}.

\bibitem{Bauer:2010jv}
C.~W. Bauer, N.~D. Dunn, and A.~Hornig, {\em {On the effectiveness of threshold
  resummation away from hadronic endpoint}\/},
\href{http://arxiv.org/abs/1010.0243}{{\tt arXiv:1010.0243 [hep-ph]}}.

\bibitem{Collins:1977jy}
P.~Collins,
{\em {An Introduction to Regge Theory and High-Energy Physics}\/}, .

\bibitem{Abarbanel:1969eh}
H.~Abarbanel, M.~Goldberger, and S.~Treiman, {\em {Asymptotic properties of
  electroproduction structure functions}\/},
\href{http://dx.doi.org/10.1103/PhysRevLett.22.500}{Phys.Rev.Lett. {\bf 22}
  (1969)  500--502}.

\bibitem{Bonciani:2007ex}
R.~Bonciani, G.~Degrassi, and A.~Vicini, {\em {Scalar particle contribution to
  Higgs production via gluon fusion at NLO}\/},
  \href{http://dx.doi.org/10.1088/1126-6708/2007/11/095}{JHEP {\bf 0711} (2007)
   095},
\href{http://arxiv.org/abs/0709.4227}{{\tt arXiv:0709.4227 [hep-ph]}}.

\bibitem{Marzani:2008az}
S.~Marzani, R.~D. Ball, V.~Del~Duca, S.~Forte, and A.~Vicini, {\em {Higgs
  production via gluon-gluon fusion with finite top mass beyond next-to-leading
  order}\/},
  \href{http://dx.doi.org/10.1016/j.nuclphysb.2008.03.016}{Nucl.Phys. {\bf
  B800} (2008)  127--145},
\href{http://arxiv.org/abs/0801.2544}{{\tt arXiv:0801.2544 [hep-ph]}}.

\bibitem{vanNeerven:1991gh}
W.~van Neerven and E.~Zijlstra, {\em {The O$(\alpha_s^2)$ corrected Drell-Yan
  $K$ factor in the DIS and MS scheme}\/},
  \href{http://dx.doi.org/10.1016/0550-3213(92)90078-P,
  10.1016/0550-3213(92)90078-P}{Nucl.Phys. {\bf B382} (1992)  11--62}.

\bibitem{admp}
C.~Anastasiou, L.~J. Dixon, K.~Melnikov, and F.~Petriello, {\em {High precision
  QCD at hadron colliders: Electroweak gauge boson rapidity distributions at
  NNLO}\/},  \href{http://dx.doi.org/10.1103/PhysRevD.69.094008}{Phys.Rev. {\bf
  D69} (2004)  094008}, \href{http://arxiv.org/abs/hep-ph/0312266}{{\tt
  arXiv:hep-ph/0312266 [hep-ph]}}.

\bibitem{Catani:2010en}
S.~Catani, G.~Ferrera, and M.~Grazzini, {\em {W boson production at hadron
  colliders: the lepton charge asymmetry in NNLO QCD}\/},
  \href{http://dx.doi.org/10.1007/JHEP05(2010)006}{JHEP {\bf 1005} (2010)
  006}, \href{http://arxiv.org/abs/1002.3115}{{\tt arXiv:1002.3115 [hep-ph]}}.
  * Temporary entry *.

\bibitem{Grazzini:2009nd}
M.~Grazzini, {\em {Transverse-momentum resummation at hadron colliders}\/},
  \href{http://arxiv.org/abs/0908.1338}{{\tt arXiv:0908.1338 [hep-ph]}}.

\bibitem{e866_1}
{FNAL E866/NuSea Collaboration}, R.~Towell et al., {\em {Improved measurement
  of the anti-d / anti-u asymmetry in the nucleon sea}\/},
  \href{http://dx.doi.org/10.1103/PhysRevD.64.052002}{Phys.Rev. {\bf D64}
  (2001)  052002}, \href{http://arxiv.org/abs/hep-ex/0103030}{{\tt
  arXiv:hep-ex/0103030 [hep-ex]}}.

\bibitem{e866_2}
J.~C. Webb, {\em {Measurement of continuum dimuon production in 800-GeV/C
  proton nucleon collisions}\/},
  \href{http://arxiv.org/abs/hep-ex/0301031}{{\tt arXiv:hep-ex/0301031
  [hep-ex]}}. Ph.D. Thesis (Advisor: Vassili Papavassiliou).

\bibitem{e866_3}
{NuSea Collaboration}, J.~Webb et al., {\em {Absolute Drell-Yan dimuon
  cross-sections in 800 GeV / c pp and pd collisions}\/},  Phys.Rev.Lett.
  (2003)  , \href{http://arxiv.org/abs/hep-ex/0302019}{{\tt
  arXiv:hep-ex/0302019 [hep-ex]}}.

\bibitem{Martin:2009iq}
A.~Martin, W.~Stirling, R.~Thorne, and G.~Watt, {\em {Parton distributions for
  the LHC}\/},
  \href{http://dx.doi.org/10.1140/epjc/s10052-009-1072-5}{Eur.Phys.J. {\bf C63}
  (2009)  189--285}, \href{http://arxiv.org/abs/0901.0002}{{\tt arXiv:0901.0002
  [hep-ph]}}.

\bibitem{Lai:2010vv}
H.-L. Lai, M.~Guzzi, J.~Huston, Z.~Li, P.~M. Nadolsky, et al., {\em {New parton
  distributions for collider physics}\/},
  \href{http://dx.doi.org/10.1103/PhysRevD.82.074024}{Phys.Rev. {\bf D82}
  (2010)  074024}, \href{http://arxiv.org/abs/1007.2241}{{\tt arXiv:1007.2241
  [hep-ph]}}.

\bibitem{Balossini:2009sa}
G.~Balossini, G.~Montagna, C.~M. Carloni~Calame, M.~Moretti, O.~Nicrosini, et
  al., {\em {Combination of electroweak and QCD corrections to single W
  production at the Fermilab Tevatron and the CERN LHC}\/},
  \href{http://dx.doi.org/10.1007/JHEP01(2010)013}{JHEP {\bf 1001} (2010)
  013}, \href{http://arxiv.org/abs/0907.0276}{{\tt arXiv:0907.0276 [hep-ph]}}.

\bibitem{Catani:2009sm}
S.~Catani, L.~Cieri, G.~Ferrera, D.~de~Florian, and M.~Grazzini, {\em {Vector
  boson production at hadron colliders: A Fully exclusive QCD calculation at
  NNLO}\/},
  \href{http://dx.doi.org/10.1103/PhysRevLett.103.082001}{Phys.Rev.Lett. {\bf
  103} (2009)  082001}, \href{http://arxiv.org/abs/0903.2120}{{\tt
  arXiv:0903.2120 [hep-ph]}}.

\bibitem{PDF4LHC}
{PDF4LHC working group}.
\newblock \url{https://wiki.terascale.de/index.php?title=PDF4LHC_WIKI}.

\bibitem{PDG}
{Particle Data Group}, K.~Nakamura et al., {\em {Review of particle
  physics}\/},
  \href{http://dx.doi.org/10.1088/0954-3899/37/7A/075021}{J.Phys.G {\bf G37}
  (2010)  075021}.

\bibitem{Bethke:2009jm}
S.~Bethke, {\em {The 2009 World Average of alpha(s)}\/},
  \href{http://dx.doi.org/10.1140/epjc/s10052-009-1173-1}{Eur.Phys.J. {\bf C64}
  (2009)  689--703}, \href{http://arxiv.org/abs/0908.1135}{{\tt arXiv:0908.1135
  [hep-ph]}}.

\bibitem{Lai:2010nw}
H.-L. Lai, J.~Huston, Z.~Li, P.~Nadolsky, J.~Pumplin, et al., {\em {Uncertainty
  induced by QCD coupling in the CTEQ global analysis of parton
  distributions}\/},
  \href{http://dx.doi.org/10.1103/PhysRevD.82.054021}{Phys.Rev. {\bf D82}
  (2010)  054021}, \href{http://arxiv.org/abs/1004.4624}{{\tt arXiv:1004.4624
  [hep-ph]}}.

\bibitem{Aaltonen:2010zza}
{CDF Collaboration}, T.~A. Aaltonen et al., {\em {Measurement of $d\sigma/dy$
  of Drell-Yan $e^+e^-$ pairs in the $Z$ Mass Region from $p\bar{p}$ Collisions
  at $\sqrt{s}=1.96$ TeV}\/},
  \href{http://dx.doi.org/10.1016/j.physletb.2010.06.043}{Phys.Lett. {\bf B692}
  (2010)  232--239}, \href{http://arxiv.org/abs/0908.3914}{{\tt arXiv:0908.3914
  [hep-ex]}}.

\bibitem{Aaltonen:2009ta}
{CDF Collaboration}, T.~Aaltonen et al., {\em {Direct Measurement of the $W$
  Production Charge Asymmetry in $p\bar{p}$ Collisions at $\sqrt{s} = 1.96$
  TeV}\/},
  \href{http://dx.doi.org/10.1103/PhysRevLett.102.181801}{Phys.Rev.Lett. {\bf
  102} (2009)  181801}, \href{http://arxiv.org/abs/0901.2169}{{\tt
  arXiv:0901.2169 [hep-ex]}}.

\bibitem{vanRitbergen:1997va}
T.~van Ritbergen, J.~Vermaseren, and S.~Larin, {\em {The Four loop beta
  function in quantum chromodynamics}\/},
  \href{http://dx.doi.org/10.1016/S0370-2693(97)00370-5}{Phys.Lett. {\bf B400}
  (1997)  379--384}, \href{http://arxiv.org/abs/hep-ph/9701390}{{\tt
  arXiv:hep-ph/9701390 [hep-ph]}}.

\bibitem{Czakon:2004bu}
M.~Czakon, {\em {The Four-loop QCD beta-function and anomalous dimensions}\/},
  \href{http://dx.doi.org/10.1016/j.nuclphysb.2005.01.012}{Nucl.Phys. {\bf
  B710} (2005)  485--498}, \href{http://arxiv.org/abs/hep-ph/0411261}{{\tt
  arXiv:hep-ph/0411261 [hep-ph]}}.

\bibitem{talbot}
J.~Abate and P.~P. Valk\'o, {\em {Multi-precision Laplace transform
  inversion}\/},  \href{http://dx.doi.org/10.1002/nme.995}{International
  Journal for Numerical Methods in Engineering {\bf 60} (2004) no.~5,
  979--993}.

\bibitem{Hamberg:1990np}
R.~Hamberg, W.~van Neerven, and T.~Matsuura, {\em {A Complete calculation of
  the order $\alpha-s^{2}$ correction to the Drell-Yan $K$ factor}\/},
  \href{http://dx.doi.org/10.1016/0550-3213(91)90064-5,
  10.1016/0550-3213(91)90064-5}{Nucl.Phys. {\bf B359} (1991)  343--405}.

\bibitem{Chetyrkin:1997un}
K.~Chetyrkin, B.~A. Kniehl, and M.~Steinhauser, {\em {Decoupling relations to
  $O(\alpha_s^3)$ and their connection to low-energy theorems}\/},
  \href{http://dx.doi.org/10.1016/S0550-3213(97)00649-4}{Nucl.Phys. {\bf B510}
  (1998)  61--87},
\href{http://arxiv.org/abs/hep-ph/9708255}{{\tt arXiv:hep-ph/9708255
  [hep-ph]}}.

\bibitem{Harlander:2001is}
R.~V. Harlander and W.~B. Kilgore, {\em {Soft and virtual corrections to $pp\to
  H + X$ at NNLO}\/},
  \href{http://dx.doi.org/10.1103/PhysRevD.64.013015}{Phys.Rev. {\bf D64}
  (2001)  013015},
\href{http://arxiv.org/abs/hep-ph/0102241}{{\tt arXiv:hep-ph/0102241
  [hep-ph]}}.

\bibitem{Dawson:1990zj}
S.~Dawson, {\em {Radiative corrections to Higgs boson production}\/},
\href{http://dx.doi.org/10.1016/0550-3213(91)90061-2}{Nucl.Phys. {\bf B359}
  (1991)  283--300}.

\bibitem{Dawson:1993qf}
S.~Dawson and R.~Kauffman, {\em {QCD corrections to Higgs boson production:
  nonleading terms in the heavy quark limit}\/},
  \href{http://dx.doi.org/10.1103/PhysRevD.49.2298}{Phys.Rev. {\bf D49} (1994)
  2298--2309},
\href{http://arxiv.org/abs/hep-ph/9310281}{{\tt arXiv:hep-ph/9310281
  [hep-ph]}}.

\bibitem{Anastasiou:2002yz}
C.~Anastasiou and K.~Melnikov, {\em {Higgs boson production at hadron colliders
  in NNLO QCD}\/},
  \href{http://dx.doi.org/10.1016/S0550-3213(02)00837-4}{Nucl.Phys. {\bf B646}
  (2002)  220--256},
\href{http://arxiv.org/abs/hep-ph/0207004}{{\tt arXiv:hep-ph/0207004
  [hep-ph]}}.

\bibitem{Catani:2003zt}
S.~Catani, D.~de~Florian, M.~Grazzini, and P.~Nason, {\em {Soft gluon
  resummation for Higgs boson production at hadron colliders}\/},  JHEP {\bf
  0307} (2003)  028,
\href{http://arxiv.org/abs/hep-ph/0306211}{{\tt arXiv:hep-ph/0306211
  [hep-ph]}}.

\bibitem{Aglietti:2006tp}
U.~Aglietti, R.~Bonciani, G.~Degrassi, and A.~Vicini, {\em {Analytic Results
  for Virtual QCD Corrections to Higgs Production and Decay}\/},
  \href{http://dx.doi.org/10.1088/1126-6708/2007/01/021}{JHEP {\bf 0701} (2007)
   021},
\href{http://arxiv.org/abs/hep-ph/0611266}{{\tt arXiv:hep-ph/0611266
  [hep-ph]}}.

\bibitem{mvv}
S.~Moch, J.~Vermaseren, and A.~Vogt, {\em {Higher-order corrections in
  threshold resummation}\/},
  \href{http://dx.doi.org/10.1016/j.nuclphysb.2005.08.005}{Nucl.Phys. {\bf
  B726} (2005)  317--335}, \href{http://arxiv.org/abs/hep-ph/0506288}{{\tt
  arXiv:hep-ph/0506288 [hep-ph]}}.

\bibitem{Moch:2005ky}
S.~Moch and A.~Vogt, {\em {Higher-order soft corrections to lepton pair and
  Higgs boson production}\/},
  \href{http://dx.doi.org/10.1016/j.physletb.2005.09.061}{Phys.Lett. {\bf B631}
  (2005)  48--57}, \href{http://arxiv.org/abs/hep-ph/0508265}{{\tt
  arXiv:hep-ph/0508265 [hep-ph]}}.

\bibitem{Ball:1999sh}
R.~D. Ball and S.~Forte, {\em {The Small x behavior of Altarelli-Parisi
  splitting functions}\/},
  \href{http://dx.doi.org/10.1016/S0370-2693(99)01013-8}{Phys.Lett. {\bf B465}
  (1999)  271--281},
\href{http://arxiv.org/abs/hep-ph/9906222}{{\tt arXiv:hep-ph/9906222
  [hep-ph]}}.

\bibitem{hardy}
G.~Hardy, {\em Divergent series}.
\newblock Chelsea Publishing Series. American Mathematical Society, 1991.

\bibitem{Weniger}
E.~J. Weniger, {\em Nonlinear sequence transformations for the acceleration of
  convergence and the summation of divergent series\/},
  \href{http://arxiv.org/abs/math/0306302v1}{{\tt arXiv:math/0306302v1}}.

\bibitem{borel1}
S.~Graffi, V.~Grecchi, and B.~Simon, {\em {Borel summability: Application to
  the anharmonic oscillator}\/},
  \href{http://dx.doi.org/10.1016/0370-2693(70)90564-2}{Phys.Lett. {\bf B32}
  (1970)  631--634}.

\bibitem{borel2}
U.~D. Jentschura, {\em {The Resummation of nonalternating divergent
  perturbative expansions}\/},
  \href{http://dx.doi.org/10.1103/PhysRevD.62.076001}{Phys.Rev. {\bf D62}
  (2000)  076001}, \href{http://arxiv.org/abs/hep-ph/0001135}{{\tt
  arXiv:hep-ph/0001135 [hep-ph]}}.

\end{thebibliography}\endgroup

\end{document}